\documentclass[12pt]{report}

\pdfoutput=1

\usepackage{xcolor}
\usepackage{appendix}
\usepackage{hyperref}

	\usepackage{epsfig}
	\usepackage{amssymb}
	\usepackage{amsfonts}
	\usepackage{amsmath}
	\usepackage{euscript}
	\usepackage{verbatim}
	\usepackage{latexsym}
	\usepackage{graphicx}
	\usepackage[utf8]{inputenc}
	\usepackage{graphicx}	
	\usepackage{subfigure}
	\usepackage{wrapfig}
	\usepackage[T1]{fontenc}
	\usepackage{mathtools}
	\usepackage{float}
	\usepackage{calligra}
	\usepackage{braket}
	\usepackage{caption}
	\usepackage{url}
	
	\usepackage{tikz}
	\usetikzlibrary{shapes.geometric, arrows,patterns,snakes}
	\tikzstyle{ellip} = [ellipse, minimum width=3cm, minimum height=1cm,text centered, draw=black]

	\newcommand{\beq}{\begin{equation}}
	\newcommand{\eeq}{\end{equation}}
	\newcommand{\bea}{\begin{eqnarray}}
	\newcommand{\eea}{\end{eqnarray}}
	
	\def\e{\epsilon}
	\def\cM{{\mathcal M}}
	\def\pd{\partial}
	\def\bA{\bar{A}}
	\def\ba{\bar{a}}
	\def\e{\epsilon}
	\def\calM{{\mathcal M}}
	\def\cL{\mathcal{L}}
	\def\cW{\mathcal{W}}
	\newcommand{\calL}[1] {\mathcal{L}^{(#1)}}
	\newcommand{\calW}[1] {\mathcal{W}^{(#1)}}
	\newcommand{\mui}[1] {\mu^{(#1)}}
	\newcommand{\nui}[1] {\nu^{(#1)}}
	\newcommand{\bcalL} {\overline{\cL}}
	\newcommand{\bcalW} {\overline{\cW}}
	\def\bmu{\overline{\mu}}
	\def\bnu{\overline{\nu}}
	\newcommand{\ei}[1] {\epsilon^{(#1)}}
	\newcommand{\si}[1] {\sigma^{(#1)}}
	\newcommand{\bei}[1] {\overline{\epsilon}^{(#1)}}
	\newcommand{\bsi}[1] {\overline{\sigma}^{(#1)}}
	\def\pp{\partial_{\phi}}
	\def\pt{\partial_t}

	\def\d{\hbox{\rm d}}
	\def\Tr{ \hbox{\rm Tr}}
	\def\rme{ \hbox{\rm e}}

	\def\bra{\langle}
	\def\ket{\rangle}

	\def\bA{{\widetilde A}}

	\def\cL{\mathcal{L}}
	
	\def\bR{\mathbb{R}}
	\def\bC{\mathbb{C}}
	\def\cN{\mathcal{N}}

	\def\gi{g^{-1}}
	\def\tA{A'}
	\def\br{\nonumber \\}
	
	\newcommand{\fsl}[1]{\ensuremath{\mathrlap{\!\not{\phantom{#1}}}#1}}
	\newcommand{\overbar}[1]{\mkern 1.5mu\overline{\mkern-1.5mu#1\mkern-1.5mu}\mkern 1.5mu}

	\newcommand{\bpsi}{\overbar{\psi}}
	\newcommand{\bV}{\overbar{V}}
	\newcommand{\bphi}{\overbar{\phi}}

	\hypersetup{colorlinks=false, linkcolor=blue, citecolor=red}

\begin{document}
	
	\pagenumbering{gobble}

\begin{titlepage}
\pagenumbering{gobble}

\begin{center}

	\null \vspace{1.6cm}
	
	{\LARGE \emph{Aspects of Higher Spin Theories, Conformal Field Theories and Holography}}
	
	\vskip 1cm

	\large{A thesis submitted for the degree of \\ Doctor of Philosophy \\ in the Faculty of Sciences}
	
	\vspace{3cm}
	{\Large\textbf{\textsf{{Avinash Raju}}}}
	
	\vspace{1.7cm}
	\begin{figure}[h]
		\centering
		\includegraphics[scale=0.25]{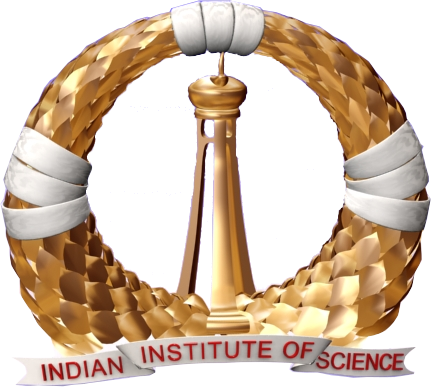}
	\end{figure}
	
	\vspace{.5cm}
	\textit{Centre for High Energy Physics\\
		Indian Institute of Science \\
		\ Bangalore - 560012. India. \\}

\end{center}

\end{titlepage}

\pagebreak
\
\pagebreak

\vskip 2cm

\begin{flushleft}
\textsf{\textbf{\LARGE Declaration}}
\end{flushleft}

\vskip 2cm

{\large I hereby declare that this thesis "Aspects of Higher Spin Theories, Conformal Field Theories and Hologrpahy" is based on my own research work, that I carried out with my collaborators in the Centre for High Energy Physics, Indian Institute of Science, during my tenure as a PhD student under the supervision of Dr. Chethan Krishnan. The work presented here does not feature as the work of someone else in obtaining a degree in this or any other institute. Any other work that may have been quoted or used in writing this thesis has been duly cited and acknowledged.

\vskip 2cm

 \ \\
Date:   \hspace{11cm} Avinash Raju

\vskip 1.5cm

 \ \\
Certified by:

\vskip 1.5cm

 \ \\
Dr. Chethan Krishnan\\
Centre for High Energy Physics\\
Indian Institute of Science\\
Bangalore: 560012\\
India}

\pagebreak

\pagebreak
\
\pagebreak

\vskip 1cm

\begin{center}
\textsf{\textbf{\LARGE Acknowledgement}}
\end{center}

The work presented in this dissertation and my PhD life in general would not have been possible without the support and companionship of people I have been surrounded by all these years. This is my humble attempt to acknowledge their help and support.\\
 
I am greatly indebted my research supervisor Chethan Krishnan for his time, invaluable advice on physics and life, and for the tremendous amount of physics he taught me. He graciously took me as his student while I was still new to the department and tolerated my whims for 5 years. It was a great pleasure to work with him and was the best choice I could have made for an adviser.\\

I would also like to express my gratitude to my early collaborators Shubho Roy and Somyadip Thakur. Their help and support were especially useful in getting me started with research. I have enjoyed many late night discussions with them. I have learned a lot from Shubho and without him my early publications would not have been possible. \\

Thanks to all my teachers, Prof. B. Ananthanarayanan, Prof. Rohini Godbole, Prof. Aninda Sinha, Prof. Sachindeo Vaidya and Prof. Justin David. I owe them for most of what I know about high energy physics.\\

I am also grateful to Mr. Keshava and Mr. Saravana for being the lifeline of CHEP. Over years I have made quiet a lot of friends in IISc. My special gratitude to Ver\'{o}nica Errasti D\'{i}ez, P.N. Bala Subramanian, Abhiram Kaushik, Anustuv Pal, Jagriti Pal, Jatin Panwar and Drew Morrill for lightning up my life. I would also like to thank my colleagues Parveen Kumar, K.V. Pavan Kumar, Aradhita Chattopadhyaya, Shouvik Datta, Gaurav Mendiratta and V Suryanarayana Mummidi. Thanks is also due to Pankaj Ajit, Anand Sateendhran, Vishnu Ananthnarayanan, Ashwin K.P. and Anirudh Rajan. \\

Finally, I would like to thank my parents and my sister for their constant encouragement, support and help. Words fail to truly describe my gratitude and love for them.

\pagebreak
\
\pagebreak

\begin{center}
\textbf{\large \underline{ABSTRACT}}
\end{center}

\vspace{0.5cm}

\noindent This dissertation consist of three parts. The first part of the thesis is devoted to the study of gravity and higher spin gauge theories in 2+1 dimensions. We construct cosmological solutions of higher spin gravity in 2+1 dimensional de Sitter space. We show that a consistent thermodynamics can be obtained for their horizons by demanding appropriate holonomy conditions. This is equivalent to demanding the integrability of the Euclidean boundary CFT partition function, and reduces to Gibbons-Hawking thermodynamics in the spin-2 case. By using a prescription of Maldacena, we relate the thermodynamics of these solutions to those of higher spin black holes in AdS$_3$. For the case of negative cosmological constant we show that interpreting the inverse AdS$_3$ radius $1/l$ as a Grassmann variable results in a formal map from gravity in AdS$_3$ to gravity in flat space. The underlying reason for this is the fact that ISO(2,1) is the Inonu-Wigner contraction of SO(2,2). We show how this works for the Chern-Simons actions, demonstrate how the general (Banados) solution in AdS$_3$ maps to the general flat space solution, and how the Killing vectors, charges and the Virasoro algebra in the Brown-Henneaux case map to the corresponding quantities in the BMS$_3$ case. Our results straightforwardly generalize to the higher spin case: the flat space higher spin theories emerge automatically in this approach from their AdS counterparts. We also demonstrate the power of our approach by doing singularity resolution in the BMS gauge as an application. Finally, we construct a candidate for the most general chiral higher spin theory with AdS$_3$ boundary conditions. In the Chern-Simons language, the left-moving solution has Drinfeld-Sokolov reduced form, but on the right-moving solution all charges and chemical potentials are turned on. Altogether (for the spin-3 case) these are 19 functions. Despite this, we show that the resulting metric has the form of the ``most general'' AdS$_3$ boundary conditions discussed by Grumiller and Riegler. The asymptotic symmetry algebra is a product of a ${\cal W}_3$ algebra on the left and an affine $sl(3)_k$ current algebra on the right, as desired. The metric and higher spin fields depend on all the 19 functions.\\

The second part is devoted to the problem of Neumann boundary condition in Einstein's gravity. The Gibbons-Hawking-York (GHY) boundary term makes the Dirichlet problem for gravity well defined, but no such general term seems to be known for Neumann boundary conditions. In our work, we view Neumann boundary condition {\em not} as fixing the normal derivative of the metric (``velocity'') at the boundary, but as fixing the functional derivative of the action with respect to the boundary metric (``momentum''). This leads directly to a new boundary term for gravity: the trace of the extrinsic curvature with a specific dimension-dependent coefficient. In three dimensions this boundary term reduces to a ``one-half'' GHY term noted in the literature previously, and we observe that our action translates precisely to the Chern-Simons action with no extra boundary terms. In four dimensions the boundary term vanishes, giving a natural Neumann interpretation to the standard Einstein-Hilbert action without boundary terms. We also argue that a natural boundary condition for gravity in asymptotically AdS spaces is to hold the {\em renormalized} boundary stress tensor density fixed, instead of the boundary metric. This leads to a well-defined variational problem, as well as new counter-terms and a finite on-shell action. We elaborate this in various (even and odd) dimensions in the language of holographic renormalization. Even though the {\em form} of the new renormalized action is distinct from the standard one, once the cut-off is taken to infinity, their {\em values} on classical solutions coincide when the trace anomaly vanishes. For AdS$_4$, we compute the ADM form of this renormalized action and show in detail how the correct thermodynamics of Kerr-AdS black holes emerge. We comment on the possibility of a consistent quantization with our boundary conditions when the boundary is dynamical, and make a connection to the results of Compere and Marolf. The difference between our approach and microcanonical-like ensembles in standard AdS/CFT is emphasized. 

In the third part of the dissertation, we use the recently developed CFT techniques of Rychkov and Tan to compute anomalous dimensions in the O(N) Gross-Neveu model in $d=2+\epsilon$ dimensions. To do this, we extend the ``cow-pie contraction'' algorithm of Basu and Krishnan to theories with fermions. Our results match perfectly with Feynman diagram computations.\\

\pagebreak
\
\pagebreak

\vspace*{7cm}

\begin{center}
	\textit{To my parents}
\end{center}

\pagebreak

\
\vspace{2cm}

\pagenumbering{arabic}

\tableofcontents

\newpage

\chapter*{\textbf{{ List of Publications}}}
\addcontentsline{toc}{chapter}{{{ List of Publications}}}

\begin{flushleft}
	\textsf{\textbf{\large This thesis is based on the following articles}}
\end{flushleft}

\begin{flushleft}
	\textsf{\textbf{\large Review article}}
\end{flushleft}

\begin{flushleft}
	\textsf{\textbf{\large Other publications during  PhD, not included in thesis}}
\end{flushleft}

\pagebreak

\

\pagebreak

\chapter{Introduction}

This thesis is devoted towards understanding gravity, higher spin theories and conformal field theories. In view of developments in string theory, these topics are intimately related, especially in the light of holography. The best known example of hologrphy is the celebrated \emph{AdS/CFT correspondence}, which is a statement of equivalence between string theory in certain curved (anti-de Sitter) spacetime and conformal field theories living on the boundary of this space. In certain limits the string theory can be approximated by Einstein's gravity and most applications of AdS/CFT rely on this limit. There also exists certain limits of string theory where higher spin theories emerge. The higher spin symmetries are believed to show up in the high momentum transfer limit of string amplitudes \cite{Gross:1987kza,Gross:1988ue,Sundborg:2000wp,Lindstrom:2003mg,Bonelli:2003kh,Sagnotti:2003qa,Craps:2014wpa,Kiran:2014kca,Krishnan:2015wha}\footnote{This is only a representative list. The subject of higher spin symmetries in string theory has a long history and many authors have contributed which are omitted here.}. Formally, we can think of higher spin limit as taking $\alpha' \rightarrow \infty$, keeping fixed the number of string oscillators \cite{Bonelli:2003kh}. There also exist various versions of holography with higher spin fields propagating in the bulk. This is our motivation for studying the aforementioned topics, namely to have a better understanding of holography. \\

In this chapter we present a brief overview of higher spin theories, conformal field theories and holography. This chapter is largely motivational and we will expand on each of these topics in subsequent chapters. In Chapter \ref{GHSR}, we begin by reviewing gravity in first order (vielbein-spin-connection) formalism. We shall see its connection to the Chern-Simons theory and extend it to include towers of higher spin fields interacting with gravity. In chapter \ref{HSC}, we construct cosmological solutions of higher spin gauge theories with positive cosmological constant. We also explore the thermodynamics of our solutions and relate it to the thermodynamics of higher spin black holes using a prescription due to Maldacena. Chapter \ref{Grassmann} deals with the case of negative cosmological constant. We give a prescription, based on Inonu-Wigner contraction of AdS isometries to map relevant quantities to flat space. We show how our prescription maps the action, solutions and charges in AdS to flat space. Our prescription also works for higher spin theories and as an application, we demonstrate the power of our techniques in singularity resolution. In Chapter \ref{CHSG}, we construct a candidate solution for the most general chiral higher spin theory in AdS. Our chiral theory falls under the same metric class as the recently constructed ``most general'' AdS$_3$ boundary condition by Grumiller and Riegler and the symmetry algebra is given by ${\cal W}_3 \times sl(3)_k$. In Chapter \ref{BC_gravity}, we investigate the issue of Neumann boundary condition in Einstein's gravity. We construct the boundary action needed for a well-defined variational problem with Neumann boundary condition. We also extend this to asymptotically AdS space where Neumann boundary condition has the interpretation as fixing the boundary stress-tensor density. Using holographic renormalization, it is possible to extract counter-terms that can render the on-shell action finite as well as reproduce the charges of black holes. We do this in dimensions $d=2,3,4$, where $d$ is the dimension of the boundary. In Chapter \ref{GNCFT}, we study the $O(N)$ Gross-Neveu CFT in $2+\epsilon$ dimensions. Based on the work of Rychkov-Tan and Krishnan-Basu, we generalize their techniques to fermions and compute the leading order anomalous dimension for a class of composite operators in the Gross-Neveu model.

\section{Motivation}

Our current understanding of matter and forces, at the fundamental level, is described by quantum field theories (QFT). While this recipe is incredibly successful in describing Electromagnetism, Weak and Strong nuclear force, it fails to describe Gravity at the fundamental level. Unifying General Relativity (GR) with quantum mechanics remains one of the challenges of theoretical physics. There are strong indications that at energies of the scale of Planck mass ($\approx 10^{19}$ GeV), where Quantum Gravity (QG) effects come into play, General Relativity has to be replaced by a new theory. One such scenario is in the vicinity of the center of black holes where the spacetime curvatures become large. String theory is a strong contender for such a new theory which is consistent with quantum mechanics and general relativity.\\

The basic objects of String theory are one dimensional strings whose oscillations give a spectrum of fields with specific mass and spin in the ambient spacetime in which the string is propagating \cite{Polchinski:1998rq}. String theory first emerged in the study of strong interactions where this idea was used to describe the large number of mesons and baryons as oscillation modes of the string. Even though the string model describes some aspects of the particle spectrum very well, it was soon abandoned with the advent of Quantum Chromodynamics (QCD). QCD is a gauge theory based on $SU(3)$ group whose basic ingredients are quarks and gluons and is asymptotically free. At low energies, the theory becomes strongly interacting and is not amenable to easy calculations. It was later suggested by `t Hooft that the theory might simplify by lifting the gauge group to $SU(N)$ and taking an $N\rightarrow \infty$ limit \cite{tHooft:1973alw}. The hope was that the theory becomes exactly solvable at $N=\infty$ and one could then do an expansion in $1/N$. The relevant coupling in this theory is the so called `t Hooft coupling $\lambda \sim g^{2}_{YM}N$. Furthermore, it was also noticed that the Feynman diagram expansion of the amplitudes in this limit closely mimics string interactions. This explains the earlier success of string theory in describing some of the aspects of strong interaction. \\

One of the major successes of string theory comes from the study of black holes. Semi-classical physics in black hole background reveals the thermodynamic aspects of black holes, as demonstrated by Bekenstein and Hawking \cite{Bekenstein:1973ur, Hawking:1974sw}. In particular, it was found that black holes have a finite temperature and entropy given by the universal formula

\begin{equation}
T_{BH} = \dfrac{\kappa}{2\pi},\qquad S_{BH} = \frac{A}{4G_N}
\end{equation}
where $\kappa$ is the surface gravity, $A$ is the area of the horizon and $G_N$ is the Newton's constant. This suggests that black holes should be understood as a thermodynamic ensemble and the job of a putative quantum gravity theory is then to explain this entropy in terms of its microstates. This idea has found some success in string theory where the entropy of a class of (supersymmetric) black holes are understood in terms of states in string theory (see \cite{Sen:2007qy,David:2002wn} and references therein). Another interesting fact about the above formula is that the entropy scales as the area unlike in a quantum field theory where entropy scales like the volume. This has led to the celebrated holographic principle of `t Hooft and Susskind \cite{tHooft:1993dmi,Susskind:1994vu} which is a property of quantum gravity. \\

Roughly speaking, the holographic principle is a statement of equivalence between quantum gravity in $d+1$ dimensions and an ordinary quantum field theory in $d$ dimensions. This is rather surprising because we typically encounter systems whose number of degrees of freedom scales as the volume of the system. String theory admits one of the concrete realizations of the holographic principle, called {\em AdS/CFT correspondence}. 

\section{$2+1$-D gravity and Higher Spins}
In $2+1$ dimensions, pure gravity has no (perturbative) dynamics because curvature is completely rigid. But despite the lack of any gravitational attraction, gravity in 2+1 dimensions is non-trivial -- black holes solutions were discovered by Banados, Teitelboim and Zanelli (BTZ) \cite{Banados:1992wn} as quotients of AdS$_{3}$ \cite{Banados:1992gq}. This fact makes 2+1 D gravity an excellent theoretical laboratory for testing a variety nonperturbative issues in quantum gravity, without the added complications of curvature dynamics which play a huge role in higher dimensions. However, effort in this direction did not begin in earnest until the work of Witten \cite{Witten:1988hc} (see also \cite{Achucarro:1987vz}). He demonstrated that $2+1$-d gravity can be recast as a Chern-Simons gauge theory,

\begin{eqnarray}
S_{EH} = I_{CS}[A] - I_{CS}[\bA]
\end{eqnarray}
where 

\begin{equation}
I_{CS} [A] = \frac{k}{4\pi} \int_{\cM} \bra A \wedge dA + \frac{2}{3} A \wedge A \wedge A \ket 
\end{equation}
with the gauge gauge group $SL(2,R)\times SL(2,R)$ when the cosmological constant $\Lambda$ is $<0$, the gauge group $SL(2,C)$ when $\Lambda$ is $>0$ and $ISO(2,1)$ when $\Lambda=0$.\\

The negative cosmological constant case drew a lot of attention, partly because that was the context in which the above mentioned BTZ black holes were discovered, but also because of the earlier work of Brown and Henneaux \cite{Brown:1986nw} who showed that the asymptotic symmetry algebra of AdS$_{3}$ gravity is a Virasoro algebra. In fact, this latter result is now widely recognized as a precursor to the celebrated AdS-CFT duality \cite{Maldacena:1997re} where a fully quantum theory of gravity in AdS$_{d+1}$ is conjectured to have an equivalent description in terms of a conformal field theory supported on the boundary of AdS$_{d+1}$.\\

On an entirely different theme, theories of interacting gauge fields with an infinite tower of higher spins ($s\geq2$) have been studied as a toy version of a full string theory \footnote{Since the latter has infinite dimensional gauge invariance, see the work by Sundborg \cite{Sundborg:2000wp}.} by Fradkin and Vasiliev \cite{Fradkin:1986ka,Fradkin:1987ks,Vasiliev:1990cm,Vasiliev:1990en,Vasiliev:1992av,Vasiliev:1999ba}, building on the early work of Fronsdal \cite{Fronsdal:1978rb}. Higher spin theories in three dimensions, as demonstrated in \cite{Blencowe:1988gj} are considerably simpler than theories in higher dimensions due to absence of any local propagating degrees of freedom \textit{i.e.} they are topological. In addition, it is possible to truncate the infinite tower of higher spin fields to spin, $s\leq N$. The complicated nonlinear interactions of the higher spin fields can be reformulated in terms of an $SL(N,R)\times SL(N,R)$ Chern-Simons gauge theory (for AdS$_{3}$ case) or an $SL(N,C)$ Chern-Simons (for dS$_{3}$). Therefore 2+1 dimensional higher spin theories are a generalization of Chern- Simons gravity -- one gets back to the spin-2 pure gravity theory when one sets $N=2$.\\

\section{AdS Space}
We begin by reviewing the geometry of anti-de Sitter space in $d+2$ dimensions. AdS$_{d+2}$ can be thought of as a hyperboloid embedded in flat $(d+3)$-dimensional space $\bR^{d+1,2}$ with Minkowski metric

\begin{equation}\label{embed_metric}
ds^2 = -dX_0^2 - dX^2_{d+2} + \sum_{i=1}^{d+1} dX_i^2.
\end{equation}
where $X^M$ are the coordinates of $\bR^{d+1,2}$. AdS space is now defined by the relation

\begin{equation}\label{hyperboloid}
X_0^2 + X_{d+2}^2 - \sum_{i=1}^{d+1}X_i^2 = L^2
\end{equation}
where $L$ is called AdS radius. Both \eqref{embed_metric} and \eqref{hyperboloid} has an $SO(d+1,2)$ isometry. We can solve \eqref{hyperboloid} by \cite{Kiritsis:2007zza}

\begin{eqnarray}
X_0 &=& L\cosh \rho \cos \tau,\qquad X_{d+2} = L\cosh \rho \sin \tau \br X_i &=& L \sinh \rho \:\Omega_i,\;\;\; \left(i=1,2,\cdots,d+1;\;\;\sum_{i=1}^{d+1}\Omega_i = 1 \right)
\end{eqnarray}
where $\Omega_i$ are the coordinates on unit $d$-sphere, S$^d$. With this parametrization, the induced metric on the hyperboloid is given by

\begin{equation}
ds^2 = L^2\left(-\cosh^2 \rho \;d\tau^2 + d\rho^2 + \sinh^2 \rho\; d\Omega_d^2\right)
\end{equation}
The hyperboloid is covered once by this parametrization for $0<\rho<\infty$ and $0\leq \tau < 2\pi$. Topologically, AdS$^{d+2}$ is S$^1 \times \bR^{d+1}$, where S$^1$ is time-like and the AdS space thus defined has closed timelike curves. We can get a causal spacetime by taking the universal cover of this space by extending $\tau$ to $-\infty<\tau<\infty$. There are no timelike curves on this cover and this is usually taken as the AdS space here. 

The causal structure of the spacetime is invariant under conformal transformations. To understand the causal structure of AdS space it is convenient to first introduce a new coordinate $\theta$ such that $\tan \theta = \sinh \rho$, $\theta \in [0,\frac{\pi}{2})$, under which the metric takes the form

\begin{equation}
ds^2 = \frac{L^2}{\cos^2 \theta}\left(-d\tau^2 + d\theta^2 + \sin^2\; \theta d\Omega_d^2\right).
\end{equation}
The metric in the parenthesis is the metric of Einstein static universe with the difference being the $\theta$ coordinate taking values in $[0,\frac{\pi}{2})$ instead of $[0,\pi)$. The equator $\theta=\pi/2$ is the boundary of the space with topology S$^d$. The boundary extends in the $\tau$ direction and we must specify a boundary condition on it to have a well-defined Cauchy problem.\\

There is another useful set of coordinates on AdS called the \emph{Poincare coordinates}, which are given by

\begin{eqnarray}
X_0 &=& \frac{z}{2}\left(1 + \frac{1}{z^2}(L^2 + \vec{x}^2 - t^2)\right), \quad X_{d+1} = \frac{z}{2}\left(1 - \frac{1}{z^2}(L^2 + \vec{x}^2 - t^2)\right), \br X_{d+2} &=& \frac{Lt}{z},\qquad\qquad X_i = \frac{Lx^i}{z}
\end{eqnarray}
with $\vec{x}\in \bR^d$. The metric in this coordinates is given by

\begin{equation}
ds^2 = \frac{L^2}{z^2}\left[dz^2 - dt^2 + d\vec{x}^2 \right]
\end{equation}
The boundary is now at $z=0$ and this coordinates only cover half of the hyperboloid. The Euclidean AdS can be obtained from the above metric by an analytic continuation $\tau \rightarrow i\tau_E$ and can be obtained by embedding hyperboloid in $\bR^{d+2,1}$.

\section{Conformal Field Theories}
In this section we present the conformal algebra in general Minkowski spacetime in $d$ dimensions. Conformal field theories are ubiquitous in many areas of physics. They appear as the theory describing the end points of renormalization group flow, \textit{i.e.} fixed points. In statistical physics, they appear as theory that describes the physics of systems undergoing a second order phase transition. CFTs appear as critical theories at the point of phase transition. CFTs are also important in string theory, especially due to AdS/CFT. AdS/CFT provides a non-perturbative definition of quantum gravity in AdS space and thus CFTs can teach us about quantum gravity.\\

Most of the material present here is based on \cite{magoo}. The conformal group in general $d$ dimensional flat space is the set of transformations that preserves the form of the metric up to a scale factor, \textit{i.e.}

\begin{equation}
\eta_{\mu \nu} \rightarrow \rme^{\omega(x)}\eta_{\mu \nu}
\end{equation}
It is easy to check that Poincare group is a subgroup of the conformal group with $\omega(x)=0$. Therefore, the conformal algebra will contain the generators of Poincare group $M_{\mu \nu}$ of Lorentz transformations and $P_{\mu}$ of translations. Together, they generate the Poincare algebra

\begin{eqnarray}\label{poincare_algebra}
\left[M_{\mu \nu},M_{\rho\sigma} \right] &=& -i\left( \eta_{\mu \rho}M_{\nu \sigma} - \eta_{\nu \rho}M_{\mu \sigma} + \eta_{\nu \sigma}M_{\mu \rho} - \eta_{\mu \sigma}M_{\nu \rho} \right) \br
\left[M_{\mu \nu},P_{\rho} \right] &=& -i\left( \eta_{\mu \rho}P_{\nu}- \eta_{\nu \rho}P_{\mu} \right) \\
\left[P_{\mu},P_{\nu} \right] &=& 0 \nonumber
\end{eqnarray}
In addition to the Poincare transformations, the conformal group also admits $(i)$ the scale transformations

\begin{equation}
x^{\mu} \rightarrow \lambda x^{\mu}
\end{equation}
and $(ii)$ the special conformal transformations

\begin{equation}
x^{\mu} \rightarrow \frac{x^{\mu} + b^{\mu}x^2}{1 + 2x^{\nu}b_{\nu} + a^2 x^2}
\end{equation}
Together, they obey the conformal algebra, which apart from the Poincare algebra \eqref{poincare_algebra}, also contains generators of scale transformation $D$ and special conformal generators $K_{\mu}$. Their algebra is given by \cite{Mack}

\begin{eqnarray}\label{conformal_algebra}
\left[M_{\mu\nu},K_{\rho} \right] &=& -i\left( \eta_{\mu \rho}K_{\nu}- \eta_{\nu \rho}K_{\mu} \right) \br
\left[P_{\mu},K_{\nu} \right] &=& 2i(M_{\mu\nu}-\eta_{\mu\nu}D) \br
\left[M_{\mu\nu},D \right] &=& 0 \\
\left[D,K_{\mu} \right] &=& i K_{\mu} \br
\left[D,P_{\mu} \right] &=& -iP_{\mu} \nonumber
\end{eqnarray}
This algebra is isomorpic to the $SO(2,d)$ Lie algebra\footnote{In the Euclidean space this becomes $SO(1,d+1)$.} which can be seen by defining generators $J_{ab}$, where $a,b = 0,1,\cdots,d+1$ such that

\begin{eqnarray}
J_{\mu\nu} &=& M_{\mu\nu}, \quad J_{\mu d} = \frac{1}{2}(K_{\mu} - P_{\mu}), \br J_{\mu (d+1)} &=& \frac{1}{2}(K_{\mu} + P_{\mu}), \quad J_{(d+1)d} = D 
\end{eqnarray}
Note that $d=2$ is a special case as the group of conformal transformations gets enhanced to all holomorphic transformations. This leads to some remarkable simplifications in $d=2$. \\

In a field theory the basic constituents are field operators which should now be in a representation of the conformal group. For physical applications we are often interested in classifying fields according to their scaling dimension $-i\Delta$ which are eigenvalues of the scaling operator $D$. Using \eqref{conformal_algebra} we see that $P_{\mu}$ raises the $D$ eigenvalues one unit and $K_{\mu}$ lowers them. In unitary theories there is a lower bound for the scaling dimension of any operator, which leads us to the conclusion that there exists a lowest dimension operator which is annihilated by $K_{\mu}$. Such lowest dimensional operators are called \textit{primary operators}. For a primary operator $\Phi(x)$, the conformal group acts as

\begin{eqnarray}
\left[P_{\mu}, \Phi(x)\right] &=& i\pd_{\mu}\Phi(x) \br
\left[M_{\mu\nu},\Phi(x) \right] &=& \left[i(x_{\mu}\pd_{\nu}-x_{\nu}\pd_{\mu}) + \Sigma_{\mu\nu} \right]\Phi(x) \\
\left[D,\Phi(x)\right] &=& i(-\Delta + x^{\mu}\pd_{\mu})\Phi(x) \br 
\left[K_{\mu},\Phi(x) \right] &=& \left[i(x^2 \pd_{\mu} - 2x_{\mu}x^{\nu}\pd_{\nu} + 2x_{\mu}\Delta) - 2x^{\nu}\Sigma_{\mu\nu} \right]\Phi(x) \nonumber
\end{eqnarray} 
where $\Sigma_{\mu\nu}$ is a finite dimensional representation of $M_{\mu\nu}$. The representations of the conformal group is labelled by its Lorentz transformation and its scaling dimensions. Thus the representations can be broken down into primary fields and all the fields obtained by acting with $P_{\mu}$ on the primary fields. \\ 

Since the conformal group is larger than the Poincare group, it restricts the form of the correlation functions. For two and three point functions, the conformal symmetry completely fixes the functional form of the correlator up to few constants. For instance, the 2-point function of two scalar primary operators is given by

\begin{equation}
\bra O_1 (x_1) O_2(x_2)\ket = \frac{c_{12}}{(x_{12})^{2\Delta}}
\end{equation}
where $x_{12} = x_1 - x_2$, $c_{12}$ is a constant and $\Delta_1 = \Delta_2 = \Delta$. The correlation function vanish for $\Delta_1 \neq \Delta_2$. Similarly, the 3-point function of three scalar primary operators is fixed by the conformal symmetry to be

\begin{eqnarray}
\bra O_1(x_1)O_2(x_2)O_3(x_3)\ket = \frac{c_{123}}{(x_{12})^{\Delta_1 + \Delta_2 - \Delta_3}(x_{13})^{\Delta_1 + \Delta_3 - \Delta_2}(x_{23})^{\Delta_2 + \Delta_3 - \Delta_1}}
\end{eqnarray}
Once again, the power of conformal symmetry fixes the functional form of the 3-point function up to a constant $c_{123}$. In general, we can use global conformal invariance to restrict the n-point function using

\begin{eqnarray}
0 &=& \bra G O_1 \cdots O_n\ket \br &=& \bra \left[G,O_1\right]O_2 \cdots O_n \ket + \bra O_1 \left[G,O_2\right]\cdots O_n \ket+ \cdots \bra O_1 \cdots \left[G,O_n\right]\ket
\end{eqnarray}
where $G$ is the generator of the global conformal group. Using \eqref{poincare_algebra} and \eqref{conformal_algebra}, the commutators can be replaced by differential operators giving differential equations for the correlator. Later on we will use this property to obtain the three point function of two scalar primaries with a primary fermion.\\

We conclude this section by pointing out one of the general properties of local operators in a field theory called \textit{operator product expansion} (OPE). OPE states that a product of two local operators $O_1(x_1)O_2(x_2)$ at short distances can be expressed as a sum of local operators acting at that point. 

\begin{eqnarray}
O_1(x_1)O_2(x_2) = \sum_n c^{n}_{12}(x_1 - x_2)O_n(x_2)
\end{eqnarray}
All operators with same global quantum numbers as $O_1 O_2$ appear in the sum in general. The coefficients $c^{n}_{12}$ does not depend on the positions of other operator insertions. The above sum should be understood in the sense of operators inside correlation function. In a CFT, the conformal symmetry restricts the form of the coefficients.

\section{AdS/CFT Correspondence}

The archetypal example of AdS/CFT correspondence is the equivalence between type $IIB$ string theory compactified on AdS$_5 \times \mathrm{S}^5$ which is dual to four-dimensional ${\cal N} = 4$ supersymmetric Yang-Mills (SYM) theory \cite{Maldacena:1997re}, proposed by Maldacena in 1997. Since then many more examples of such dualities have been worked out (see \cite{Aharony:2008ug},for example).\\

Maldacena's conjecture was borne out of studying low energy limit of a stack of $N$ coincidental $D3$-branes in 10-dimensional type $IIB$ supergravity. $D3$ branes are 3 dimensional objects propagating through spacetime. The worldvolume theory for this stack of $D3$ branes is given by ${\cal N} = 4$, $SU(N)$ supersymmetric Yang-Mills (SYM) theory. On the other hand $D3$ branes act as sources in the supergravity and contributes to the stress tensor, so the geometry becomes curved. It turs out, the near horizon geometry of the stack of $D3$ branes is AdS$_5$ while far away from the horizon the metric asymptotes to flat space. The key point here is that the 4 dimensional flat spacetime can be viewed as the boundary of 5 dimensional AdS space. \\

Since AdS/CFT is an equivalence between theories, there must exist a dictionary that relates the observables in one theory to observables in the other. One also expects the global symmetries on both sides to match. This is indeed true in the case of AdS$_5 \times \mathrm{S}^5$ and ${\cal N} = 4$, $SU(N)$ supersymmetric Yang-Mills (SYM) theory. The isometries group of AdS$_5 \times \mathrm{S}^5$ is $SO(4,2)\times SO(6)$ while on the gauge theory side $SO(4,2)$ arises due to conformal invariance. The gauge theory also has an R-symmetry given by $SU(4)\simeq SO(6)$ (see \cite{magoo} for a convenient review). In addition to the above bosonic symmetries, ${\cal N} = 4$ SYM also has 32 supercharges, enhanced due to conformal invariance. This enhanced supersymmetry also appears in the near horizon limit from string theory in AdS$_5 \times \mathrm{S}^5$. The parameters on the gauge theory side such as $g_{YM}$, $\theta$ and $N$ are all dimensionless as expected of a CFT. They are related to string coupling through

\begin{equation}
	\tau \equiv \frac{4\pi i}{g^{2}_{YM}} + \frac{\theta}{2\pi} = \frac{i}{g_s} + \frac{\chi}{2\pi}
\end{equation}
where $g_s$ is the string coupling and $\chi$ is the expectation value of the R-R axion scalar of type $IIB$ supergravity. There are three dimensionful parameter in string theory on AdS$_5$, namely, string length $l_s$, AdS radius $L$ and 10 dimensional Newton's constant $G_{10}$ (we can scale the S$^5$ radius to 1). However it is related to parameters in the gauge theory side only through dimensionless ratios. They are given by \cite{Kiritsis:2007zza}

\begin{equation}
	\frac{L^4}{l^4_s} = 4\pi g_s N = \lambda,\quad \frac{16\pi G_{10}}{l^8_s} = (2\pi)^7 g^2_s = 8\pi^5 \frac{\lambda^2}{N^2}
\end{equation}
In fact the dictionary goes way beyond the symmetries and couplings and can be extended to other gauge theory observables such as correlation functions, Wilson loops, entanglement entropy, etc. On the string theory side, we notice that $G_N \sim 1/N^2$, so quantum effects are suppressed when $N>>1$. The stringy effects are suppressed when the background curvature are small, \textit{i.e.} $L>>l_s$. This implies $\lambda >> 1$ in the gauge theory side. We thus reach the conclusion that strongly coupled ($\lambda>>1$) gauge theory at large N is well captured by supergravity on AdS$_5 \times \mathrm{S}^5$. \\

We can now write the statement of AdS/CFT correspondence in its most useful way. Suppose $\phi(x,r)$ is a bulk field whose boundary value is $\phi_0(x)$. The interaction of this bulk field with the branes suggests that the boundary value of the bulk fields acts as sources for appropriate operators in the CFT \cite{Witten:1998qj,GKP}. The correspondence can be symbolically expressed as

\begin{eqnarray}\label{adscft}
\bra \rme^{\int d^4x \phi_0(x) O(x)}\ket_{CFT_4} = Z_{String}[\phi_0]
\end{eqnarray}

AdS/CFT correspondence, in principle, gives a precise, non-perturbative definition of quantum gravity in an asymptotically AdS spacetime. Moreover the utility of the correspondence comes about because it is a strong/weak coupling duality, which makes it very useful for applications to strongly coupled systems. Below we will describe how to use the correspondence to compute 2 and 3 point function in the strongly coupled CFT.

\subsection{Correlation Functions}
In \eqref{adscft} we have equated two partition functions, namely the gauge theory partition function in presence of sources to the string partition function. We begin by first noticing that given a partition function, the logarithm of the partition function is the generator of all connected correlation functions.

\begin{equation}
\bra \rme^{\int d^4x \phi_0(x) O(x)}\ket_{CFT_4} = \rme^{-W[\phi_0]}
\end{equation}
The string partition function is not exactly solvable in AdS$^5 \times {\rm S}^5$ and we need to resort to approximation techniques. As we argued previously, at large $N, \lambda$ we can ignore the contribution of loops and stringy effects and use supergravity action to compute the partition function. At leading order we then have the saddle point evaluation of the partition function

\begin{equation}
Z_{gauge}[\phi_0] = \rme^{-I_{SUGRA}[\phi_0]}
\end{equation} 
The on-shell action $I_{SUGRA}$ is evaluated on solution that satisfy the equations of motion with the boundary condition 

\begin{eqnarray}
\phi(z,x)|_{z=0} = z^{4-\Delta}\phi_0(x)
\end{eqnarray}
Usually the on-shell supergravity actions are beset with infrared divergences coming from $z\rightarrow0$ limit of the integral, so the boundary conditions are prescribed on the surface $z=\e$ with $\e \rightarrow 0$ taken at the end of calculations.\\

We are now in a position to apply the procedure outlined above to compute correlation functions. For simplicity, we describe the computation of 2-point function for a massive scalar field in some detail. To the leading order, we can forget about the interactions and start with the action

\begin{equation}
	I = \frac{1}{2}\int d^{d+2}x\sqrt{g}\left[(\pd \phi)^2 + m^2 \phi^2 \right]
\end{equation}
To compute the on-shell action, we need to first solve the equations of motion 

\begin{eqnarray}
(\Box - m^2)\phi = 0
\end{eqnarray}
subject to the boundary conditions

\begin{equation}
\phi(z,x)|_{z=0} = z^{d+1-\Delta}\phi_0(x)
\end{equation}
where $\phi_0$ is an arbitrary function on $\bR^{d+1}$. Fourier transforming the boundary coordinates, we have

\begin{eqnarray}
\phi(z,x) = \int \frac{d^{d+1} q}{(2\pi)^{d+1}}\phi(z,q)\rme^{iq.x}
\end{eqnarray}
In the Poincare coordinates, the wave equation then takes the form

\begin{equation}
\left(\pd_z^2 - \frac{q}{u}\pd_z - q^2 - \frac{m^2 L^2}{z^2}\right)\phi(z,q) = 0	
\end{equation}
To solve the equation we first need to construct the bulk to boundary propagator which is defined as

\begin{equation}
(\Box - m^2)K(z,x;x') = 0, \quad K(z,x;x')|_{z=0} = \delta^{d+1}(x-x')
\end{equation}
Fouries transforming the boundary coordinates, we have

\begin{equation}
	K(z,q)|_{z=0} = 1
\end{equation}
We can now solve the Fourier transformed wave equation as done in \cite{magoo}. For convenience we will Wick rotate to the Euclidean space and work in the configuration space instead of Fourier space. The solution in the configuration space is given by

\begin{eqnarray}\label{propsolution}
\frac{z^{\Delta}}{(z^2 + |x-x'|^2)^{\Delta}},
\end{eqnarray}
where $\delta$ is the larger root of 

\begin{eqnarray}\label{delta_eqn}
\Delta(\Delta - d -1) = m^2 L^2
\end{eqnarray}
The solution \eqref{propsolution} vanishes as $z\rightarrow 0$ for $x- x'\neq 0$. Regularity of the solution demands we choose $\Delta = \Delta_+$, the larger root of \eqref{delta_eqn}. For this choice, it is now easy to see that the solution vanishes as $z\rightarrow 0$ for $|x-x'|\neq 0$, \textit{i.e.}

\begin{eqnarray}
\lim_{z\rightarrow 0} \frac{z^{\Delta_+}}{(z^2 + |x-x'|^2)^{\Delta_+}} = C_d \delta^{d+1}(x-x')
\end{eqnarray}
where

\begin{eqnarray}
C_d = \pi^{\frac{d+1}{2}}\frac{\Gamma[\Delta_+ - \frac{d+1}{2}]}{\Gamma[\Delta_+]}
\end{eqnarray}
The normalized propagator is therefore given by

\begin{eqnarray}
K_{\Delta}(z,x;x') = \frac{\Gamma[\Delta]}{\pi^{\frac{d+1}{2}}\Gamma[\Delta - \frac{d+1}{2}]}\frac{z^{\Delta}}{(z^2 + |x-x'|^2)^{\Delta}}
\end{eqnarray}
and the solution to massive wave equation is given by

\begin{eqnarray}\label{normalizable_sol}
\phi(z,x) = \frac{1}{C_d} \int d^{d+1}x'\; \frac{z^{\Delta_+}}{(z^2 + |x-x'|^2)^{\Delta_+}} \phi_0(x')
\end{eqnarray}
so that 

\begin{eqnarray}
\lim_{z\rightarrow 0} \phi(z,x) \sim z^{\Delta_-}\phi_0(x)
\end{eqnarray}
For $d=3$, \emph{i.e.} in AdS$_5$ the bulk-boundary propagator near $z=0$ is given by

\begin{eqnarray}
\lim_{z\rightarrow 0} K_{\Delta}(z,x;x') \simeq z^{4-\Delta} \left[\delta^{(4)}(x-x') + O(z^2) \right] + z^{\Delta}\left[ \frac{C_3^{-1}}{|x-x'|^{2\Delta}} + O(z^2) \right]
\end{eqnarray}
The part of the solution proportional to $z^{4-\Delta}$ is the source while the other part corresponds to the normalizable mode and is determined by the source. We can now compute the on-shell action by plugging in \eqref{normalizable_sol} into the action

\begin{eqnarray}
I_{on-shell} &=& -\frac{1}{2}\int d^4x_1 d^4 x_2 \int d^4x \left. \frac{K(z,x;x_1)\pd_z K(z,x;x_2)}{z^3}\right|_{z=0}
\end{eqnarray}
From the near boundary expansion of $K$, we then have

\begin{eqnarray}
\int d^4 x \frac{K(z,x;x_1)\pd_z K(z,x;x_2)}{z^3} \simeq \Delta_- z^{2\Delta_- - 4}\delta(x_1 - x_2) + \frac{4}{C_3}\frac{1}{|x-x'|^{2\Delta}} + \cdots
\end{eqnarray}
The first term is a contact term that diverges as we approach boundary and can be removed by adding appropriate counter-terms in the action. The systematic procedure for finding counter-terms is called \emph{Holographic renormalization} and will be dicussed in detail in Chapter 7. The second terms gives a finite contribution and indicates that the 2-point function of dual operator is

\begin{eqnarray}
\bra O(x_1)O(x_2)\ket = \frac{1}{|x-x'|^{2\Delta_+}}
\end{eqnarray} 
where $O(x)$ is a CFT operator. Similar philosophy also holds for the computation of higher n-point functions. This concludes our discussion of AdS/CFT correspondence.

\chapter{A Review of $(2+1)$-D Gravity and Higher Spin theories}\label{GHSR}

This chapter is an elementary review of gravity and higher spin theories in $2+1$ dimensions. Most of the material presented here is based on the review article \cite{Kiran}\footnote{See also \cite{Afshar:2014rwa}}. The organization of this chapter is as follows. In Sec. \ref{2one}, we recap the first order formulation of Einstein's gravity in 2+1-dimensions and its recasting as a Chern-Simons theory with a non-compact gauge group. We fix all notations and conventions for the map between the second order variables (metric) and the gauge connection here. In section \ref{two}, we review the basics of higher spin gauge theories in $2+1$ dimension. We show how Chern-Simons formulation of pure gravity can be easily generalized to include towers of interacting higher spin fields.

\section{Chern-Simons Formulation of Gravity}\label{2one}
In this section we will briefly review the Chern-Simons formulation of 3D Einstein's gravity, first explored in the context of supergravity by \cite{Achucarro:1987vz}, later rediscovered and revived by \cite{Witten:1988hc}. The aim of this section is to orient the reader and give a flavor of 3D gravity and its topological nature. See \cite{Carlip:1998uc} and references therein for a comprehensive survey. \\

The Chern-Simons formulation is a classical equivalence of Einstein's gravity in 3 dimensions to the Chern-Simons action with appropriate gauge group. To begin with, the (bulk) Einstein-Hilbert action is given by

\begin{equation}\label{normal_action}
S_{G} = \frac{1}{16 \pi G }\int_{M} d^3 x {\sqrt{-g}} (R - 2 \Lambda) + I_{matter}.
\end{equation} 
where $R$ is the Ricci scalar, constructed out of the metric $g_{\mu\nu}$ and $g$ is the matrix determinant of the metric and we have set the cosmological constant to zero. Equations of motion for the action \eqref{normal_action} are,
 
\begin{equation}
R_{\mu \nu} + \Lambda g_{\mu \nu} - \frac{1}{2} g_{\mu \nu} R = -8 \pi G T_{\mu \nu} \label{eom}
\end{equation}
The number of independent components of Ricci and Riemann tensors in $d$ dimensions are 

\[ \frac{d(d+1)}{2} \quad \& \quad \frac{d(d-1)}{4} \left( \frac{d(d-1)}{2} +1 \right) \]
respectively. In $2+1$ dimensions it is interesting to note that both these tensors have six independent components. Hence, Riemann tensor can be written completely in terms of Ricci tensor and vice versa. Using this and the symmetries of Riemann tensor, it is easy to show that 
\begin{equation}
R_{\mu \nu \rho \sigma} = g_{\mu \rho} R_{\nu \sigma} + g_{\nu \sigma} R_{\mu \rho} - g_{\mu \sigma} R_{\nu \rho} - g_{\nu \rho} R_{\mu \sigma} - \frac{1}{2} (g_{\mu \rho} g_{\nu \sigma} - g_{\mu \sigma} g_{\nu \rho}) R \label{rmn}.
\end{equation}
There is no traceless part, i.e. Weyl curvature tensor is zero. The above equation also means that in vacuum($T_{\mu \nu}=0$), the solutions of Einstein equation are flat for $\Lambda = 0$, and for $\Lambda \neq 0$ they have constant curvature (\textit{i.e.} the Ricci scalar is a constant). That  is, in vacuum when $\Lambda = 0 $, \eqref{eom} becomes
\begin{equation}
R_{\mu \nu} = \frac{1}{2} g_{\mu \nu} R 
\end{equation}
which upon taking trace implies $R=0=R_{\mu \nu}$ and hence from \eqref{rmn}, Riemann tensor vanishes and solution is locally flat. Similarly, when $\Lambda \neq 0$ vacuum solutions of Einstein equation have constant curvature. This means that $2+1$ dimensional space-time does not have local degrees of freedom. It has curvature only where there is matter, and there are no gravitational waves. \\

The fact that there are no local degrees of freedom in this case can also be seen by looking at the number of independent parameters in the phase space of GR. The independent parameters we have here are independent components of spatial metric on a constant time hypersurface, which is $d(d-1)/2$ in $d$ dimensions, and their time derivatives(conjugate momenta) which are again $d(d-1)/2$ in number. Einstein field equations act as $d$ constraints on initial conditions and furthermore coordinate choice eliminates $d$ degrees of freedom. This leaves us with $d(d-1) - 2d = d(d-3)$ degrees of freedom, which is zero for $d=3$.\\

\subsection{First Order Formulation of Gravity}
The Chern-Simons formulation rests heavily on the so called first order formalism, where the fundamental variables are vielbein, $e^{a}=e^{a}_{\mu}dx^{\mu}$ and spin connection, $\omega^{a}_{\; b} = \omega^{a}_{\mu\;b}dx^{\mu}$\;\footnote{The Latin alphabets $a,b,\cdots$ stand for tangent space indices while the manifold indices are denoted by Greek alphabets $\mu, \nu, \cdots$. The Latin indices are raised and lowered using the Minkowski metric \[ \eta_{ab} = {\rm diag}(-1,1,\cdots,1) \]}. The vielbein satisfies 

\begin{eqnarray}
g^{\mu \nu} e^{a}_{\mu} e^{b}_{\nu} &=& \eta^{ab}  \br 
\eta_{ab} e^{a}_{\mu} e^{b}_{\nu} &=& g_{\mu \nu}.
\end{eqnarray} 
Collection of all possible vielbeins at every point on $M$ is called a frame/vielbein bundle. For any spacetime vector $V_{\mu}$, now we can work with $V^a = V^\mu e^a_\mu$ instead. Covariant derivative of $V^a$ would be, 

\begin{equation}
D_\mu V^a = \partial_\mu V^a + \omega^a_{\mu b} V^b \label{cov}
\end{equation}
where, $\omega^a_{\mu b}$ is a connection in vielbein basis, it is called  the spin connection. The choice of $\omega^a_{\mu b}$ can be fixed by demanding the net parallel transport of $e^a_\mu$ to give a vanishing covariant derivative (see \cite{Ortin:2004ms} or section (12.1) of \cite{Green:1987mn} for a clear discussion)

\begin{equation}
D_{\mu} e_{\nu}^a = \partial_\mu e_\nu^a - \Gamma^\rho_{\mu \nu} e^a_\rho + \e^{abc} \omega_{\mu b}e_{\nu c} = 0.
\end{equation}
If the connection $\Gamma^{\rho}_{\mu \nu}$ is torsion free, then

\begin{equation}
T^a = D_{\omega} e^{a} = \mathrm{d} e^{a} + \omega^{a}_{b} \wedge e^{b} = 0 \label{c1}
\end{equation}
Equation \eqref{c1} is called Cartan's first structure equation. For torsion free case, expression for $\omega^{a}_{\mu b}$ can be written explicitly in terms of frame 1-forms by inverting them, we will see this for 2+1 dimensional case later. 

The curvature tensor can be defined using the usual expression for gauge field strength, adapted to the present case \cite{Ortin:2004ms}

\begin{equation}
[D_\mu, D_\nu]V^a = R_{\mu \nu b}^aV^b .
\end{equation}
Using \eqref{cov}, Riemann tensor now takes the form,

\begin{eqnarray}
dx^\mu \wedge dx^\nu R_{\mu \nu b}^a &=& (\partial_{[\mu}\omega_{\nu] a}^b - \omega_{[\mu | a}^c \omega_{| \nu ] c}^b) dx^\mu \wedge dx^\nu \\
&=& d \omega^{b}_{a} + \omega^{b}_{c} \wedge \omega^{c}_{a}
\end{eqnarray}
which is analogous to the familiar gauge theory expression

\begin{equation}
F = dA + A\wedge A .
\end{equation}
We can now use this along with metric and spin connection to write Einstein action in first order formalism,

\begin{equation}
I = k \int  \left[  \e_{a_1 a_2 \dots a_D} R^{a_1 a_2} \wedge e^{a_3} \wedge \dots e^{a_D} + 
\frac{\Lambda}{D ! } \e_{a_1 a_2 \dots a_D} e^{a_1} \wedge e^{a_2} \dots \wedge e^{a_D} \right] \label{foeh} . 
\end{equation}
where $R^{a_1 a_2}$ is a curvature two form i.e., $R^{a_1 a_2}\equiv R^{a_1 a_2 a_3 a_4}e_{a_3}\wedge e_{a_4}$.\\

Now it can be seen that in $(2+1)$-dimensions the first order action \eqref{foeh} can be written as:

\begin{equation}\label{3d_first_ord_act}
I = \frac{1}{8\pi G} \int_{\cM} \left[e^a \wedge (\mathrm{d}\omega_a + \frac{1}{2} \e_{abc} \omega^b \wedge \omega^c) + 
\frac{\Lambda}{6} \e_{abc} e^a \wedge e^b \wedge e^c \right] . 
\end{equation}
where we have $\omega^a = 1/2 \e^{abc}\omega_{\mu bc}dx^{\mu}$. One of the equations of motion is obtained by varying the above action with respect to $\omega_a$: 

\begin{equation}
T_a = \mathrm{d}e_a + \e_{abc} \omega^b \wedge e^c = 0 \label{eqn1}. 
\end{equation}
If triads $e^a_\mu$ are invertible, \eqref{eqn1} can be solved to obtain the following expression for spin connection, 

\begin{equation}
\omega^a_\mu = \e^{abc}e^\nu_c(\partial_\mu e_{\nu b}-\partial_\nu e_{\mu b}) -\frac{1}{2} \e^{bcd}(e^\nu_b e^\rho_c \partial_\rho e_{\nu d}) e_\mu^a \label{eqn2}
\end{equation}
invertibility of triad is important as the solution \eqref{eqn2} is a second order equation while \eqref{eqn1} is a first order equation\;\footnote{Non-invertible triads may find its way in quantum theory but we will only deal with classical theories}. Varying the action \eqref{3d_first_ord_act} with respect to $e^a$ gives, 

\begin{equation}
\mathrm{d}\omega_a + \frac{1}{2} \e_{abc} \omega^b \wedge \omega^c + \frac{\Lambda}{2} \e_{abc} e^b \wedge e^c = 0
\end{equation}
which can also be expressed as, 

\begin{equation}
R_a = \mathrm{d}\omega_a + \frac{1}{2} \e_{abc} \omega^b \wedge \omega^c = - \frac{\Lambda}{2} \e_{abc} e^b \wedge e^c 
\end{equation}
which is the Einstein's equation in vielbein-spin-connection language. 

Up to boundary terms, action \eqref{3d_first_ord_act} is invariant under two sets of gauge symmetries, (a) Local Lorentz Transformations (LLT),
\begin{align}
\delta_l e^a &= \e^{abc} e_b \tau_c  \\
\delta_l \omega^a &= d\tau^a + \e^{abc} \omega_b \tau_c \label{local}
\end{align}
where $\tau_a$ is a local function, and (b) Local Translations (LT), 

\begin{align}
\delta_t e^a &= d\rho^a + \e^{abc} \omega_b \rho_c  \\
\delta_t \omega^a &= -\Lambda \e^{abc} e_b \rho_c. \label{trans}
\end{align}
The subscripts $t$ and $l$ above on $\delta$ are labels that stand for LLT and LT respectively. These are called local Lorentz transformations and local translations because the number of components of $\tau$ and $\rho$ are precisely equal to the number of parameters of Lorentz transformations ($d(d-1)/2$ of them in $d$ dimensions) and translations ($d$ of them in $d$ dimensions) respectively. The Einstein-Hilbert action in the second order (\emph{i.e.}, in the metric) formulation is invariant under space-time diffeomorphisms which is related to LLT and LT in the first order formulation \cite{Carlip:1998uc}.

\subsection{Connection to Chern Simons Theory}
Gravity in 2+1 dimensions behaves like a gauge theory in many ways because its first order action \eqref{3d_first_ord_act} is that of a gauge theory: the so-called Chern-Simons theory. We demonstrate this by taking $A = A^a_\mu T_a dx^\mu$ to be a connection one form of group $G$ on a 3-manifold $\cM$, i.e $A$ is the vector potential of a gauge theory whose gauge group is $G$, the generators of whose Lie algebra are $T_a$. Chern-Simons action for $A$ is then:

\begin{equation}\label{gen_CS_action}
I_{CS} [A] = \frac{k}{4\pi} \int_{\cM} \bra A \wedge dA + \frac{2}{3} A \wedge A \wedge A \ket 
\end{equation}
Here $k$ is coupling constant and $\bra \cdots \ket$ is the non-degenerate invariant bilinear form on Lie algebra of G, which we will define more concretely below for the various cases. Euler-Lagrange equations of motion coming from \eqref{gen_CS_action} are

\begin{equation}
F[A] = dA + A \wedge A = 0 \label{flat}
\end{equation}
hence, $A$ is a flat connection(\emph{i.e.} field strength of $A$ vanishes). This does not mean that $A$ is always trivial, as potential with vanishing field strength might give rise to non-abelian Aharanov-Bohm effect. \\

We will start with the flat space theory. Our goal is to reproduce the Einstein-Hilbert action in the first order formulation from a Chern-Simons gauge theory. The triad and the spin connection are taken in the form
\bea
e^a=e^a_\mu \ dx^\mu, \ \ \omega^a=\frac{1}{2} \epsilon^{abc} \omega_{\mu bc}\ dx^{\mu}. \label{dualspincon}
\eea
The tangent space indices are raised and lowered using the 2+1 Minkowski metric ${\rm diag}(-1,1,1)$.
Now the claim is that the Chern-Simons action
\bea
I_{CS}[{\cal A}]=\frac{k}{4 \pi}\int \bra{\cal A} \wedge d {\cal A}+\frac{2}{3}{\cal A} \wedge {\cal A} \wedge {\cal A}\ket
\eea
with 
\bea
{\cal A}\equiv e^a\ P_a +\omega^a\ J_a
\eea
is the Einstein-Hilbert-Palatini action (with zero cosmological constant) in the first order formulation, if the generators satisfy the $ISO(2,1)$ algebra
\begin{equation}
\left[P_{a},P_{b}\right]=0,\qquad\left[J_{a},J_{b}\right]=\epsilon_{abc}J^{c},\qquad\left[J_{a},P_{b}\right]=\epsilon_{abc}P^{c},\label{eq: ISO(2,1) algebra}
\end{equation}
with $\epsilon^{012}=1$ and the invariant non-degenerate bilinear form
is defined by,
\begin{equation}
\bra J_{a} \ P_{b}\ket =\eta_{ab}, \ \ \bra J_{a} \ J_{b} \ket = 0 = \bra P_{a} P_{b} \ket. \label{eq: ISO(2,1) metric}
\end{equation}
Here, the level $k$ of the Chern-Simons theory is related to Newton's constant by
\bea
k=\frac{1}{4G}.
\eea
Once crucial ingredient here worthy of note is the choice of the trace form. For all components of the gauge field to have appropriate kinetic terms, it is necessary that the trace form is non-degenerate.\\ 

Now we turn to gravity with a negative cosmological constant $\Lambda\equiv -\lambda<0$. In this case, Witten's  observation is that again the Einstein-Hilbert action (this time including the cosmological constant piece) can be obtained from the Chern-Simons action and identical definitions as above, if one simply changes the algebra of the $P_a$ and $J_a$ to the $SO(2,2)$ algebra:
\bea
\left[P_{a},P_{b}\right]=\lambda \epsilon_{abc}J^c,\qquad\left[J_{a},J_{b}\right]=\epsilon_{abc}J^{c},\qquad\left[J_{a},P_{b}\right]=\epsilon_{abc}P^{c}.\label{eq: SO(2,2) algebra}
\eea 
In particular, the trace form is the same as before. 

There is a slightly different way of writing the latter (negative cosmological constant) case, that is often used in the literature and we will find convenient. One first introduces the generators
\bea
J_a^{\pm}=\frac{1}{2}\left(J^a\pm l\ P^a \right), \label{pmJ}
\eea
where $l=\frac{1}{\sqrt{\lambda}}$.
It is easy to check that (\ref{eq: SO(2,2) algebra}) now takes the form
\bea
\left[J_{a}^+,J_{b}^-\right]=0,\qquad\left[J_{a}^+,J_{b}^+\right]=\epsilon_{abc}J^{c+},\qquad\left[J_{a}^-,J_{b}^-\right]=\epsilon_{abc}J^{c-}. \label{sl2 factorized form}
\eea
The first of the above commutators implies that the algebra is a direct sum: what we have essentially shown is that $SO(2,2)\sim SL(2,\bR)\times SL(2,\bR)$, and that its algebra can be written as a direct sum of two copies of ${\bf sl}(2,\bR)$. In particular this means that we can introduce $T^a$ and $\tilde T^a$ via
\bea
J_{a}^+=\left(\begin{array}{cc}
	T^a&0\\
	0 & 0
\end{array}\right),\qquad J_{a}^-=\left(\begin{array}{cc}
0&0\\
0 & \tilde T^a
\end{array}\right)
\eea
so that if $T^a$ and $\tilde T^a$ each satisfy the $SL(2,\bR)$ algebra,
\bea
\left[T_{a},T_{b}\right]=\epsilon_{abc}T^{c},\qquad\left[\tilde T_{a},\tilde T_{b}\right]=\epsilon_{abc}\tilde T^{c}. \label{T-tilde-T form}
\eea
then (\ref{sl2 factorized form}), and therefore (\ref{eq: SO(2,2) algebra}), are satisfied. An important point to note is that from the trace form (\ref{eq: ISO(2,1) metric}) one finds that the trace form in terms of  $T$ and $\tilde T$ are
\bea
\bra T_a T_b\ket =\frac{l}{2} \eta_{ab}, \ \ \bra \tilde T_a \tilde T_b \ket =-\frac{l}{2} \eta_{ab}, 
\eea
In terms of $T$ and $\tilde T$, the gauge field now takes the form
\bea
{\cal A}_\mu=\left(\begin{array}{cc}
	\left(\omega_\mu^a+\frac{1}{l}e_\mu^a\right)\ T_a&0\\
	0 & \left(\omega_\mu^a-\frac{1}{l}e_\mu^a\right)\ \tilde T_a
\end{array}\right)\equiv \left(\begin{array}{cc}
A_\mu^a T_a&0\\
0 & \tilde A_\mu^a \tilde T_a
\end{array}\right)
\eea
so that the Einstein-Hilbert action with a cosmological constant can be written as the sum of two pieces now:
\bea
\frac{k}{4 \pi}\int \bra { A} \wedge d { A}+\frac{2}{3}{ A} \wedge { A} \wedge { A}\ket + \frac{k}{4 \pi}\int \bra { \tilde A} \wedge d {\tilde A}+\frac{2}{3}{\tilde A} \wedge {\tilde A} \wedge {\tilde A}\ket
\eea
Since the algebra of both $T$'s and $\tilde T$'s is identical (namely the $SL(2,\bR)$ algebra), what is typical in the literature is to identify the generator matrices $T^a = \tilde T^a$. This means that their trace forms are also identical, which one takes to be
\bea
\bra T_a T_b \ket =\frac{1}{2} \eta_{ab}. \label{simpltr}
\eea
Note that this trace form does not have the factor of $l$ as before, so that the missing $l$ has to be incorporated into the Chern-Simons level by hand for the action to reduce to the Einstein-Hilbert form. So now
\bea
k=\frac{l}{4G}.
\eea
Also, the negative sign in the trace form of the $\tilde T$ should also be incoprorated into the action by hand, so that now the AdS Einstein-Hilbert action takes the final form
\bea
I_{EH_{AdS}}=\frac{k}{4 \pi}\int \bra { A} \wedge d { A}+\frac{2}{3}{ A} \wedge { A} \wedge { A}\ket - \frac{k}{4 \pi}\int \bra { \tilde A} \wedge d {\tilde A}+\frac{2}{3}{\tilde A} \wedge {\tilde A} \wedge {\tilde A} \ket  \label{adsaction}
\eea 
where now the $A$ and $\tilde A$ are understood to be expanded in a basis of $T^a$'s (and no $\tilde T^a$'s):
\bea
A_\mu=\left(\omega_\mu^a+\frac{1}{l}e_\mu^a\right)\ T_a, \ \ \tilde A_\mu=\left(\omega_\mu^a-\frac{1}{l}e_\mu^a\right)\ T_a \label{gaugef}
\eea
with trace form (\ref{simpltr}).\\

When $\Lambda = - 1/l^2 > 0$, we will be looking at de-sitter gravity. The generators are that of $SL(2,\bC)$, 
\beq
\left[T_a,T_b \right] = \e_{abc}T^{c} \label{alg2}
\eeq
with invariant bilinear form 

\begin{eqnarray}\label{IB}
\bra T_a T_b \ket = \frac{1}{2} \eta_{ab}
\end{eqnarray}
is same as before and the connection one forms are 

\begin{align}
A &= \left( \omega + \frac{i}{l} e \right) \\
\tilde{A} &= \left( \omega - \frac{i}{l}e \right).
\end{align}
with \eqref{IB} as an invariant bilinear form and $k= -il/4G$, the Chern-Simons action $$I[A,\tilde{A}] = I_{CS}[A] - I_{CS}[\tilde{A}]$$ is same as first order action \eqref{3d_first_ord_act}. Note that the two gauge connections $A$ and $\tilde{A}$ are not independent in this case $\tilde{A}_aT^a = A^\ast_a T^a$ (to make action real). This condition also tells us that, unlike in previous cases, $A$ and $\tilde{A}$ are not independent of one another, hence, the gauge group of the theory is $SL(2,\bC)$ and not $SL(2,\bC) \times SL(2,\bC)$. As $F[A]$ of \eqref{flat} is the only gauge covariant local object, hence flat connection implies no local observables. The algebras \eqref{T-tilde-T form} and \eqref{alg2} are identical: they are the $SL(2)$ algebra. The coefficients which the field takes values in decides weather it is $SL(2,\bR) \times SL(2,\bR)$ or $SL(2,\bC)$. \\

Before we end this section, let us look at gauge symmetries and boundary terms that come up during gauge transformation of \eqref{gen_CS_action}. The Chern-Simons action depends directly on the gauge-variant $A$ and not on the gauge-invariant field strength $F$ in a conventional gauge theory. Let us look at its behaviour under gauge transformation of \eqref{gen_CS_action}, 
\beq
A = \gi \d g + \gi \tA g.
\eeq
Substituting this in action and simplifying, we get 
\begin{align}
I_{CS}[A] &= I_{CS}[\tA] - \frac{k}{4\pi}\int_{\partial \cM}\bra (\d g \gi)\wedge \tA \ket \\ \nonumber &+ \frac{k}{4\pi}\int_{\cM} \bra (\gi \d g)\wedge(\d \gi)\wedge(\d g) + \frac{2}{3}(\gi \d g)\wedge(\gi \d g)\wedge(\gi \d g) \ket.
\end{align}
The last term upon simplification yields,

\begin{equation}
-\frac{k}{12\pi} \int_{\cM} \bra (\gi \d g)\wedge(\gi \d g)\wedge(\gi \d g)  \ket.
\end{equation}
Thus under a finite gauge transformation of the Chern-Simons action, we have

\begin{align*}
I_{CS}[A] = I_{CS}[\tA] &- \frac{k}{4\pi}\int_{\partial \cM}\bra (\d g \gi)\wedge \tA \ket \\ \nonumber &- \frac{k}{12\pi} \int_{\cM} \bra  (\gi \d g)\wedge(\gi \d g)\wedge(\gi \d g)  \ket.
\end{align*}
The boundary term in RHS vanishes if the space is compact. If the gauge group $G$ is also compact, the last term is related to the winding number of the gauge transformation \cite{Nakahara}, its value will be $ 2 \pi n$($n$ is integer) for appropriate $k$. Hence $\exp(iI_{CS}) $ which occurs in path integral is gauge invariant. If $M$ is not closed, then we need to add boundary contributions to \eqref{gen_CS_action} to make the variational principle well defined(as the expression is for a closed manifold). This can be seen explicitly by varying $A$ in Chern-Simons action \eqref{gen_CS_action}, 

\beq
\delta I_{CS}[A] = {\rm Eqs.\;of\;motion} - \frac{k}{4\pi} \int_{\pd \cM} \bra A \wedge \delta A \ket
\eeq 
The boundary term does not vanish if $\cM$ is not closed and action will not have an extrema. 

A similar simple exercise can be done with Chern Simons action, if we choose a complex structure and look at AdS theory \cite{Campoleoni:1}, 

\begin{align}
\partial M &= R \times S^1  \\
x^\pm &= \frac{t}{l} \pm \phi.
\end{align} 
We then have

\begin{align}
\delta I_{CS}[A] &= - \frac{k}{4\pi} \int_{\partial M} \bra A \wedge \delta A \ket \\
&= - \frac{k}{4\pi} \int_{R \times S^1} dx^+ dx^- \bra A_+ \delta A_- - A_- \delta A_+ \ket
\end{align} 
apart from cases like $A_+ = 0$ or$A_- = 0$ at boundary (which can also be worthy of study), to define a general boundary value problem, we can add
\beq
I_{\partial \cM} [A] = \frac{k}{2\pi} \int_{\pd \cM} dx^+ dx^- \bra A_+ A_- \ket.
\eeq 
Depending on weather $A_+$ is held constant or $A_-$, final action will be
\beq
I_{CS}'[A] = I_{CS}[A] \pm I_{\partial \cM} [A]. 
\eeq 
If we keep $A_+$ fixed, the modified Chern Simons action $I_{CS}'[A] = I_{CS}[A] + I_{\partial \cM} [A]$ transforms under gauge transformation $A = \gi \d g + \gi \tilde{A} g$ as 
\beq
I_{CS}'[A] = I_{CS}'[\tilde{A}] + k I^+_{WZW} [g,\tilde{A}_+]. 
\eeq 
where $I^+_{WZW} [g,\tilde{A}_+]$ is the action of a chiral Wess-Zumino-Witten model on the boundary $\partial M$,

\bea
I^+_{WZW} [g,\tilde{A}_+] &=& \frac{1}{4\pi} \int_{\pd \cM}\bra (\gi \pd_+ g) (\gi \partial_-g) - 2 \gi A_- g \tilde{A}_+ \ket \br &+& \frac{1}{12\pi} \int_{\cM} \bra \gi \d g \ket^3 
\eea
This implies that number of physical degrees of freedom of Chern-Simons theory, ($2+1$)-D gravity in particular, depends on whether space time has a boundary. If there is boundary, gauge invariance is broken on it and these gauge degrees of freedom are dynamical, each broken symmetry adding infinite dimensional space of solutions that are not equivalent. We will address the issue of boundary conditions and variational principle in more detail in Chapter \ref{BC_gravity}.

\section{Chern-Simons formulation of Higher Spin Fields}\label{two}
In quantum field theory we have fields that carry irreducible representations of the Poincare group. Such fields are labelled by their mass and spin, the two Casimirs of the Poincare group. For instance, the gauge fields in Standard Model are massless fields th spin-$1$. Similarly, the matter content is made of massive fermionic fields which carry spin-$1/2$. Another familiar example is the gravitational field which is spin-$2$. In fact the representations of Poincare group can be labelled by arbitrary integer or half-integer spins. \\

While we can write free theories of higher spin fields, interacting theories suffer various no-go theorems \cite{Weinberg:1964ew,Grisaru:1977kk,Weinberg:1980kq,Grisaru:1976vm,Porrati:2008rm,Boulanger:2008tg,Porrati:2012rd} in flat space which rule out higher spin fields interacting with the Maxwell field or gravity. This is unacceptable if we believe that at low energies gravity couple to all fields universally. One of the ways to evade such no-go theorems is by constructing theories in backgrounds with a non-zero cosmological constant. On such backgrounds, it is possible to have a large higher spin gauge symmetry (which generates consistent interaction vertices) while avoiding no-go results such as that of Coleman and Mandula \cite{Coleman:1967ad}. Such an interacting theory was in fact proposed by Vasiliev \cite{Vasiliev:1990en} which consists of a set of non-linear equations of motion with an infinite tower of higher spin gauge fields interacting with each other (see \cite{Rahman:2015pzl} for a review). Vasiliev's approach rests on a description of higher spin fields that mimic the first order approach to gravity discussed before. On the other hand,Vasiliev theory still lacks a metric-like action description in 4 or higher dimensions.\\

In three dimensions however, things simplify to a great extent, owing to the topological nature of gravity and other massless higher spin fields. In fact, in 3 dimensions, it is possible to truncate the infinte tower of higher spin fields to any desired spin-N \cite{Aragone:1983sz}. Another simplification that is unique to $D=3$ is the fact that a spin-s field can be described by a pair of gauge fields $e_{\mu}^{\;\;a_1 a_2 \cdots a_{s-1}}$ and $\omega_{\mu}^{\;\;a_1 a_2 \cdots a_{s-1}}$ carrying the same index structure (see \cite{Vasiliev:1995dn} ). This is reminiscent of the fact that for pure gravity, one can work with dualized spin connection

\begin{equation}
\omega_{\mu}^{\;\; a} = \frac{1}{2}\e^{abc}\omega_{\mu\;bc}
\end{equation}
For a free field $\varphi_{\mu_1\cdots \mu_s}$, in $D=3$, the Fronsdal equation reduces to a flatness condition on the field. The above two observations are the starting point of Chern-Simons formulation of higher spin gauge fields. Based on the above requirements Blencowe \cite{Blencowe:1988gj} proposed an interacting theory of higher spins based on Chern-Simons theory.\\

The structrue of higher spin Chern-simons theory is as follows. In the CS formulation of gravity we have gauge potentials 

\begin{eqnarray}
a_{\mu}^{\;\;a} = \omega_{\mu}^a + \frac{e_{\mu}^{a}}{l},\quad\quad \tilde{a}_{\mu}^{\;\;a} = \omega_{\mu}^a - \frac{e_{\mu}^{a}}{l} 
\end{eqnarray}
formed out of linear combinations of vielbein and the spin-connection. We then define a similar linear combination of the higher spin gauge potentials

\begin{eqnarray}\label{hs_potential}
b_{\mu}^{\;\;a_1\cdots a_{s-1}} = \omega_{\mu}^{\;\;a_1\cdots a_{s-1}} + \frac{e_{\mu}^{\;\;a_1\cdots a_{s-1}}}{l},\quad \tilde{b}_{\mu}^{\;\;a_1\cdots a_{s-1}} = \omega_{\mu}^{\;\;a_1\cdots a_{s-1}} - \frac{e_{\mu}^{\;\;a_1\cdots a_{s-1}}}{l}. 
\end{eqnarray}
We then introduce higher spin generators $T_{\;\;a_1\cdots a_{s-1}}$ which can be now added to the $sl(2,\bR)$ generators to yield gauge potentials of the form

\begin{eqnarray}
A &=& \left(a_{\mu}^{\;\;a}T_a + b_{\mu}^{\;\;a_1\cdots a_{s-1}}T_{\;\;a_1\cdots a_{s-1}} \right)dx^{\mu} \\
\bA &=& \left(\tilde{a}_{\mu}^{\;\;a}\tilde{T}_a + \tilde{b}_{\mu}^{\;\;a_1\cdots a_{s-1}}\tilde{T}_{\;\;a_1\cdots a_{s-1}} \right)dx^{\mu}
\end{eqnarray}
Since $T_{\;\;a_1\cdots a_{s-1}}$ is contracted with the higher spin potential \eqref{hs_potential}, it should transform as an irreducible $sl(2,\bR)$ tensor and should be traceless \cite{Campoleoni:1}

\begin{eqnarray}\label{generator_traceless}
T^{a}_{\;\;a\; a_3\cdots a_{s-1}} = 0
\end{eqnarray} 
and must satisfy

\begin{equation}\label{generator_commutation}
	\left[T_b, T_{a_1\cdots a_{s-1}}\right] = \e^{c}_{\;\;b(a_1}T_{a_2\cdots a_{s-1})c}
\end{equation}
If the generators $T_a$ and $T_{\;\; a_1\cdots a_{s-1}}$ generate a Lie algebra with a non-degenerate bilinear form, we can form the action

\begin{eqnarray}
S = S_{CS}[A] - S_{CS}[\bA]
\end{eqnarray}
where $S_{CS}$ is the Chern-Simons action given by \eqref{gen_CS_action}. In \cite{Campoleoni:1}, it was shown that the linearized equations of motion which follows from the CS action is nothing but the Fronsdal equation. Here the metric and higher spin fields are related to their frame fields by the relation

\begin{eqnarray}
e_{\mu} = \frac{1}{2}(A_{\mu}-\bA_{\mu})
\end{eqnarray}
and
\begin{eqnarray}\label{metric}
g_{\mu\nu} = \frac{1}{2}\bra e_{\mu}e_{\nu} \ket, \quad \varphi_{\mu_1 \cdots \mu_s} = \frac{1}{s!}\bra e_{(\mu_1}\cdots e_{\mu_s)} \ket
\end{eqnarray}
where $\bra \cdots \ket$ refers to the bilinear form defined on the algebra.\\

So the problem of finding consistent theories of higher spin gauge fields reduces to finding a semisimple Lie algebra that satisfies \eqref{generator_traceless} and \eqref{generator_commutation}. The commutator of two generators of spin $s_1$ and $s_2$ in principle can produce a generator of spin different from either and therefore we may have to add other spins in the theory\footnote{In fact we can choose $	\left[T_{a_1\cdots a_{s-1}}, T_{b_1\cdots b_{s-1}}\right] =0$. This however leads to a free theory.}. In \cite{Bergshoeff:1989ns}, it was shown that $sl(n)$ generators admit all the above properties and thus provides a candidate theory.\\

Let us illustrate this with the simplest possible higher spin theory, a spin-3 field coupled to gravity. The gauge group that we consider is $SL(3,\bR)\times SL(3,\bR)$ with generators satisfying 

\begin{eqnarray}\label{sl(3)_lie_algebra}
\left[T_a, T_b \right] &=& \e_{abc}T^c \br
\left[T_a,T_{bc} \right] &=& \e^{d}_{\;\;a(b}T_{c)d}\\
\left[T_{ab},T_{cd}\right] &=& -(\eta_{a(c}\e_{d)be} + \eta_{b(c}\e_{d)ae})T^e \nonumber
\end{eqnarray}
This algebra is isomorphic to $sl(3)$ Lie algebra and also admits a non-degenerate bilinear form which is guaranteed by the existence of a quadratic Casimir

\begin{eqnarray}
C = T_a T^a + \frac{1}{2}T_{ab}T^{ab}
\end{eqnarray}
Now we can use the tracelessness condition on the generators and express \eqref{sl(3)_lie_algebra} in a more convenient basis \cite{Campoleoni:1}

\begin{eqnarray}
T_0 &=& \frac{1}{2}(L_1 + L_{-1}),\quad T_1 =  \frac{1}{2}(L_1 - L_{-1}), \quad T_2 = L_0, \br
T_{00} &=& \frac{1}{4}(W_2 + W_{-2} + 2W_0),\quad T_{01} = \frac{1}{4}(W_2 - W_{-2}), \\
T_{11} &=& \frac{1}{4}(W_2 + W_{-2} - 2W_0),\quad T_{02} = \frac{1}{2}(W_1 + W_{-1}), \br
T_{22} &=& W_0, \quad T_{12} = \frac{1}{2}(W_1 - W_{-1})
\end{eqnarray} 
where the new set of generators $\{L_m,W_n\}$ satisfy the commutation relation

\begin{eqnarray}
\left[L_m,L_n\right] &=& (m-n)L_{m+n}, \br
\left[L_m,W_n\right] &=& (2m-n)W_{m+n}, \\
\left[W_m,W_n\right] &=& -\frac{1}{3}(m-n)(2m^2 + 2n^2 -mn - 8)L_{m+n} \nonumber
\end{eqnarray}
With this basis, it is now easy to check that the tracelessness condition is trivially satisfied

\begin{equation}
-T_{00} + T_{11} + T_{22} = 0
\end{equation}
In the appendix we give a $3\times 3$ representation for the generators $\{L_m,W_n\}$.

It has to be noted that our arguments also hold for the case of positive as well as zero cosmological constant. In the case of $\Lambda > 0$, the appropriate gauge group is $sl(N,\bC)$, while for $\Lambda=0$, the higher spin generators are appropriate generalization of Poincare algebra \cite{Gonzalez:2013oaa,Afshar:2013vka}.

\pagebreak

\chapter{Higher Spin Cosmology}\label{HSC}

\section{Introduction}
We are now in a position to construct higher spin generalizations of de Sitter (dS) cosmology. de Sitter metric is the vacuum solution to Einstein's equation  with a positive cosmological constant. Although 3D gravity has no propagating degrees of freedom, the counterpart to the BTZ black hole quotients were constructed in \cite{Park:1998qk} (also see \cite{Balasubramanian:2001nb}). The fact that de Sitter is a cosmological spacetime with a spacelike boundary \cite{Spradlin:2001pw} has made the development of a consistent dS/CFT proposal much more confusing. Various interesting attempts were made in \cite{Witten:2001kn,Strominger:2001gp,Maldacena:2002vr}, but there seems to be a fundamental difficulty in realizing de Sitter space in $any$ kind of unitary quantum set up as a stable vacuum \cite{Polyakov:2007mm,Banks:2005bm,Fischler:2001yj,Dyson:2002nt,Dyson:2002pf,Krishnan:2006bq}. \\

The aim of this chapter is to construct cosmological solutions in higher spin dS$_{3}$ gravity. We work specifically with the case where the rank of the gauge group, $N=3$. The solutions we construct are the higher spin generalizations of dS$_{3}$ quotients such as Kerr-dS$_{3}$ and quotient cosmology \cite{Park:1998qk,Balasubramanian:2001nb,Krishnan:2013cra}
and should be thought of as the de Sitter counterparts of the spin-3 charged AdS$_{3}$ black hole solutions of \cite{Gutperle:2011kf,Ammon:2011nk}. It has been shown that big-bang type singularities contained in quotient cosmologies in the purely $SL(2)$ sector of this higher spin theory can be removed by performing a spin-3 gauge transformation \cite{Krishnan:2013cra}. But the problem of constructing spin-3 charged cosmologies was left open. In this work, we fill this gap and discuss the thermodynamics of their cosmological horizons. 

The plan of this chapter is as follows. We fix all notations and conventions for the map between the second order variables (metric and spin-3 field) and the gauge connection here. We also review and discuss the variational principle for asymptotically de Sitter like connections in the gauge theory formulation. In Sec. \ref{sec:hs dS cosmologies}, we first review pure gravity i.e. $SL(2,C)$ sector solutions, namely the Kerr de Sitter universe and the quotient cosmology. We do this both in the metric and gauge theory set-up. Then we construct higher spin extensions of these geometries by modifying their gauge connection and adding spin-3 charges in a manner consistent with the triviality of gauge connection holonomies along contractible cycles. These solutions are shown to contain cosmological horizons and, in the case of quotient cosmology, higher spin big bang/ big crunch like causal singularities. In the final section, Sec. \ref{sec: dS TD}, these holonomy conditions are shown to be necessary for the consistency of thermodynamics associated with cosmological horizons. These consistency conditions turn out to be identical to demanding integrability of a ``boundary CFT partition function''. Using a prescription of Maldacena \cite{Maldacena:2002vr}, we relate thermodynamics of our solutions to those of higher spin AdS$_{3}$ black holes. Our formulation gives the same results as the Gibbons-Hawking results when we restrict to spin-2 and work in the metric language.\\

In the static coordinates, the de Sitter metric is given by

\begin{eqnarray}
ds^2 = -\left(1-\frac{r^2}{l^2} \right)dt^2 +\frac{dr^2}{\left(1-\frac{r^2}{l^2} \right)} + r^2 d\Omega^2_{d-2}, \quad 0\leq r <l
\end{eqnarray}
in $d$ dimensions where $d\Omega^{2}_{d-2}$ is the metric on a unit $(d-2)$-sphere $S^{d-2}$. Another interesting set of coordinates is the so called ``flat slicing'' coordinates

\begin{eqnarray}
ds^2= -dt^2 + \rme^{2t/l}dy^2,\quad dy^2 = \delta_{\mu\nu}dy^{\mu}dy^{\nu}
\end{eqnarray}

As we discussed earlier, the connection to the Chern-Simons formulation of gravity is achieved by considering $sl(2,\bC)$ valued connections. In order to introduce higher spin fields, we can simply lift the gauge group from $sl(2,\bC)$ to $sl(N,\bC)$ which describes an interacting theory of fields of spin $1,2,\cdots N$ (see \cite{Ouyang:2011fs,Krishnan:2013cra}). We will study the simplest case of $sl(3,\bC)$ theory. One simply defines the higher spin (up to spin 3) theory i.e. an interacting theory of gravity and a spin 3-field, by the action \cite{Blencowe:1988gj} 

\begin{equation}
I_{CS}[A]=\frac{k}{4\pi }\int_{\cM}\bra AdA+\frac{2}{3}A^{3}\ket-\frac{k}{4\pi }\int_{\cM}\bra \bar{A}d\bar{A}+\frac{2}{3}\bar{A}^{3}\ket.\label{eq: CS action for dS_3}
\end{equation}
The general $sl(3,\bC)$ gauge potential can be expressed as

\begin{eqnarray}
A & = & \left(\omega_{\mu}^{A} + \frac{i}{l}e_{\mu}^{A}\right){\cal T}_{A}dx^{\mu} \\
\bA & = & \left(\omega_{\mu}^{A} - \frac{i}{l}e_{\mu}^{A}\right){\cal T}_{A}dx^{\mu}
\end{eqnarray}
where ${\cal T}_{A}$ now stands for both $L$ and $W$ generators. Then, the more familiar metric and spin-3 field can be extracted from the (imaginary parts of the basis coefficients) of the gauge field: 

\begin{equation}
g_{\mu\nu}=\frac{1}{2!}\bra e_{\mu}e_{\nu}\ket, \quad \phi_{\mu\nu\lambda}=\frac{1}{3!}\bra e_{(\mu}e_{\nu}e_{\lambda)}\ket,\label{eq: metric and spin-3 field definition}
\end{equation}
while the three-dimensional Newton's constant (in units of the dS radius, $l$) is given by the Chern-Simons level number, 

\begin{equation}
k = -i\frac{l}{4 G_{3}}.\label{eq: Relation between CS level and Newton's constant}
\end{equation}
We work in the prevalent general relativity convention where, $8G_{3}=1.$ Since the gauge group $SL(3,\bC)$ is non-compact the Chern-Simons level number is $not$ quantized.

\section{Variational Principle and Boundary Conditions}
Now lets consider the variation of the action (\ref{eq: CS action for dS_3}). Generically a variation has a bulk (volume) piece, proportional to the equation of motion and boundary pieces supported on temporal and spatial boundaries, 

\begin{eqnarray}
\delta I=\int d^{3}x\:(E.O.M)+\int d^{2}x\,\pi_{\mu}\delta A^{\mu}|_{t_{i}}^{t_{f}}+\int dtdx^{j}\;\left.\pi_{\mu}^{j}\delta A^{\mu}\right|_{x_{i,min}}^{x_{i,max}}
\end{eqnarray}
To have a good variational principle one has to ensure that these boundary pieces vanish (on-shell) by prescribing initial and final conditions and spatial boundary conditions. If the prescribed conditions do not lead to vanishing contribution for the boundary pieces of the variation, then one has to add supplementary boundary terms to the action to cancel these. One crucial point to be noted here in contrast with the AdS case is that the action (\ref{eq: CS action for dS_3}) already defines a good variational principle without any supplementary boundary terms. This is because asymptotically de Sitter spaces have $closed$ spatial sections and the only boundary contributions are from future infinity ($t_{f}\rightarrow\infty$) and at some time coordinate in the past ($t_{i}$ $={\rm const}$). As the variational principle is usually defined with vanishing variations at the initial and final times, 

\begin{eqnarray}
\delta A|_{t_{i},t_{f}}=0,
\end{eqnarray}
these boundary pieces vanish. However we shall $not$ demand the future data to be fixed (i.e. $\delta A|_{t_{f}\rightarrow\infty}\neq0$) and look to set up a variational principle by demanding instead the conjugate momentum vanishes 

\begin{eqnarray}
\pi_{\mu}|_{t_{f}\rightarrow\infty}\rightarrow0.
\end{eqnarray}
Such a variational principle will be made to appear natural in Sec.\ref{sec: dS TD} where the close parallel between de Sitter and Anti-de Sitter cases is brought out. This will often restrict us to a subclass of solutions which are specified by their future fall-off behaviors (which close under gauge transformations), 

\begin{eqnarray}
\lim_{t\rightarrow\infty}A_{\mu}\sim t^{\alpha_{\mu}}
\end{eqnarray}
for some real $bounded$ exponent $\alpha_{\mu}$. This is the analogue of non-normalizable fall-offs in AdS. These fall-off behaviors are
fixed by conducting the asymptotic (future/past) symmetry analysis in a manner closely parallel to the AdS$_{3}$ counterpart \cite{Campoleoni:1, Henneaux:2010xg} as was done in \cite{Ouyang:2011fs}. By demanding that the asymptotic symmetries of this larger theory still contain the Virasoro algebras already present in the $SL(2,C)$ case, it was found that the suitable fall-offs behaviors at future infinity for the $SL(3,C)$ gauge connections are 
\begin{equation}
A_{\bar{w}}=0,\qquad A_{\rho}=b^{-1}\partial_{\rho}b,\qquad A-A_{dS_{3}}\stackrel{\tau\rightarrow\infty}{\longrightarrow}\mathcal{O}(1).\label{eq: asymptotic future dS_3}
\end{equation}
Here $b$ is a gauge transformation $\in SL(3,C)$. However as we shall see in the next section, in order to construct gauge field configurations with non-vanishing higher spin charges, one has to violate the asymptotic fall-offs (\ref{eq: asymptotic future dS_3}) and hence one has to supplement the action (\ref{eq: CS action for dS_3}) with boundary terms. Again this is parallel to the situation for higher spin AdS black hole solutions \cite{Gutperle:2011kf} for which the boundary counter terms were worked out in \cite{Banados:2012ue, deBoer:2013gz}.

\section{Higher Spin de Sitter Cosmologies\label{sec:hs dS cosmologies}}

We are interested in constructing solutions of the $SL(3,C)$ gauge theory describing spacetimes of positive cosmological constant which have non-zero spin-3 charges in addition to the spin-2 charges i.e. energy and angular momentum. Since these are higher spin extensions of the pure gravity solutions or $SL(2)$ sector, let us first review the solutions of the $SL(2)$ sector obtained by taking quotients of pure three-dimensional de Sitter space \cite{Balasubramanian:2001nb}.

\subsection{Kerr-dS$_{3}$ universe}

The first class of $SL(2)$ quotients of pure de Sitter space is the so called Kerr - de Sitter universe ($KdS_{3}$). These are very similar to de Sitter space itself, in the sense that these solutions have two regions bounded by cosmological horizons, and have future and past infinite regions outside the cosmological horizons. However the topology of the past and future infinities of $KdS_{3}$ is that of a cylinder, $S^{1}\times R$, in contrast to the de Sitter space, for which they have topology of a sphere, $S^{2}$. 

In static Schwarzschild-like coordinates, the $KdS_{3}$ metric \cite{Park:1998qk,Balasubramanian:2001nb} reads like, 

\begin{eqnarray}
ds^{2}&=&-N^{2}(r)dt^{2} + N^{-2}(r)dr^{2} + r^{2}\left(d\phi+N_{\phi}dt\right)^{2},\\ N^{2}(r)&=& M - \frac{r^{2}}{l^{2}} + \frac{J^{2}}{4r^{2}},\quad N_{\phi}=-\frac{J}{2r^{2}}.\nonumber \label{eq: Kerr-dS_3 in Sch}
\end{eqnarray}
Introducing, the outer and inner radii 

\begin{equation}
r_{\pm}^{2}=Ml^{2}\left(\sqrt{1+\left(J/M\, l\right)^{2}}\pm1\right)/2,\label{eq: r_plus and r_minus}
\end{equation}
one can rewrite Eq.(\ref{eq: Kerr-dS_3 in Sch}) as 

\begin{equation}
ds^{2}=-\frac{\left(r^{2}+r_{-}^{2}\right)\left(r_{+}^{2}-r^{2}\right)}{r^{2}l^{2}}dt^{2}+\frac{r^{2}l^{2}}{\left(r^{2}+r_{-}^{2}\right)\left(r_{+}^{2}-r^{2}\right)}dr^{2}+r^{2}\left(d\phi+\frac{r_{+}r_{-}}{r^{2}}\frac{dt}{l}\right)^{2},r<r_{+}\label{eq:Kerr-dS_3 inside}
\end{equation}
and we note that this geometry has a horizon at $r=r_{+}$. This metric can be analytically continued across outside i.e. for $r>r_{+}$:

\begin{equation}
ds^{2}=-\frac{r^{2}l^{2}}{\left(r^{2}+r_{-}^{2}\right)\left(r^{2}-r_{+}^{2}\right)}dr^{2}+\frac{\left(r^{2}+r_{-}^{2}\right)\left(r^{2}-r_{+}^{2}\right)}{r^{2}l^{2}}dt^{2}+r^{2}\left(d\phi+\frac{r_{+}r_{-}}{r^{2}}\frac{dt}{l}\right)^{2}.\label{eq:Kerr-dS_3 outside}
\end{equation}
In this region $r$ is timelike while $t$ is spacelike. To make contact with the gauge theory we write down the $SL(2,C)$ connections for the two regions. Introducing, $\mathcal{N}^{2}(r)\equiv\frac{\left(r^{2}+r_{-}^{2}\right)\left(r^{2}-r_{+}^{2}\right)}{r^{2}l^{2}}=-N^{2}(r)$,
the gauge field expressions are,

\begin{eqnarray}
A^{0} & = & N(r)\left(d\phi+i\frac{dt}{l}\right),\quad A^{1}=\frac{lN_{\phi}-i}{N(r)}\frac{dr}{l}\quad,A^{2}=\left(rN_{\phi}+i\frac{r}{l}\right)\left(d\phi+i\frac{dt}{l}\right)\br &&{\rm for}\;\; r<r_{+} \br
A^{0} & = & -\frac{lN_{\phi}-i}{\mathcal{N}(r)}\frac{dr}{l},\quad A^{1}=\mathcal{N}(r)\left(d\phi+i\frac{dt}{l}\right),\quad A^{2}=\left(rN_{\phi}+i\frac{r}{l}\right)\left(d\phi+i\frac{dt}{l}\right); \br &&{\rm for}\;\; r>r_{+} .\label{eq: SL(2) gauge connection for KdS_3 in Sch.}
\end{eqnarray}
\\

In the exterior region, $r>r_{+}^{2}$, on can transform to Fefferman-Graham like coordinates ($\rho,w,\bar{w}$) defined by 

\begin{equation}
\rho=\ln\left(\frac{\sqrt{r^{2}-r_{+}^{2}}+\sqrt{r^{2}+r_{-}^{2}}}{2l}\right),w=\phi+it/l,\bar{w}=\phi-it/l\label{eq: FG radius in terms of Sch. radius}
\end{equation}
and obtain the form of the metric, 

\begin{equation}
ds^{2}=-l^{2}d\rho^{2}+\frac{1}{2}\left(Ldw^{2}+\bar{L}d\bar{w}^{2}\right)+\left(l^{2}e^{2\rho}+\frac{L\bar{L}}{4}e^{-2\rho}\right)dwd\bar{w},\label{eq: Kerr-dS_3 in FG}
\end{equation}
where, the zero modes, $L,\bar{L}$ are defined by, 

\begin{equation}
L+\bar{L}=Ml,\qquad L-\bar{L}=iJ.\label{eq: Zero Modes}
\end{equation}
Note that $\rho$ here is a time coordinate, eg. \cite{deBuyl:2013ega}. This coordinate system is better suited than the Schwarzschild one for conducting the asymptotic symmetry analysis of dS$_{3}$ and its identification with Euclidean Virasoro algebra and its charges \cite{Balasubramanian:2001nb}. As noted in \cite{Krishnan:2013cra}, the corresponding $SL(2,C)$ gauge field is, 

\begin{equation}
A=i\, T_{0}\, d\rho+\left[\left(e^{\rho}-\frac{L}{2l}e^{-\rho}\right)\, T_{1}+i\,\left(e^{\rho}+\frac{L}{2l}e^{-\rho}\right)T_{2}\right]dw\label{eq: SL(2,C) connection for general metric}
\end{equation}
One can obtain the above Kerr-dS$_{3}$ connection from a primitive connection, $a$ given by 

\begin{equation}
a=\left[\left(1-\frac{L}{2l}\right)T_{1}+i\left(1+\frac{L}{2l}\right)T_{2}\right]dw\label{eq: Kerr-dS3 curly}
\end{equation}
free of any $\rho$ dependence, by performing a single valued gauge transformation on $a$: 

\begin{eqnarray}
A=\mathcal{B}^{-1}a\:\mathcal{B}+\mathcal{B}^{-1}d\mathcal{B},
\end{eqnarray}
for 

\begin{equation}
\mathcal{B}=\exp\left(i\rho T_{0}\right)=\exp(\rho L_{0})\label{eq: Radial gauge transformation}
\end{equation}
(because $\mathcal{B}$ being a sole function of $\rho$ is single valued in the $\phi$ direction).

\subsection{Quotient Cosmology}

One can also construct de Sitter quotients containing (spinning) big bang/big crunch singularities \cite{Balasubramanian:2001nb} (also reviewed in \cite{Castro:2012gc} ). These quotients are locally given by the same exterior Kerr- de Sitter metric (\ref{eq:Kerr-dS_3 outside}). But since $t$ and $r$ switch their roles and become spacelike and timelike respectively, we are better off switching their roles in the metric itself, 

\begin{equation}
ds^{2}=-\frac{t^{2}l^{2}}{\left(t^{2}+r_{-}^{2}\right)\left(t^{2}-r_{+}^{2}\right)}dt^{2}+\frac{\left(t^{2}+r_{-}^{2}\right)\left(t^{2}-r_{+}^{2}\right)}{t^{2}l^{2}}dr^{2}+t^{2}\left(d\phi+\frac{r_{+}r_{-}}{t^{2}}dr\right)^{2}.\label{eq: Quotient Cosmology metric}
\end{equation}
The quotient cosmology arises when we compactify $r$ into a circle. With, $r$ and $\phi$ both being periodic the future and past infinity of this quotient cosmology have the topology of a torus, $S^{1}\times S^{1}$ as opposed to $R\times S^{1}$ for the case of the Kerr de Sitter universe. Also with a periodic $r$, this metric cannot be extended to $-r_{+}<t<r_{+}$ where $g_{rr}<0$ and one has closed timelike curves. Removing this region then leaves us with a big bang (big crunch) like solution for $t>r_{+}$ ($t<r_{+}$) with the $r$-$\phi$ torus degenerating to a circle \cite{Balasubramanian:2001nb}. This is an example of a causal structure singularity \cite{Banados:1992gq}, and these are the analogues of higher dimensional curvature singularities in 2+1 dimensions.  These singularities were shown to be removable via a higher spin gauge transformation when embedded into a spin-3 $SL(3)$ theory in \cite{Krishnan:2013cra}.\\

Since the quotient cosmology is metrically identical to the exterior regions of the Kerr de Sitter universe, the Fefferman-Graham gauge metric expression (\ref{eq: Kerr-dS_3 in FG}) and the gauge connection expressions (\ref{eq: SL(2,C) connection for general metric},\ref{eq: Kerr-dS3 curly},\ref{eq: Radial gauge transformation}) also carry over with the coordinate changes, 

\begin{equation}
\rho=\ln\left(\frac{\sqrt{t^{2}-r_{+}^{2}}+\sqrt{t^{2}+r_{-}^{2}}}{2l}\right),w=\phi+ir/l,\bar{w}=\phi-ir/l.\label{eq: FG time in terms of Schwarzschild time}
\end{equation}

\subsection{The Higher Spin cosmological gauge fields}

In the $SL(3)$ theory, the general primitive connection that satisfies asymptotic (future) de Sitter fall off conditions is 

\begin{equation}
a'=\left[\left(1-\frac{L}{2l}\right)T_{1}+i\left(1+\frac{L}{2l}\right)T_{2}+\frac{W}{8l}W_{-2}\right]dw\label{eq:Asymptotic de Sitter form}
\end{equation}
$L$ and $W$ can be functions of $z$, but we will consider the constant case in analogy with \cite{Gutperle:2011kf}. Explicit forms
for the generators can be found in \cite{Krishnan:2013cra}. We can transform from the primitive connection $a'$, to $A'$,
the fully $\rho$-dependent form by applying the transformation (\ref{eq: Radial gauge transformation})
 
\begin{equation}
A'=iT_{0}d\rho+\left[\left(e^{\rho}-\frac{L}{2l}e^{-\rho}\right)\, T_{1}+i\,\left(e^{\rho}+\frac{L}{2l}e^{-\rho}\right)T_{2}+\frac{W}{8l}e^{-2\rho}W_{-2}\right]dw,\label{eq: dS fall-off with spin-3}
\end{equation}
which we call the Fefferman-Graham gauge because it manifests the proper $\rho\rightarrow\infty$ fall-offs behaviors Eq. (\ref{eq: asymptotic future dS_3}) as derived in \cite{Campoleoni:1,Ouyang:2011fs}. \\

As the trace of $W_{-2}$ with any $SL(3)$ generator is zero, we find that metric obtained from $A',\bar{A}'$ is same as
(\ref{eq: Kerr-dS_3 in FG}). But the spin-3 field now attains a non-zero value. These non-vanishing components of spin-3 fields are given by

\begin{align}
\varphi_{www} & =-\frac{i}{8}l^{2}W,\nonumber \\
\varphi_{ww\bar{w}} & =-\frac{i}{24}l\bar{L}We^{-2\rho}+\frac{i}{24}l^{2}\overline{W},\nonumber \\
\varphi_{w\bar{w}\bar{w}} & =-\frac{i}{96}\bar{L}^{2}We^{-4\rho}+\frac{i}{24}l\bar{L}\overline{W}e^{-2\rho},\nonumber \\
\varphi_{\bar{w}\bar{w}\bar{w}} & =\frac{i}{32}\bar{L}^{2}\overline{W}e^{-4\rho}.\label{eq:Spin-3 field components}
\end{align}
In order to construct metrics (cosmologies) with \textit{non-vanishing} spin-3 charges (which will necessarily violate the asymptotically dS fall-offs), we propose the following ansatz for the primitive connection corresponding to a general spin-3 cosmology

\begin{eqnarray}
a'&=&\left[\left(1-\frac{L}{2l}\right)T_{1}+i\left(1+\frac{L}{2l}\right)T_{2}+\frac{W}{8l}W_{-2}\right]dw \br &+& \mu\left[W_{2}+w_{0}W_{0}+w_{-2}W_{-2}+t\;(T_{1}-iT_{2})\right]d\bar{w},\label{eq: primitive spin-3 cosmology}
\end{eqnarray}
where $\mu,w_{0},w_{-2},$ and $t$ are constants. The motivation for this comes from the fact that under a suitable set of analytically continuations of the charges and sign of the cosmological constant (which will be elaborated in the following sections) de Sitter higher spin cosmologies turn into the Euclidean sections of AdS higher spin black hole solutions of \cite{Gutperle:2011kf} (much like in the case of pure gravity or $SL(2)$ sector, Kerr-dS$_{3}$ solutions continue on to Euclidean $BTZ$ black holes). \\

Now, the connection (\ref{eq: primitive spin-3 cosmology}) is an off-shell object and contains too many independent parameters. Restricting on-shell, we find that the connection has to be of the form

\begin{eqnarray}
a'&=&\left[\left(1-\frac{L}{2l}\right)T_{1}+i\left(1+\frac{L}{2l}\right)T_{2}+\frac{W}{8l}W_{-2}\right]dw \br &+& \mu\left[W_{2}-\frac{L}{2l}W_{0}+\frac{L^{2}}{16l^{2}}W_{-2}+\frac{W}{l}(T_{1}-iT_{2})\right]d\bar{w}.\label{eq: On-shell primitive spin-3}
\end{eqnarray}
Now although the connection is on-shell, it is still arbitrary in the sense that one does $not$ know whether such solutions make a regular or singular contribution to the Hartle-Hawking wave-function (or equivalently, when continued to Euclidean AdS, the corresponding Gibbons-Hawking partition function, $Z_{ECFT}$). Just as in the second-order or metric formulation of gravity, this is guaranteed by demanding the regularity of the Euclidean section of the metric, in case of the first-order or connection formulation, it is fixed by demanding \emph{triviality} of the gauge-connection $A$ along contractible circle(s). The non-trivial topology of the connection is captured by the holonomy matrix or the Wilson loop operator along any contractible circle, $\mathcal{C}$

\begin{eqnarray}
\mathrm{Hol}_{\mathcal{C}}(A)\equiv\mathcal{B}^{-1}\exp\left[\oint_{\mbox{\ensuremath{\mathcal{C}}}}\; dx^{\mu}a_{\mu}\right]\;\mathcal{B}=e^{H_{\mathcal{C}}}.
\end{eqnarray}
The triviality of the connection is ensured when this holonomy matrix is identity. Equivalently, this means that the matrix, $H_{\mathcal{C}}$ has eigenvalues $(0,-2\pi i,2\pi i)$. In the case of Kerr-dS$_{3}$ and its spin-3 generalizations one has a contractible thermo-angular circle, 

\begin{eqnarray}
\left(t,\phi\right)\sim\left(t+i\beta,\phi+i\beta\Omega\right).
\end{eqnarray}
Equivalently, if one defines $\tau=\frac{\beta}{2\pi}\left(1-i\Omega l\right)$, this thermo-angular circle can be reexpressed as $\left(w,\bar{w}\right)\sim\left(w+2\pi\tau/l,\bar{w}+2\pi\bar{\tau}/l\right).$\\

The associated holonomy is, 

\begin{eqnarray}
\mbox{Hol}(a) & = & \mathcal{B}^{-1}\exp\left[\int_{0}^{i\beta}\; dt\, a_{t}+\int_{0}^{i\beta\Omega}\: d\phi\: a_{\phi}\right]\;\mathcal{B}\nonumber \\
& = & \mathcal{B}^{-1}\exp\left[\int_{0}^{i\beta}\; dt\, i\left(a_{w}-a_{\bar{w}}\right)/l+\int_{0}^{i\beta\Omega}\: d\phi\:\left(a_{w}+a_{\bar{w}}\right)\right]\;\mathcal{B}\nonumber \\
& = & \mathcal{B}^{-1}\exp\left[-\left(2\pi\tau\: a_{w}-2\pi\bar{\tau}\: a_{\bar{w}}\right)/l\right]\;\mathcal{B}\nonumber \\
& = & e^{W(a)}.\label{eq: Holonomy around the thermo-angular circle}
\end{eqnarray}
This means that the matrix $W(a)$ should have eigenvalues $(0,-2\pi i,2\pi i)$\footnote{Of course, one could consider the eigenvalues to be integer multiples of $\pm2\pi i$, in general, to get trivial holonomy. However, the choice of $(0,-2\pi i,2\pi i)$ is motivated by the fact that the time component of the vielbein should also be regular (finite and single-valued) at the horizon \cite{Banados:2016nkb}}. These two (complex) conditions entirely fix the charges $L,W$ in terms of the potentials $\beta,\mu$.\\

For generic gauge connections it is nontrivial to compute the holonomy matrix exactly. Since all we need are its eigenvalues,
we are perfectly fine to work with the matrix, $\exp(\tilde{w}(a)),\:\tilde{w}(a)\equiv-\left(2\pi\tau\, a_{w}-2\pi\bar{\tau}\: a_{\bar{w}}\right)$, instead, since it is related to the Holonomy matrix, $\exp(W(a))$ by a single-valued gauge transformation (similarity transformation), $\mathcal{B}$ and hence has the same eigenvalue spectrum. Demanding the eigenvalues of the $\tilde{w}_{z}$ be $(0,-2\pi i,2\pi i)$ implies

\begin{equation}
\det\left(\tilde{w}(a)\right)=0,\quad\text{Tr}\left[\tilde{w}(a)^{2}\right]=-8\pi^{2},\quad\text{Tr}\left[\tilde{w}(a)\right]=0.
\end{equation}
which translate to the following relations determining the charges $L,W$ in terms of the potentials $\tau,\mu$,
 
\begin{eqnarray}
27l^{2}\tau^{3}W-36l\tau^{2}\alpha L^{2}-54l\tau\alpha^{2}LW-54l\alpha^{3}W^{2}-8\alpha^{3}L^{3} & = & 0\nonumber \\
1-\frac{2\tau^{2}L}{l^{3}}-\frac{6\tau\alpha W}{l^{3}}+\frac{4}{3}\frac{\alpha^{2}L^{2}}{\tau^{2}l^{2}}\qquad\qquad\qquad & = & 0\label{eq:Holonomy constraints}
\end{eqnarray}
where $\alpha=\mu\bar{\tau}$. Since we are already familiar with the exact solution for purely spin-2 charges i.e. mass and angular momentum, we can now obtain a solution to the charges in the presence of spin-3 potentials in a perturbation series in the spin-3 chemical potential, $\mu$:

\begin{equation}
W=\sum_{i=1}^{\infty}a_{i}\mu^{i}\qquad L=\frac{l}{2\tau^{2}}+\sum_{j=1}^{\infty}b_{j}\mu^{j}
\end{equation}
Substituting this in Eq.(\ref{eq:Holonomy constraints}) and solving both equations to quadratic order $\mu$, we get following perturbative solution for $L$ and $W$ 

\begin{equation}
\frac{L}{l}=\frac{l^{2}}{2\tau^{2}}-\frac{5\alpha^{2}l^{4}}{6\tau^{6}}+\cdots,\qquad\frac{W}{l}=\frac{\alpha l^{4}}{3\tau^{5}}-\frac{20\alpha^{3}l^{6}}{27\tau^{9}}+\cdots\label{eq: Solutions to holonomy conditions}
\end{equation}
These solutions satisfy the \emph{integrability conditions} (as can be checked order by order), 

\begin{eqnarray}
\frac{\partial L}{\partial\alpha}=\frac{\partial W}{\partial\tau}, \label{names}
\end{eqnarray}
pointing out to the existence of a bulk (Euclidean) action, $I$

\begin{eqnarray}
\delta I\sim \delta \tau \  L+\delta \alpha \  W
\end{eqnarray}
with $L$ and $W$ being functions of $\tau,\alpha $. The basic reason why eqn (\ref{names}) arises is because we are demanding that there be an underlying partition function description for the system (the exponential of the action   being the semi-classical partition function). The integrability condition is the statement that the double derivatives of  the partition function (with respect to $\alpha$ and $\tau$) commute. A closely related discussion can be found in section (5.2) of \cite{Gutperle:2011kf}. The precise form of the action functional requires taking care of various subtleties, see \cite{David:2012iu}. We will make use of their results when we make comparisons with the AdS case.\\

For the (higher spin) AdS case, the integrability conditions were understood \cite{Gutperle:2011kf} to be integrability conditions of a \emph{boundary} CFT partition function, $Z_{CFT}$ dual to the higher spin AdS bulk theory (vide AdS/CFT). The on-shell bulk action, $I^{\mbox{on-shell}}$ is the saddle-point contribution to $Z_{CFT}$, corresponding to the classical higher
spin Black hole configuration. Similarly it will be shown later, the integrability conditions Eq. (\ref{eq:Holonomy constraints},\ref{eq: Solutions to holonomy conditions}) for the case of (higher spin) de Sitter connections apply to that a putative dual $Euclidean$ CFT partition function, $Z_{CFT*}$. It will also be shown that two partition functions ($Z_{CFT}$, $Z_{CFT*}$) are related by a suitable ``Wick-rotation'' of the Cherns-Simons level number (cosmological constant) and gauge theory charges (mass, spin, spin-3 charges).\\

Finally we apply the radial gauge transformation, (\ref{eq: Radial gauge transformation}) to obtain full radial dependence, 
\begin{eqnarray}
A' & =iT_{0}d\rho+\left[\left(e^{\rho}-e^{-\rho}\frac{L}{2l}\right)T_{1}+i\left(e^{\rho}+e^{-\rho}\frac{L}{2l}\right)T_{2}+e^{-2\rho}\;\frac{W}{8l}W_{-2}\right]dw\nonumber \\
& +\mu\left[e^{2\rho}\; W_{2}-\frac{L}{2l}W_{0}+e^{-2\rho}\;\frac{L^{2}}{16l^{2}}\; W_{-2}+e^{-\rho}\frac{W}{l}(T_{1}-iT_{2})\right]d\bar{w},\nonumber \\
\bar{A}' & =-iT_{0}d\rho+\left[\left(e^{\rho}-e^{-\rho}\frac{\bar{L}}{2l}\right)T_{1}-i\left(e^{\rho}+e^{-\rho}\frac{\bar{L}}{2l}\right)T_{2}+e^{-2\rho}\;\frac{\bar{W}}{8l}W_{-2}\right]d\bar{w}\nonumber \\
& +\bar{\mu}\left[e^{2\rho}\; W_{2}-\frac{L}{2l}W_{0}+e^{-2\rho}\;\frac{L^{2}}{16l^{2}}\; W_{-2}+e^{-\rho}\frac{\bar{W}}{l}(T_{1}+iT_{2})\right]dw.
\end{eqnarray}
\\
The corresponding metric expression is 
\begin{eqnarray}
ds^{2} & = & -l^{2}d\rho^{2} + \left( \frac{lL}{2} + \frac{lW\bar{\mu}}{2} - \frac{1}{2}e^{-2\rho}L\overline{W}\bar{\mu} - \frac{\bar{L}^{2}\bar{\mu}^{2}}{3}\right)dw^{2} \br &+& \left( \frac{l\bar{L}}{2} + \frac{l\bar{W}\mu}{2} - \frac{1}{2}e^{-2\rho} \bar{L}W\mu - \frac{L^{2}\mu^{2}}{3}\right)d\bar{w}^{2} \\
& + & \left( \frac{1}{2}e^{2\rho}l^{2} - \frac{3lW\mu}{4} - \frac{3l\overline{W}\bar{\mu}}{4} + \frac{L^{2}\mu\bar{\mu}}{8} + \frac{L\bar{L}\mu\bar{\mu}}{12} + \frac{\bar{L}^{2}\mu\bar{\mu}}{8} \right.\br  && \left. \hspace{3cm} + \frac{1}{8}e^{-2\rho}\left(L\bar{L} + 4W\overline{W}\mu\bar{\mu}\right)\right)dwd\bar{w} \nonumber\label{eq: hs KdS_3 metric}
\end{eqnarray}
while the expressions for the non-vanishing spin-3 field components are, 

\begin{eqnarray}
\psi_{\rho\rho w} & = & \frac{1}{18}il^{2}\bar{L}\bar{\mu},\qquad\qquad\psi_{\rho\rho\bar{w}}=-\frac{1}{18}il^{2}L\mu,\nonumber \\
\psi_{www} & = & -\frac{1}{8}il^{2}W + \frac{1}{16}ilL^{2}\bar{\mu} + \frac{1}{24}ilL\bar{L}\bar{\mu} + \frac{1}{16}il\bar{L}^{2}\bar{\mu} - \frac{1}{12}il\bar{L}W\bar{\mu}^{2} + \frac{1}{27}i\bar{L}^{3}\bar{\mu}^{3}\br & + & \frac{1}{4}ie^{-2\rho}\left( lW\overline{W}\bar{\mu} - \frac{1}{6}L\bar{L}\overline{W}\bar{\mu} - \frac{1}{2}\bar{L}^{2}\overline{W}\bar{\mu}^{2}\right) - \frac{1}{8}ie^{-4\rho}W\overline{W}^{2}\text{\ensuremath{\bar{\mu}}}^{2} + \frac{ie^{-4\rho}\bar{L}^{2}\overline{W}^{2}\bar{\mu}^{3}}{16l},\br
\psi_{\bar{w}\bar{w}\bar{w}} & = & -\frac{1}{4}ie^{4\rho}l^{3}\mu+\frac{1}{2}ie^{2\rho}l^{2}W\mu^{2}-\frac{1}{24}ilL\bar{L}\mu+\frac{1}{12}ilL\overline{W}\mu^{2}-\frac{1}{27}iL^{3}\mu^{3}-\frac{1}{4}ilW^{2}\mu \br
& + & \frac{1}{24}ie^{-2\rho}L\bar{L}W\mu^{2} + \frac{1}{32}ie^{-4\rho}\left(\bar{L}^{2}\overline{W} - \frac{L^{2}\bar{L}^{2}\mu}{2l}\right), \\
\psi_{ww\bar{w}} & = & \frac{1}{12}ie^{2\rho}\left(l^{2}L\bar{\mu}+\frac{1}{3}il^{2}\bar{L}\bar{\mu}\right) + \frac{1}{24}il^{2}\overline{W} - \frac{1}{18}ilL^{2}\mu - \frac{1}{18}ilLW\mu\bar{\mu} - \frac{1}{72}iL^{2}\bar{L}\mu\bar{\mu}^{2} \br &-& \frac{1}{108}iL\bar{L}^{2}\mu\bar{\mu}^{2} - \frac{1}{72}i\bar{L}^{3}\mu\bar{\mu}^{2} + \frac{ie^{-2\rho}}{12}\left(\frac{1}{12}L\bar{L}^{2}\bar{\mu}+\frac{1}{4}\bar{L}^{3}\bar{\mu}-l\overline{W}^{2}\bar{\mu}-\frac{1}{2}l\bar{L}W \right. \br &+& \left. \frac{2}{3}L^{2}\overline{W}\mu\bar{\mu} + \frac{1}{36}\bar{L}W\overline{W}\mu\bar{\mu}^{2}\right)-\frac{ie^{-4\rho}}{24}\left(\frac{\bar{L}^{3}\overline{W}\bar{\mu}^{2}}{2l}-\bar{L}W\overline{W}\bar{\mu}-\overline{W}^{3}\bar{\mu}^{2}+\frac{L^{2}\overline{W}^{2}\mu\bar{\mu}^{2}}{2l}\right),\nonumber \\
\psi_{w\bar{w}\bar{w}} & = & \frac{1}{12}ie^{4\rho}l^{3}\bar{\mu} - ie^{2\rho}l^{2}\left(\frac{1}{9}L\mu + \frac{1}{6}W\mu\bar{\mu}\right) + \frac{1}{12}i\left(lLW\mu^{2} + \frac{1}{6}l\bar{L}^{2}\bar{\mu} - \frac{1}{3}l\bar{L}\overline{W}\mu\bar{\mu} \right. \br &+& \left.  \frac{1}{6}L^{3}\mu^{2}\bar{\mu} + \frac{1}{9}L^{2}\bar{L}\mu^{2}\bar{\mu} + \frac{1}{6}L\bar{L}^{2}\mu^{2}\bar{\mu} + lW^{2}\mu^{2}\bar{\mu}\right) + \frac{1}{12}ie^{-2\rho}\left( l\bar{L}\overline{W} - \frac{1}{3}L^{2}\bar{L}\mu \right. \br &-& \left. \frac{1}{6}\bar{L}^{2}W\mu\bar{\mu} - \frac{1}{3}LW\overline{W}\mu^{2}\bar{\mu}\right) - \frac{1}{24}ie^{-4\rho}\left( \bar{L}^{2}W-\frac{\bar{L}^{4}\bar{\mu}}{8l} + \bar{L}\overline{W}^{2}\bar{\mu} - \frac{L^{2}\bar{L}\overline{W}\mu\bar{\mu}}{2l}\right).\nonumber \label{eq: Spin-3 field for hs KdS_3}
\end{eqnarray}

\subsection{Schwarzschild Gauge}

The Fefferman-Graham (FG) gauge expressions only cover a part of the spacetime outside the horizon. In this section, we describe solution of the gauge connection in Schwarzschild gauge. For simplicity, we will consider the purely non-rotating case from
now on, 

\begin{eqnarray}
L=\bar{L},\qquad\bar{W}=-W\qquad\text{and}\qquad\bar{\mu}=-\mu.
\end{eqnarray}
The metric in FG gauge is then, 

\begin{eqnarray}
g_{\rho\rho} & = & -l^{2},\nonumber \\
g_{tt} & = & \left(e^{\rho}-\frac{L+2W\mu}{2l}e^{-\rho}\right)^{2},\nonumber \\
g_{\phi\phi} & = & l^{2}\left(e^{\rho}+\frac{L-2W\mu}{2l}e^{-\rho}\right)^{2}+\frac{4L^{2}|\mu|^{2}}{3}-2l\, W\mu.
\end{eqnarray}
We observe that there is a horizon i.e. $g_{tt}$ vanishes, at 

\begin{equation}
\rho_{+}=\frac{1}{2}\ln\left[\frac{L+2W\mu}{2l}\right].\label{sl3hr}
\end{equation}
Now, we can introduce the Schwarzschild radial coordinate $r$ (motivated from the definition of the Schwarzschild radial coordinate for the pure $SL(2)$ case), 

\begin{eqnarray}
\rho=\ln\left[\frac{r+\sqrt{r^{2}-r_{+}^{2}}}{2l}\right],
\end{eqnarray}
where $r_{+}^{2}$ is 

\begin{equation}
r_{+}^{2}=2l(L+2W\mu).
\end{equation}
In the limit $\mu=0$, the above equation reduces to the pure $SL(2)$ case, (\ref{eq: FG radius in terms of Sch. radius}). In the Schwarzschild like gauge the metric is given by, 

\begin{eqnarray}
ds^{2} & = & -\frac{l^{2}}{r^{2}-r_{+}^{2}}dr^{2}+\frac{2(L+2W\mu)}{l\, r_{+}^{2}}\left(r^{2}-r_{+}^{2}\right)dt^{2} + \\
&  & \left[\left(\frac{r\, L}{L+2W\mu}+\frac{2l\mu W\sqrt{r^{2}-r_{+}^{2}}}{L+2W\mu}\right)^{2}+\frac{4L^{2}|\mu|^{2}}{3}-2l\, W\mu\right]d\phi^{2} \nonumber \label{eq:Sch. gauge higher spin}
\end{eqnarray}
We also note that $g_{\phi\phi}>0$ as it is a sum of manifestly positive quantities ($W$ and $\mu$ are imaginary quantities with
same sign vide (\ref{eq: Solutions to holonomy conditions})) and there are no closed timelike curves in the $\phi$ direction.\\

Now that we have the metric expressions for the higher spin versions of the Kerr de Sitter universe in Schwarzschild gauge (\ref{eq:Sch. gauge higher spin}) outside the cosmological horizon, one can now write down metric for higher spin generalizations of the quotient cosmologies (\ref{eq: Quotient Cosmology metric}) by simply swapping $r$ and $t$. 

\begin{eqnarray}
ds^{2}&=&-\frac{l^{2}}{t^{2}-r_{+}^{2}}dt^{2}+\frac{2(L+2W\mu)}{l\, r_{+}^{2}}\left(t^{2}-r_{+}^{2}\right)dr^{2} \\ &+& \left[\left(\frac{L\, t+2\mu W\sqrt{t^{2}-r_{+}^{2}}}{L+2W\mu}\right)^{2}+\frac{4L^{2}|\mu|^{2}}{3}-2l\, W\mu\right]d\phi^{2}. \nonumber \label{eq: hs quotient cosmology}
\end{eqnarray}
Just as in the case for the $SL(2)$ quotient cosmology, $r$ is now compactified into a circle and this metric cannot be continued
inside the horizon, $r_{+}$. As a result this it contains big bang/big crunch like singularities at $t=\pm r_{+}$ when the
$r$-circle degenerates to a point, exactly like its $SL(2)$ cousin. It will be interesting to consider the resolution of these singularities along the lines of \cite{Krishnan:2013cra}, but we will not pursue it here.

\section{Thermodynamics of asymptotically de Sitter connections\label{sec: dS TD}}

The aim of this section to derive a consistent thermodynamics for asymptotically dS$_{3}$ spin-2 connections in the Chern-Simons language. (See \cite{Perez:2013xi} for an explicit expression for the entropy in metric-like variables.) In a metric (second order) formalism of gravity, more precisely spin-2 gravity, thermodynamics of spacetimes containing horizons of any kind
is provided by the Gibbons-Hawking generalization \cite{Gibbons:1977mu} of the black hole thermodynamics of Bardeen, Carter and Hawking \cite{Bardeen:1973gs}.  However, as we shall see, in the Chern-Simons or first order set-up, a consistent thermodynamics is obtained extremely efficiently by first mapping de Sitter solutions to Euclidean AdS (EAdS) solutions and then demanding integrability conditions on free energy (equivalently partition function) of a putative Euclidean CFT located on the future infinity of the asymptotic de Sitter (same as conformal boundary of the analytically continued EAdS solution). Maldacena \cite{Maldacena:2002vr} notes that the conformal patch of dS \footnote{ie., the upper quadrant in the dS Penrose diagram containing the infinite future at $\eta=0$ and bounded by the horizon at $\eta=1$. No light rays from the infinite past can reach this region.},

\begin{eqnarray}
ds^{2}=\frac{-d\eta^{2}+d{\bf x}_{d}^{2}}{\eta^{2}/l^{2}}
\end{eqnarray}
goes over to the Poincare patch of the EAdS, under, $l^{2}\rightarrow-l^{2}$ and, $\eta^{2}\rightarrow-z^{2}$, 

\begin{eqnarray}
ds^{2}=\frac{dz^{2}+d{\bf x}_{d}^{2}}{z^{2}/l^{2}},
\end{eqnarray}
and then he proposes that for any asymptotic (in time) de Sitter space, 

\begin{eqnarray}
\Psi_{Hartle-Hawking}=Z_{CFT^{*}},
\end{eqnarray}
since for EAdS one has the celebrated AdS-CFT conjecture $Z_{ESUGRA}=Z_{CFT}.$ Under the identifications, the Euclidean path-integral in AdS becomes the Hartle-Hawking wave function of dS. Next we construct a similar map between the $exterior$ regions of Kerr deSitter and and Euclidean BTZ black hole and then generalize to the higher spin case where the bulk action would be a first order action instead of second order (metric) action.

\subsection{``Wick-rotation'' from Kerr de Sitter to $EBTZ$}

We simply write down these identifications for the FG gauge, 

\begin{eqnarray}
\rho_{dS} & \rightarrow & \rho_{EAdS}+i\:\frac{\pi}{2},\nonumber \\
t_{ds} & \rightarrow & i\; t_{EAdS}\\
l_{dS} & \rightarrow & i\: l_{EAdS},\nonumber \\
L_{dS},\bar{L}_{dS} & \rightarrow & -iL_{EAdS},-i\bar{L}_{EAdS}\nonumber \\
M_{dS},J_{dS} & \rightarrow & -M_{AdS},\;-J_{AdS}.\label{eq: dS to AdS identifications}
\end{eqnarray}
Under these identifications, the KdS$_{3}$ metric Eq. (\ref{eq:Kerr-dS_3 outside}) goes over to 

\begin{equation}
ds^{2}\rightarrow d\tilde{s}^{2}=l^{2}d\rho^{2}+\frac{l}{2}\left(Ldw^{2}+\bar{L}d\bar{w}^{2}\right)+\left(l^{2}e^{2\rho}+\frac{L\bar{L}}{4}e^{-2\rho}\right)dwd\bar{w}\label{eq:Wick-rotated dS in FG}
\end{equation}
but now with, 

\begin{eqnarray}
L=\frac{Ml+J}{2},\bar{L}=\frac{Ml-J}{2}.
\end{eqnarray}
This is an evidently an Euclidean metric. To determine whether this is the Euclidean $BTZ$ metric ($EBTZ$), we write the $EBTZ$ metric expressions directly from Euclideanizing the Lorentzian $BTZ$, 

\begin{eqnarray}
ds^{2}&=&\frac{l}{2}\left(L^{+}\: dw^{+2}+L^{-}\: dw^{-2}\right)+\left(l^{2}e^{2\rho}+\frac{L^{+}L^{-}}{4}e^{-2\rho}\right)dw^{+}\: dw^{-}+l^{2}d\rho^{2},\br w^{\pm}&=&\phi\pm\frac{t}{l},\label{eq: BTZ metric in FG}
\end{eqnarray}
where, the ``zero modes'' $L^{+},L^{-}$ are defined in terms of the mass and the spin by, 

\begin{equation}
L^{+}=\frac{Ml+J}{2},\qquad L^{-}=\frac{Ml-J}{2}.\label{eq: BTZ zero modes}
\end{equation}
Upon a replacing $t\rightarrow it_{E}$, we obtain the $EBTZ$ metric,

\begin{eqnarray}
ds^{2}&=&\frac{l}{2}\left(L^{+} dw^{2} + L^{-} d\bar{w}^{2}\right) + \left(l^{2}e^{2\rho} + \frac{L^{+}L^{-}}{4}e^{-2\rho}\right)dw d\bar{w} +  l^{2}d\rho^{2},\br w &=& \phi+\frac{it_{E}}{l},\bar{w}=\phi-\frac{it_{E}}{l},\label{eq:EBTZ in FG}
\end{eqnarray}
Clearly, this is identical to the Wick-rotated $KdS_{3}$ metric eq. (\ref{eq:Wick-rotated dS in FG}).\\

In Schwarzschild-like coordinates, the Kerr-dS$_{3}$ metric \ref{eq: Kerr-dS_3 in Sch}, on using the identifications, \ref{eq: dS to AdS identifications}, becomes, 

\begin{eqnarray}
d\tilde{s}^{2}&=&N^{2}dt^{2}+N^{-2}dr^{2}+r^{2}\left(i\, N^{\phi}dt+d\phi\right)^{2},\br N^{2}&=&-M+\frac{r^{2}}{l^{2}}+\frac{J^{2}}{4r^{2}},\qquad N^{\phi}=\frac{J}{2r^{2}}.\label{eq:wick-rotated dS}
\end{eqnarray}
The Lorentzian exterior $BTZ$ metric Eq. (\ref{eq: BTZ metric in FG}) reads, 

\begin{eqnarray}
ds^{2}&=&-N^{2}dt^{2}+N^{-2}dr^{2}+r^{2}\left(N^{\phi}dt+d\phi\right)^{2},\br N^{2}&=&-M+\frac{r^{2}}{l^{2}}+\frac{J^{2}}{4r^{2}},\qquad N^{\phi}=\frac{J}{2r^{2}}.\label{eq:BTZ in Sch}
\end{eqnarray}
which upon Euclideanizing i.e. $t\rightarrow it_{E},$ 

\begin{eqnarray}
ds^{2}&=&N_{E}^{2}dt_{E}^{2}+dr^{2}/N_{E}^{2}+r^{2}\left(i\: N_{E}^{\phi}dt_{E}+d\phi\right)^{2},\br N_{E}^{2}&=&-M+\frac{r^{2}}{l^{2}}+\frac{J^{2}}{4r^{2}},N_{E}^{\phi}=\frac{J}{2r^{2}}.\label{eq:EBTZ in Sch}
\end{eqnarray}
One can write a metric expression in terms of outer and inner horizons for the $BTZ$ along the lines of \ref{eq:Kerr-dS_3 outside}, 

\begin{eqnarray}
ds^{2}&=& - \frac{\left(r^{2}-r_{+}^{2}\right)\left(r^{2}-r_{-}^{2}\right)}{r^{2}l^{2}}dt^{2} + \frac{r^{2}l^{2}}{\left(r^{2}-r_{+}^{2}\right)\left(r^{2}-r_{-}^{2}\right)}dr^{2}\\ && \hspace{5cm} +r^{2}\left(d\phi+\frac{r_{+}r_{-}}{r^{2}}\frac{dt}{l}\right)^{2}, r>r_{+} \nonumber \label{eq: BTZ outside in Sch}
\end{eqnarray}
with, 

\begin{eqnarray}
r_{\pm}^{2}=\frac{Ml^{2}}{2}\left(1\pm\sqrt{1-\left(\frac{J}{Ml}\right)^{2}}\right).
\end{eqnarray}
So the identifications are, 

\begin{equation}
r_{+}\rightarrow r_{+},\left(r_{-}\right)_{KdS_{3}}\rightarrow-i\,\left(r_{-}\right)_{BTZ}.\label{eq:r_plus and r_minus identifications}
\end{equation}
For $KdS_{3}$ note that in terms of $L,\bar{L}$, 

\begin{eqnarray}
r_{+}=\frac{\sqrt{Ll}+\sqrt{\bar{L}l}}{\sqrt{2}},r_{-}=\frac{\sqrt{Ll}+\sqrt{\bar{L}l}}{\sqrt{2}}
\end{eqnarray}

\begin{eqnarray}
r_{+}^{2}+r_{-}^{2}=2\sqrt{L\bar{L}l^{2}}.
\end{eqnarray}
So, the temperature inverse of $KdS_{3}$ in terms of $L,\bar{L}$,

\begin{equation}
\frac{\beta}{2\pi}=\frac{l^{2}}{2}\left(\frac{1}{\sqrt{2Ll}}+\frac{1}{\sqrt{2\bar{L}l}}\right).\label{eq: Inverse temperature as a function of L and Lbar}
\end{equation}
For non-rotating $KdS_{3}$, $\tau=\beta/2\pi$ and we have 

\begin{eqnarray}
\frac{L}{l}=\frac{l^{2}}{2\tau^{2}}.
\end{eqnarray}
Again, this metric (\ref{eq:EBTZ in Sch}) is exactly that of the Wick-rotated dS metric in Schwarzschild coordinates, Eq. (\ref{eq:wick-rotated dS}).

\subsection{dS-AdS ``Wick rotation'' at work: Equivalence of thermodynamics in the metric formulation}

In order to further solidify our heuristic identifications, we show that under these identifications the Gibbons-Hawking thermodynamics \cite{Gibbons:1977mu}, including the temperature and entropy of the Kerr-dS$_{3}$ solution, maps onto to those the ``Wick-rotated'' EBTZ solutions.

\begin{enumerate}
	\item The entropies for either geometry are the same since entropy of either cosmological or Black-hole horizons in the Gibbons-Hawking framework is given by 
	\begin{equation}
	S=\frac{1}{4G}(\mbox{Horizon Area})=2\left(2\pi r_{+}\right)=4\pi r_{+}.\label{eq: KdS_3 or BTZ entropy}
	\end{equation}
	This is borne out by our heuristic identifications, since $r_{+}\rightarrow r_{+}$.
	\item The temperature of KdS$_{3}$ is given by Gibbons-Hawking thermodynamics by the conical singularity trick, 
	\begin{equation}
	T_{KdS_{3}}=\frac{r_{+}^{2}+r_{-}^{2}}{2\pi l^{2}r_{+}}.\label{eq: Hawking temperature of Kerr-dS_3}
	\end{equation}
	Using the identification Eq. (\ref{eq:r_plus and r_minus identifications}) and the additional identification $T_{ds}\rightarrow-T_{AdS}$\footnote{This temperature sign flip is a direct result of the flip in the sign of mass parameter or ``internal energy'' $M$ in identification \ref{eq: dS to AdS identifications}. The conjugacy relation 
	\begin{eqnarray}
	T^{-1}=\frac{\partial S}{\partial M},
	\end{eqnarray}
	directs you that one needs to perform, 
	\begin{eqnarray}
	T_{dS}\rightarrow-T_{AdS}
	\end{eqnarray}
	in consonance with 
	\begin{eqnarray}
	M_{dS}\rightarrow-M_{AdS}.
	\end{eqnarray}
	} this temperature continues to the Hawking temperature of the corresponding
	BTZ black hole! 
	\begin{equation}
	T_{BTZ}=\frac{r_{+}^{2}-r_{-}^{2}}{2\pi l^{2}r_{+}}.\label{eq: Hawking Temperature of BTZ}
	\end{equation}
	
	\item The chemical potential conjugate to angular momentum is, 
	\begin{equation}
	\Omega_{KdS_{3}}=-T\frac{\partial S}{\partial J}=-\frac{r_{-}}{r_{+}l}\label{eq: KdS_3 angular chemical potential}
	\end{equation}
	Again under the identifications, we obtain the expected behavior $\Omega_{dS}\rightarrow\Omega_{AdS}$ since $J_{dS}\rightarrow-J_{AdS}$ \footnote{Since, going over from dS to AdS implies the replacements $S\rightarrow S$, $J\rightarrow-J$ and $\beta\rightarrow-\beta$, we must have $\Omega\rightarrow\Omega$ in order to reproduce the correct thermodynamic relation, 
	\begin{eqnarray}
	\frac{\partial S}{\partial J}=-\Omega\beta
	\end{eqnarray}
	}. We note that, $\Omega_{BTZ}=\frac{r_{-}}{r_{+}l}.$ Parenthetically, we note that when we move to Euclidean BTZ, we need to define, $J_{EAdS}=-iJ_{AdS}$ and consequently the new conjugate $\Omega_{EAdS}=i\Omega_{AdS}$, so that respective identifications are, $J_{dS}\rightarrow-iJ_{EAdS}$ and $\Omega_{KdS_{3}}\rightarrow-i\Omega_{EBTZ}$. 
\end{enumerate}

Since under the identifications, one can successfully map any dS thermodynamic quantities like entropy, internal energy, angular charges and their respective conjugates to AdS quantities, the laws of thermodynamics will continue as well. When higher spin charges are added, we will demand a similar statement to hold with higher spin charges and chemical potential added to the
thermodynamical relations.

\subsection{Thermodynamics in the Chern-Simons formulation}

So far everything we discussed was in the $SL(2,C)$ sector of the theory with just metric or spin-2 fields turned on, but we extend
this analogy to the full $SL(3,C)$ sector i.e. when both metric and spin-3 field are present. In that case though we do not know the generalization of the Gibbons-Hawking thermodynamics \cite{Bardeen:1973gs}. However, taking dS/CFT as a principle, we can propose that the thermodynamics of a dS-connection is identical to that of a suitably continued Euclidean AdS-connection i.e. a higher spin AdS black hole \cite{Gutperle:2011kf}. The thermodynamics of $SL(3,C)$ valued Euclidean AdS$_{3}$-connections for higher spin black-holes (connected to BTZ, i.e. the so called ``BTZ'' branch) has been shown to be dictated by the integrability conditions of the free energy of a dual CFT \cite{Gutperle:2011kf}. These conditions which can be cast in a gauge-invariant form by the $holonomy$ conditions \cite{Gutperle:2011kf}. Under the correct identifications of charges and potentials, the integrability or holonomy conditions of a dS connection should continue to those of an AdS connection. Or turning this fact around, we expect the charges we obtain functions of the potentials $\mu$ and $T$ on solving the integrability conditions on the dS side, (\ref{eq: Solutions to holonomy conditions}) to reproduce the respective solution of AdS integrability conditions \emph{i.e.} AdS charges as a function of AdS potentials \cite{David:2012iu} upon making the dS-to-AdS identifications. For AdS, the solution to the holonomy conditions is,

\begin{equation}
\frac{L^{+}}{l}=\frac{l^{2}}{2\tau^{2}}+\frac{10}{3}\frac{\alpha^{2}l^{4}}{\tau^{6}}+\dots,\qquad\frac{W^{+}}{l}=-\frac{4}{3}\frac{\alpha l^{4}}{\tau^{5}}+\dots.\label{eq: Equation of State for AdS}
\end{equation}
We have the identifications for the spin-3 charges when going from $dS$ to $AdS$, 

\begin{eqnarray}
\alpha_{AdS}=\frac{\alpha_{dS}}{2},W_{AdS}^{+}=-2iW_{dS},
\end{eqnarray}
or, 
\begin{equation}
\mu_{AdS}=-\frac{\mu_{dS}}{2},W_{AdS}^{+}=-2iW_{dS}.\label{eq:Identifications for the spin-3 charge}
\end{equation}

To compute, $Z_{CFT}$ from the bulk gauge theory, we make use of the saddle point approximation, 

\begin{eqnarray}
Z_{CFT}=Z_{(E)SUGRA}=e^{I_{E}^{On-shell}}
\end{eqnarray}
where, $I_{E}$ is the Euclidean bulk action, defined in terms of the original action, $I$ by 

\begin{eqnarray}
I_{E}[F({\bf x},t)]=iI[F({\bf {\bf x}},it_{E})].
\end{eqnarray}
The Chern-Simons action without any supplementary boundary terms,

\begin{eqnarray}
I_{CS}=\frac{k}{4\pi}\int\bra AdA+\frac{2}{3}A^{3}\ket,
\end{eqnarray}
is the right action for the $SL(2)$ sector. On-shell this becomes \cite{Banados:1998ta}, 

\begin{equation}
I_{CS}[A]=-\frac{k}{4\pi}\int dt\, d\mbox{\ensuremath{\phi}}\; \bra A_{t}A_{\phi}\ket.\label{eq: On shell CS action}
\end{equation}
For $SL(3)$ sector one needs to add new boundary terms as formulated in \cite{Banados:2012ue,deBoer:2013gz}. But it is easy to see that a similar map as we are presenting below will also hold for the boundary terms, so in the following, we will illustrate it only for the bulk terms. Using, (5.1) of \cite{Gutperle:2011kf}

\begin{equation}
I_{\mbox{on-shell}}^{\mbox{EAdS}}=-\frac{2\beta L}{l}+\frac{16\beta\mu^{2}L^{2}}{3l^{2}}.\label{eq:On-shell action for AdS}
\end{equation}
The higher spin de Sitter on-shell action turns out to be\footnote{To get this on-shell action, we perform integration over $t$-circle $(0,i\beta)$ and over $\phi$-circle $(0,2\pi)$ 
\begin{eqnarray}
\tilde{I}_{\mbox{on-shell }}^{\mbox{dS}}=i\left(-\frac{k}{4\pi\epsilon_{R}}\right)\int_{0}^{-\beta}\left(i\, dt\right)\int_{0}^{2\pi}d\phi\left[-i\:\mbox{Tr}\left(A_{t}A_{\phi}-\bar{A}_{t}\bar{A}_{\phi}\right)\right],
\end{eqnarray}
with $k=-2il$. }, 

\begin{equation}
\tilde{I}_{\mbox{on-shell}}^{\mbox{dS}}=-\left(\frac{2\beta L}{l}+\frac{4}{3}\frac{\beta\mu^{2}L^{2}}{l^{2}}\right).\label{eq: On-shell action}
\end{equation}
Again, using the identifications, we see that the $dS$ on-shell action reproduces the $EAdS$ on-shell action, (\ref{eq:On-shell action for AdS})

\begin{eqnarray}
\tilde{I}_{On-shell}^{dS}=I_{On-shell}^{EAdS}.
\end{eqnarray}
Thus we have demonstrated that the higher spin generalizations of Kerr de Sitter universes are related to (higher spin) AdS black holes just as they were in the pure gravity (spin-2) case in the metric formulation. However this on-shell action is not yet equal to $-\beta \Phi $, where $\Phi$ is the grand ``higher spin'' canonical potential $ \Phi = E- TS - \mu W$. But it is possible to add boundary/supplementary terms and change the action $\tilde{I}_{on-shell}^{dS}$ to a new action $I_{On-shell}$ such that,
\begin{eqnarray}
-I_{On-shell}= \beta \Phi.
\end{eqnarray}
Such a procedure was conducted in the anti de Sitter case in \cite{David:2012iu}, see their section (2.2). For our de Sitter case, the necessary extra terms can be obtained from their expressions by the AdS-dS identifications, in exact analogy with our computation here for the bulk terms. The match between our entropy and the higher spin AdS$_{3}$ black hole entropy \cite{Kraus:2013esi} is a natural consequence, and we have explicitly checked this. This concludes our discussion about the connection between the thermodynamics of the Kerr-dS$_{3}$ solution and that of higher spin black holes in AdS$_{3}$.\\

\chapter{A Grassmann Path from AdS$_3$ to Flat Space}\label{Grassmann}
\section{Introduction}

The AdS/CFT correspondence has provided us with substantial insight into the nature of quantum gravity when there is a negative cosmological constant. This includes the possibility of a resolution of the black hole information paradox, and potential exact candidates for quantum gravity in terms of non-gravitational quantum gauge theories. \\

Eventually, one would like to understand flat space quantum gravity as well, but taking the vanishing comsological constant limit of the AdS/CFT correspondence in order to accomplish this has remained a challenge. Some progress in this direction has been made by Barnich and collaborators \cite{Barnich:2010eb, Barnich:2012aw, Barnich:2012xq}\footnote{A very recent work on this topic is \cite{Fareghbal:2013ifa}. See also \cite{Strominger:2013jfa} for some recent interesting thoughts on the asymptotics of flat space.} in the AdS$_3$ case. Specifically, Barnich and Compere \cite{Barnich:2006av} showed that the asymptotic symmetry algebra of AdS$_3$ (the Virasoro algebra of Brown-Henneaux) turns into that of flat 2+1 dimensional space (namely, the centrally extended version \cite{Barnich:2006av,Barnich:2012aw} of the so-called BMS$_3$ \cite{Ashtekar:1996cd} algebra) in a certain scaling limit where the cosmological constant is sent to zero. \\

In this chapter, we will show that there is a simple {\em algebraic} way to relate semi-classical gravity in flat space to that in AdS when the spacetime is 2+1 dimensional. The starting point is the fact that 2+1 dimensional gravity can be thought of as a Chern-Simons gauge theory. The gauge group of the theory is $SO(2,2)$ when there is a cosmological constant $\Lambda \equiv -\lambda < 0$, but when $\Lambda = 0$ the gauge group is $ISO(2,1)$. It turns out that an Inonu-Wigner contraction on the  $SO(2,2)$ algebra gives us the $ISO(2,1)$ algebra. This Inonu-Wigner contraction and its connection the BMS/GCA correspondence has been studied in \cite{Bagchi:2010eg, Bagchi:2012cy, Afshar:2013bla}.\\

Our simple observation is that this Inonu-Wigner contraction of the algebras can be realized at the level of the theories, by taking the inverse AdS$_3$ radius $\epsilon \equiv 1/l =\sqrt{\lambda}$ to be a Grassmann parameter such that $\epsilon^2=0$, \cite{KRNEW}\footnote{Recently, this approach has been furthur generalized to various kinds of graded alegebras which appear in physics, in \cite{Krishnan:2017zux}.}. 
We show that this trick can be used to map the actions, the solutions and the asymptotic symmetry algebras. Specifically, the general Fefferman-Graham solution for AdS$_3$ gravity written down by Banados goes over into the general flat space solution in BMS gauge, and the Virasoro algebra with the Brown-Henneaux central charge goes over into the BMS$_3$ algebra of flat space with the correct central charge. \\

We also show that this approach generalizes to higher spin theories which are essentially Chern-Simons theories with higher rank gauge groups. The recently constructed flat-space higher spin theories emerge very simply and straightforwardly from this approach.  As an illustration of the usefulness of our approach, we show how we can resolve singularities in flat space gravity using higher spins in a BMS-like gauge. We claim that our construction is more ``advantageous" than various other implementations of the limit/contraction, in particular, in the case of higher spin gravity. One reason for this is the fact that our approach can be implemented algebraically. Another (technical) reason is that our approach automatically provides us with a useful trace form in  the Chern-Simons formulation of flat space theory. Since the observables are nonlocal gauge theory objects like holonomies of Wilson loops our approach provides an instantly readable/executable map to read them off, unlike in the previous approaches.

\section{Inonu-Wigner Contraction}
For the case of a negative cosmological constant, the 3D Einstein-Hilbert action in Chern-Simons form is given by  \eqref{adsaction}. The basic reason why we have set up these constructions carefully in Chapter 1 is because the precise chain of logic in writing down the action in the form (\ref{adsaction}) is often not discussed in the literature, but is crucial for what we are about to discuss. One of our basic observation in this work is that if one makes the replacement 
\bea
\frac{1}{l} \rightarrow \epsilon
\eea
(where $\epsilon$ is a Grassmann parameter so that $\epsilon^2=0$), in (\ref{gaugef}), then the AdS Einstein-Hilbert action (\ref{adsaction}) turns into the flat space Einstein-Hilbert action, but multiplied by an overall factor of $\epsilon$. In other words we will see that the quantity multiplying the $\epsilon$, after the above replacement, is the flat space gravitational action. This makes sure that the Newton's constant and Chern-Simons level after this replacement are related by

\bea
k =\frac{1}{4G}. \label{grassG}
\eea
Even though we will not do so here, we can absorb the overall factor of $\epsilon$ into the definition of the $G$ and formally treat $G$ as a Grassmann parameter: since we are mostly interested in classical equations of motion where $G$ is merely an overall factor, this will not make any difference at the level of the solutions.  
These claims are easy to check by direct computation, and we have done so. 

For most purposes we will be using this map from AdS to flat space as a useful technical tool for dealing with various aspects of classical solutions, so for the purposes of this thesis, we will think of it as a formal tool. But the simplifications that happen are sufficiently drastic, that it is tempting to speculate that there is more to this story than a mere trick.

The fundamental reason why the above replacement works is because of the fact that $ISO(2,1)$ is an Inonu-Wigner contraction of $SO(2,2)$. For the specific case here, Inonu-Wigner contraction is the statement that if one scales the generators $P^a$ in the $SO(2,2)$ algebra (\ref{eq: SO(2,2) algebra}) by a (non-Grassmann) parameter $\epsilon$ (that is $P^a \rightarrow \epsilon P^a$) and then takes $\epsilon \rightarrow 0$, one is left with the $ISO(2,1)$ algebra (\ref{eq: ISO(2,1) algebra}). But instead of taking the {\em analytic} limit $\epsilon \rightarrow 0$ to implement the Inonu-Wigner contraction, one can also treat $\epsilon$ as a Grassmann parameter and end with the same (\ref{eq: SO(2,2) algebra}). This is an {\em algebraic} realization of the contraction and that is what we are putting to use here. 

An explicit way in which both the norms and the algebras of $ISO(2,1)$ can be realized in terms of the $T^a$ and $\tilde T^a$ generators of $SL(2,\bR)$ is to define:
\bea
P^a=\left(\begin{array}{cc}
	\epsilon \ T^a&0\\
	0 & -\epsilon \ \tilde T^a
\end{array}\right), \ &&\ J^a=\left(\begin{array}{cc}
T^a&0\\
0 &  \tilde T^a
\end{array}\right)
\eea
If one identifies $T^a$ with $J^{a+}=\Big(\begin{array}{cc}
T^a&0\\
0 & 0
\end{array}\Big)$ and $\tilde T^a$ with $J^{a-}=\Big(\begin{array}{cc}
0&0\\
0 & \tilde T^a
\end{array}\Big)$, then this can be thought of as another way to write
\bea
P^a=\epsilon(T^a-\tilde T^a), \ \ J^a=(T^a+\tilde T^a)
\eea
which in turn follows from (\ref{pmJ}) upon $1/l \rightarrow \epsilon$. This generalizes very straightforwardly to higher spin theories as well, as we will briefly discuss later.

Another (non-Grassmann) way to think of the mapping from one theory to other is to think of it as the scaling limit where $1/l \rightarrow 0$ but with $k/l$ is held fixed. Even though it is not couched there in this language, this is essentially what BGG have done \cite{Barnich:2012aw}. We will find this useful in our discussion of the Brown-Henneaux algebra.

It is trivial to check that the diffeomorphisms and local Lorentz transformations, written in terms of the triads and the spin connections, also go over from the AdS to the flat case without any difficulty when we set $1/l \rightarrow \epsilon$. The explicit expressions can be found in Witten's paper and the check is trivial, so we will not repeat them here.

Often, in what follows we will use generators $T^a$ that have the trace form
\bea
{\rm Tr}(T^aT^b)=2 \eta^{ab},
\eea
following the conventions of \cite{Castro:2011fm}, where they are working with 3 $\times$ 3 generators, which are more convenient from the perspective of generalizations to higher spin theories. 
This implies that we should take
\bea
k=\frac{l}{16 G},
\eea
in the AdS case.

\section{AdS$_{3}$ in BMS-like gauge}

Our goal is first to show the transition from general locally AdS$_3$ 
solution \cite{Banados:1998gg} to the general asymptotically flat solution using the Grassmann approach\footnote{In this section, we have chosen to set $8G=1$ in agreement with the general convention in $2+1$-D general relativity literature.%
}. 

Following \cite{Barnich:2012aw}, we first write down the general locally AdS solution in a BMS-like gauge to ease the transition to flat space. The general asymptotically AdS$_3$ line element that satisfies the Einstein equations with a negative cosmological constant can be written in the form
\begin{equation}
ds^{2}=\left(\mathcal{M}-\frac{r^{2}}{l^{2}}\right)du^{2}-2dudr+2\mathcal{N}\: dud\phi+r^{2}d\phi^{2}\label{eq: AdS_3 metric in BMS gauge}
\end{equation}
provided 
\begin{equation}
\partial_{u}\mathcal{M}=\frac{2}{l^{2}}\partial_{\phi}\mathcal{N},\qquad2\partial_{u}\mathcal{N}=\partial_{\phi}\mathcal{M}.\label{eq: neccessary conditions on M,N}
\end{equation}
This solution, and these conditions on the arbitrary functions are merely a re-writing of the general Fefferman-Graham solution in AdS$_3$ \cite{Banados:1998gg}. 
This is easily checked by noting that $\mathcal{M},\mathcal{N}\equiv \mathcal{M}(\phi,u),\mathcal{N}(\phi,u)$ satisfying the above conditions
can be expressed in terms of the usual left and right moving functions,
$\mathcal{L}(x^{+}),\bar{\mathcal{L}}(x^{-})$, 
\begin{equation}
\mathcal{M}(u,\phi)=2\left(\mathcal{L}(x^{+})+\bar{\mathcal{L}}(x^{-})\right), \ \ \mathcal{N}(u,\phi)=l\left(\mathcal{L}(x^{+})-\bar{\mathcal{L}}(x^{-})\right),\label{eq: BMS gauge AdS parameter}
\end{equation}
with $x^{\pm}=\frac{u}{l}\pm\phi$.\\

We take the triad for this locally AdS$_{3}$ solution (in BMS like coordinates) to be

\bea
e=-\frac{1}{\sqrt{2}}\left[\left(\frac{\mathcal{M}}{2}-1-\frac{r^{2}}{2l^{2}}\right)\: du-dr+\mathcal{N}\: d\phi\right]T_{0}+\hspace{1in}\nonumber \\ \hspace{1in}-\frac{1}{\sqrt{2}}\left[\left(\frac{\mathcal{M}}{2}+1-\frac{r^{2}}{2l^{2}}\right)\: du-dr+\mathcal{N}\: d\phi\right]T_{1}
-rd\phi T_{2}.\label{eq: BMS gauge AdS triad}
\eea
The (dualized) spin-connection (\ref{dualspincon}) can be computed directly from the triads, and the result is:

\begin{eqnarray}
\omega^{0} & = & -\frac{1}{\sqrt{2}}\left(\frac{\partial_{u}\mathcal{N}}{r}-\frac{\partial_{\phi}\mathcal{M}}{2r}+\frac{\mathcal{N}}{l^{2}}\right)du-\frac{1}{\sqrt{2}}\left(\frac{\mathcal{M}}{2}-1-\frac{r^{2}}{2l^{2}}\right)d\phi,\\
\omega^{1} & = & -\frac{1}{\sqrt{2}}\left(\frac{\partial_{u}\mathcal{N}}{r}-\frac{\partial_{\phi}\mathcal{M}}{2r}+\frac{\mathcal{N}}{l^{2}}\right)du-\frac{1}{\sqrt{2}}\left(\frac{\mathcal{M}}{2}+1-\frac{r^{2}}{2l^{2}}\right)d\phi,\\
	\omega^{2} & = & -\frac{r}{l^{2}}du.
\end{eqnarray}
These expressions have been checked to satisfy the torsion-free condition,
\bea
\partial_{\mu}e_{\nu}^{a}-\partial_{\nu}e_{\mu}^{a}+\epsilon^{a}\,_{bc}\left(e^{b}\,_{\mu}\omega^{c}\,_{\nu}-e^{b}\,_{\nu}\omega^{c}\,_{\mu}\right)=0,
\eea
and the Einstein equation \cite{Witten:1988hc} (provided conditions
(\ref{eq: neccessary conditions on M,N}) hold),
\[
\partial_{\mu}\omega_{\nu}^{a}-\partial_{\nu}\omega_{\mu}^{a}+\epsilon^{a}\,_{bc}\left(\omega^{b}\,_{\mu}\omega^{c}\,_{\nu}+\frac{1}{l^{2}}e^{b}\,_{\mu}e_{\nu}^{c}\right)=0.
\]
Note that the asymptotic AdS$_3$ fall off conditions went into the construction of the Fefferman-Graham form: they are implicit in our starting point. So the constraints (\ref{eq: neccessary conditions on M,N}) came purely from imposing the AdS$_3$ Einstein equations.

Using the triads and the spin connection, now we can immediately write down the explicit gauge field corresponding to the general asymptotically AdS$_3$ solution via (\ref{gaugef}).

\section{Grassmann Path to Flat Space}

Now we turn to the general locally flat solution. In the ``BMS-gauge'' \cite{Barnich:2010eb}, where asymptotic analysis
is easiest (akin to Fefferman-Graham gauge in the case of $AdS$),
the most general solution in $2+1$-d is,
\begin{equation}
ds^{2}=\mathcal{M}(\phi)du^{2}-2dudr+2\left[\mathcal{J}(\phi)+\frac{u}{2}\partial_{\phi}\mathcal{M}(\phi)\right]dud\phi+r^{2}d\phi^{2}.\label{eq: most general asymptotically flat}
\end{equation}
(Later we will specialize to the case when $\mathcal{M}(\phi)=M$
and $\mathcal{J}(\phi)=J/2$ are constants, which has a cosmological interpretation).

Now, the gauge field from the last section, upon the Grassmann replacement of $1/l \rightarrow \epsilon$ gives us the explicit form
\begin{eqnarray}
A & = & -\frac{1}{\sqrt{2}}\left[\epsilon\left(\frac{\mathcal{M}}{2}-1\right)\: du-\epsilon dr+\epsilon\left(\mathcal{J}+\frac{u}{2}\mathcal{M}'\right)\: d\phi+\left(\frac{\mathcal{M}}{2}-1\right)\: d\phi\right]T_{0}\nonumber \\
&  & \qquad-\frac{1}{\sqrt{2}}\left[\epsilon\left(\frac{\mathcal{M}}{2}+1\right)\: du-\epsilon dr+\epsilon\left(\mathcal{J}+\frac{u}{2}\mathcal{M}'\right)\: d\phi+\left(\frac{\mathcal{M}}{2}+1\right)\: d\phi\right]T_{1}\nonumber \\
&  & \qquad\qquad\qquad\qquad\qquad\qquad\qquad\qquad\qquad\qquad\qquad\qquad\qquad\qquad\qquad-\epsilon\: r\: d\phi\; T_{2}.\hspace{0.5in}\label{eq: BMS gauge full connection-1}
\end{eqnarray}
Note that the Grassmann replacement gives a simple interpretation for the form of the functions now because of the constraints (\ref{eq: neccessary conditions on M,N}):
\bea
\mathcal{M} \equiv \mathcal{M}(\phi), \ \ \mathcal{N} \equiv \mathcal{J}(\phi)+\frac{u}{2}\mathcal{M}'(\phi).
\eea
Our claim from section 2 is that the $ISO(2,1)$ theory can be reformulated as a Grassmann valued $SO(2,2)$
gauge theory with the connection\footnote{We will exclusively work with ``holomorphic" part. The ``anti-holomorphic" $\tilde A=\left(\omega^{a}-\epsilon e^{a}\right)T_{a}$ part is entirely analogous.}
\[
A=\left(\omega^{a}+\epsilon e^{a}\right)T_{a}
\]
where, $T_{a}\in SO(2,1)=SL(2,R)$ and $\epsilon$ is a Grassman parameter.
This means that we can read off the flat space triad and spin connection from this gauge field. 

Indeed, it is easy to check that the triad and spin connection one obtains this way, can reproduce the most general flat metric  (\ref{eq: most general asymptotically flat}). The natural expressions for the flat space triads are \cite{Gonzalez:2013oaa}%
\footnote{However in contrast to the convention of \cite{Gonzalez:2013oaa}, we choose a convention where $\eta_{ab}=\mbox{diag}\left(-1,1,1\right)$.},

\begin{eqnarray}
e^{0} & = & -\frac{1}{\sqrt{2}}\left[\left(\frac{\mathcal{M}(\phi)}{2}-1\right)\: du-dr+\left(\mathcal{J}(\phi)+\frac{u}{2}\frac{d\mathcal{M}(\phi)}{d\phi}\right)\: d\phi\right],\nonumber \\
e^{1} & = & -\frac{1}{\sqrt{2}}\left[\left(\frac{\mathcal{M}(\phi)}{2}+1\right)\: du-dr+\left(\mathcal{J}(\phi)+\frac{u}{2}\frac{d\mathcal{M}(\phi)}{d\phi}\right)\: d\phi\right],\nonumber \\
e^{2} & = & -r\: d\phi,\label{eq: BMS gauge triad}
\end{eqnarray}
We can compute the spin-connection from Cartan's torsion-free condition and this also matches the result obtained from Grassmann replacement from AdS:

\begin{eqnarray}
\omega^{0} & = & -\frac{1}{\sqrt{2}}\left(\frac{\mathcal{M}(\phi)}{2}-1\right)\: d\phi,\nonumber \\
\omega^{1} & = & -\frac{1}{\sqrt{2}}\left(\frac{\mathcal{M}(\phi)}{2}+1\right)\: d\phi,\nonumber \\
\omega^{2} & = & 0.\label{eq: BMS gauge spin-connection}
\end{eqnarray}

For later use we write down the $ISO(2,1)$ connection in terms of the $ISO(2,1)$ generators as well (\ref{eq: ISO(2,1) algebra}):%

\begin{eqnarray}
A & = & -\frac{1}{\sqrt{2}}\left[\left(\frac{\mathcal{M}}{2}-1\right)\: du-dr+\left(\mathcal{J}+\frac{u}{2}\mathcal{M}'\right)\: d\phi\right]P_{0}\nonumber \\
&  & -\frac{1}{\sqrt{2}} \left[\left(\frac{\mathcal{M}}{2}+1\right)\: du-dr+\left(\mathcal{J}+\frac{u}{2}\mathcal{M}'\right)\: d\phi\right]P_{1} \\
&  & -r\: d\phi\; P_{2}-\frac{1}{\sqrt{2}}\left(\frac{\mathcal{M}}{2}-1\right)\: d\phi\: J_{0}-\frac{1}{\sqrt{2}}\left(\frac{\mathcal{M}}{2}+1\right)\: d\phi\: J_{1},\nonumber \label{eq: BMS gauge full connection}
\end{eqnarray}
where, $\mathcal{M}'=\partial_{\phi}\mathcal{M}(\phi)$.

\section{Asymptotic Charge Algebra}

We restore all factors of $8G$ for this section to facilitate a consistent derivation of the flat space from a grassmanian AdS expressions. The AdS charges (as derived in BMS looking gauge) were written down by \cite{Barnich:2012aw}, 

\begin{equation}
Q_{f,Y}=\frac{1}{16\pi G}\int_{0}^{2\pi}d\phi\left[f\left(\mathcal{M}+1\right)+2Y\mathcal{N}\right],\label{eq: AdS charges}
\end{equation}
associated with the killing vector,

\bea
\xi_{f,Y}=f\: du-Y-l\partial_{\phi}f\left(\int_{r}^{\infty}dr'r'^{-2}e^{2\beta}\right)\: d\phi-r\left(\partial_{\phi}\xi^{\phi}-U\partial_{\phi}f\right)
\eea
with $f,Y$ defined in terms of purely holomorphic and antiholomorphic (arbitrary) functions, $Y^{\pm}$: 

\bea
f=\frac{l}{2}\left(Y^{+}(x^{+})+Y^{-}(x^{-})\right),Y=\frac{1}{2}\left(Y^{+}(x^{+})-Y^{-}(x^{-})\right).
\eea
$U,\beta,V$ are metric paramaters,

\bea
ds^{2}=e^{2\beta}\frac{V}{r}du^{2}-2e^{2\beta}dudr+r^{2}\left(d\phi-Udu\right)^{2}.
\eea
In fact looking at the metric (\ref{eq: AdS_3 metric in BMS gauge}), we have,

\bea
\beta=0,\frac{V}{r}+r^{2}U^{2}=-\frac{r^{2}}{l^{2}}+\mathcal{M},\mathcal{N}=-r^{2}U.
\eea
First we do a mode decomposition \cite{Barnich:2012aw},

\bea
\mathcal{L}(\bar{\mathcal{L}})=-\frac{1}{4}+\sum_{m}\frac{1}{2l}L^{\pm}e^{-imx^{\pm}},
\eea
we have,

\bea
\mathcal{M}&=&-1+\sum_{m}\frac{8G}{l}\left(L_{m}^{+}e^{-imu/l}+L_{-m}^{-}e^{imu/l}\right)e^{-im\phi},\\
\mathcal{N}&=&4G\sum_{m}\left(L_{m}^{+}e^{-imu/l}-L_{-m}^{-}e^{imu/l}\right)e^{-im\phi}.
\eea
So replacing,

\bea
\frac{1}{l}\rightarrow\epsilon,
\eea
these mode expansions, become

\begin{eqnarray}
\mathcal{M} & = & -1+8G\epsilon\left(L_{0}^{+}+L_{0}^{-}\right)+8G\epsilon\sum_{m\neq0}\left(L_{m}^{+}(1-\epsilon\: imu)+L_{-m}^{-}(1+\epsilon\: imu)\right)e^{-im\phi}\nonumber \\
& = & -1+8G\epsilon\sum_{m}\left(L_{m}^{+}+L_{-m}^{-}\right)e^{-im\phi},\nonumber \\
\mathcal{N} & = & 4G\left(L_{0}^{+}-L_{0}^{-}\right)+4G\sum_{m\neq0}\left[\left(L_{m}^{+}-L_{-m}^{-}\right)-\epsilon u\: im\left(L_{m}^{+}+L_{-m}^{-}\right)\right]e^{-im\phi}.\label{eq:Intermediate stage M, N}
\end{eqnarray}
Next we define,

\begin{equation}
P_{m}=\frac{1}{l}\left(L_{m}^{+}+L_{-m}^{-}\right),J_{m}=L_{m}^{+}-L_{-m}^{-},\label{eq: New ISO generators}
\end{equation}
which after the Grassman replacement turns into

\begin{equation}
\mathcal{P}_{m}=\epsilon\left(L_{m}^{+}+L_{-m}^{-}\right),\mathcal{J}_{m}=L_{m}^{+}-L_{-m}^{-},\label{eq: New ISO generators}
\end{equation}
These definitions can be motivated in two ways. One is by taking a cue from \cite{Barnich:2012xq} and making the replacement $1/l \rightarrow \epsilon$ in the expressions there. Another way to motivate this definition is as follows. These modes are to be thought of as capturing the infinite dimensional extension of $SL(2, \bR)$ (or $ISO(2,1)$ after the flat space limit). The zero mode part of these generators is $SL(2, \bR)$ (respectively $ISO(2,1)$). The definitions, restricted to the zero mode sector is precisely what is needed to make the transition from $SL(2, \bR)$  to $ISO(2,1)$, so it is natural extend the definitions to the higher modes as well. Either way, ultimately the only thing that matters is that this definition ends up giving us BMS$_3$ from Virasoro as we show presently.

With the above definitions,
\begin{equation}
\mathcal{M}(\phi)=-1+8G\sum_{m}P_{m}e^{-im\phi},\label{eq: familiar BMS flat M}
\end{equation}
\begin{equation}
\mathcal{N}=8G\left(\mathcal{J}(\phi)+\frac{u}{2}\partial_{\phi}\mathcal{M}(\phi)\right),\mathcal{J}(\phi)\equiv\frac{1}{2}\sum_{m}J_{m}e^{-im\phi}.\label{eq:familiar BMS flat N}
\end{equation}
Similarly for the killing vector parameters after making the replacements,

\begin{eqnarray*}
	\epsilon\: f & = & \frac{1}{2}\sum_{m}\left(Y_{m}^{+}+Y_{-m}^{-}\right)e^{-im\phi}-\epsilon\: u\sum_{m\neq0}im\left(Y_{m}^{+}-Y_{-m}^{-}\right)e^{-im\phi},\\
	Y & = & \frac{1}{2}\sum_{m}\left(Y_{m}^{+}-Y_{-m}^{-}\right)e^{-im\phi}-\epsilon\: u\sum_{m\neq0}im\left(Y_{m}^{+}+Y_{-m}^{-}\right)e^{-im\phi}
\end{eqnarray*}
Analogous to (\ref{eq: New ISO generators}) we have

\begin{equation}
\epsilon T_{m}\equiv\frac{1}{2}\left(Y_{m}^{+}+Y_{-m}^{-}\right),Y_{m}\equiv\frac{1}{2}\left(Y_{m}^{+}-Y_{-m}^{-}\right).\label{eq: Definition for Killing modes to go from SL2 to ISO}
\end{equation}
This leads to the expressions,

\begin{equation}
f=T(\phi)+u\partial_{\phi}Y(\phi),Y=Y(\phi)\label{eq: Killing vector parameters for ISO theory}
\end{equation}
Now finally we can plug equations (\ref{eq: familiar BMS flat M}), (\ref{eq:familiar BMS flat N}), and (\ref{eq: Killing vector parameters for ISO theory}) in the expression (\ref{eq: AdS charges}) for the AdS charges to obtain,

\begin{equation}
Q_{T,Y}=\frac{1}{16\pi G}\int_{0}^{2\pi}\left(T(\phi)\:\mathcal{M}(\phi)+2Y(\phi)\mathcal{J}(\phi)\right).\label{eq: BMS charges}
\end{equation}
This is exactly the expression of $ISO$ charges obtained in \cite{Barnich:2010eb} upon conducting a Henneaux-Teitelboim like asymptotic symmetry analysis for flat space (BMS/CFT correspondence).

To conlude this section we show how the Virasoro algebra with Brown-Henneaux central charge goes over to the BMS algebra with the correct central charge\footnote{See \cite{Henkel:2005zu} for a related discussion in a different context.}. 
The latter has central charges $c^{\pm}=\frac{3l}{2G}$:
\begin{equation}
\left[L_{m}^{\pm},L_{n}^{\pm}\right]=\left(m-n\right)L_{m+n}^{\pm}+\frac{c^{\pm}}{12}m(m^{2}-1)\delta_{m+n},\qquad\left[L_{m}^{\pm},L_{n}^{\mp}\right]=0.\label{eq: Virasoro algebra}
\end{equation}
To this end use a more convenient version of the Virasoro for our
contraction purpose, 
\begin{eqnarray}
	\left[J_{m},J_{n}\right] & = & \left(m-n\right)J_{m+n},\\
	\left[P_{m},P_{n}\right] & = & \frac{1}{l^{2}}\left(m-n\right)J_{m+n},\\
	\left[J_{m},P_{n}\right] & = & (m-n)P_{m+n}+\frac{k}{12}m\left(m^{2}-1\right)\delta_{m+n}.
\end{eqnarray}
where 

\bea
k\equiv\frac{c^{+}+c^{-}}{l}=\frac{3}{G}
\eea 
Now we arrrive at the $\mathfrak{bms}_{3}$ algebra by the simple replacement, $\frac{1}{l}\rightarrow\epsilon,$
(and accordingly $P_{m}\rightarrow\mathcal{P}_{m},J_{m}\rightarrow\mathcal{J}_{m}$)

\begin{eqnarray}
	\left[\mathcal{J}_{m},\mathcal{J}_{n}\right] & = & \left(m-n\right)\mathcal{J}_{m+n},\\
	\left[\mathcal{P}_{m},\mathcal{P}_{n}\right] & = & 0,\\
	\left[\mathcal{J}_{m},\mathcal{P}_{n}\right] & = & (m-n)\mathcal{P}_{m+n}+\frac{k}{12}m\left(m^{2}-1\right)\delta_{m+n}.
\end{eqnarray}

\section{Higher Spin Extension}

For the higher spin version, one has to extend the $ISO(2,1)$ algebra by including a new set of spin-3 generators, $J_{ab},P_{ab}$ \cite{Gonzalez:2013oaa,Afshar:2013vka}. The algebra takes the form

\begin{eqnarray}
	\left[J_{ab},J_{cd}\right] & = & -\left(\eta_{a(c}\epsilon_{d)bm}+\eta_{b(c}\epsilon_{d)am}\right)J^{m},\br
	\left[J_{ab},P_{cd}\right] & = & -\left(\eta_{a(c}\epsilon_{d)bm}+\eta_{b(c}\epsilon_{d)am}\right)P^{m},\br
	\left[P_{ab},P_{cd}\right] & = & 0,\br
	\left[J_{a},J_{bc}\right] & = & \epsilon^{m}\,_{a(b}J_{c)m},\\
	\left[J_{a},P_{bc}\right] & = & \epsilon^{m}\,_{a(b}P_{c)m},\br
	\left[P_{a},J_{bc}\right] & = & \epsilon^{m}\,_{a(b}P_{c)m},\br
	\left[P_{a},P_{bc}\right] & = & 0. \nonumber
\end{eqnarray}
We will call this the $hsf_{3}$ algebra. The invariant nondegenerate bilinear product is given by (the only non-vanishing pieces),

\bea
\bra P_{a},J_{b}\ket &=&\eta_{ab},\\ \bra P_{ab},J_{ab}\ket &=& \eta_{ac}\eta_{bd} + \eta_{ad}\eta_{bc} - \frac{2}{3}\eta_{ab}\eta_{cd}. \label{hstrace}
\eea
This algebra can be realized as as Inonu-Wigner contraction of $SL(3, \bR) \times SL(3, \bR)$ algebra analogous to the spin-2 case. In terms of the two copies of the $SL(3)$ generators 
\bea
&&[T_a,T_b] = \epsilon_{abc} T^c, \\
&&[T_a,T_{bc}] = \epsilon^d_{\ \ a(b} T_{c)d}, \\ 
&&[T_{ab},T_{cd}] = \sigma \left(\eta_{a(c} \epsilon_{d)be}+\eta_{b(c} \epsilon_{d)ae}\right)T^e.
\eea
it can be straightforwardly checked that one can define the $hsf_3$ generators via
\bea
P^a=\left(\begin{array}{cc}
	\epsilon \ T^a&0\\
	0 & -\epsilon \ T^a
\end{array}\right), \ &&\ J^a=\left(\begin{array}{cc}
T^a&0\\
0 &  T^a
\end{array}\right) \\
P^{ab}=\left(\begin{array}{cc}
	\epsilon \ T^{ab}&0\\
	0 & -\epsilon \ T^{ab}
\end{array}\right), \ && \ J^{ab}=\left(\begin{array}{cc}
T^{ab}&0\\
0 &  T^{ab}
\end{array}\right).
\eea
This is the Grassmann realization of Inonu-Wigner and our point is that this can be used to interpret a Grassmann valued $SL(3, \bR) \times SL(3, \bR)$ gauge field as an $hsf_3$ (that is, flat space higher spin) gauge field. The Grassmann approach immediately enables us to get to the above result from the generators of \cite{Campoleoni:1}. The traces of the $SL(3,\bR) \times SL(3,\bR)$ are designed so that it reproduces (\ref{hstrace}). As in the spin-2 case, the flat space higher spin gauge field can be expressed via Grassmann parameter by its natural generalization
\bea
A =\sum_{a=0}^{2}\left(e^{a}+\epsilon\:\omega^{a}\right)T_{a} +\sum_{a,b=0}^{2}\left(e^{ab}+\epsilon\:\omega^{ab}\right)T_{ab}
\eea
Again we expect the actions and the asymptotic symmetries to work out exactly analogously, but we leave the details.

\section{Application: Singularity Resolution in the BMS Gauge}

So far what we have done is to merely repeat known results (but from a new and perhaps a simpler and more elegant point of view). Now we will show that this new technology makes certain computations tractable and show that certain singularity resolution questions become anwerable in this frame work. The reason for this is that constructing a set of explicit matrix generators that satisfy the $hsf_3$ algebra {\em while having the non-degenerate trace form (\ref{hstrace}) } is non-trivial. But one can bypass this problem while having a non-degenerate trace form by working within the Grassmann technology.

Other discussions on singularity resolution in higher spin theories can be found in \cite{Castro:2011fm, Krishnan:2013cra, Krishnan:2013zya, Burrington:2013dda, Krishnan:2013tza}. The basic idea of singularity resolution in this set up is to consider a singular solution of the spin-2 theory, embed it in the higher spin theory, and then to look for gauge transformations that retain the holonomy within the same conjugacy class. If there exists a gauge transformation that gives rise to metric and higher spin fields that are regular while not changing the conjugacy class, we have resolved the singularity.

\subsection{Metric Formulation of the Singular Cosmology}

\begin{figure}[H]
	\begin{center}
		\includegraphics[height=0.4\textheight]{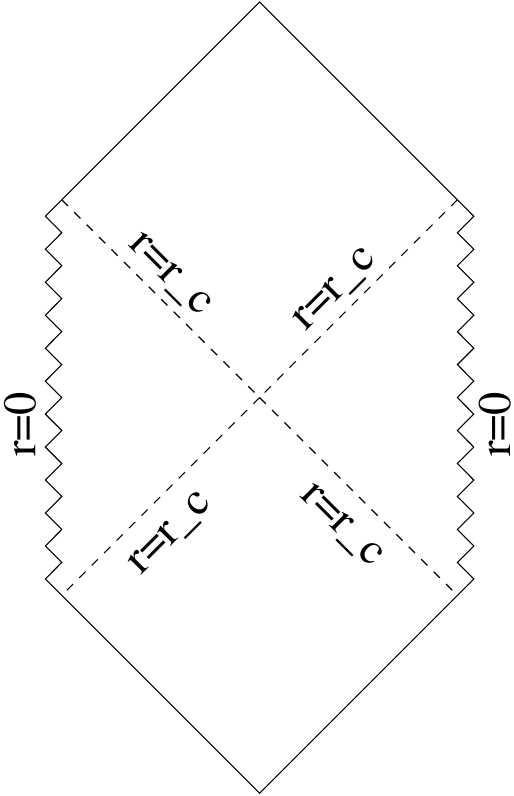}
		\caption{2D slice of Penrose diagram of the Shifted Boost orbifold.}
		\label{Penrose}
	\end{center}
\end{figure}

We start with the boost-shifted orbifold cosmology \cite{Bagchi:2012xr} which has a sigularity we intend to resolve. To obtain this solution, we can start with the general flat space BMS-gauge solution that we considered previously and specialize to the case when $\mathcal{M}(\phi)=M$ and $\mathcal{J}(\phi)=J/2$ are constants. As pointed out in \cite{Barnich:2012aw,Barnich:2012xq} this BMS gauge metric could be thought of as an expression in outgoing null coordinate, 

\begin{equation}
u=t-\int dr/N^{2}(r),\label{eq: null coordinate}
\end{equation}
and a new angular coordinate,

\begin{equation}
\varphi=\phi-\int dr\; N^{\varphi}/N{}^{2}\label{eq: new angular coordinate}
\end{equation}
to correspond to a Schwarzschild-type metric,

\begin{equation}
ds^{2}=-N^{2}(r)dt^{2}+N^{-2}(r)dr^{2}+r^{2}\left(d\varphi+N^{\varphi}dt\right)^{2},\label{eq: Sch. gauge metric}
\end{equation}
where
\bea
N^{2}(r)=-M+\frac{J^{2}}{4r^{2}}=\frac{M}{r^{2}}\left(r_{C}^{2}-r^{2}\right),N^{\varphi}=\frac{J}{2r^{2}}.
\eea
Note that $\varphi=\phi-\int dr\; N^{\varphi}/N{}^{2}$ and hence does {\em not} parametrize a compact direction. However one can identify $\varphi\sim\varphi+2\pi$ and construct quotient spaces \cite{Cornalba:2002fi,Cornalba:2003kd,Barnich:2012aw} with cosmological (Cauchy) horizons at $r=r_{C}$. However these spaces contain pathological regions with closed time-like curves, $r<0$
and such regions are excised. $r=0$ thus becomes a causal structure singularity. These have been dubbed \emph{shifted boost orbifolds} \cite{Cornalba:2002fi,Cornalba:2003kd, Berkooz:2002je}\footnote{The reason for the name is the fact that the metric can be understood as an orbifold of flat space under shifts and boosts, but we will not need that connection, so we will not elaborate on it.} when they were discovered and discussed in the context of string theory. However, we shall refer to these as flat quotient cosmologies. A Penrose diagram of the flat quotient cosmology is provided in Fig. \ref{Penrose}. 

\subsection{Gauge Theory Formulation of the Singular Cosmology}

For completeness, we present the expressions for the set of triads and the dual spin connection for the flat quotient cosmology in Schwarzschild gauge (\ref{eq: Sch. gauge metric}):

\begin{eqnarray}
e^{0}&=&N(r)dt,\qquad e^{1}=-N^{-1}(r)dr,\qquad e^{2}=rN^{\varphi}(r)\; dt+rd\varphi,\label{eq: triads} \\
\omega^{0}&=&N(r)d\varphi,\qquad\omega^{1}=\frac{N^{\varphi}(r)}{N(r)}\; dr,\qquad\omega^{2}=r\; N^{\varphi}(r)d\varphi.\label{eq: dual spin connections}
\end{eqnarray}
The triad and the spin connection for the general BMS gauge solution has been written down before. We will need the full gauge connection that we wrote down in (\ref{eq: BMS gauge full connection}). It turns out that one can express this full connection, $A$ in terms of a \emph{primitive} connection, $a$ which is stripped-off of any $r$-dependence,

\begin{eqnarray}
a(u,\phi) & = & -\frac{1}{\sqrt{2}}\left[\left(\frac{\mathcal{M}}{2}-1\right)\: du + \left(\mathcal{J} + \frac{u}{2}\mathcal{M}'\right)\: d\phi\right]P_{0}\nonumber \\
&-& \frac{1}{\sqrt{2}}\left[ \left(\frac{\mathcal{M}}{2}+1\right)\: du + \left(\mathcal{J}+\frac{u}{2}\mathcal{M}'\right)\: d\phi\right]P_{1} \\ &-& \frac{1}{\sqrt{2}}\left(\frac{\mathcal{M}}{2}-1\right)\: d\phi\: J_{0} - \frac{1}{\sqrt{2}}\left(\frac{\mathcal{M}}{2} + 1\right)\: d\phi\: J_{1},\nonumber\label{eq:primitive connection}
\end{eqnarray}
using a radial gauge transformation, $b(r)=\exp\left(\frac{rP_{0}+rP_{1}}{\sqrt{2}}\right)$,

\bea
A=b^{-1}\: a\: b+b^{-1}\partial_{r}b.
\eea
This will be useful to us in resolving the singularity.

\subsection{Holonomy of the flat cosmology}

For the flat quotients, $\mathcal{M}(\phi)=M$ and $2\mathcal{J}(\phi)=J$ are constants, and the Wilson loop operator along a constant $u$, $\phi$-circle around $r=0$ is,

\bea
W=\exp\left(\int_{0}^{2\pi}d\phi\; A\right)=\exp\left(b^{-1}2\pi a_{\phi}b\right)=b^{-1}\exp\left(2\pi a_{\phi}\right)b,
\eea
where,

\bea
a_{\phi}=-\frac{1}{\sqrt{2}}\left[\frac{J}{2}\left(P_{0}+P_{1}\right)+\left(\frac{M}{2}-1\right)\: J_{0}+\left(\frac{M}{2}+1\right)\: J_{1}\right].
\eea
Under a trivial gauge transformation, $U$,

\bea
W=UWU^{-1}=\rme^{UwU^{-1}}
\eea
where, $w\equiv2\pi a_{\phi}$. 

\subsection{$ISO(2,1)$ Solution as a Grassmann Valued $SO(2,2)$ Solution\label{sec: From Grassmann SL(2,R) to ISO(2,1)}}

We haven't introduced explicit matrices for the $P^a$ and $J^a$, but we do not need to. This is because the same holonomy information can be captured equivalently in the Grassmann language.  To this end, we first note that the full connection can also be written as the Grassmann valued connection (\ref{eq: BMS gauge full connection-1}). \\

This connection can also be written in terms of a primitive connection. The radial dependence is contained in the term,

\bea
A_{r}=\epsilon\frac{T_{0}+T_{1}}{\sqrt{2}}dr.
\eea
We can try to gauge away $r$-dependence and construct a primitive connection by gauge transforming with 

\bea
U=\exp(\epsilon T_{+}r), 
\eea
with $ T_{\pm}=\frac{T_{0}\pm T_{1}}{\sqrt{2}}$, so that $A=U^{-1}aU+U^{-1}dU$. This $r$-independent primitive connection is

\begin{equation}
a(u,\phi)= \left[ \epsilon T_{-} - \epsilon \frac{M}{2} T_{+} \right]du + \left[ -\left( \frac{M}{2} + \epsilon \left(J + \frac{u}{2}M' \right) \right)T_{+} + T_{-} \right]d\phi
\label{eq:Primitive connection in SL(2) sector}
\end{equation}
We are interested in computing the holonomy matrix of this connection $a$ along a $\phi$-circle of constant $u,r$ in the special case when $M$, $J$ are constant. The eigenvalues of the holonomy are given by 

\begin{equation}
w=2\pi a_{\phi} = \{0,\ -2\pi \sqrt{M} -\pi \frac{J}{\sqrt{M}}\epsilon,\ 2\pi \sqrt{M} +\pi \frac{J}{\sqrt{M}}\epsilon \} \label{eq:grassmann valued holonomy matrix}
\end{equation}

For more generic cases determining the eigenvalues is hard, so instead we use the characteristic polynomial theorem for $3\times3$ square matrices. For a square matrix $M$, the characteristic equation is
\begin{equation}
M^{3}= \mbox{M} \mathbb{I}_{3}+\frac{1}{2}\left(\mbox{tr}(M^{2})-(\mbox{tr}(M))^{2}\right)M+\mbox{tr}(M)\: M^{2}.\label{eq: Characteristic Polynomial Expansion}
\end{equation}
The eigen-values of two holonomy matrices are identical, iff the coefficients of their characteristic polynomials agree. It is easy to check that this theorem is valid even when the matrix has Grassmann valued matrix elements. 

For the flat cosmology, and $M=w$, the left and right holonomy matrices give rise to
\begin{eqnarray}
\mbox{Det}w & = & 0,\nonumber \\
\mbox{tr}\left(w\right) & = & 0,\nonumber \\
\mbox{tr}\left(w_{\pm}^{2}\right) & = & 8\pi^{2} \left(M + 2\epsilon\: J\right) \label{eq: Holonomy conditions}
\end{eqnarray}
Of course, since $\exp(w_{\pm})\in SL(3)$, $\mbox{tr}(w_{\pm})=0$ is automatically ensured.%

\subsection{Singularity Resolution}

We extend the $SL(2,R)$ connection (\ref{eq:Primitive connection in SL(2) sector}) by adding Grassmann valued $SL(3)$ generators,
\begin{equation}
a' = a + \sum_{a,b=0}^{2}\left(c_{ab}+\epsilon\: d_{ab}\right)T_{ab}.\label{eq: SL(3,R) X SL(3,R) primitive}
\end{equation}
After gauge transforming to include the radial dependence we will have a form,
\bea
A'=A + \sum_{a,b=0}^{2}\left(e_{ab}+\epsilon\:\omega_{ab}\right)T_{ab} 
\eea
This is a Grassmann valued $SL(3,R)\times SL(3,R)$ connection and equivalently a connection in the higher spin theory in asymptotically flat space. The metric and the higher spin fields can be obtained from the gauge field by identifying the triad (and its higher spin version)\cite{Gonzalez:2013oaa}. The correction to metric takes the explicit form
\bea
ds^{2}=\left(\eta_{ab}e^{a}\,_{\mu}e^{b}\,_{\nu}+2\eta_{ac}\eta_{bd}e^{ab}{}_{\mu}e^{cd}\,_{\nu}\right)dx^{\mu}dx^{\nu}.\label{eq:metric components after higher spin corrections}
\eea

Actually, instead of using the generators, $T_{ab}$ which do not constitute a linearly independent set, we will use the set, $W_{a}$ \cite{Castro:2011fm}
\bea
a'=a+\sum_{a=-2}^{2}\left(c^{a}+\epsilon\: d^{a}\right)W_{a}  \\ A'=A+\sum_{a=-2}^{2}\left(C^{a}+\epsilon\: D^{a}\right)W_{a} 
\eea
We may look at the simplest case of singularity resolution where we only turn on $W$ generators in $a_{\phi}$ component of the primitive connection.
\begin{equation}
a'_{\phi} = a_{\phi} + \sum_{a=-2}^{2} (c_a + \epsilon d_a)W_a
\end{equation}
Next we need to satisfy the equations of motion i.e flatness of the connection, 
\bea 
da'+a'\wedge a'=0 
\eea
\begin{itemize}
	\item Demanding $c_{r},d_{r}=0$ i.e., no radial components, flatness implies the coefficients $c^{a}$ and $d^{a}$ are independent of $r$. This should not be surprising as this is still in ``radial'' gauge or a primitive connection, where radial dependence has been gauged away just like in the $SL(2)$ sector, 
	\bea
	c_{\mu}^{a}=c_{\mu}^{a}(u,\phi) \qquad d_{\mu}^{a}=d_{\mu}^{a}(u,\phi) 
	\eea
	But the surprising fact that higher spin contribution to the metric is \emph{$r$ independent} as evident from Eq. (\ref{eq:metric components after higher spin corrections}).
	
	\item We futhermore assume all coefficients in the primitive connection to be constants. This is justified because we care to find {\em some} resolution, not the most general resolution of the singularity. The equation of motion for these coefficients are then given by
	\begin{equation}
	[a_u,a_{\phi}] = 0
	\end{equation}
	This gives us following conditions
	\begin{eqnarray}
	c_1 = 0, \ \ c_{-1} =0, \ \ c_0 + Mc_{-2} = 0, \ \ M c_0 + 4c_2 = 0 . 
	\end{eqnarray}
	Coefficients $d_1$ and $d_{-1}$ are not determined by any equation and can be freely choosen to be zero.
\end{itemize}
Next, we impose the holonomy constraints. As before we want the eigenvalues of
$w=2\pi a'_{\phi}$ to be
same as that of Eq. (\ref{eq: Holonomy conditions}).

\begin{enumerate}
	\item The trace condition gives,
	\begin{eqnarray}
	&&\frac{8c_0^2}{3} + 32c_2 c_{-2} =  0, \hspace{1.5in} \\
	&&\frac{16c_0 d_0}{3} + 32 c_{-2} d_2 + 32 c_2 d_{-2} = 0.\nonumber
	\end{eqnarray}
	\item Determinant condition gives,
	
	\begin{eqnarray}
	&&-\frac{16 c_0^3}{27} + 4c_2 + \frac{64c_0 c_2 c_{-2}}{3} - \frac{2c_0 M}{3} + c_{-2}M^2 = 0,\hspace{4cm} \\
	&&-\frac{16c_0^2 d_0}{9} + \frac{64 c_2 c_{-2} d_0}{3} + 4d_2 + \frac{64c_0 c_{-2} d_{-2}}{3} - \frac{4c_0 J}{3} \\ && \hspace{5cm}- \frac{2d_0 M}{3} + 4c_{-2}JM + d_{-2}M^2=0. \nonumber
	\end{eqnarray}
\end{enumerate}
These equations can be consistently solved for various coefficients $c_a$ and $d_a$. Here we list one particular solution which helps in singularity resolution.
\begin{eqnarray}
c_2 = 0, \ \ c_{-2} = 0, \ \ c_0 = 0,
\end{eqnarray}
together with coefficients $d_0$, $d_{2}$ and $d_{-2}$ which are now constrained to obey following relation
\bea\label{d_constraint}
4d_2 - \frac{2d_0 M}{3} + d_{-2}M^2 = 0
\eea
Transforming back to full $r$-dependent gauge, we obtain the metric to be
\begin{equation}
ds^{2} = \mathcal{M}du^{2}-2dudr+2\mathcal{J}dud\phi + \left[ \frac{12d_0^2}{9} + 16d_2 d_{-2} + r^2 \right] d\phi^{2}
\end{equation}
The fact that the collapsing $\phi$-cycle is now stabilized at finite radius is evident from the metric. More concretely, it can also be seen explicitely from the form of Ricci scalar
\begin{equation}\label{resolved_ricci}
R = \frac{24\left(d_0^2 + 12d_2 d_{-2} \right)M}{\left( 4d_0^2 + 48d_2 d_{-2} + 3r^2  \right)}
\end{equation}
which is a non-constant, but everywhere non-singular function of $r$. It can be checked that there exists values of $d_0$, $d_2$ and $d_{-2}$ consistent with \eqref{d_constraint} which makes \eqref{resolved_ricci} non-singular everywhere. The higher spin fields that result from the gauge transformation are also regular everywhere, even though we will not present the details. 

In any event, singularity resolution was only illustrative for our purposes here: our goal was to demonstrate that the Grassmann approach can be a useful technical tool and not merely a curiosity.

\chapter{Chiral Higher Spin Gravity}\label{CHSG}
\section{Introduction}

In gravity theories with a negative cosmological constant, it is natural to consider Anti-deSitter space as the vacuum, and to think of solutions with asymptotically AdS$_3$ boundary conditions as states built on that vacuum. However, the precise choice of the fall-offs one allows in one's definition of ``asymptotically AdS$_3$'' has turned out to be somewhat arbitrary and many consistent choices are known in the literature \cite{Compere:2013bya,Afshar:2016wfy,Troessaert:2013fma,Avery:2013dja,Compere:2015knw,Grumiller:2016kcp}. For example, the most famous of these is the Brown-Henneaux \cite{Brown:1986nw} boundary conditions which corresponds in a certain radial gauge to the choice of left (right) Chern-Simons gauge field with (anti-)holomorphic dependence on the boundary coordinates, along with some restrictions on the charges and chemical potentials that show up in these gauge fields.\\

Recently, Grumiller and Riegler \cite{Grumiller:2016pqb} have written down what is arguably the ``most general'' such AdS$_3$ boundary condition for gravity. By this they mean that {\em all} the charges and chemical potentials that are visible in the Chern-Simons formulation are also visible in the metric formulation. Yet in the asymptotic limit the metric has an AdS$_3$ form, albeit with fall-offs that are more general than the ones found in Fefferman-Graham. They showed that the asymptotic symmetry algebra in this case is two copies of the $sl(2)_k$ current algebra. They accomplished this by working with a choice of radial gauge that was different from the standard radial ``Banados'' gauge. We will call this the Grumiller-Riegler radial gauge.\\

A very natural question to ask in this context is to see whether this can be generalized to higher spins. Can one work with a higher spin theory, turn on all the charges and chemical potentials (including higher spin ones) in the Chern-Simons language and be lead to an asymptotically AdS$_3$ metric, perhaps in the generalized Fefferman-Graham gauge of \cite{Grumiller:2016pqb}? We will see that this is not possible. If we keep all the charges, the metric fall-offs are no longer of the generalized Fefferman-Graham type. This means that we have to remove some of the charges/chemical potentials in a consistent way\footnote{The jargon for this activity is ``reduction''.} and see whether one can get a consistent asymptotic symmetry algebra with acceptable AdS$_3$ fall-offs. In the Banados gauge, one such choice is well known: this is the so-called Drinfeld-Sokolov reduction \cite{Campoleoni:1, Campoleoni2} which when done on both sides truncates left and right sides equally and leads to an asymptotically AdS$_3$ metric in the Fefferman-Graham gauge and a $\cW_3 \times \cW_3$ asymptotic symmetry algebra. Can one allow more general charges and chemical potentials by working with the Grumiller-Riegler radial gauge and allowing the generalized Fefferman-Graham metric?\\

Indeed, in this chapter we will see that if one allows this, one can do {\em much} better. In fact, one can make a Drinfeld-Sokolov reduction on only one side and (self-consistently) set a certain higher spin chemical potential to zero, while letting the other side be fully general: this still leads to the generalized Fefferman-Graham metric. This is a total count of 19 unknown functions between the charges and chemical potentials. One can also check that all of these functions show up in the metric/higher spin fields side of the story as well. Furthermore, we can calculate the asymptotic symmetry algebra and we find that the result is a copy of the $\cW_3$ algebra on the left and an affine $sl(3)_k$ algebra on the right. We suspect that this is the most general chiral higher spin gravity that satisfies these requirements, even though we do not prove it\footnote{One way to disprove this claim is to have an algebra that is ``bigger'' than $\cW_3$ but ``smaller'' than $sl(3)_k$ on the left, and show that there exists a radial gauge where this, together with the right side, leads to an asymptotically AdS$_3$ fall-off. We have a suspicion that an $sl(2)_k$ on the left might also be allowed while having $sl(3)_k$ on the right, but this is a different class of chiral higher spin theory than the one we are looking at here: it does not have higher spin excitations at all on the left. Note that the right side is already as general as can be.}. 

In the final section, we will offer some comparison with the closely related results of \cite{nemani}.

\section{Chiral Spin-3 Gravity}

Taking a cue from the AdS/dS solutions discussed in previous chapters, we can choose a radial gauge of the form

\begin{equation}\label{radial_bc}
A = b^{-1}\mathrm{d}b + b^{-1}a(t,\phi)b, \quad \bA = b\mathrm{d}b^{-1} + b\ba(t,\phi)b^{-1}
\end{equation}
where all the $\rho$ dependence now comes from the group element $b(\rho)$. For empty AdS there are many choices of radial gauge, the usual one being $b = \exp(\rho L_0)$, which we will call the Banados gauge. In \cite{Grumiller:2016pqb}, the group element $b$ was instead chosen to be

\begin{equation}\label{radial_gauge}
b= \exp(L_{-1})\exp(\rho L_0)
\end{equation}
This radial gauge (which we will call the Grumiller-Riegler gauge) manifests all the $sl(2)$ charges and potentials present in the gauge field language, in the metric language as well. For our case, $a$ and $\ba$ are the $sl(3,R)$ Lie algebra valued fields which takes the general form

\begin{eqnarray}\label{gen_connection}
a_t(t,\phi) &=& \mui{-1}L_{-1} + \mui{0}L_0 + \mui{1}L_{1} + \nui{-2}W_{-2} + \nui{-1}W_{-1} \br &\quad \quad &+ \nui{0}W_0 + \nui{1}W_1 + \nui{2}W_2 \br 
a_{\phi}(t,\phi) &=&  \calL{-1}L_{-1} + \calL{0}L_0 + \calL{1}L_1 + \calW{-2}W_{-2} + \calW{-1}W_{-1} \br &\quad \quad &  +\; \calW{0}W_0 + \calW{1}W_1 + \calW{2}W_2 \br
\ba_t(t,\phi) &=& \bmu^{(-1)}L_{-1} + \bmu^{(0)}L_{0} + \bmu^{(1)}L_{1} + \bnu^{(-2)}W_{-2} + \bnu^{(-1)}W_{-1} \br &+& \bnu^{(0)}W_{0} + \bnu^{(1)}W_{1} + \bnu^{(2)}W_{2}\\ 
\ba_{\phi}(t,\phi) &=&  \frac{\pi}{8k}\left( \bcalL^{(-1)} L_{-1} - 2\bcalL^{(0)} L_{0} +  \bcalL^{(1)} L_{1} - \frac{1}{4} \bcalW^{(-2)}W_{-2} + \bcalW^{(-1)}W_{-1} \right. \br &\quad \quad & - \; \left. \frac{3}{2} \bcalW^{(0)}W_{0} + \bcalW^{(1)}W_{1} - \frac{1}{4} \bcalW^{(-2)}W_{-2}  \right) \nonumber
\end{eqnarray}
where $\{L_i,W_j \}$ are the $sl(3,R)$ generators whose algebra is given in Appendix A Eq. \eqref{generators}. The normalization of coefficients of $\ba_{\phi}$ have been chosen for later convenience. The equations of motion impose relationship between the set of potentials $\{\mui{i}, \nui{j} \}$ and the charges $\{\calL{i}, \calW{j} \}$ and is given in the appendix for the unbarred sector while it is identical for the barred sector also. Following \cite{Grumiller:2016pqb}, we hold chemical potentials as fixed functions $\delta a_t = \delta \ba_t =0$ while

\begin{eqnarray}
\delta a_{\phi}(t,\phi) &=& \sum_{i=-1}^{1} \delta \calL{i}(t,\phi)L_i + \sum_{j=-2}^{2} \delta \calW{j}(t,\phi)W_j \br
\delta \ba_{\phi}(t,\phi) &=& \sum_{i=-1}^{1} \delta \bcalL{i}(t,\phi)L_i + \sum_{j=-2}^{2}\delta \bcalW{j}(t,\phi)W_j \br
\end{eqnarray}
The time-like boundary of AdS gives rise to an infinite dimensional phase space with infinitely many global charges. The algebra of charges can be then determined by the Regge-Teitelboim approach \cite{Regge,Banados:1994tn}.

In \cite{Grumiller:2016pqb}, a generalization of Fefferman-Graham gauge for asymptotically AdS metrics was introduced. The motivation of this gauge was to capture all independent $sl(2)$ charges in the metric formulation as well. This metric takes the general form

\begin{eqnarray}\label{ads_metric}
ds^2 &=& d\rho^2 + 2\left(e^{\rho}N^{(0)}_i + N^{(1)}_i + e^{-\rho}N^{(2)}_i + O(e^{-2\rho}) \right)d\rho dx^i \br &+& \left( e^{2\rho}g^{(0)}_{ij} + e^{\rho }g^{(1)}_{ij} + g^{(2)}_{ij} + O(e^{-\rho}) \right)dx^i dx^j
\end{eqnarray}
We call this the \textit{generalized Fefferman-Graham} gauge and various expansion coefficients capture all independent combinations of the chemical potentials and charges in the $sl(2)$ Chern-Simons language. This was the motivation for the choice of radial gauge \eqref{radial_gauge}. 

We take \eqref{ads_metric} as our definition of \textit{asymptotically AdS} metric and construct a higher spin theory which preserves this form of metric. Using \eqref{radial_gauge}, \eqref{radial_bc} and \eqref{metric}, we find that the most general gauge connection \eqref{gen_connection} would violate the metric form \eqref{ads_metric} and therefore the coefficients of the gauge connection have to be restricted. We begin by imposing the Drinfeld-Sokolov (DS) condition on the left gauge connection $a_{\phi}$ in order to further restrict the coefficients. DS reduction in the highest weight gauge amounts to fixing the coefficients of $a_{\phi}$ such that \cite{Campoleoni:1,Campoleoni2,Coussaert:1995zp,Henneaux:2010xg}

\begin{equation}
a_{\phi}(\phi) = L_1 + a^{(-)}_{\phi}(\phi)
\end{equation}
where 

\begin{equation}
[L_{-1}, a^{(-)}_{\phi}(\phi)] = 0
\end{equation}
The above equation fixes the form of $a^{(-)}$ and the reduced gauge field can be now written as

\begin{equation}
a_{\phi}(\phi) = L_1 + \frac{2\pi}{k}{\cal L}L_{-1} -\frac{\pi}{2k} {\cal W}W_{-2}
\end{equation}
As shown by \cite{Campoleoni:1, Campoleoni2}, this restriction helps getting a $\cW_3$ charge algebra. With this restriction on $a_{\phi}$, the equations of motion takes the form

\begin{eqnarray}
\mui{0} + \pp \mui{1} &=& 0 \br 
\cL \mui{1} + 2\cW \nui{2} - \frac{k}{2\pi} \mui{-1} - \frac{k}{4\pi}\pp \mui{0} &=& 0 \br 
\cL \mui{0} + \cW \nui{1} - \frac{k}{2\pi} \pp \mui{-1} + \pt \cL &=& 0 \br 
2\cL \nui{-1} - \cW \mui{0} - \frac{k}{\pi}\pp \nui{-2} - \frac{1}{2}\pt \cW &=& 0 \br 
\cL \nui{0} - \frac{1}{2}\cW \mui{1} - \frac{k}{\pi}\nui{-2} - \frac{k}{4\pi}\pp \nui{-1} &=& 0 \\ 
\cL \nui{1} - \frac{k}{2\pi}\nui{-1} - \frac{k}{6\pi}\pp \nui{0} &=& 0 \br 
\cL \nui{2} - \frac{k}{4\pi}\nui{0} - \frac{k}{8\pi}\pp \nui{1} &=& 0 \br 
\nui{1} + \pp \nui{2} &=& 0 \nonumber
\end{eqnarray}
However, to get the right fall-offs for the metric it turns out that the DS reduction is not enough. But it can be accomplished by specifying the chemical potentials. This is consistent, because chemical potentials are fixed functions that we are allowed to specify. For the metric to be have correct fall-off we need $\nui{2} =0$. This fact combined with the equations of motion (partially) fix other chemical potentials. It turns out that the chemical potentials are now specified by a single function $\mu$. The final left gauge connection can be written as

\begin{eqnarray}\label{gauge_sol_unbarred}
a &=& \left(\mu L_1 - \mu' L_0 + \left( \frac{2\pi}{k}{\cal L}\mu +\frac{1}{2}\mu'' \right)L_{-1} -\frac{\pi}{2k} {\cal W}\mu W_{-2} \right)dt \br &+& \left( L_1 + \frac{2\pi}{k}{\cal L}L_{-1} - \frac{\pi}{2k}{\cal W} W_{-2} \right)d\phi \\
\end{eqnarray}
and the left over equations of motion for $a(t,\phi)$ now gives

\begin{eqnarray}
\dot{\cL} - \mu \cL' - 2\cL \mu' - \frac{k}{4\pi}\mu''' &=& 0 \\
-\dot{\cW} + \mu \cW' + 3\cW \mu' &=& 0
\end{eqnarray}
where dot and prime refers to derivatives with respect to time and $\phi$ respectively. Note that setting $\mu=1$ results in holomorphic dependence of ${\cal L}$ and ${\cal W}$ on the boundary coordinates \cite{Campoleoni:1}.

For the barred sector, no further restrictions are needed to stay in the generalized Fefferman-Graham form of the metric, and therefore $\ba(t,\phi)$ stays in the most general form:

\begin{eqnarray}\label{gauge_sol_barred}
\ba(t,\phi) &=& \left( \bmu^{(-1)}L_{-1} + \bmu^{(0)}L_{0} + \bmu^{(1)}L_{1} + \bnu^{(-2)}W_{-2} + \bnu^{(-1)}W_{-1} \right. \br &+& \left. \bnu^{(0)}W_{0} + \bnu^{(1)}W_{1} + \bnu^{(2)}W_{2}\right) dt + \frac{\pi}{8k}\Bigg( \bcalL^{(-1)} L_{-1} - 2\bcalL^{(0)} L_{0} +  \bcalL^{(1)} L_{1} \br  &-& \frac{1}{4} \bcalW^{(-2)}W_{-2} + \bcalW^{(-1)}W_{-1} - \frac{3}{2} \bcalW^{(0)}W_{0} + \bcalW^{(1)}W_{1} - \frac{1}{4} \bcalW^{(-2)}W_{-2}  \Bigg) \nonumber
\end{eqnarray}
The equations of motion for $\ba(t,\phi)$ is substantially more complicated because of more number of terms and is listed in the appendix (where we suppress the bar's however). The metric is explicitly given by:

\begin{eqnarray} \label{full_metric}
g_{\rho \rho} &=&  1, \quad g_{\rho t} = \frac{\rme^{\rho}}{2}\mu + \left( \mu - \frac{1}{2}\bmu^{(0)} - \frac{1}{2}\mu' \right) + \frac{\rme^{-\rho}}{2}\bmu^{(1)} \br 
g_{\rho \phi} &=& \frac{\rme^{\rho}}{2} + \left( 1- \frac{\pi}{8k}\bcalL^{(0)} \right) +  \frac{\rme^{-\rho}\pi}{16k}\bcalL^{(1)} \br 
g_{tt} &=& \rme^{2\rho} \mu \bmu^{(-1)} - \rme^{-\rho} \mu \bmu^{(0)} + \left[ -\frac{2\pi}{k}\cL \mu^2 - \mu\bmu^{(0)} + \frac{1}{4}\bmu^{(0)2} + \mu \bmu^{(1)}  - \bmu^{(1)}\bmu^{(1)} + \frac{1}{3}\bnu^{(0)2} \right. \br &-& \left. \bnu^{(1)}\bnu^{(-1)} + 4\bnu^{(2)}\bnu^{(-2)} + \frac{1}{2}\bmu^{(0)}\mu' + \frac{1}{4}\mu'^{2} - \frac{1}{2}\bmu^{(1)}\mu''  \right] + \rme^{-\rho} \bmu^{(1)}(2\mu - \mu') \br &+& \rme^{-2\rho}\left( \bmu^{(1)}\mu + \frac{2\pi}{k}\cL \mu \bmu^{(1)} - \bmu^{(1)}\mu' + \frac{1}{2}\bmu^{(1)}\mu''   \right) + \rme^{-4\rho}\frac{2\pi }{k}\cW \mu \bnu^{(2)} \br
g_{t\phi} &=& \frac{\rme^{2\rho}}{2}\left( \frac{1}{4}\bcalL^{(-1)}\mu + \bmu^{(-1)} \right) - \frac{\rme^{\rho}}{2}\left( \frac{1}{4}\bcalL^{(0)}\mu - \bmu^{(0)}   \right) + \frac{\pi}{16k}\left[ - 32\cL \mu + 2\bcalL^{(0)}\mu \right. \br &+& \left. \bcalL^{(1)}\mu - \frac{8k}{\pi}\bmu^{(0)} - \bcalL^{(0)}\bmu^{(0)} + \frac{8k}{\pi}\bmu^{(1)} - \bcalL^{(-1)}\bmu^{(1)} -\bcalL^{(1)}\bmu^{(-1)}- \bcalW^{(0)}\bnu^{(0)} \right. \br &-& \left. \bcalW^{(-1)}\bnu^{(1)} - \bcalW^{(-2)}\bnu^{(2)} - \bcalW^{(1)}\bnu^{(-1)} - \bcalW^{(2)}\bnu^{(-2)} - \bcalL^{(0)}\mu' - \frac{4k}{\pi}\mu'' \right] \br &+& \frac{\rme^{-\rho}\pi}{2k} \left[\frac{2k}{\pi}\bmu^{(1)} + \frac{\bcalL^{(1)}}{4}\left(\mu - \frac{1}{2}\mu' \right)   \right] + \frac{\rme^{-2\rho}\pi}{4k}\left[ \frac{1}{4}\bcalL^{(1)}\mu + \frac{\pi}{2k}\bcalL^{(1)}\cL \mu + \frac{2k}{\pi}\bmu^{(1)} \right. \br &+& \left. 4\cL\bmu^{(1)} - \frac{1}{4}\bcalL^{(1)}\mu' + \frac{1}{8}\bcalL^{(1)}\mu''  \right] + \rme^{-4\rho}\frac{\pi}{k}\cW \left(  \bnu^{(2)} - \frac{\pi}{32k}\bcalW^{(2)}\mu  \right) \\ 
g_{\phi \phi} &=& \rme^{2\rho}\frac{\pi}{8k}\bcalL^{(-1)} + \rme^{\rho}\frac{\pi}{4k}\bcalL^{(0)} + \frac{\pi}{4k}\left[ \bcalL^{(0)}+ \frac{\pi}{16k}\bcalL^{(0)2} + \frac{1}{2}\bcalL^{(1)} - \frac{\pi}{16k}\bcalL^{(1)}\bcalL^{(-1)} \right. \br &-& \left.  8\cL + \frac{3\pi}{64k}\bcalW^{(0)2} -\frac{\pi}{16k}\bcalW^{(1)}\bcalW^{(-1)} + \frac{\pi}{16k}\bcalW^{(2)}\bcalW^{(-2)} \right] + \rme^{-\rho}\frac{\pi}{4k}\bcalL^{(1)} \br &+& \rme^{-2\rho}\frac{\pi}{8k^2}\bcalL^{(1)}\left( k + 2\pi \cL \right) - \rme^{-4\rho}\frac{\pi^2}{16k^2}\bcalW^{(2)}\cW \nonumber
\end{eqnarray}
We omit writing the spin-3 field here to avoid clutter. However we like to remark that the metric and the spin-3 field combined has all the 19 independent functions that appeared in the gauge field. In the next section, we present the asymptotic symmetry algebra for our solution.

\section{Global Charges and their Algebra}
In this section we elucidate the approach of Regge-Teitelboim \cite{Regge}, applied to Chern-Simons theories \cite{Banados:1994tn} to compute the algebra of global charges. Asymptotically AdS spaces have a time-like boundary and the gauge transformations that act non-trivially on the boundary (\textit{i.e.} gauge transformations that do not become identity as $\rho \rightarrow \infty$) are not true gauge transformations but are rather genuine symmetry transformations. On the time-like boundary, these transformations map one solution to another giving rise to a non-trivial boundary phase space.

We begin with a space + time decomposition of the CS gauge field\footnote{Our discussion in this section is for the pure Chern-Simons theory with no extra boundary terms. This corresponds to a Neumann boundary condition in the metric language, see \cite{latest,Pavan,Bala}, which we address in Chapter \ref{BC_gravity}. The variational principle with $a_t$ held fixed that we have used in the previous sections requires the addition of an additional boundary term to the CS action, to be well-defined. But this point will not affect our discussion in this section because it relies only on the Lagrange multiplier term that generates the gauge algebra, which arises from the bulk piece in \eqref{lagrange}.}. We begin by assuming that the three manifold $\calM$ is topologically a solid cylinder with a time-like boundary which is topologically $\partial \calM \simeq \mathbb{R}\times S^1$. The gauge connection can be expressed as 

\begin{equation}
A_{\mu}dx^{\mu} = A_t dt + A_{\rho} d\rho + A_{\phi} d\phi
\end{equation}
The action now takes the form

\begin{equation}
I_{CS} = \frac{k}{4\pi}\int_{\calM}dt d\rho d\phi \bra A_{\phi}\dot{A}_{\rho} - A_{\rho}\dot{A}_{\phi} + 2A_t F_{\rho \phi} \ket + \frac{k}{4\pi}\int_{\partial \calM}dt d\phi \bra A_t A_{\phi} \ket \label{lagrange}
\end{equation}
In the above action we see that $A_t$ is a Lagrange multiplier enforcing the constraint $F_{\rho \phi}=0$ while $A_{\rho}$ and $A_{\phi}$ are the dynamical variables (with the caveat that this theory is topological). Similar story will also hold for the barred sector $\bA_{\mu}$. We first define the smeared generators of gauge transformations

\begin{equation}
G[\Lambda] = \frac{k}{4\pi}\int_{\Sigma}dtd\phi \bra \Lambda F_{t\phi} \ket + Q[\Lambda]
\end{equation} 
where $\Sigma$ is the spatial hypersurface and $Q[\Lambda]$ is a boundary term that is added to make $G[\Lambda]$ a differentiable functional of $A_i, i=\rho, \phi$. For a state independent gauge parameter $\Lambda$, $Q[\Lambda]$ takes the form

\begin{equation}
Q[\Lambda] = -\frac{k}{2\pi} \int_{\partial \Sigma}d\phi \bra \Lambda A_{\phi} \ket
\end{equation}
The generators $G(\Lambda)$ satisfies the Poisson bracket relation

\begin{equation}\label{G-algebra}
\{ G[\Lambda], G[\Gamma] \} = G[[\Lambda, \Gamma]] + \frac{k}{2\pi}\int_{\partial \Sigma}d\phi \bra \Lambda \partial_{\phi}\Gamma \ket
\end{equation}
The second term is a central extension and comes out as a consequence of the surface term $Q[\Lambda]$. $Q[\Lambda]$ does not vanish on-shell, \textit{i.e.} when $F_{ij}=0$ and any transformation for which $Q[\Lambda]\neq 0$ generate global symmetries through 

\begin{equation}\label{gauge_var}
\delta_{\Lambda}F = \{Q[\Lambda],F[A_i] \}
\end{equation}
for any phase space functional $F[A_i]$. The gauge transformations $\Lambda$ that preserve the boundary condition \ref{radial_bc} and \ref{gen_connection} can be written as

\begin{equation}\label{gauge_param}
\Lambda = b^{-1}\lambda(t,\phi) b, \quad \lambda = \ei{i}(t,\phi)L_i + \si{j}(t,\phi)W_j
\end{equation}
where the summation is implied over indices $i$ and $j$. This gives a non-vanishing charge $Q[\Lambda]$

\begin{equation}\label{surface_charge}
Q[\Lambda] = -\frac{k}{2\pi}\int d\phi \bra \lambda a_{\phi} \ket
\end{equation}

Now we are in a position to compute the Poisson algebra of the charges. We begin with the unbarred sector whose coefficients are restricted by the DS highest weight gauge condition. Under the gauge variation \ref{gauge_param}, the connection $a$ transforms by

\begin{equation}
\delta_{\lambda}a = {\rm d}\lambda + [a,\lambda]
\end{equation}

Since the form of the gauge field is fixed, the boundary condition preserving transformations $\Lambda$, given by \eqref{gauge_param},  are now characterized by two functions $\e$ and $\sigma$. The exact form of the gauge transformation parameter and variation of charges can be found in \cite{Campoleoni:1}. The expressions the global charge associated with the transformation is given by

\begin{equation}
Q[\Lambda] = \oint d\phi \left( \e(\phi){\cal L}(\phi) + \sigma(\phi){\cal W}(\phi) \right)
\end{equation}
The charges generate an algebra through the Poisson brackets which is nothing but the classical ${\cal W}_3$ algebra 

\begin{eqnarray}
\{{\cal L}(\phi), {\cal L}(\phi') \} &=& -\left( \delta(\phi-\phi'){\cal L}'(\phi) + 2\delta'(\phi-\phi'){\cal L}(\phi) \right) - \frac{k}{4\pi}\delta'''(\phi-\phi') \\
\{{\cal L}(\phi), {\cal W}(\phi') \} &=& -\left( 2\delta(\phi-\phi'){\cal W}'(\phi) + 3\delta'(\phi-\phi'){\cal W}(\phi) \right) \\
\{{\cal W}(\phi), {\cal W}(\phi') \} &=& -\frac{1}{3} \left( 2\delta(\phi-\phi'){\cal L}'''(\phi) + 9\delta'(\phi-\phi'){\cal L}''(\phi) + 15\delta''(\phi-\phi'){\cal L}'(\phi)\right. \br &+& \left. 10\delta'''(\phi-\phi'){\cal L}(\phi) + \frac{k}{4\pi}\delta^{(5)}(\phi-\phi') \right. \br &+& \left. \frac{64\pi}{k}\left( \delta(\phi-\phi'){\cal L}{\cal L}'(\phi) + \delta'(\phi-\phi'){\cal L}^2(\phi)  \right)   \right)
\end{eqnarray}
with the central charge

\begin{equation}
c= 6k = \frac{3l}{2G_N}
\end{equation}
In terms of the Fourier modes

\begin{equation}
{\cal L}(\phi) = - \frac{1}{2\pi}\sum L_n {\rm e}^{-in\phi}, \quad {\cal W}(\phi) =  \frac{1}{2\pi}\sum W_n {\rm e}^{-in\phi}
\end{equation}
and shifting the zero mode of ${\cal L}$ by

\begin{equation}
L_0 \rightarrow L_0 - \frac{k}{4}.
\end{equation}
The algebra now takes the familiar form

\begin{eqnarray}
i\{ L_m, L_n \} &=& (m-n)L_{m+n} + \frac{c}{12}\delta_{m+n,0} \\
i\{ L_m, W_n \} &=& (2m-n)W_{m+n} \\
i\{ W_m, W_n \} &=& -\frac{1}{3} \left( (m-n)(2m^2 + 2n^2 -mn -8)L_{m+n} + \frac{96}{c}(m-n)\Lambda_{m+n} \right. \br &+& \left. \frac{c}{12}m(m^2 -1)(m^2 -4)\delta_{m+n,0} \right)
\end{eqnarray}
where $\Lambda_n$ is defined as

\begin{equation}
\Lambda_n = \sum_{n \in \mathbb{Z}}L_{m+n}L_{-m}
\end{equation}

Now we return to the barred sector and compute the symmetry algebra. Since there are no restrictions on the coefficients of $\ba_{\phi}$ all the charges transform under the gauge transformation. The boundary preserving gauge transformation is given by
\begin{equation}\label{gauge_param}
\overline{\Lambda} = b \overline{\lambda}(t,\phi) b^{-1}, \quad \overline{\lambda} = \bei{i}(t,\phi)L_i + \bsi{j}(t,\phi)W_j
\end{equation}
The transformation of charges under the above gauge transformation is then given by

\begin{eqnarray}\label{charge_var}
\delta \bcalL^{(0)} &=& \frac{1}{4}\Bigg(\bcalL^{(-1)}\bei{1} -\bcalL^{(1)}\bei{-1}+ \bcalW^{(-1)}\bsi{1} +2\bcalW^{(-2)}\bsi{2} - \bcalW^{(1)}\bsi{-1} \br && \hspace*{5cm} -2\bcalW^{(2)}\bsi{-2}  -\frac{k}{\pi}\partial_{\phi}\bei{0} \Bigg) \br
\delta \bcalL^{(-1)} &=& \frac{1}{4} \Bigg(-\bcalL^{(-1)}\bei{0} - 2\bcalL^{(0)}\bei{-1} - 2\bcalW^{(-1)}\bsi{0} - \bcalW^{(-2)}\bsi{1} - 3\bcalW^{(0)}\bsi{-1} \br && \hspace*{5cm} - 4\bcalW^{(1)}\bsi{-2} + \frac{2k}{\pi}\partial_{\phi}\bei{-1} \Bigg) \br
\delta \bcalL^{(1)} &=& \frac{1}{4} \Bigg( \bcalL^{(1)}\bei{0} + 2\bcalL^{(0)}\bei{1} + 2\bcalW^{(1)}\bsi{0} + 3\bcalW^{(0)}\bsi{1} + 4\bcalW^{(-1)}\bsi{2} \br && \hspace*{5cm} + \bcalW^{(2)}\bsi{-1} + \frac{2k}{\pi}\partial_{\phi}\bei{1} \Bigg) \br
\delta \bcalW^{(2)} &=& \frac{1}{4}\Bigg( 2\bcalW^{(2)}\bei{0} + 4\bcalW^{(1)}\bei{1} - 4\bcalL^{(1)}\bsi{1} - 16\bcalL^{(0)}\bsi{2} - \frac{8k}{\pi}\partial_{\phi}\bsi{2} \Bigg) \\
\delta \bcalW^{(1)} &=& \frac{1}{4} \Bigg( \bcalW^{(1)}\bei{0} + 3\bcalW^{(0)}\bei{1} - \bcalW^{(2)}\bei{-1}  + 2\bcalL^{(1)}\bsi{0} + 2\bcalL^{(0)}\bsi{1} \br && \hspace*{5cm} - 4\bcalL^{(-1)}\bsi{2} + \frac{2k}{\pi}\partial_{\phi}\bsi{1} \Bigg) \br
\delta \bcalW^{(0)} &=& \frac{1}{4}\Bigg( 2\bcalW^{(-1)}\bei{1} - 2\bcalW^{(1)}\bei{-1} + 2\bcalL^{(-1)}\bsi{1} - 2\bcalL^{(1)}\bsi{-1} - \frac{4k}{3\pi}\partial_{\phi}\bsi{0} \Bigg)\br
\delta \bcalW^{(1)} &=& \frac{1}{4}\Bigg(-\bcalW^{(-1)}\bei{0} + \bcalW^{(-2)}\bei{1} - 3\bcalW^{(0)}\bei{-1} - 2\bcalL^{(-1)}\bsi{0} - 2\bcalL^{(0)}\bsi{-1} \br && \hspace*{5cm} + 4\bcalL^{(1)}\bsi{-2} + \frac{2k}{\pi}\partial_{\phi}\bsi{-1} \Bigg) \br
\delta \bcalW^{(2)} &=& \frac{1}{4}\Bigg(-2\bcalW^{(-2)}\bei{0} - 4\bcalW{(-1)}\bei{-1} + 4\bcalL^{(-1)}\bsi{-1} + 16\bcalL^{(0)}\bsi{-2} - \frac{8k}{\pi}\partial_{\phi}\bsi{-2} \Bigg) \nonumber
\end{eqnarray}
Similarly, the $\ba_t$ is fixed, $\delta_{\lambda}\ba_t = 0$ gives the time-evolution of the gauge parameters $\{\bei{i},\bsi{j} \}$. From \eqref{surface_charge}, the charge associated with the above gauge transformation can be computed to be

\begin{eqnarray}
\bar{Q}[\overline{\lambda}] &=& \oint d\phi \left(\bcalL^{(0)}\bei{0} + \bcalL^{(1)}\bei{-1} + \bcalL^{(-1)}\bei{1} + \bcalW^{(2)}\bsi{-2} + \bcalW^{(1)}\bsi{-1}  + \bcalW^{(0)}\bsi{0} \right. \br  &+& \left. \bcalW^{(-1)}\bsi{1} + \bcalW^{(-2)}\bsi{2}  \right)
\end{eqnarray}
Using eq. \eqref{gauge_var} and eq. \eqref{charge_var}, the Poisson algebra of the connections can be determined to be an $sl(3)_{k}$ Kac-Moody algebra \cite{Campoleoni:1}.

\begin{equation}
\{ \ba^{A}_{\phi}(\phi),\ba^{B}_{\phi}(\phi') \} = -\frac{2\pi}{k}\left(\delta(\phi - \phi')f^{AB}_{\;\;\;\;\; C}\ba^{C}_{\phi}(\phi) - \delta'(\phi-\phi')\gamma^{AB}  \right) .
\end{equation}
$f^{AB}_{\;\;\;\;\; C}$ are the structure constants of $sl(3)$ and $\gamma^{AB}$ is the inverse of the $sl(3)$ Killing metric. The $sl(3)_k$ algebra can be written in a more familiar form by Fourier decomposing the gauge connection 

\begin{eqnarray}
\ba^{A}(\phi) = \frac{1}{k}\sum_{n \in \mathbb{Z}}\ba^{A}_n \mathrm{e}^{-i n\phi}
\end{eqnarray}
which gives

\begin{equation}
\{ \ba^{A}_n, \ba^{B}_m \} = -f^{AB}_{\;\;\;\;\; C} \ba^{C}_{n+m} + in\gamma^{AB}\delta_{n+m,0}
\end{equation}
Thus our solution presents a $\cW_3 \times sl(3)_k$ as its asymptotic symmetry algebra.

\section{Comments}

In this concluding section, we briefly contrast our work with previous results. In an interesting paper \cite{nemani} Poojary and Suryanarayana made the following choice of the bare gauge field:
\begin{eqnarray}\label{nemani_gauge1}
a &=& \left( L_1 - \kappa L_{-1} - \omega W_{-2} \right)dt + \left( L_1 - \kappa L_{-1} - \omega W_{-2} \right)d\phi \\
\ba &=& \left(  -L_{-1} + \tilde{\kappa} L_{1} + \tilde{\omega} W_{2} + \sum^{1}_{a=-1}f^a L_a + \sum_{b=-2}^{2} g^b W_b \right)dt \br &+& \left( L_{-1} - \tilde{\kappa} L_{1} - \tilde{\omega} W_{2} + \sum^{1}_{a=-1}f^a L_a + \sum_{b=-2}^{2} g^b W_b \right)d\phi\label{nemani_gauge2}
\end{eqnarray}
In our language, this amounts to the following restrictions on the charges and the chemical potential: 
\begin{eqnarray}
\mu &=& 1,\;\; \cL = -\frac{k}{2\pi}\kappa,\;\; \cW = \frac{2k}{\pi}\omega,\;\; \br 
\bmu^{(1)} &=& \tilde{\kappa} + f^{1},\;\; \bmu^{(-1)} = f^{-1} -1,\;\;\bmu^{(0)} = f^0 \\ 
\bnu^{(2)} &=& g^2 + \tilde{\omega},\quad \bnu^{(a)} = \bcalW^{(a)} = g^a, a\neq 2 \br 
\bcalL^{(1)} &=& \frac{8k}{\pi}(f^1 - \tilde{\kappa}),\;\; \bcalL^{(-1)} = \frac{8k}{\pi}(1+f^{-1}), \;\; , \bcalL^{(0)} = -\frac{4k}{\pi}f^0, \;\; \br \bcalW^{(2)} &=& -\frac{32k}{\pi}(g^2 - \tilde{\omega}),\;\; \bcalW^{(-2)} = -\frac{32k}{\pi}g^{-2}, \;\; \bcalW^{(0)} = -\frac{16k}{3\pi} g^0, \br \bcalW^{(\pm1)} &=& \frac{8k}{\pi}g^{\pm 1}
\end{eqnarray}
Note that there are 12 independent functions in this case as opposed to our 19. Furthermore imposing the equations of motion makes $\kappa$ and $\omega$ holomorphic in the boundary coordinates. They showed that with this choice the asymptotic charge algebra is $\cW_3 \times sl(3)_k$. 

Since the charge algebra depends only on the bare gauge field, our construction is an explicit demonstration that the restriction of \cite{nemani} is not necessary if one's goal is to reproduce the $\cW_3 \times sl(3)_k$ algebra: the most general gauge field on the right side, together with a somewhat more general gauge field on the left (see Section 3 for details), will still do the job\footnote{We would also like to bring to attention the work of \cite{Ponomarev:2016lrm} who have considered chiral higher spin fields in flat space}.

Another comment worth making is that even with the restricted form \eqref{nemani_gauge1}, \eqref{nemani_gauge2} of \cite{nemani}, one can check that the metric does {\em not} have the typical Fefferman-Graham fall-off, neither in the Banados radial gauge that \cite{nemani} are working with (we show this in an Appendix), nor in the Grumiller-Riegler radial gauge that we use (which is a corollary of our results in section 3).  To make sense as an asymptotically AdS space, one {\em must} think of the metric in the generalized Fefferman-Graham gauge\footnote{This issue is possibly moot however if one views asymptotic AdS$_3$ in higher spin theories as a not-very-meaningful idea in the metric formulation. If one adopts such a point of view, there is nothing stopping one from turning on all the charges on either side and the resulting charge algebra would be two copies of $sl(3)_k$. It should be kept in mind however, that part of the motivation for the Drinfeld-Sokolov reduction choice in \cite{Campoleoni:1} was that the metric had the usual (asymptotic) AdS$_3$ form even with higher spins turned on. This is our motivation for taking the metric (somewhat) seriously even with higher spins turned on.}. We have checked that our general field configuration leads to this metric in both Banados and Grumiller-Riegler radial gauges. 

\chapter{Neumann Boundary Condition in Gravity and Holography}\label{BC_gravity}
\section{Introduction}

To derive equation of motion from an action using a variational principle (see e.g., \cite{Hint}), we need to make sure that the boundary terms arising from the variation vanish. For two-derivative theories, the boundary terms that arise from the variation of the action typically contain variations of both the field and its normal derivative at the boundary. Holding both fixed at the boundary will trivially get rid of these terms, but it will also remove most of the interesting dynamics. Instead, what one tries to do is to add {\em boundary terms} to the action such that the total boundary variation after the addition of these new terms depends either only on the field variation, or only on the normal derivative variation. When we can find a boundary term to accomplish this, we say that we have a Dirichlet problem (in the former case) or a Neumann problem (in the latter). In such a situation, we have a well-defined variational problem upon demanding that the field (for Dirichlet) or its normal derivative (for Neumann) be held fixed at the boundary.\\

In the case of gravity, it has been known since \cite{GH, Y} that there exists a boundary term one can add to the bulk Einstein-Hilbert action to make the Dirichlet problem for the metric well-defined. This is the Gibbons-Hawking-York (GHY) term. But the Neumann problem as we have stated in the above paragraph does not seem to be well-defined for gravity. This is because, (to the best of our knowledge) no one has written down a boundary term to be added to the Einstein-Hilbert action such that {\em just} holding the normal derivatives fixed at the boundary kills all boundary terms that arise from the variation of the total action \footnote{See for example \cite{Stefan} for some explicit expressions for the variations in the context of three dimensions; the structure in general dimensions is similar.}. This is striking, since the GHY term has been around for forty years. 

In our work, we take an alternative view on the Neumann problem. Instead of holding the normal derivative of the metric fixed at the boundary, we will seek a variational problem where the functional derivative of the action with respect to the boundary metric is held fixed. In the case of classical particle mechanics, these two formulations are equivalent because the former corresponds to holding the velocity fixed at the boundary, whereas the latter corresponds to holding the momentum fixed. But we will see that in the case of gravity, the latter formulation results in a drastic simplification, and we can indeed write down a boundary term that makes this type of a Neumann problem well-defined.\\

On a related issue, in the usual AdS/CFT correspondence \cite{Maldacena:1997re, GKP, Witten:1998qj}, the boundary values of fields on the gravity side are identified as the sources of the fields in the field theory. Thus AdS/CFT correspondence is formulated as a Dirichlet problem as well (on the gravity side). Typically, to get a finite action on solutions, one has to take care of infrared divergences of the Einstein-Hilbert action in both flat space and in AdS. This is true even with the addition of boundary terms that make the variational problem well-defined. In flat space, this was done for the GHY boundary term in \cite{GH,Y} and for the Neumann term in \cite{latest} via appropriate background subtraction procedures. In AdS however, for the Dirichlet problem, there exists a well-defined and quite natural way to get finite actions by the addition of counter-terms \cite{Balasubramanian:1999re, deHaro}, which have a very natural interpretation in the dual field theory as canceling UV divergences. Such counter-terms lead to a finite action and a finite (renormalized) stress tensor. The existence of this finite stress tensor suggests that in AdS, one can define the Neumann variational problem to be one where we hold the renormalized stress tensor density fixed, and one should get a well-defined variational principle and finite Neumann action. We can do this in two ways: we can do this via starting from the renormalized Dirichlet action in AdS (which is well-known from, say, \cite{deHaro}) and do a Legendre transform on the boundary metric, or we can start from a Fefferman-Graham expansion as the definition of asymptotically AdS space, and systematically construct counter-terms for the un-renormalized Neumann action by demanding vanishing of divergences. In the next section, we will adopt the latter strategy and write down explicit renormalized Neumann actions in AdS$_{d+1}$ with $d=2,3,4$. Remarkably, we will find that both these approaches yield the same results.\\

The organization of this chapter is as follows. In Sec. \ref{Sec:Dirichlet} we recapitulate the standard Dirichlet problem in gravity, including the Gibbons-Hawking-York (GHY) term and variational principle. In Sec. \ref{Sec:Neumann}, we proceed with constructing the Neumann boundary term through a boundary Legendre transform. In Sec. \ref{Sec:VariousD} and \ref{Sec:Microcanonical} we comment about the form of Neumann action in various dimensions and its relation to a previous work of Brown and York \cite{Brown}.

\section{Dirichlet problem}\label{Sec:Dirichlet}

We will start by reviewing the Dirichlet problem for gravity. Everything in this section is well-known, but we want to write the Dirichlet action in a form that is suitable for moving to Neumann. The Einstein-Hilbert action in $(d+1)$-dimensions with a cosmological term is given by\footnote{Our notations and conventions are that of \cite{poisson}}

\begin{equation}\label{eh_action}
S_{EH} = \frac{1}{2\kappa}\int_{\cM}d^{d+1}x \sqrt{-g}(R-2\Lambda)
\end{equation}
where $\kappa=8\pi G$. Variation of Einstein-Hilbert action yields

\begin{equation}
\delta S_{EH} = \frac{1}{2\kappa}\int_{\cM}d^{d+1}x \sqrt{-g}(G_{ab}+\Lambda g_{ab})\delta g^{ab} -\frac{1}{\kappa}\int_{\partial \cM}d^{d}y \sqrt{|\gamma|} \varepsilon \left( \delta \Theta + \frac{1}{2}\Theta^{ij}\delta \gamma_{ij} \right)
\end{equation}
where $G_{ab} = R_{ab}-\frac{1}{2}R g_{ab}$ is the Einstein tensor, $\gamma_{ij} = g_{ab}e^{a}_{i}e^{b}_{j}$ is the induced metric on the boundary $\pd \cM$ and $e^{a}_{i} = \frac{\partial x^a}{\partial y^i}$ is the coordinate transformation relating the boundary coordinates $y^i$ to the bulk coordinates $x^a$, and $\Theta = \gamma^{ij}\Theta_{ij}$ is the trace of the extrinsic curvature. $\varepsilon$ distinguishes the space-like and time-like hypersurfaces  and takes values $\varepsilon = \pm1$ for time-like and space-like boundaries respectively. We also assume that the boundaries are not null. The extrinsic curvature is defined as

\begin{equation}
\Theta_{ij} = \frac{1}{2}(\nabla_a n_b + \nabla_b n_a) e^{a}_{i}e^{b}_{j}
\end{equation}
where $n_a$ is the unit normal to the boundary. The variational principle is spoiled by the offending surface term which does not vanish for a fixed boundary metric $\gamma_{ij}$. Therefore we need to add a boundary term to \eqref{eh_action} to make the variational principle well defined. This boundary piece is the Gibbons-Hawking-York term
 
\begin{equation}
S_{GHY} = \frac{1}{\kappa}\int_{\partial \cM}d^{d}y \sqrt{|\gamma|} \varepsilon \Theta
\end{equation}
whose variation is given by 
 
\begin{equation}
\delta S_{GHY} = \frac{1}{\kappa}\int_{\partial \cM}d^{d}y \sqrt{|\gamma|}\varepsilon \left(\delta \Theta + \frac{1}{2}\Theta \gamma^{ij}\delta \gamma_{ij} \right)
\end{equation}
Therefore the variation of total gravitational action yields
 
\begin{eqnarray}\label{dirichlet_variation}
\delta S_{D} &=& \delta S_{EH} + \delta S_{GHY} = \frac{1}{2\kappa}\int_{\cM}d^{d+1}x \sqrt{-g}(G_{ab}+\Lambda g_{ab})\delta g^{ab} \\ \nonumber &-& \frac{1}{2\kappa}\int_{\partial \cM}d^{d}y \sqrt{|\gamma|} \varepsilon  \left( \Theta^{ij} - \Theta \gamma^{ij}  \right)\delta \gamma_{ij}
\end{eqnarray}
Thus we find that the action $S_{D}$ is stationary under arbitrary variations of the metric in the bulk provided we satisfy the bulk equations of motion and the variations vanish on the boundary. 
 
For use in the next subsection, we define the canonical conjugate of the boundary metric as,
 
\begin{equation}
\pi^{ij} \equiv \frac{\delta S_{D}}{\delta \gamma_{ij}} = -\frac{\sqrt{|\gamma|}}{2\kappa} \varepsilon (\Theta^{ij}-\Theta \gamma^{ij})
\end{equation}
The variation of \eqref{dirichlet_variation} can thus be expressed as
 
\begin{equation}\label{variation}
\delta S_{D} = \frac{1}{2\kappa}\int_{\cM}d^{d+1}x \sqrt{-g}(G_{ab}+\Lambda g_{ab})\delta g^{ab} + \int_{\partial \cM}d^{d}y\; \pi^{ij} \delta \gamma_{ij}
\end{equation}

\section{Neumann problem}\label{Sec:Neumann}
We want to write down a variational principle where instead of holding the metric $\gamma_{ij}$ fixed at the boundary, we can hold $\pi^{ij}$ fixed \footnote{Note that in particle mechanics, holding $\dot q$ fixed and holding $p$ fixed at the boundary are identical because of the simple ${\dot q}^2$ form of the kinetic term, which guarantees that $p(q,\dot q)=\dot q$. Note that both in particle mechanics as well as in our case for gravity, we are using the suggestive symbols $p$ and $\pi^{ij}$, but thinking of them as functions of $(q, \dot q)$ and $(\gamma_{ij}, \partial_a \gamma_{ij})$ respectively.}. 
It is easy to see from (\ref{variation}) that this can be easily accomplished by adding yet another term to the Dirichlet action of the previous section. The form suggested by (\ref{variation}) is

\begin{equation}
S_{N} = S_{D} -\int_{\partial {\cM}} d^{d}y \ \pi^{ij}\gamma_{ij}
\end{equation}
It is trivial now to check that the variation of $S_{N}$ is

\begin{eqnarray}
\delta S_{N} =  \frac{1}{2\kappa}\int_{{\cM}}d^{d+1}x \sqrt{-g}(G_{ab}+\Lambda g_{ab})\delta g^{ab} - \int_{\pd {\cM}}d^{d}y\  \delta \pi^{ij} \gamma_{ij}\nonumber
\end{eqnarray}
guaranteeing that holding $\pi^{ij}$ fixed at the boundary yields a well-defined variational problem. Explicitly, our Neumann action is given by

\begin{eqnarray}
S_{N} & = & S_{EH}+S_{N_b} \\
& \equiv & S_{EH} 
+ \frac{(3-d)}{2\kappa}\int_{\partial {\cal M}}d^{d}y \sqrt{|\gamma|}\varepsilon K \label{NeumannA}
\end{eqnarray}
We will discuss features, applications and extensions of this action. We will see that it is natural one to consider from many different angles.

\section{d=1, 2, 3}\label{Sec:VariousD}

In two dimensions, our boundary term is identical to the GHY boundary term. This might seem puzzling, but is easy to understand by writing the metric in conformal gauge. The only degree of freedom is the conformal factor in the metric $ ds^2=e^{2 \phi (t,r)} (-dt^2+dr^2)$
and in terms of $\phi$, the Einstein-Hilbert action takes the form 

\begin{eqnarray}
S_{EH} \sim \int dt\ dr (\ddot \phi-\phi'')
\end{eqnarray}
Variation of this results in boundary terms of the form $\int dt \ \delta \phi'|_{\Lambda}$. The GHY term is designed precisely to cancel this when varied, and unlike in higher dimensions this form is trivial enough that its variation kills off the entire boundary piece. We have taken the boundary to be timelike and put it at $r=\Lambda$ for definiteness, but the discussion is obviously analogous for a spacelike boundary. \\

In more than two dimensions, the Dirichlet and Neumann boundary terms are always distinct. In three dimensions, our Neumann boundary term reduces to $S_{N_b}=\frac{1}{2\kappa}\int_{\partial {\cal M}}d^{2}y \sqrt{|\gamma|}\varepsilon K$, which is ``one-half" the Gibbons-Hawking term.  Remarkably, this boundary term has previously appeared in the literature for other reasons both in the flat case \cite{Bagchi:2013lma,Stefan} as well as  in AdS \cite{Banados:1998ys, Miskovic:2006tm}. 

Especially in light of holography in AdS$_3$ and flat 2+1 dimensional space, there is much to be said here. But we will leave that for future work, and make only one brief comment: the Einstein-Hilbert action leads to a boundary piece

\begin{eqnarray}
S^{CS}_{B}\sim \int_{\partial {\cal M}} {\rm Tr} (A \wedge \bar A)
\end{eqnarray}
in the Chern-Simons formulation (see eg., \cite{Kiran, KRNEW} for an elementary discussion of Chern-Simons gravity in 2+1 dimensions; we work with the AdS case for concreteness). It is straightforward to check that the ``one-half" Gibbons-Hawking term precisely gets rid of this boundary piece \footnote{See for example \cite{Apolo}, the discussion at the beginning of section 2.1. The ``one-half" GHY term is not discussed there, but our claims about it follow trivially from the expressions there.}. Therefore in 2+1 dimensions, our boundary term makes the metric formulation of gravity translate precisely into the bulk Chern-Simons action, with {\em no} boundary term at all. A natural choice of boundary conditions \cite{Banados:1998ys} in the Euclidean geometry is then to set $A_z=0$ or $A_{\bar z}=0$ at the boundary (here $z \sim (t+ix)$ where $t$ and $x$ are boundary coordinates). These boundary conditions are satisfied by AdS$_3$ and the BTZ black hole.  

In four dimensions, our boundary term identically vanishes. Thus, we come to a perhaps surprising conclusion: standard Einstein-Hilbert gravity in four dimensions, {\em without} boundary terms, has an interpretation as a Neumann problem. 

\section{Microcanonical Gravity}\label{Sec:Microcanonical}

In \cite{Brown}, a microcanonical definition of the gravitational path integral was proposed. The basic idea there was to add new boundary terms to the gravitational field, so that the energy surface \footnote{The term ``surface" arises because \cite{Brown} work in four dimensions, but the approach is more general.} density and the momentum surface density are held fixed at the boundary, in the definition of the variational principle. One can then use this action to define a microcanonical path integral for gravity which has some pleasing properties. 

Even though \cite{Brown} does not emphasize it, it is easy to see that the surface energy/momentum densities that they hold fixed are just some of the components of the energy momentum surface density, which is the quantity we have held fixed in defining our Neumann variational problem. The approach of \cite{Brown} results in the somewhat awkward action, eqn. (3.13) in their paper. However, despite this, since charges are best defined globally in general relativity, we feel that the approach of \cite{Brown} is a very interesting one. This is one of the motivations for defining a more ``covariant" variational problem where it is the whole stress-energy tensor density on the boundary  that is held fixed. Happily, this also turns out to have a Neumann interpretation. 

The work of \cite{Brown} was before the era of AdS/CFT, and in hindsight, we believe this approach is an even more interesting one in the AdS/CFT context. Fixing the boundary stress energy tensor gives us a natural definition for a microcanonical approach to AdS/CFT. We will be reporting on this in greater detail elsewhere \cite{Bala}, but we briefly outline some of the results here. 

The natural boundary stress tensor in AdS/CFT is the one introduced by \cite{Balasubramanian:1999re} (see also, for example \cite{deHaro}). These stress tensors are obtained by adding further boundary terms (which have a natural interpretation as counter-terms in the holographic language) while demanding that the on-shell action be finite \footnote{The original action has a bulk IR divergence.}. This gives rise to the finite {\em renormalized boundary stress tensors} found in \cite{Balasubramanian:1999re, deHaro}. 

Together with the results of Sec. \ref{Sec:Neumann}, this suggests that a natural object to hold fixed while doing a Neumann variation in the AdS/CFT context is the renormalized stress tensor density, not the bare stress tensor density. Indeed, it is possible to show \cite{Bala} that one can add appropriate counter-terms to our Neumann action (\ref{NeumannA}) so that:

(a) the variation of the total action leads to the {\em renormalized} stress tensor of \cite{Balasubramanian:1999re, deHaro} (up to an ambiguity in odd dimensions \cite{Bala}), 

(b) the variational principle is well-defined when one holds this renormalized boundary stress tensor density fixed, and 

(c) the total on-shell action is finite. \\

A detailed discussion of these issues in various dimensions will be presented in upcoming sections following \cite{Bala}. There, we will also argue that AdS/CFT is the natural context for discussing, extending and finding applications for the (stress-tensor version of the) microcanonical path integral approach of \cite{Brown}

\section{Holographic Renormalization of Neumann Gravity}\label{Sec:Holo_Ren}

In this section we will derive the renormalized Neumann action by directly dealing with the Fefferman-Graham expansion (\ref{FG}) and demanding that the action be finite. Typically in Dirichlet theory one imagines that the boundary conditions are set by the leading part of the FG expansion, in our case it is a combination of the $g_i$'s (see (\ref{FG}) that is getting fixed via the renormalized boundary stress tensor. A standard review is \cite{skn}.

\subsection{Regularized Action in Fefferman-Graham Coordinates}

By asymptotically AdS$_{d+1}$ space, we mean a metric that solves the Einstein equation with a negative cosmological constant, that can be expressed asymptotically (\textit{i.e.}, as $z \rightarrow 0$) by a general Fefferman-Graham expansion given by

\begin{equation}
ds^2 = G_{\mu \nu}dx^{\mu} dx^{\nu} = \frac{l^2}{z^2}\left( dz^2 + g_{ij}(x,z)dx^i dx^j \right)
\end{equation}
where

\begin{equation}
g(x,z) = g_{0} + z^2 g_2 + \cdots + z^d g_d + z^d \log z^2\; h_d + O(z^{d+1}).\label{FG}
\end{equation}
Only even powers of $z$ appear up to $O(z^{[d-1]})$. The log term appears only for even $d$. In all the discussions that follow, we set $l=1$. The cosmological constant is related to the AdS radius through the relation $\Lambda = - \frac{d(d-1)}{2l^2}$. Since only even powers appear in the expansion, we introduce a new coordinate $\rho = z^2$ in which the metric takes the form

\begin{eqnarray}\label{asymptotic_metric}
ds^2 &=& \frac{d\rho^2}{4\rho^2} + \frac{1}{\rho}g_{ij}(x,\rho)dx^i dx^j \\ \nonumber
g(x,\rho) &=& g_0 + \rho g_2 + \cdots + \rho^{d/2} g_d + \rho^{d/2}\;\log \rho \; h_d 
\end{eqnarray}
Note that the condition that this metric solves the Einstein equation means that the higher order $g_{(m)ij}$ can be determined in terms of the lower order ones, and explicit formulas can be written down for them. We present explicit expressions in an Appendix. We can compute the Neumann action \cite{latest,Pavan} (note that \cite{Pavan} worked with the bulk dimension, so our $d=D-1$ in the notation there),
\begin{equation}\label{neumann_action}
S_{N} = \frac{1}{2\kappa}\int_{\cM}d^{d+1}x \sqrt{-g}(R-2\Lambda) - \frac{(d-3)}{2\kappa}\int_{\partial \cM}d^{d}y \sqrt{|\gamma|}\varepsilon \Theta
\end{equation}
for (\ref{asymptotic_metric}) and we immediately sees that it diverges. This is not a surprise: the same thing happens for the Dirichlet action as well, and the process of adding counter-terms to the Dirichlet action to make it finite is known as holographic renormalization \cite{deHaro}. We can adopt a similar approach here. The first step is to cut-off the radial integration at a finite $\rho=\epsilon$, to regulate the action. After this regularization, the Neumann action (\ref{neumann_action}) is given by
\begin{eqnarray}
S_{N}^{reg} &=& -\frac{d}{2\kappa}\int d^d x \int_{\epsilon}d\rho \; \frac{1}{\rho^{d/2 +1}}\sqrt{-g} \br  &-&\frac{(d-3)}{2\kappa}\int d^d x \left. \frac{1}{\rho^{d/2}}\left( d\sqrt{-g} - 2 \rho \partial_{\rho}\sqrt{-g} \right) \right|_{\rho = \epsilon}
\end{eqnarray}

Our goal is to add counter-terms so that the Neumann action becomes finite. We will find that this is indeed a natural construction and for standard black hole solutions it leads to the same on-shell action as the Dirichlet theory.

\subsection{AdS$_3$ ($d=2$)}

In $d=2$ the regularized Neumann action takes the form,

\begin{eqnarray}
S_{N}^{reg} &=& -\frac{1}{\kappa}\int d^2 x \left[ \int_{\epsilon} d\rho \frac{\sqrt{-g}}{\rho^2} + \left. \left( -\frac{\sqrt{-g}}{\epsilon} + \partial_{\rho}\sqrt{-g} \right) \right|_{\rho = \epsilon} \right]
\end{eqnarray}
Using the expansion for the determinant \eqref{detexpansion} and doing the $\rho$ integral, we arrive at following final form for the regulated action

\begin{equation}
S^{reg}_N = \frac{1}{2\kappa}\int d^2 x \sqrt{-g_0} \log \epsilon \; \Tr\; g_{2} \label{2log}
\end{equation}
In this work, we will ignore this Logarithmic divergence, because it will not be relevant for the situations we consider, like black holes. This is similar to the approach of \cite{Balasubramanian:1999re} and we would like to write down counter-terms parallel to theirs in terms of the induced metric. The logarithmic divergence in the Dirichlet case were presented later in \cite{deHaro}. We emphasize however that even though we do not use them, our presentation of logarithmic divergences is complete: the expressions for the quantities involving $g_2$ in \eqref{2log}, \eqref{4log} in terms of curvatures of the boundary metric $g_0$ are presented in an Appendix. Note however that unlike the other counter-terms, we cannot absorb the cut-off dependence of the logarithmic divergence entirely into expressions involving the induced metric; a logarithmic cut-off dependence will remain. This is unavoidable, and this is the form in which \cite{deHaro} also leave their results, see their equation (B.4), last term. The renormalized quantities are of course cut-off independent by construction. 

Once we ignore the logarithmic term, the renormalized Neumann action is therefore identical to the original Neumann action $S_{N}$ in three dimensions: no counter-terms are required to render the action finite.  
\bea
S^{ren}_N=S_N
\eea
This was an observation that was already made in a slightly different language in \cite{Stefan, Banados:1998ys}, as a special observation about three dimensions. From our perspective, the fact that the bare action is already finite in 2+1 dimensions is the crucial reason why their construction works.

Now we come to one crucial observation. The renormalized stress-tensor in 2+1 dimensions is given by \cite{Balasubramanian:1999re}:

\begin{equation}
T^{ren}_{ab} = \frac{1}{\kappa}\left[ \Theta_{ab} - \Theta \gamma_{ab} + \gamma_{ab}\right]
\end{equation}
We will now show that the renormalized Neumann action (which coincidentally happens to be the same as the bare Neumann action in 2+1 dimensions\footnote{This coincidence of the renormalized and the bare Neumann actions is a feature of 2+1 dimensions and does not hold in higher dimensions, but the statements we make about the renormalized action apply in higher dimensions as well.}) gives rise to a well-defined variational principle when we demand that the renormalized boundary stress tensor density is held fixed. This means that, given the renormalized stress-tensor as our boundary data, we have a well defined variational principle. 

To show this, first note that in three dimensions,
\begin{eqnarray}\label{ads3_Nvariation}
\delta S^{ren}_N &=& \delta S_N=\delta \left[ \frac{1}{2\kappa}\int_{\mathcal{M}} d^3 x\;\sqrt{-g}(R-2\Lambda) +\frac{1}{2\kappa}\int_{\partial \mathcal{M}}\;d^2 x\;\sqrt{-\gamma} \Theta \right] \\ \nonumber &=& \frac{1}{2\kappa}\int_{\mathcal{M}} d^3 x\;\sqrt{-g}(G_{ab}+\Lambda g_{ab})\delta g^{ab} - \int d^2 x \left[ \delta \left( -\frac{\sqrt{-\gamma}}{2\kappa}(\Theta^{ab}-\Theta \gamma^{ab}) \right)\gamma_{ab} \right]
\end{eqnarray}
The bare stress-tensor is defined as

\begin{equation}
T^{bare}_{ab}=\frac{1}{\kappa}\left[ \Theta_{ab} - \Theta \gamma_{ab} \right]
\end{equation}
The surface term in \eqref{ads3_Nvariation} can be thus expressed as

\begin{equation}
\delta \left( -\frac{\sqrt{\gamma}}{2\kappa}(\Theta^{ab}-\Theta \gamma^{ab}) \right)\gamma_{ab} = \delta \left(\frac{\sqrt{-\gamma}}{2}T^{bare\;ab} \right)\gamma_{ab}
\end{equation}
Now by an explicit calculation, we can see that

\begin{equation}
\delta \left(\frac{\sqrt{-\gamma}}{2}T^{bare}_{ab} \right)\gamma^{ab} = \delta \left(\frac{\sqrt{-\gamma}}{2}T^{ren}_{ab} \right)\gamma^{ab}\label{ambig}
\end{equation}
This shows that the Neumann variational problem of the renormalized action might as well be formulated by holding the renormalized boundary stress tensor density fixed. This arises because in formulating the variational problem one has the freedom to add a $\chi_{ab}$ to the stress tensor that one is holding fixed at the boundary as long as it satisfies
\bea
\delta\Big(\sqrt{-\gamma} \chi_{ab}\Big) \ \gamma^{ab}=0\label{ambig1}
\eea
We will see that in odd $d$ dimensions, this ambiguity in practice does not arise because the variational problem of Neumann type for the renormalized action essentially automatically leads to the renormalized stress tensor. We turn now to demonstrate this in four dimensions. 

\subsection{AdS$_4$ ($d=3$)}

In $d=3$, the singular part of regularized action evaluates to

\begin{eqnarray}\label{ads4_ren_action}
S_{N}^{reg} &=& -\frac{3}{2\kappa}\int d^3 x \int_{\epsilon} d\rho \frac{\sqrt{-g}}{\rho^{5/2}} \\ \nonumber &=&  -\frac{1}{\kappa}\int d^3 x \sqrt{-g_0} \left( \frac{1}{\epsilon^{3/2}} + \frac{3}{2 \epsilon^{1/2}}\; \Tr\; g_2    \right)
\end{eqnarray}
where we have once again used the determinant expansion \eqref{detexpansion}. The determinant of the induced metric $\gamma_{ab}$ can be expressed as

\begin{equation}\label{h-g_relation}
\sqrt{-\gamma} = \frac{\sqrt{-g}}{\epsilon^{d/2}}
\end{equation}
This, together with \eqref{g2traces} allows us to write the counter-term action

\begin{equation}\label{threed_counterterm}
S^{ct} = \frac{1}{\kappa}\int d^3 x\; \sqrt{-\gamma}\left(1- \frac{1}{4}R[\gamma] \right)
\end{equation}
The fact that this is the correct counter-term can be checked by expanding \eqref{threed_counterterm} in the Fefferman-Graham expansion order by order and using \eqref{detexpansion} and \eqref{g2traces}. The renormalized Neumann action, in a notation analogous to that in \cite{Balasubramanian:1999re}, is thus given by

\begin{equation}
S^{ren}_N = \frac{1}{2\kappa}\int_{\mathcal{M}} d^4 x \sqrt{-g}\;(R-2\Lambda) + \frac{1}{\kappa}\int d^3 x\; \sqrt{-\gamma}\left(1- \frac{1}{4}R[\gamma] \right)
\end{equation}
Including this counter-term and doing variations, we also reproduce the stress-tensor of \cite{Balasubramanian:1999re, deHaro}

\begin{equation}
T^{ren}_{ab} = \frac{1}{\kappa}\left[ \Theta_{ab} - \Theta \gamma_{ab} + 2 \gamma_{ab} - G_{ab} \right] \label{4dRenT}
\end{equation}
where $G_{ab}=R_{ab}[\gamma]-\frac{1}{2}R[\gamma] \gamma_{ab}$ is the Einstein tensor of the induced metric\footnote{More precisely, what we reproduce is $\delta T^{ren}_{ab}$ from the variational problem for the renormalized Neumann action. But unlike in odd $d$, this leads directly to (\ref{4dRenT}) and we do not need to use the ambiguity of the type (\ref{ambig1}).}. This stress tensor is known for empty AdS and AdS black hole to be finite and also has the right leading fall-offs to reproduce the correct finite charges for the AdS black hole.

This shows again that the renormalized Neumann action leads to a well-defined variational problem when holding the renormalized boundary stress tensor fixed.

\subsection{AdS$_5$ ($d=4$)}

For the case of $d=4$, the divergent part of the action evaluates to

\begin{eqnarray}\label{ads5_onshellaction}
S_{N}^{reg} &=&  -\frac{2}{\kappa}\int d^4 x \sqrt{-g_0} \left( \frac{3}{2\epsilon^{2}} + \frac{3}{4 \epsilon}\; \Tr\; g_2  - \log \epsilon \;\frac{1}{8} \left( (\Tr (g_2))^2 - \Tr (g_2)^2 \right) \right) \label{4log}
\end{eqnarray}
Barring the log term, all other divergences in \eqref{ads5_onshellaction} can be cancelled by adding a counter-term given by 

\begin{equation}\label{fived_counterterm}
S^{ct}_{N} = \frac{3}{\kappa} \int d^4 x \; \sqrt{-\gamma}
\end{equation}
Once again, this can be explicitly checked by expanding \eqref{fived_counterterm} in Fefferman-Graham expansion and using the relations \eqref{detexpansion} and \eqref{g2traces}. The renormalized Neumann action is given by

\begin{eqnarray}\label{fived_Neumann_action}
S^{ren}_N &=&   \frac{1}{2\kappa}\int_{\mathcal{M}} d^5 x \sqrt{-g}\;(R-2\Lambda) -\frac{1}{2\kappa}\int_{\partial \mathcal{M}}\;d^4 x \sqrt{-\gamma}\; \Theta + \frac{3}{\kappa} \int d^4 x \; \sqrt{-\gamma}
\end{eqnarray}
As in the case of $d=2$, there is an ambiguity in the stress-tensor. The renormalized stress-tensor we hold fixed for the variational principle is given by

\begin{equation}
T^{ren}_{ab} = \frac{1}{\kappa}\left[ \Theta_{ab} - \Theta \gamma_{ab} + 3\gamma_{ab} -\frac{1}{2}G_{ab} \right]
\end{equation}
Once again this shows that the renormalized Neumann action \eqref{fived_Neumann_action} gives a well defined variational principle with renormalized stress-tensor.  We also note that \eqref{fived_Neumann_action}, being an even $d$ case has an ambiguity similar to $d=2$ case, and we have used the fact that 
\bea
\delta \Big(\sqrt{-\gamma}G_{ab}\Big)\ \gamma^{ab}=0.
\eea

In what follows, we will often suppress the superscript ${ren}$ when there is no source of ambiguity that we are indeed working with the renormalized action. 

\subsection{Comparison With Standard Holographic Renormalization}

How does all this compare with the standard discussion of holographic renormalization in the Dirichlet case? 

One difference is that the counter-terms that are added in the Dirichlet case do not change the variational problem: before and after their addition, the boundary metric that is held fixed is identical. This is not true in our case. Before renormalization, the quantity that is held fixed is the {\em unrenormalized} stress tensor density, but at the end it is the {\em renormalized} stress tensor density. It is of course not surprising that added terms can change the variational problem, what is worthy of remark here is the philosophy behind it: we demanded the finiteness of the Neumann action, and that leads to a well-defined variational problem with the {\em renormalized} quantity held fixed. Satisfyingly, this same object can also be obtained as the Legendre transform of the {\em renormalized} Dirichlet action, see Appendix B. Note that the unrenormalized actions are merely a crutch and the renormalized actions are the physically relevant objects.  

Let us also note that the {\em total} action/partition function (including counter-terms {\em and} everything else) can only be a functional of the quantity fixed at the boundary. This is guaranteed at the level of the action because again, the Neumann action is a Legendre transform of Dirichlet and therefore (by construction) depends only on the conjugate variable. In equations, as we discuss in an Appendix, we can view our action as
\begin{equation}\label{Neu_leg_action}
S^{ren}_N [\pi^{ren}_{ab}]= S^{ren}_D[\gamma^{ab}] - \int_{\partial \mathcal{M}}d^{D-1}x\;\pi^{ren}_{ab}\gamma^{ab}
\end{equation}
where 
\bea
\pi^{ren}_{ab} = \frac{\delta S^{ren}_D}{\delta \gamma^{ab}}.
\eea 
This can be viewed as the semi-classical version\footnote{We will briefly discuss the existence of a full quantum theory further in Section 5 and 6, as well as in more detail in \cite{CKCG}.} of a Legendre transform at the level of partition functions:
\bea
\Gamma[\delta W/\delta \gamma^{ab}]=W[\gamma^{ab}]- \int_{\partial \mathcal{M}}d^{D-1}x\; \frac{\delta W}{\delta \gamma^{ab}}\gamma^{ab}
\eea
At the level of the semi-classical saddle, this translates to the statement that the variational principle (while holding the conjugate quantity fixed at boundary) is well-defined, which we checked explicitly earlier in this section. The separate terms (including counter-terms) in the action which are  integrated over can have complicated dependences, but they conspire to satisfy the above demands. 

As an aside, we also note some papers in the literature which deal with related set-ups. In particular, in \cite{Geoffrey} the boundary metric fluctuates but they arrange that the variational principle with the Dirichlet action works, by setting $T^{ij}=0$.  There are other papers, especially in three dimensions, which deal with similar set-ups  \cite{Grumiller:2016pqb,KRNEW,Apolo, Compere:2013bya, Afshar:2016wfy, Troessaert:2013fma, Avery:2013dja, KRChiralHS, Myers}. In fact, our approach can be thought of in many ways as a general framework for dealing with some of these situations. The work of \cite{Geoffrey} treats the boundary stress tensor to be a fixed given {\em value} (namely, zero), so their partition function is a {\em number}, so they do not discuss the points we emphasize in the previous paragraph. Our work can be thought of as a generalization of theirs and our partition function is a proper functional, where instead of setting the stress tensor (density) to be zero, we treat it as arbitrary but fixed\footnote{The ``arbitrariness" of the boundary stress tensor should of course still satisfy the requirement that the Fefferman-Graham expansion should satisfy the bulk equations of motion, see the discussion in \cite{deHaro} for details.}.

\section{Finite On-shell Action}

In this section we present the results of on-shell action and stress-energy tensor for the Neumann action in various dimensions. We also draw comparison of our on-shell action with the on-shell Dirichlet action. Note that the precise value of the action is sensitive to the infrared cutoff of the action integral. So one cannot work abstractly at the level of the Fefferman-Graham expansion like we did so far, because we need to know the metric finitely deep into the geometry and not merely as an expansion at the boundary. So we will consider explicit solutions like black holes.

\subsection{AdS$_3$}

The Dirichlet action for gravity in AdS$_3$ is given by \cite{Balasubramanian:1999re}

\begin{equation}
S_D = \frac{1}{2\kappa}\int_{\mathcal{M}}d^3x\;\sqrt{-g}(R-2\Lambda)+ \frac{1}{\kappa}\int_{\partial\mathcal{M}}\sqrt{-\gamma}\Theta -\frac{1}{\kappa}\int_{\partial\mathcal{M}}\sqrt{-\gamma}
\end{equation}
We evaluate the above action on the BTZ metric

\begin{equation}
ds^2 = -\frac{(r^2 - r^{2}_{+})(r^2 - r^{2}_{-})}{r^2}dt^2 + \frac{r^2\;dr^2}{(r^2 - r^{2}_{+})(r^2 - r^{2}_{-})} + r^2 \left(d\phi - \frac{r_{+}r_{-}}{r^2}dt \right)^2
\end{equation}
where $r_{+}$ and $r_{-}$ are the outer and inner horizons respectively and are related to the charges through the relation $M=r_{+}^{2}+r_{-}^{2}$ and $J=2r_{+}r_{-}$. In the above metric we have set $l=1$. Evaluating the action between time $-T$ to $T$ and $r_{+}<r<R$ on this solution yields

\begin{equation}
S_{D}^{BTZ} = \frac{2\pi(r_{+}^2+r_{-}^2)T}{\kappa} + O\left(\frac{1}{R^2}\right)
\end{equation}
The on-shell Neumann action for the BTZ solution yields

\begin{equation}
S_{N}^{BTZ} = \frac{2\pi(r_{+}^2+r_{-}^2)T}{\kappa}
\end{equation}
which matches with the Dirichlet action in the limit $R\rightarrow \infty$. The stress-energy tensor similarly takes the form

\begin{equation}
T_{ab} = \left( \begin{array}{cc}
-\frac{r_{+}^2+r_{-}^2}{2\kappa}  & \frac{r_{+}r_{-}}{\kappa}  \\
\frac{r_{+}r_{-}}{\kappa}  & -\frac{r_{+}^2+r_{-}^2}{2\kappa} 
\end{array} \right)+ O\left(\frac{1}{R^2}\right)
\end{equation}
This stress tensor has the right fall-offs to reproduce finite charges $M$ and $J$ through the relation \cite{Brown, quasilocal}

\begin{equation}
Q_{\xi} = -\int_{\Sigma}d^{D-1}x\;\sqrt{\sigma}(u^{a}T_{ab}\xi^{b})
\end{equation}
where $\xi^{a}$ is the Killing vector generating the isometry of the boundary metric and $u^{a}$ is the unit time-like vector. We see that the counter-term action that was chosen to make the on-shell Neumann action finite also produces a finite stress tensor.  This was shown for the Dirichlet case by \cite{Balasubramanian:1999re}.

\subsection{AdS$_4$}

The (renormalized) Dirichlet action in $D=4$ takes the form

\begin{eqnarray}
S_D &=& \frac{1}{2\kappa}\int_{\mathcal{M}}d^4x\;\sqrt{-g}(R-2\Lambda)+ \frac{1}{\kappa}\int_{\partial\mathcal{M}}d^3x\;\sqrt{-\gamma}\Theta \br &-&\frac{2}{\kappa}\int_{\partial\mathcal{M}}d^3x\; \sqrt{-\gamma}\left(1+\frac{{}^{(3)}R}{4}\right)
\end{eqnarray}
The AdS-Schwarzschild black hole metric is given by

\begin{equation}
ds^2 = -(1-\frac{2M}{r}+r^2)dt^2 + \frac{dr^2}{(1-\frac{2M}{r}+r^2)} + r^2d\Omega^2
\end{equation}
The horizon is obtained by the real root of

\begin{equation}
1-\frac{2M}{r_H}+r_{H}^2 = 0
\end{equation}
Evaluating the action for this metric yields (integrated in the region $-T<t<T$ and $r_{H}<r<R$)

\begin{equation}
S_{D}^{AdS-BH} = -\frac{8\pi (M-r_{H}^3) T}{\kappa} + O\left(\frac{1}{R}\right)
\end{equation}
The stress tensor computed for this metric is given by

\begin{equation}
T_{ab}=\left(
\begin{array}{ccc}
-\frac{2 M }{\kappa R } & 0 & 0 \\
0 & -\frac{M}{\kappa R } & 0 \\
0 & 0 & -\frac{M \sin ^2(\theta )}{\kappa R} \\
\end{array}
\right)+O\left(1/R^2 \right)
\end{equation}
which once again has the right fall-offs to obtain finite charges as described in the previous section. The Neumann action in $D=4$ takes the form

\begin{equation}
S_{N} = \frac{1}{2\kappa}\int_{\mathcal{M}}d^4x\;\sqrt{-g}(R-2\Lambda)+\frac{1}{\kappa}\int_{\partial\mathcal{M}}d^3x\;\sqrt{-\gamma}\left(1-\frac{{}^{(3)}R}{4} \right)
\end{equation}
which evaluates to

\begin{equation}
S_{N}^{AdS-BH} = -\frac{8\pi (M-r_{H}^3) T}{\kappa} + O\left(\frac{1}{R}\right)
\end{equation}
The sub-leading term here differs from the sub-leading term in the Dirichlet action and the two actions are same only in the $R\rightarrow \infty$ limit.

\subsection{AdS$_5$}

In $D=5$ the Dirichlet action takes the form

\begin{eqnarray}
S_D &=& \frac{1}{2\kappa}\int_{\mathcal{M}}d^5x\;\sqrt{-g}(R-2\Lambda)+ \frac{1}{\kappa}\int_{\partial\mathcal{M}}d^4x\;\sqrt{-\gamma}\Theta \br &-& \frac{3}{\kappa} \int_{\partial\mathcal{M}}d^3x\;\sqrt{-\gamma}\left(1+ \frac{{}^{(4)}R}{12}\right)
\end{eqnarray}
Evaluating this action for the black hole metric

\begin{equation}
ds^2 = -f(r)dt^2 + \frac{dr^2}{f(r)}+r^2d\Omega_{3}^{2}
\end{equation}
where

\begin{equation}
f(r) = r^2 + 1 -\frac{2M}{r^2}
\end{equation}
The horizon is once again determined by the largest positive root of

\begin{equation}
r_{H}^2 + 1 -\frac{2M}{r_{H}^2} = 0
\end{equation} 
The action evaluates to
\begin{equation}
S_{D}^{BH} = -\frac{2\pi^2T}{\kappa}(2M+\frac{3}{4}-2r_{H}^{4}) + O\left(1/R^4 \right)
\end{equation}
The stress tensor takes the form

\begin{equation}
T_{ab} = \left(
\begin{array}{cccc}
-\frac{3 (8M+1) }{8R^2 \kappa } & 0 & 0 & 0 \\
0 & -\frac{(8M+1)}{8R^2 \kappa } & 0 & 0 \\
0 & 0 & -\frac{\left((8M+1) \sin ^2(\psi )\right)}{8R^2
	\kappa } & 0 \\
0 & 0 & 0 & -\frac{\left((8M+1) \sin ^2(\theta ) \sin
	^2(\psi )\right) }{8R^2 \kappa } \\
\end{array}
\right)+O\left(1/R^4\right)
\end{equation}
The Neumann action in this case can be written as

\begin{equation}
S_N = \frac{1}{2\kappa}\int_{\mathcal{M}}d^5x\;\sqrt{-g}(R-2\Lambda)- \frac{1}{\kappa}\int_{\partial\mathcal{M}}d^4x\;\sqrt{-\gamma}\Theta +\frac{3}{\kappa}\int_{\partial\mathcal{M}}d^4x\;\sqrt{-\gamma}
\end{equation}
which evaluates to

\begin{equation}
S_{N}^{BH} = -\frac{2\pi^2T}{\kappa}(2M+\frac{3}{4}-2r_{H}^{4}) + O\left(1/R^2 \right)
\end{equation}
Again, we find agreement when the radial cutoff is taken to infinity.

\section{ADM Formulation of Renormalized AdS$_4$ Action}

The ADM formulation of GR works by singling out the time direction from the spatial direction and re-expressing the content of GR in terms of ADM variables. Thus the spacetime is thought of as foliated by spatial slices $\Sigma_t$ which are the hypersurfaces of constant $t$. The spacetime metric can be expressed as

\begin{equation}\label{adm_metric_full}
ds^2 \equiv g_{\alpha \beta}dx^{\alpha} dx^{\beta} = -N^2 dt^2 + h_{ab}(dy^a + N^{a}dt)(dy^b + N^{b}dt) 
\end{equation}
where $N$ is the Lapse function, $N^{a}$ is the shift vector and $h_{ab}$ is the induced metric on the hypersurface $\Sigma_t$. In what follows, we assume that the manifold is a box with finite spatial extend such that the boundary is time-like, denoted ${\cal B}$. The spatial section of ${\cal B}$ is denoted $B$. We will also ignore the space-like boundaries at initial and final times and work with coordinates such that the time-like boundary is orthogonal to the spatial hypersurfaces, $\Sigma_t$. Under the ADM split of the bulk metric, \eqref{adm_metric_full}, the induced metric on the boundary ${\cal B}$, also undergoes a decomposition

\begin{equation}\label{gamma_ADM}
ds^2 \equiv \gamma_{ij}dx^{i} dx^{j} = -N^2 dt^2 + \sigma_{AB}(d\theta^A + N^{A}dt)(d\theta^{B} + N^{B}dt)
\end{equation} 
where $\sigma_{AB}$ is the induced metric on $B$. We will also need the expression for the decomposition of Ricci scalar

\begin{equation}\label{ricci_decomp}
{}^{(D)} R = {}^{(D-1)}R + K^{ab}K_{ab} - K^2 - 2\nabla_{\alpha}\left( u^{\beta} \nabla_{\beta}u^{\alpha} - u^{\alpha} \nabla_{\beta} u^{\beta} \right)
\end{equation}
where $K_{ab}$ is the extrinsic curvature of the spatial hypersurface $\Sigma_t$ (not to be confused with the boundary). The point about ADM split is that $N$ and $N^a$ are not dynamical fields and therefore their conjugates are constraint relations. The dynamical field is the spatial metric $h_{ab}$ and the canonical conjugate momentum is given by

\begin{equation}
p^{ab} \equiv \frac{\partial}{\partial \dot{h}_{ab}}(\sqrt{-g}\mathcal{L}_G) = \frac{\sqrt{h}}{2\kappa}(K^{ab}-Kh^{ab})
\end{equation}
where $K_{ab}$ is the extrinsic curvature of $\Sigma_t$. The details of the ADM decomposition of gravitational action can be found in \cite{poisson, Pavan}. We will work with AdS$_4$ in what follows, for convenience.

\subsection{Dirichlet action}
In this section, we return to the case of ADM decomposition, for the renormalized Dirichlet action in AdS$_4$. The renormalized action in the covariant form is given by \cite{deHaro,Balasubramanian:1999re}

\begin{eqnarray}\label{D_ads_cov_action}
S_D &=& \frac{1}{2\kappa}\int_{\mathcal{M}}d^4x\;\sqrt{-g}(R-2\Lambda)+ \frac{1}{\kappa}\int_{\cal B}d^3x\;\sqrt{-\gamma}\Theta \\ \nonumber &+& \frac{1}{\kappa}\int_{\cal B}d^3x\;\sqrt{-\gamma}\left(-\frac{2}{l} \right) \left[1+ \frac{{}^{(3)}R}{4} \right]
\end{eqnarray}
The first two terms are the Einstein-Hilbert and the GHY piece, and can be written in terms of the ADM variables following the steps of \cite{Pavan}. This gives us the following form for the action \cite{poisson}

\begin{eqnarray}
S_D &=& S_{EH} + S_{GHY} + S_{ct} \\ \nonumber &=& \int_{\cal M}d^D x \left( p^{ab}\dot{h}_{ab} - NH - N_a H^a \right) +  \int_{\cal B}d^{D-1}y \sqrt{\sigma}\left(N\varepsilon - N^a j_a \right) + S_{ct}
\end{eqnarray}
where $H$ and $H^a$ are the Hamiltonian and momentum constraints,

\begin{eqnarray}
H &=& \frac{\sqrt{h}}{2\kappa}\left( K^{ab}K_{ab} - K^2 - {}^{(3)}R + 2\Lambda \right) \\ \nonumber
H^a &=& -\frac{\sqrt{h}}{\kappa}D_b(K^{ab}-Kh^{ab}) \\ \nonumber
\end{eqnarray}
$\sqrt{\sigma}\varepsilon$, $\sqrt{\sigma}j_a$ and $N\sqrt{\sigma}s^{ab}/2$ are the momenta conjugate to $N$, $N^{a}$ and $\sigma_{ab}$. and are given by

\begin{eqnarray}\label{canonical_variables}
\varepsilon &=& \frac{k}{\kappa}, \quad  j_a = \frac{2}{\sqrt{h}}r_b p_{\;a}^{b} \\ \nonumber s^{ab} &=& \frac{1}{\kappa}\left[k^{ab} - \left(\frac{r^a \partial_a N}{N} + k\right)\sigma^{ab} \right]
\end{eqnarray} 
where $k^{ab}$ is the extrinsic curvature of $B$ embedded in $\Sigma_t$ and $k=k^{ab}\sigma_{ab}$. The counter-term action is given in the covariant form by

\begin{equation}
S_{ct} = \frac{1}{\kappa}\int_{\cal B}d^3x\;\sqrt{-\gamma}\left(-\frac{2}{l} \right) \left[1+ \frac{{}^{(3)}R}{4} \right]
\end{equation}
Using \eqref{ricci_decomp} and the expression for the determinant $\sqrt{-\gamma}=N\sqrt{\sigma}$, we obtain the counter-term action as

\begin{equation}
S_{ct} = \frac{1}{\kappa}\int_{\cal B}d^3x\left(-\frac{2}{l} \right)\left[1 + \frac{l^2}{4}\left( {}^{(2)}R + \hat{k}_{ab}\hat{k}^{ab} - \hat{k}^2 \right) \right]
\end{equation}
where $\hat{k}_{ab}$ is the extrinsic curvature of $B$ as a hypersurface embedded in ${\cal B}$. For black hole geometries, we also get a contribution from the horizon which is given by decomposing the covariant Neumann action with a boundary at the horizon where no data is specified \cite{Brown,Pavan,martinez}. The action then takes the form

\begin{eqnarray}
S_D &=& \int_{\cal M}d^D x \left( p^{ab}\dot{h}_{ab} - NH - N_a H^a \right) \\ \nonumber &+& \int_{\cal H}d^{D-1}y\;\sqrt{\sigma}\left(\frac{r^a \partial_a N}{\kappa} + \frac{2r_a N_b p^{ab}}{\sqrt{h}} \right) + \int_{\cal B}d^{D-1}y \sqrt{\sigma}\left(N\varepsilon - N^a j_a \right) \\ \nonumber &+& \frac{1}{\kappa}\int_{\cal B}d^3x\left(-\frac{2}{l} \right)\left[1 + \frac{l^2}{4}\left( {}^{(2)}R + \hat{k}_{ab}\hat{k}^{ab} - \hat{k}^2 \right) \right]
\end{eqnarray}
We can further express the above action in terms of the renormalized parameters thereby absorbing the counter-term into the renormalized quantities $\varepsilon^{ren}=\varepsilon+\varepsilon^{ct}$, $j_{a}^{ren}=j_{a}+j^{ct}_{a}$ and $s_{ab}^{ren}=s_{ab}+s_{ab}^{ct}$. To do so, we do a canonical decomposition of the tensor using normal and tangential projections \cite{quasilocal}. The expressions for renormalized quantities are given by

\begin{eqnarray}
\varepsilon^{ren} &=& u_a u_b T^{ab} \\ \nonumber
j_{a}^{ren} &=& -\sigma_{ab}T^{bc}u_c \\ \nonumber
s_{ab}^{ren} &=& \sigma_{ac}\sigma_{bd} T^{cd} \nonumber
\end{eqnarray} 
where $T^{ab}$ is the renormalized stress tensor given by

\begin{equation}
T^{ab} = \frac{1}{\kappa}\left( \Theta^{ab}-\Theta \gamma^{ab} + \frac{2}{l}\gamma^{ab} -l G^{ab} \right)
\end{equation}
Using the above expressions, we get

\begin{eqnarray}
\varepsilon^{ren} &=& \varepsilon - \frac{1}{\kappa}\left[ \frac{2}{l} + \frac{l}{2}\left({}^{(2)}R - \hat{k}_{ab}\hat{k}^{ab} + \hat{k}^2 \right) \right] \\ \nonumber
j_{a}^{ren} &=& j_a + \frac{l}{\kappa}\left(d_{a}\hat{k} - d_{b}\hat{k}^{b}_{\; a} \right) \\ \nonumber
s_{ab}^{ren} &=& s_{ab} + \frac{1}{\kappa}\left[ \frac{2}{l}\sigma_{ab} + \frac{l}{2}\left( {}^{(2)}R + \hat{k}^{ab}\hat{k}_{ab} - \hat{k}^2 \right) \right. \\ \nonumber &-& \left. l\left( -\frac{1}{N}{\cal L}_{m}\hat{k}_{ab}- \frac{1}{N}d_a d_b N + {}^{(2)}R_{ab}+ \hat{k}\hat{k}_{ab}-2 \hat{k}_{ac}\hat{k}^{c}_{b} \right) \right]
\end{eqnarray}
In writing the above expressions, we have made use of Gauss-Codazzi relations whose exact expressions are given in the Appendix. Thus, the renormalized action can be expressed as

\begin{eqnarray}\label{D_ads_adm_action}
S_D &=& \int_{\cal M}d^D x \left( p^{ab}\dot{h}_{ab} - NH - N_a H^a \right) \\ \nonumber &+& \int_{\cal H}d^{D-1}y\;\sqrt{\sigma}\left(\frac{r^a \partial_a N}{\kappa} + \frac{2r_a N_b p^{ab}}{\sqrt{h}} \right) + \int_{\cal B}d^{D-1}y \sqrt{\sigma}\left(N\varepsilon^{ren} - N^a j_{a}^{ren} \right)
\end{eqnarray}

\subsubsection{Kerr-AdS: Covariant}

As an illustration of our construction, we can evaluate the action on the Kerr-AdS metric in $D=4$. Rotating black holes are better defined in AdS, than flat space (see eg., \cite{CKRotation, CKTomogram}). The metric in Boyer-Lindquist type coordinates is given by

\begin{eqnarray}\label{kerr_metric}
ds^2 &=& \rho^2\left( \frac{dr^2}{\Delta}+\frac{d\theta^2}{\Delta_{\theta}}\right)+\frac{\Delta_{\theta}\sin^2\theta}{\rho^2}\left(adt - \frac{r^2 + a^2}{\Sigma}d\phi \right)^2 \br &-& \frac{\Delta}{\rho^2}\left( dt - \frac{a\sin^2 \theta}{\Sigma}d\phi \right)^2
\end{eqnarray}
where

\begin{eqnarray} 
\rho^2 &=& r^2 + a^2\cos^2 \theta, \qquad \Delta = (r^2 + a^2)\left(1+\frac{r^2}{l^2} \right)-2Mr \\ \nonumber \Delta_{\theta} &=& 1 - \frac{a^2}{l^2}\cos^2 \theta, \qquad \Sigma = 1- \frac{a^2}{l^2}
\end{eqnarray}
The horizon is at the largest positive root of $\Delta(r_H)=0$. The angular velocity of the black hole (for $r \geq r_H$) is given by
\begin{equation}
\omega = a\Sigma \left( \frac{\Delta_{\theta}(r^2 + a^2)- \Delta}{(r^2 + a^2)^2 \Delta_{\theta} - a^2 \Delta \sin^2 \theta} \right)
\end{equation}
The angular velocity at the horizon is given by

\begin{equation}
\Omega_H  = \frac{a\Sigma}{r_H^{2}+a^2}
\end{equation}
while the angular velocity at the boundary ($r \rightarrow \infty$), is given by $\Omega_{\infty} = -a/l^2$. The angular velocity relevant for the thermodynamics is given by $\Omega = \Omega_H - \Omega_{\infty}$ \cite{Klemm, HHT}.
Given the metric, the ADM variables can be read off by comparing \eqref{kerr_metric} with the ADM form of the metric. The Lapse, Shift and spatial metric is given by
\begin{eqnarray}
N &=& \sqrt{\frac{\rho^2 \Delta \Delta_{\theta}}{(r^2 + a^2)^2 \Delta_{\theta}-a^2 \Delta \sin^2 \theta}} \\ \nonumber
N^{\phi} &=& a\Sigma \frac{\left( \Delta - \Delta_{\theta}(r^2 + a^2) \right)}{(r^2 + a^2)^2 \Delta_{\theta}-a^2 \Delta \sin^2 \theta} \\ \nonumber
h_{ab} &=& \left( \begin{array}{ccc}
\frac{\rho^2}{\Delta} & 0 & 0 \\
0 & \frac{\rho^2}{\Delta_\theta} & 0 \\
0 & 0 & \frac{\left( (r^2 + a^2)^2 \Delta_\theta - a^2 \Delta \sin^2 \theta  \right)}{\rho^2 \Sigma^2}
\end{array} \right) \nonumber
\end{eqnarray}

For thermodynamic interpretation we must work with the complex metric associated with the black hole, which is given by the identification $N \rightarrow -i\tilde{N}, N^{\phi} \rightarrow -i\tilde{N}^{\phi}$ \cite{Brown, Pavan}. The periodicity of time circle can be estimated by evaluating $r^a\partial_a \tilde{N} \equiv 2\pi/\beta$ term on the horizon. This gives the time periodicity, $\beta$, to be

\begin{equation}\label{beta}
\beta = \frac{4\pi (r^{2}_{H} + a^2)}{r_H\left(1+ \frac{a^2}{l^2} + \frac{3r_{H}^2}{l^2}-\frac{a^2}{r_{H}^2}\right)}
\end{equation}
The expressions for various terms in the covariant action are:

\begin{eqnarray}
R &=& -\frac{12}{l^2} \\ \nonumber
\Theta &=& \frac{3}{l} + \frac{(-3a^2 + 2l^2 -5a^2 \cos 2\theta)}{4l R_c^2} + O(1/R_c^4) \\ \nonumber
{}^{(3)}R &=& \frac{2l^2 -3a^2 -5a^2 \cos 2\theta}{l^2 R_c^2} + O(1/R_c^4)
\end{eqnarray}
Evaluating the complex metric on the covariant action \eqref{D_ads_cov_action}, and using the expression \eqref{beta} for the periodicity, we get

\begin{equation}
S_D = -i \frac{\pi l^2 (r_{H}^{2}+a^2)^2 (l^2 - r_{H}^{2})}{(l^2 - a^2)\left(a^2 l^2-(a^2 + l^2)r_{H}^2 - 3r_{H}^4 \right)}
\end{equation}
This is related to the free energy through the relation

\begin{equation}\label{D_ads_free_energy}
-\beta F_D \equiv \log Z_D \approx iS_D
\end{equation}
where $\beta$ is the inverse temperature which can be identified with the periodicity of the time circle. This gives the free energy of the black hole to be

\begin{equation}\label{cov_Dir_FE}
F_D = \frac{(r_{H}^2 + a^2)(l^2 - r_{H}^{2})}{4(l^2-a^2)r_H}
\end{equation}

\subsubsection{Kerr-AdS: ADM}
Evaluating the complex metric on the ADM decomposed action, the bulk term vanishes because the metric is stationary and satisfies Einstein's equation. The horizon term gives a contribution of

\begin{equation}
S_{\cal H} = -i \frac{A}{4} - i\Omega_H P J
\end{equation}
On the boundary we can see that the renormalized $\varepsilon$, $j^\phi$ and $s^{AB}$ have correct fall-offs so as to give finite results for the integral,

\begin{eqnarray}
\varepsilon_{ren} &=& \left( \frac{M(a^2 - 4l^2 + 3a^2 \cos 2\theta)}{l\Sigma \kappa}  \right)\frac{1}{R_c^3} + O(\frac{1}{R_c^4}) \\ \nonumber
j^{\phi}_{ren} &=& \frac{3aM}{\kappa}\sqrt{\frac{\Delta_{\theta}}{\Sigma}}\frac{1}{R_c^4} + O(\frac{1}{R_c^5}) \\ \nonumber
\end{eqnarray}
Evaluating the boundary integrals, we have
\begin{equation}
S_{\cal B} = iEP + i \Omega_{\infty}P J
\end{equation}
where $E$ and $J$ are calculated as

\begin{equation}
E = \frac{M}{\Sigma^2},\qquad J = \frac{Ma}{\Sigma^2}
\end{equation}
which are the ADM charges of the Kerr black hole. Using \eqref{D_ads_free_energy}, we have

\begin{equation}\label{adm_Dir_FE}
F_D = E - TS -\Omega J
\end{equation}
Now, by an explicit computation, we can verified that the free energy, $F_D$, in \eqref{cov_Dir_FE} can be expressed as

\begin{equation}\label{Dir_cov_FE_new}
F_D = -T \frac{A}{4} -\Omega J + g(A,J)
\end{equation}
where
\begin{equation}
g(A,J) = \sqrt{\frac{A}{16\pi}+\frac{4\pi}{A}J^2 + \frac{J^2}{l^2} + \frac{A}{8\pi l^2}\left(\frac{A}{4\pi} + \frac{A^2}{32\pi^2 l^2} \right)}
\end{equation}
Equating \eqref{Dir_cov_FE_new} to the free energy computed using ADM approach, \eqref{adm_Dir_FE}, we get the generalized Smarr formula (see eq.(41) of \cite{Klemm}).

\begin{equation}
E^2 = \frac{A}{16\pi}+\frac{4\pi}{A}J^2 + \frac{J^2}{l^2} + \frac{A}{8\pi l^2}\left(\frac{A}{4\pi} + \frac{A^2}{32\pi^2 l^2} \right)
\end{equation}
Following \cite{Klemm} we can also relate these calculations to the first law, which we will not repeat.

\subsection{Neumann action}

The renormalized Neumann action in AdS$_4$ is given by

\begin{equation}
S_N = \frac{1}{2\kappa}\int_{\mathcal{M}}d^4x\;\sqrt{-g}(R-2\Lambda) + \frac{1}{\kappa}\int_{\cal B}d^3x \sqrt{-\gamma}\left(\frac{1}{l}\right) \left[1-\frac{l^2}{4}{}^{(3)}R \right]
\end{equation}
The bare part of the Neumann action in ADM was derived in \cite{latest}. In $D=4$ it can be used to write

\begin{eqnarray}
S_N &=& S_{EH} + S_{ct} \\ \nonumber &=& \int_{\mathcal{M}}d^4 x \left( p^{ab}\dot{h}_{ab} - NH - N_a H^a \right) +  \int_{\mathcal{B}}d^{3}x \sqrt{\sigma}\left(\frac{N\varepsilon}{2}  - N^a j_a + \frac{N}{2}s^{ab}\sigma_{ab} \right) + S_{ct}
\end{eqnarray}
The counter-term action can be decomposed similar to the Dirichlet case and we get

\begin{equation}
S_{ct} = \frac{1}{\kappa}\int_{\cal B}d^3x\left(\frac{1}{l} \right)\left[1 - \frac{l^2}{4}\left( {}^{(2)}R + \hat{k}_{ab}\hat{k}^{ab} - \hat{k}^2 \right) \right]
\end{equation}
For the black hole geometries, one again has a contribution from the horizon and action takes the form

\begin{eqnarray}
S_N &=& \int_{\cal M}d^4 x \left( p^{ab}\dot{h}_{ab} - NH - N_a H^a \right) \\ \nonumber &+&\int_{\mathcal{H}}d^{3}y\;\sqrt{\sigma}\left(\frac{r^a \partial_a N}{\kappa} + \frac{2r_a N_b p^{ab}}{\sqrt{h}} \right) + \int_{\mathcal{B}}d^{3}x \sqrt{\sigma}\left(\frac{N\varepsilon}{2}  - N^a j_a + \frac{N}{2}s^{ab}\sigma_{ab} \right) \\ \nonumber &+& \frac{1}{\kappa}\int_{\cal B}d^3x\left(\frac{1}{l} \right)\left[1 - \frac{l^2}{4}\left( {}^{(2)}R + \hat{k}_{ab}\hat{k}^{ab} - \hat{k}^2 \right) \right]
\end{eqnarray}
Using the expressions for renormalized parameters, the Neumann action can be expressed as

\begin{eqnarray}
S_N &=& \int_{\cal M}d^4 x \left( p^{ab}\dot{h}_{ab} - NH - N_a H^a \right) \\ \nonumber &+&\int_{\mathcal{H}}d^{3}y\;\sqrt{\sigma}\left(\frac{r^a \partial_a N}{\kappa} + \frac{2r_a N_b p^{ab}}{\sqrt{h}} \right) + \int_{\mathcal{B}}d^{3}x \sqrt{\sigma}\left(\frac{N\varepsilon^{ren}}{2}  - N^a j^{ren}_a + \frac{N}{2}s^{ren\;ab}\sigma_{ab} \right)
\end{eqnarray}

\subsubsection{Kerr-AdS: Covariant}
We can evaluate the covariant Neumann action on the Kerr-AdS complex metric, we obtain

\begin{equation}
S_N = -i \frac{\pi l^2 (r_{H}^{2}+a^2)^2 (l^2 - r_{H}^{2})}{(l^2 - a^2)\left(a^2 l^2-(a^2 + l^2)r_{H}^2 - 3r_{H}^4 \right)}
\end{equation}
Notice that unlike the asymptotically flat case, the on-shell value of the Dirichlet and Neumann action are equal. The on-shell action is related to the Neumann free energy through the relation

\begin{equation}\label{N_ads_free_energy}
-\beta F_N \equiv \log Z_N \approx iS_N
\end{equation}
which gives the free energy of the black hole to be

\begin{equation}
F_N = \frac{(r_{H}^2 + a^2)(l^2 - r_{H}^{2})}{4(l^2-a^2)r_H}
\end{equation}

\subsubsection{Kerr-AdS: ADM}
Evaluating the complex metric on the ADM decomposed action, the horizon term gives a contribution of

\begin{equation}
S_{\cal H} = -i \frac{A}{4} - i\Omega_{H}P J
\end{equation}
On the boundary we have,

\begin{eqnarray}
s^{ab}_{ren} &=& \left( \begin{array}{cc}
-\frac{lM\Delta_\theta}{\kappa} & 0 \\
0 & -\frac{M(a^2 + 2l^2 -3a^2\cos 2\theta)}{2l\kappa \sin^2\theta}
\end{array} \right)\frac{1}{R_c^3} + O(1/R_c^4) \\ \nonumber
\sigma_{ab} &=& \left( \begin{array}{cc}
\frac{\rho^2}{\Delta_\theta} & 0 \\
0 & \frac{\left( (r^2 + a^2)^2 \Delta_\theta - a^2 \Delta \sin^2 \theta  \right)}{\rho^2 \Sigma^2}
\end{array} \right) \nonumber
\end{eqnarray}
We get a contribution of $iEP/2$ from the integration over $\varepsilon^{ren}$ term and another contribution of $iEP/2$ from the integration over $s_{ab}^{ren}$ term. The $j^{ren}_{\phi}$ gives a contribution of $i\Omega_{\infty} P J$. Together we have again
\begin{equation}
S_{\cal B} = iEP - i\Omega_{\infty}P J
\end{equation}
Again using \eqref{N_ads_free_energy}, the free energy takes the form

\begin{equation}
F_N = E - TS - \Omega J
\end{equation}
where $\Omega = \Omega_H - \Omega_{\infty}$ is the potential relevant for the thermodynamics. So we end up getting the exact same expressions for $F_N$ and $F_D$ (in covariant and canonical approaches, separately).

The emergence of the canonical ensemble together with the Smarr formula implies the first law as well. This follows from the discussion in \cite{Klemm}, so we will not repeat it.

\section{Alternative Quantizations in AdS}

We would like to investigate whether these boundary conditions can define a consistent quantum gravity in AdS. If so, this will provide a set of boundary conditions that are different from the standard Dirichlet boundary conditions familiar from AdS/CFT.  So far on the other hand, a skeptic could choose to think of our discussion as merely a class of well-defined boundary conditions/terms for {\em classical} gravity in AdS. However, the fact that these boundary conditions give rise to finite actions that lead to correct thermodynamical relations is suggestive to us of an underlying quantum theory: so let us try and explore to see whether we can take these boundary conditions seriously at the quantum level. We will not prove it here (but see \cite{CKCG}) that our approach can be the starting point of a consistent quantum theory, but we will merely make some related observations. 

From the boundary theory point of view, the translation from the metric-fixed to stress-tensor-fixed point of view is a Legendre transform that takes the boundary partition function to the boundary effective action\footnote{A similar approach for scalar fields was taken in \cite{KW}, the source and condensate are dual variables in the Legendre transform sense.}. This seems to us to be a perfectly natural and consistent operation as we discussed in Section 2.5, so we believe there should be a legitimate formulation of holography in which the correspondence is phrased in the language of the effective action and not in terms of the generating functional. Note that for this, we will have to move away from the standard Dirichlet formulation of holography where the boundary values of bulk fields are interpreted as sources.

The trouble is that it is well-known that (for example) for scalars in a fixed AdS background, of the two modes (which we can call Dirichlet and Neumann) only the Dirichlet mode is typically normalizable \cite{KW, BF}. The exception to this is when the mass of the scalar falls in the Brietenlohner-Freedman window, where a Legendre transform analogous to ours takes the Dirichlet scalar theory to the Neumann scalar theory, and both are well-defined quantum mechanically \cite{KW}. When the scalar mass is not in this specific range, there is only one choice of acceptable normalizable mode and a unique quantization in a fixed AdS background.

To understand this better, let us note that the reason why we want normalizable modes is because we want them to be well-defined states in the Hilbert space of the putative quantum theory, with finite norm. This translates to a notion of finite energy: when the scalar mode has finite energy in the bulk of AdS, it can be well-defined as a state in the Hilbert space of the quantum theory. This is what happens in the case of scalar quantum field theory in a fixed AdS background \cite{Isham, BF, KW}. 

Now, lets consider the case when the background is not rigid and the metric is allowed to fluctuate. Lets start by considering scalar fields in such a set up. We note two things -- one is that a dynamical background makes the notion of energy more subtle, and secondly the notion of mass of the scalar is ambiguous because (say) a term of the form
\bea
(m^2+\lambda R) \phi^2 
\eea
where $R$ is the curvature scalar of the background will look like a usual mass term in the rigid limit. So a non-minimal coupling can sometimes be difficult to distinguish. As it happens both these issues have been addressed in \cite{MincesRivelles} (see also \cite{MincesScalar, RivellesSummary}) and it was found that once one deals with the appropriate notion of (canonical) energy both quantizations are admissible. We will take this as an encouraging fact: when dealing with the full gravity theory with appropriate counterterms etc. it is not necessarily only a Dirichlet boundary condition that can be well-defined, the notion of canonical energy needs to take into account the full theory.

Indeed, a similar conclusion was arrived at by Compere and Marolf \cite{Geoffrey}, who considered the possibility of not fixing the boundary metric, and instead considered simply integrating it over in the path integral. At the semi-classical level, the variational principle would then yield 
\begin{equation}
\delta S^{ren}_{D} = \text{ Eqs. of motion} + \frac{1}{2} \int_{\partial M} d^d x \sqrt{-g_{0}}
T^{ij} \delta g_{0_{ij}},
\end{equation}
where now there is no assumption that $\delta g_{0_{ij}}=0$ because we are letting it fluctuate. This means that to ensure that the action is stationary, now we need the boundary (renormalized) stress tensor to vanish\footnote{The boundary stress-energy tensor that we have often used in our discussions in this work is given by the relation 
	\begin{equation}
	T^{ren}_{ij}[\gamma] = -\frac{2}{\sqrt{-\gamma}}\frac{\delta S^{ren}_D}{\delta \gamma^{ij}} 
	\end{equation}
	where the boundary is placed at $\rho=\epsilon$. This is related to the CFT stress tensor (which is the true renormalized stress tensor, and the one we are using in this section) through
	\begin{eqnarray}
	T_{ij} &=& \lim_{\epsilon \rightarrow 0}\left(\frac{1}{\epsilon^{d/2 -1}} T^{ren}_{ij}[\gamma] \right) = \lim_{\epsilon \rightarrow 0}\left(-\frac{2}{\sqrt{-g(x,\epsilon)}}\frac{\delta S^{ren}_D}{\delta g^{ij}} \right) \\ \nonumber &=& -\frac{2}{\sqrt{g_{0}}}\frac{\delta S^{ren}_D}{\delta g^{ij}_{0}}.
	\end{eqnarray}
	Here, $g_0$ is the leading term in the Fefferman-Graham expansion.}.
Remarkably, Compere-Marolf found that such boundary metric fluctuations are in fact normalizable with respect to the canonical (symplectic) structure defined by the {\em full} renormalized Dirichlet action $S^{ren}_D$. Furthermore they also showed that the symplectic structure is also conserved when the boundary condition $T_{ij}=0$ holds. They further showed that if we couple the full renormalized bulk Dirichlet action above to a boundary action that is a functional of the boundary metric (\textit{i.e.}, the boundary is dynamical), so that the variation now becomes
\begin{eqnarray}
\delta S^{bndry}_D &\equiv & \delta \left(S_D + S_{bndry} \right) = \text{e.o.m} - \frac{1}{2}\int_{\partial \mathcal{M}} d^d x \sqrt{-g_{0}}T^{ij}\delta g_{0\;ij} \\ \nonumber &+& \int_{\partial \mathcal{M}} d^d x \frac{\delta S_{bndry}}{\delta g_{0\;ij}}\delta g_{0\;ij} \\ \nonumber &=& \text{e.o.m} -\frac{1}{2}\int_{\partial \mathcal{M}} d^d x \sqrt{g_{0}}\left( T^{ij} - \frac{2}{\sqrt{g_{0}}} \frac{\delta S_{bndry}}{\delta g_{0\;ij}}  \right)\delta g_{0\;ij}
\end{eqnarray}
then {\em again} the claims above hold, if instead of requiring $T^{ij}=0$ we now require 
\begin{equation}
T^{ij} - \frac{2}{\sqrt{g_{0}}} \frac{\delta S_{bndry}}{\delta g_{0\;ij}} = 0.\label{bounddyn}
\end{equation}

With that aside, let us turn to our Neumann case. We will merely discuss some connections between our work and and that of Compere-Marolf and leave it at that for now. We first note that the usual Dirichlet action plus a boundary term, after a Legendre transform of the kind we discussed, takes the form
\begin{eqnarray}
S^{bndry}_N &\equiv & S_D + S_{bndry} + \int_{\partial \mathcal{M}} d^d x \frac{\sqrt{g_{0}}}{2}\left( T^{ij} - \frac{2}{\sqrt{g_{0}}} \frac{\delta S_{bndry}}{\delta g_{0\;ij}}  \right)g_{0\;ij}
\end{eqnarray}
This has the variation
\begin{eqnarray}
\delta S^{bndry}_N &=& \text{e.o.m} + \frac{1}{2} \int_{\partial \mathcal{M}} d^d x  g_{0\;ij}\ \delta \left[ \sqrt{g_{0}} \left( T^{ij} - \frac{2}{\sqrt{g_{0}}} \frac{\delta S_{bndry}}{\delta g_{0\;ij}} \right)   \right]
\end{eqnarray}
Note that this is of the Neumann form, but now with boundary dynamics. It would be interesting to see if this leads to normalizable fluctuations, perhaps if one imposes the  condition (\ref{bounddyn}) that Compere and Marolf do. It is worth mentioning here that what \cite{Geoffrey} calls Neumann boundary condition is (as is often conventional in the gravity literature) the vanishing of  $T_{ij}$. This is the gravitational analogue of starting with the standard {\em Dirichlet} action in particle mechanics, letting the coordinate $q$ fluctuate at the boundary, but demanding that $\dot q=0$ at the boundary so that the boundary piece dies anyway, so that variational problem is well-defined. A genuinely Neumann condition is less constraining: it merely says that the normal derivative/canonical conjugate is {\em fixed}, not necessarily zero. This is what we do and therefore our Neumann case is more analogous to fixing $\dot{q}$ to a specific value at the boundary

Of course to conclusively settle this question requires further work, but we suspect that when one takes into account the full dynamics of the system instead of a fixed AdS background, more boundary conditions than what are usually considered lead to consistent quantum theories. It seems likely that one can discuss the normalizability via the symplectic structure in a covariant phase space approach, and we will report on work in this direction elsewhere.

\section{Comments and Discussions}
A standard Neumann boundary term for gravity, where one can obtain a well-defined variational problem by simply holding the normal derivative of the metric fixed at the boundary, is not known for Einstein-Hilbert gravity to the best of our knowledge. Occasionally statements of the form $K_{ij}=0$  or $K_{ij}-\alpha h_{ij}=0$ are considered as Neumann or Robin (mixed) boundary conditions in the literature. But note that these are in fact far stronger conditions than the usual Neumann or Robin boundary conditions, which merely say that these quantities are fixed, {\em not} that they are zero. Indeed, these conditions put {\em differential} constraints on the boundary surface (surfaces which admit conditions of this type are sometimes called totally umbilical surfaces \cite{Zanelli}). In hindsight, our approach provides a natural explanation why some of these boundary conditions had nice properties: they arise as natural further restrictions on the Neumann boundary condition, which puts restrictions on the boundary stress tensor (or equivalently, the extrinsic curvature). In this context, we suspect that it should be possible to define a Neumann boundary condition for higher derivative gravity theories \cite{Hint} as well, by setting the functional derivative of the action with respect to the boundary metric fixed.\\

In the context of AdS/CFT some papers with a loosely Neumann flavor have appeared, one of the more visible ones being \cite{Geoffrey}. What they do is to treat the bulk path integral (or action, when one is working semi-classically) as a functional of the boundary metric and then integrate over the boundary metric to define a new path integral. This has its interest, but its connection to the standard Neumann problem is not immediate.\\

We believe that what we have considered in this work is a natural version of the Neumann condition for gravity: our approach reduces to the standard Neumann problem when applied to particle mechanics, and it does not put any differential constraints on the boundary surface. It also puts various curiosities in the literature on boundary terms, in context. The Neumann path integral that we have considered in this work is also related to the ``microcanonical" path integral that was considered by Brown-York \cite{Brown}. Their approach amounts to holding {\em some} of the components of the quasi-local (boundary) stress tensor density fixed, whereas our approach is in some sense more covariant: we hold the entire boundary stress tensor density fixed. We saw that this has a natural interpretation as a Neumann problem, and results in a very simple Neumann action that leads to various nice features, some of which we investigated in \cite{latest,Pavan,Bala}. \\

The path integral of \cite{Brown} was called a ``microcanonical" functional integral. The motivation of \cite{Brown} for this nomenclature was that in gravity, the total charges reduce to surface integrals over the boundary. In \cite{Brown} this surface integral is not explicitly done, but we believe this surface integral actually needs to be done in order to get a true charge, and to make the path integral truly ``microcanonical" from the gravity perspective. \\

We would like to emphasize however that even keeping the integrated charge (energy) fixed on the gravity side in the sense of \cite{Brown} is not quite the same as holding the CFT energy fixed in AdS/CFT. This is because in \cite{Brown} the boundary metric is allowed to fluctuate. In AdS/CFT however, in the microcanonical ensemble when we hold the CFT energy fixed, we also hold the metric fixed. If we have infinite resolution, there is no ensemble of states in the CFT satisfying both these conditions. \\

In AdS/CFT, the natural microcanonical object to hold fixed from the CFT perspective is the total CFT energy, which should be compared to a charge (the boundary stress tensor density is a current from the CFT perspective). In the thermodynamic limit, the microcanonical density of states is a Laplace transform of the canonical partition function \cite{Chaichian}. The usual discussion of Hawking-Page transition in AdS/CFT is in the context of the canonical ensemble, but by doing this Laplace transform we can move to the microcanonical ensemble as well. The resulting discussion is guaranteed to match with the discussion of AdS thermodynamics in the microcanonical ensemble done in the Hawking-Page paper \cite{HawkPage}\footnote{This discussion is in the last section of their paper, and is not as well-known as their canonical discussion. The only thing relevant for our purposes here is that they change ensembles via the aforementioned Laplace transform.}, because the corresponding canonical discussions match.\\ 

Our construction, as we have emphasized, is different from both \cite{Brown} as well as the AdS/CFT discussion. Morally it is more similar to \cite{Brown} because we also do not pin down the metric at the boundary. Our approach could be viewed as an alternate implementation of holography in AdS where the boundary metric is allowed to fluctuate. In a follow-up paper \cite{CKCG} (unpublished), further evidence  will be provided that these boundary conditions may be consistent boundary conditions for quantum gravity in AdS: we will find that in odd $d$ the fluctuations are normalizable, and that in even $d$, normalizability of the bulk fluctuations is guaranteed when the dynamics of the  boundary metric is controlled by conformal gravity. Recently, our results were further generalized to construct Robin boundary terms for gravity in \cite{Krishnan:2017bte}. Considering the fact that the Dirichlet boundary term \cite{GH,Y} has had numerous applications since its inception more than 40 years ago, perhaps it is not surprising that the Neumann term \cite{latest} also leads to natural applications and generalizations.

\chapter{$\varepsilon$-Expansion in Gross-Neveu CFT}\label{GNCFT}
\section{Introduction}
Symmetry principles have emerged as the hallmark of modern physics. Our best understanding of nature so far is based on Standard model where (gauge) symmetries largely fixes the theory. In a quantum theory, symmetries affect the particle/field content of the theory as well as provide constraints on the correlation function of operators in the theory.

Conformal field theories have emerged as fascinating class of quantum field theories which finds many applications in physics, ranging from critical systems to Quantum gravity. Conformal symmetry can be thought of as an extension of Poincare symmetry, augmented with scale invariance and conformal invariance\footnote{The notion of conformal invariance is in conflict with the notion of asymptotic states, hence conformal field theories lack a S-matrix. However $n$-point correlation functions are well defined}.  Often times, especially in lower dimension systems, conformal invariance can be invoked as a powerful tool that allows us to go beyond perturbation theory.

In recent work, Rychkov and Tan \cite{Rychkov} have shown that the power of conformal invariance can be used to compute $\epsilon$-expansions at the Wilson-Fisher fixed point (see also \cite{Chethan-Pallab}). This approach is not reliant on Feynman diagrams (and in that sense is non-perturbative\footnote{This should be taken with a pinch of salt -- the epsilon expansion is afterall perturbative. The idea here is that the perturbative parameter in the present approach is not (at least manifestly) the coupling constant.}), and uses only conformal symmetry and analyticity in $\epsilon$ as inputs.

The results of Rychkov-Tan were generalized to other dimensions and other fixed point theories in \cite{Chethan-Pallab}. The computations require a systematic approach to handling contractions of fields in these theories, and a systematic approach for doing this was developed for scalar $O(N)$ theories \cite{Chethan-Pallab}. One of the goals of this work is to generalize this to CFTs with fermions.

Concretely, we will work with $O(N)$ Gross-Neveu model in $d=2+\epsilon$ dimensions \cite{GN} \footnote{The multiplicative renormalizability of Gross-Neveu model in $2+\e$ dimensions is discussed in \cite{Vasiliev1,Vasiliev2}, our results are unchanged for the $U(N)$ model as well.}. This theory is interesting for various reasons: there is a huge literature on this theory, and its large-N expansion and asymptotic freedom (among various other features) have been thoroughly investigated in the last decades. We generalize the approach of \cite{Rychkov,Chethan-Pallab} to this theory, and verify that the results agree with existing perturbative results in the literature, where they overlap. The contents of this chapter is based on \cite{Raju:2015fza}.

\section{$O(N)$ Gross-Neveu model in $2+\epsilon$ dimensions}\label{sec1}

The Gross-Neveu model action in $d=2+\epsilon$ dimensions is given by

\bea
S = \frac{1}{2\pi}\int d^d \sigma \left[ \bpsi^A \fsl{\partial}\psi^A +  \frac{1}{2}g\mu^{-\epsilon}(\bpsi^A \psi^A)^2 \right]
\eea
In 2 dimensions, this theory is renormalizable with a dimensionless coupling constant. The coupling constant is proportional to $\e$ and hence this theory describes a weakly coupled fixed point for small values of $\e$. We have introduces a scale $\mu$ to make the coupling constant dimensionless. \\
The engineering dimension of the fields is fixed by the action

\bea
[\psi] \equiv \delta = \frac{1+\e}{2}
\eea
The equations of motion for this theory are given by

\begin{eqnarray}
\gamma^{\mu}\partial_{\mu}\psi^A + g\mu^{-\epsilon} (\bpsi^B \psi^B)\psi^A &=& 0 \\
\partial_{\mu}\bpsi^A \gamma^{\mu} - g\mu^{-\epsilon} (\bpsi^B \psi^B)\bpsi^A &=& 0 
\end{eqnarray}

According to \cite{Rychkov}, this equation has to be seen as a conformal multiplet shortening condition, where in the free theory, the operators $(\bpsi^B \psi^B)\psi^A$ and $(\bpsi^B \psi^B)\bpsi^A$ are primaries, but in the interacting theory they are made secondary by above equations. Following \cite{Rychkov}, we formalize the relationship between operators in the free and interacting case by means of following axioms:

\begin{itemize}
	\item The interacting theory enjoys conformal symmetry.
	\item For any operator in the interacting theory, there is a corresponding operator in the free theory, which the interacting theory operator approaches to in the $\e \rightarrow 0$ limit. \\
	For definiteness, we call the interacting theory operators as $V_{2n}$, $V^{A}_{2n+1\;a}$ and $\overbar{V}^{A}_{2n+1\;a}$\footnote{A word on notations: small latin indices $a$, $b$, $\cdots$ are the spinor indices whereas $A$, $B$, etc stand for $O(N)$ indices} which in the free limit goes to 
	
	\begin{eqnarray}
	V_{2n} & \rightarrow & (\bpsi^A \psi^A)^{n}  \\ \nonumber
	V^{A}_{2n+1\;a} & \rightarrow & (\bpsi^B \psi^B)^{n} \psi^{A}_{a} \\ \nonumber
	\overbar{V}^{A}_{2n+1\;a} & \rightarrow & (\bpsi^B \psi^B)^{n} \bpsi^{A}_{a} \nonumber
	\end{eqnarray}
	\item Operators $V^{A}_{3\;a}$ and $\overbar{V}^{A}_{3\;a}$ are not primaries, instead they are related to the primaries by the multiplet shortening conditions
	
	\begin{eqnarray}\label{multiplet_short}
	\gamma^{\mu}\partial_{\mu}V^{A}_{1} = - \alpha(\e) V^{A}_{3}  \\ \nonumber
	\partial_{\mu}\bV^{A}_{1} \gamma^{\mu} = \alpha(\e) \bV^{A}_{3}
	\end{eqnarray}
	This puts restrictions on the dimensions of these operators,
	
	\bea\label{dimconstraint}
	\Delta_{3} = \Delta_{1} + 1
	\eea
	The proportionality constant $\alpha(\e)$ can be fixed later using the axioms above. All other operators $V_{m}$, $m \neq 3$, are primaries. 
	
\end{itemize}

The two-point function of two primaries of same dimension $\Delta_1$ is 

\bea
\langle V^{A}_{1\;a}(x_1) \overbar{V}^{B}_{1\;b}(x_2) \rangle = \frac{(\fsl{x}_{12})_{ab}}{(x_{12}^{2})^{\Delta_1 + \frac{1}{2}}}\delta^{AB} \label{int2pt}
\eea
where we have used the Feynman slash notation $\fsl{x} = x_{\mu}\gamma^{\mu}$. In the free limit this becomes
\bea
\langle \psi^{A}_{a}(x_1) \bpsi^{B}_{b}(x_2) \rangle = \frac{(\fsl{x}_{12})_{ab}}{x_{12}^{2}}\delta^{AB} \label{free2pt}
\eea
The anomalous dimension is defined as the difference between the actual scaling dimension of the operator and the engineering dimension, i.e, $\Delta_{n} = n\delta + \gamma_{n}$. We also make the crucial assumption that the anomalous dimensions are analytic functions of $\e$ and therefore admits a power series expansion

\bea
\gamma_n=y_{n,1} \e + y_{n,2} \e^2+ ...
\eea

 Our first task is to fix $\alpha$ in \eqref{multiplet_short}. Differentiating \eqref{int2pt} and substituting appropriate factors of $\gamma$ matrices, we obtain

\begin{eqnarray}\label{diff_correlator}
&&(\gamma^{\mu})_{ca}\langle \partial_{1\; \mu}V^{A}_{1\;a}(x_1) \partial_{2\; \nu}\bV^{B}_{1\; b}(x_2) \rangle (\gamma^{\nu})_{bd} \br &&\hspace{3cm} = (\gamma^{\mu})_{ca}\partial_{1\; \mu} \partial_{2\; \nu} \left( \frac{(\fsl{x}_{12})_{ab}}{(x_{12}^2)^{\Delta_1 + \frac{1}{2}}} \right) (\gamma^{\nu})_{bd}\delta^{AB} \\ &&\hspace{3cm} =  -(2\Delta_1 + 1)(2\Delta_1 +1 -d) \frac{(\fsl{x}_{12})_{cd}}{(x^{2}_{12})^{\Delta_1 + \frac{3}{2}}} \delta^{AB} \nonumber
\end{eqnarray}
Left hand side of \eqref{diff_correlator} takes the form

\begin{equation}
-\alpha^2 \langle V^{A}_{3\;c}(x) \bV^{B}_{3\;d}(y) \rangle 
\end{equation}
which in the free limit evaluates to

\bea
-\alpha^{2}(\e)(N-1)\frac{(\fsl{x}_{12})_{cd}}{(x^{2}_{12})^{2}}\delta^{AB}
\eea
Comparing both sides, we obtain

\begin{equation}\label{alpha}
\alpha = \sigma \sqrt{\frac{4\gamma_1}{N-1}}
\end{equation}
where $\sigma=\pm 1$. The exact sign will be determined later. Following \cite{Rychkov, Chethan-Pallab}, we consider correlators of the form

\bea
\langle V_{2n}(x_1) V^{A}_{2n+1\;a}(x_2) \bV^{B}_{1\;b}(x_3) \rangle, \quad \langle V_{2n}(x_1) V^{A}_{2n+1\;a}(x_2) \bV^{B}_{3\;b}(x_3) \rangle
\eea
which in the free limit goes to

\bea
\langle \phi_{2n}(x_1) \phi_{2n+1\; a}^{A}(x_2) \bphi^{B}_{1\;b}(x_3) \rangle, \quad  \langle \phi_{2n}(x_1) \phi_{2n+1\; a}^{A}(x_2) \bphi^{B}_{3\;b}(x_3)
\eea
 where we have introduced operators $\phi_{2n}$ and $\phi_{2n+1\; a}^{A}$ as a shorthand for $(\bpsi^{B}_{b}\psi^{B}_{b})^n$ and $(\bpsi^{B}_{b} \psi^{B}_{b})^n \psi^{A}_{a}$. The reason we are interested in these correlators is because of its sensitivity to multiplet recombination. To see this, we notice that in the free theory, $\phi_{2n} \times \phi_{2n+1\; a}^{A}$ OPE contains operators $\psi^{A}_{a}$ and $(\bpsi^{B}_{b}\psi^{b}_{b})\psi^{A}_{a}$ whereas in the interacting theory $V_{2n} \times V^{A}_{2n+1\;a}$ OPE only contains $V_1$ as the primary. The coefficients in both cases are independently computable and by Axiom:2, we expect them to match in the limit $\e \rightarrow 0$.

 In the free case, we have following OPE

\bea\label{freeOPE1}
\phi_{2n}(x_1) \times \phi_{2n+1\;a}^{A}(x_2) \supset f_{2n} (x^{2}_{12})^{-n}\left(\psi_{a}^{A} + \rho_{2n} (\fsl{x}_{12})_{ab} (\bpsi \psi)\psi^{A}_{b} \right)
\eea
The coefficients $f_{2n}$ and $\rho_{2n}$ can be determined by counting the number of Wick contractions. In next section, we provide an algorithm, based on \cite{Chethan-Pallab}, to determine these coefficients for arbitrary $n$. This is matched with the interacting theory OPE

\begin{eqnarray}\label{intOPE}
V_{2n}(x_1)\times V^{A}_{2n+1\;a}(x_2)  & \supset & \tilde{f}_{2n} (x^{2}_{12})^{-\frac{1}{2}[\Delta_{2n} + \Delta_{2n+1}-\Delta_1]} \\ \nonumber &\;& \Big[ \delta_{ac} + q_1 \delta_{ac} x_{12}^{\mu} \partial_{2\; \mu} + q_2 (\fsl{x_{12}}\fsl{\partial}_{2})_{ac}  \Big] V^{A}_{1\;c}(x_2)
\end{eqnarray}

\section{Counting contractions}\label{cowpie}

We now turn our attention to computing $f$ and $\rho$ coefficients in \eqref{freeOPE1}. Apart from \eqref{freeOPE1}, we also need OPE's of the form

\bea\label{freeOPE2}
\phi^{A}_{2n+1\;a}(x_1) \times \phi_{2n+2}(x_2) \quad \supset && f_{2n+1} (x^{2}_{12})^{-(n+1)} \times \br && \Big[(\fsl{x}_{12})_{ab}\psi_{b}^{A} + \rho_{2n+1} x^{2}_{12} (\bpsi \psi)\psi^{A}_{a} \Big]
\eea
which are used to fix the anomalous dimensions of odd operators. In \cite{Chethan-Pallab} a recursive algorithm was used to count Wick contractions, which can be adapted for the fermions. The Wick contractions can then be viewed as various ways of connecting upper and lower rows, resulting in recursive equations. In the case of fermions, the principle is essentially the same, but the contractions have a bit more structure. We use '$+$' and '$-$' to denote $\bpsi$ and $\psi$ respectively. To capture the contractions, we introduce the quantity $F^{p,r_{+},r_{-}}_{p+q,s_{+},s_{-};m_{+},m_{-}}$, where $p$ is the number of upper double cow-pies\footnote{Cow-pies is the jargon originally used in \cite{Chethan-Pallab}, which refers to the pictorial representation of individual fields that are being Wick-contracted. In our case they correspond to boxes which carry a $+$ or $-$ sign denoting $\bpsi$ and $\psi$ respectively.} which stand for $\bpsi \psi$, $r_{+}$ is the number of upper single cow-pies of '$+$' type, $r_{-}$ is the number of upper single cow-pies of '$+$' type, $p+q$ is the number of lower double cow-pies, $s_{\pm}$ is the number of lower single cow-pies of type $\pm$, $m_{\pm}$ is the number of uncontracted $\bpsi$s and  $\psi$s respectively. A contraction is always between an upper $+$ and a lower $-$ or vice-versa.

\begin{figure}[H]
	\begin{tikzpicture}[scale=0.8]
	
	\draw (2,-0.25) node[anchor=south] {\textbullet\quad\textbullet\quad\textbullet};
	\draw (7,-0.25) node[anchor=south] {\textbullet\textbullet\textbullet};
	\begin{scope}
	\draw (0,0) ellipse (1cm and 0.5cm);
	\node[draw=red] at (-0.4,0)  {$+$};
	\node[draw=red] at (0.4,0)  {$-$} ;
	\end{scope}

	\begin{scope}[xshift=4cm]
	\draw (0,0) ellipse (1cm and 0.5cm);
	\node[draw=red] at (-0.4,0)  {$+$};
	\node[draw=red] at (0.4,0)  {$-$} ;
	\end{scope}
	
	\begin{scope}[xshift=6cm]
	\draw (0,0) circle (0.5cm);
	\node[draw=red] at (0,0)  {$+$} ;
	\end{scope}
	
	\begin{scope}[xshift=8cm]
	\draw (0,0) circle (0.5cm);
	\node[draw=red] at (0,0)  {$+$} ;
	\end{scope}
	\begin{scope}[xshift=10cm]
	\draw (0,0) circle (0.5cm);
	\node[draw=red] at (0,0)  {$-$} ;
	\end{scope}
	
	\begin{scope}[xshift=12cm]
	\draw (0,0) circle (0.5cm);
	\node[draw=red] at (0,0)  {$-$} ;
	\end{scope}
	\draw (2,-2.25) node[anchor=south] {\textbullet\quad\textbullet\quad\textbullet};
	\draw (6,-2.25) node[anchor=south] {\textbullet\quad\textbullet\quad\textbullet};
	\draw (11,-2.25) node[anchor=south] {\textbullet\textbullet\textbullet};
	
	\begin{scope}[yshift=-2cm]
	\draw (0,0) ellipse (1cm and 0.5cm);
	\node[draw=red] at (-0.4,0)  {$+$};
	\node[draw=red] at (0.4,0)  {$-$} ;
	\end{scope}
	
	\begin{scope}[xshift=4cm,yshift=-2cm]
	\draw (0,0) ellipse (1cm and 0.5cm);
	\node[draw=red] at (-0.4,0)  {$+$};
	\node[draw=red] at (0.4,0)  {$-$} ;
	\end{scope}
	
	\begin{scope}[xshift=8cm,yshift=-2cm]
	\draw (0,0) ellipse (1cm and 0.5cm);
	\node[draw=red] at (-0.4,0)  {$+$};
	\node[draw=red] at (0.4,0)  {$-$} ;
	\end{scope}
	
	\begin{scope}[xshift=10cm,yshift=-2cm]
	\draw (0,0) circle (0.5cm);
	\node[draw=red] at (0,0)  {$+$} ;
	\end{scope}
	
	\begin{scope}[xshift=12cm,yshift=-2cm]
	\draw (0,0) circle (0.5cm);
	\node[draw=red] at (0,0)  {$+$} ;
	\end{scope}
	
	\begin{scope}[xshift=14cm,yshift=-2cm]
	\draw (0,0) circle (0.5cm);
	\node[draw=red] at (0,0)  {$-$} ;
	\end{scope}
	
	\begin{scope}[xshift=16cm,yshift=-2cm]
	\draw (0,0) circle (0.5cm);
	\node[draw=red] at (0,0)  {$-$} ;
	\end{scope}
	
	\begin{scope}[yshift=0.7cm]
	\draw [thick,decorate,decoration={brace,amplitude=10pt}] (-1.0,0) -- (5.0,0);
	\node[] at (2.05,0.55) {\large $p$}; 
	\end{scope}
	
	\begin{scope}[yshift=0.7cm]
	\draw [thick,decorate,decoration={brace,amplitude=10pt}] (5.5,0) -- (8.5,0);
	\node[] at (7.05,0.55) {\large $r_{+}$}; 
	\end{scope}
	
	\begin{scope}[yshift=0.7cm]
	\draw [thick,decorate,decoration={brace,amplitude=10pt}] (9.5,0) -- (12.5,0);
	\node[] at (11.0,0.55) {\large $r_{-}$}; 
	\end{scope}
	
	\begin{scope}[yshift=-2.7cm]
	\draw [thick,decorate,decoration={brace,mirror,amplitude=10pt}] (-1.0,0) -- (9.0,0);
	\node[] at (4.0,-.6) {\large $p+q$}; 
	\end{scope}
	
	\begin{scope}[yshift=-2.7cm]
	\draw [thick,decorate,decoration={brace,mirror,amplitude=10pt}] (9.5,0) -- (12.5,0);
	\node[] at (11.0,-.6) {\large $s_{+}$}; 
	\end{scope}
	
	\begin{scope}[yshift=-2.7cm]
	\draw [thick,decorate,decoration={brace,mirror,amplitude=10pt}] (13.5,0) -- (16.5,0);
	\node[] at (15.0,-.6) {\large $s_{-}$}; 
	\end{scope}
	
	\end{tikzpicture}
\end{figure}
The various coefficients $f$s and $\rho$s in our notation becomes

\begin{eqnarray}
f_{2p} = F^{p,0,0}_{p,0,1;0,1} & \quad & f_{2p}\rho_{2p} = F^{p,0,0}_{p,0,1;1,2} \\ \nonumber
f_{2p+1} = F^{p,0,1}_{p+1,0,0;0,1} & \quad & f_{2p+1}\rho_{2p+1} = F^{p,0,0}_{p+1,0,0;1,2} \nonumber
\end{eqnarray}

\subsection{$\mathbf{f_{2p}}$}

There are 3 different kinds of contractions that are possible. Of the first type, the two kernels of the $p^{\mathrm{th}}$ double cow-pie are contracted with two kernels of same lower cow-pie. This gives a factor of $Np$. The second possibility is to contract the two kernels of upper cow-pie to two different kernels of lower double cow-pie resulting in a factor of $-p(p-1)$ and the last possibility is to contract one of the kernels of upper double cow-pie with a kernel in lower double cow-pie and the second kernel of upper cow-pie with the single kernel of lower row. This gives a factor of $-p$. So, the resulting contraction can now be expressed as following recursion equation

\begin{equation}
F^{p,0,0}_{p,0,1;0,1} = (Np - p(p-1) - p)F^{p-1;0}_{p-1,1;0,1}
\end{equation}

\begin{figure}[H]
	\begin{tikzpicture}
	
	\draw (5,-0.25) node[anchor=south] {\textbullet\quad\textbullet\quad\textbullet};
	
	\begin{scope}
	\draw (0,0) ellipse (1cm and 0.5cm);
	\node[draw=red] at (-0.4,0)  {$+$};
	\node[draw=red] at (0.4,0)  {$-$} ;
	\end{scope}

	\begin{scope}[xshift=3cm]
	\draw (0,0) ellipse (1cm and 0.5cm);
	\node[draw=red] at (-0.4,0)  {$+$};
	\node[draw=red] at (0.4,0)  {$-$} ;
	\end{scope}
	
	\begin{scope}[xshift=7cm]
	\draw (0,0) ellipse (1cm and 0.5cm);
	\node[draw=red] at (-0.4,0)  {$+$};
	\node[draw=red] at (0.4,0)  {$-$} ;
	\end{scope}
	
	\draw (5,-2.25) node[anchor=south] {\textbullet\quad\textbullet\quad\textbullet};
	
	\begin{scope}[yshift=-2cm]
	\draw (0,0) ellipse (1cm and 0.5cm);
	\node[draw=red] at (-0.4,0)  {$+$};
	\node[draw=red] at (0.4,0)  {$-$} ;
	\end{scope}
	
	\begin{scope}[xshift=3cm,yshift=-2cm]
	\draw (0,0) ellipse (1cm and 0.5cm);
	\node[draw=red] at (-0.4,0)  {$+$};
	\node[draw=red] at (0.4,0)  {$-$} ;
	\end{scope}
	
	\begin{scope}[xshift=7cm,yshift=-2cm]
	\draw (0,0) ellipse (1cm and 0.5cm);
	\node[draw=red] at (-0.4,0)  {$+$};
	\node[draw=red] at (0.4,0)  {$-$} ;
	\end{scope}
	
	\begin{scope}[xshift=9cm,yshift=-2cm]
	\draw (0,0) circle (0.5cm);
	\node[draw=red] at (0,0)  {$-$} ;
	
	\end{scope}
	
	\draw (-0.4,0) -- (3.4,-2);
	\draw (0.4,0) --  (2.6,-2);
	
	
	\end{tikzpicture}
\end{figure}

\begin{figure}[H]
	\begin{tikzpicture}
	
	\draw (5,-0.25) node[anchor=south] {\textbullet\quad\textbullet\quad\textbullet};
	
	\begin{scope}
	\draw (0,0) ellipse (1cm and 0.5cm);
	\node[draw=red] at (-0.4,0)  {$+$};
	\node[draw=red] at (0.4,0)  {$-$} ;
	\end{scope}

	\begin{scope}[xshift=3cm]
	\draw (0,0) ellipse (1cm and 0.5cm);
	\node[draw=red] at (-0.4,0)  {$+$};
	\node[draw=red] at (0.4,0)  {$-$} ;
	\end{scope}
	
	\begin{scope}[xshift=7cm]
	\draw (0,0) ellipse (1cm and 0.5cm);
	\node[draw=red] at (-0.4,0)  {$+$};
	\node[draw=red] at (0.4,0)  {$-$} ;
	\end{scope}
	
	\draw (5,-2.25) node[anchor=south] {\textbullet\quad\textbullet\quad\textbullet};
	
	\begin{scope}[yshift=-2cm]
	\draw (0,0) ellipse (1cm and 0.5cm);
	\node[draw=red] at (-0.4,0)  {$+$};
	\node[draw=red] at (0.4,0)  {$-$} ;
	\end{scope}
	
	\begin{scope}[xshift=3cm,yshift=-2cm]
	\draw (0,0) ellipse (1cm and 0.5cm);
	\node[draw=red] at (-0.4,0)  {$+$};
	\node[draw=red] at (0.4,0)  {$-$} ;
	\end{scope}
	
	\begin{scope}[xshift=7cm,yshift=-2cm]
	\draw (0,0) ellipse (1cm and 0.5cm);
	\node[draw=red] at (-0.4,0)  {$+$};
	\node[draw=red] at (0.4,0)  {$-$} ;
	\end{scope}
	
	\begin{scope}[xshift=9cm,yshift=-2cm]
	\draw (0,0) circle (0.5cm);
	\node[draw=red] at (0,0)  {$-$} ;
	
	\end{scope}
	
	\draw (-0.4,0) -- (3.4,-2);
	\draw (0.4,0) --  (6.6,-2);
	
	
	\end{tikzpicture}
\end{figure}

\begin{figure}[H]
	\begin{tikzpicture}
	
	\draw (5,-0.25) node[anchor=south] {\textbullet\quad\textbullet\quad\textbullet};
	
	\begin{scope}
	\draw (0,0) ellipse (1cm and 0.5cm);
	\node[draw=red] at (-0.4,0)  {$+$};
	\node[draw=red] at (0.4,0)  {$-$} ;
	\end{scope}

	\begin{scope}[xshift=3cm]
	\draw (0,0) ellipse (1cm and 0.5cm);
	\node[draw=red] at (-0.4,0)  {$+$};
	\node[draw=red] at (0.4,0)  {$-$} ;
	\end{scope}
	
	\begin{scope}[xshift=7cm]
	\draw (0,0) ellipse (1cm and 0.5cm);
	\node[draw=red] at (-0.4,0)  {$+$};
	\node[draw=red] at (0.4,0)  {$-$} ;
	\end{scope}
	
	\draw (5,-2.25) node[anchor=south] {\textbullet\quad\textbullet\quad\textbullet};
	
	\begin{scope}[yshift=-2cm]
	\draw (0,0) ellipse (1cm and 0.5cm);
	\node[draw=red] at (-0.4,0)  {$+$};
	\node[draw=red] at (0.4,0)  {$-$} ;
	\end{scope}
	
	\begin{scope}[xshift=3cm,yshift=-2cm]
	\draw (0,0) ellipse (1cm and 0.5cm);
	\node[draw=red] at (-0.4,0)  {$+$};
	\node[draw=red] at (0.4,0)  {$-$} ;
	\end{scope}
	
	\begin{scope}[xshift=7cm,yshift=-2cm]
	\draw (0,0) ellipse (1cm and 0.5cm);
	\node[draw=red] at (-0.4,0)  {$+$};
	\node[draw=red] at (0.4,0)  {$-$} ;
	\end{scope}
	
	\begin{scope}[xshift=9cm,yshift=-2cm]
	\draw (0,0) circle (0.5cm);
	\node[draw=red] at (0,0)  {$-$} ;
	
	\end{scope}
	
	\draw (-0.4,0) -- (9.1,-2);
	\draw (0.4,0) --  (2.6,-2);
	
	
	\end{tikzpicture}
\end{figure}
This recursion equation can be solved along with the launching condition $F^{0;0}_{0,1;0,1}=1$ and we obtain

\begin{equation}\label{f2p}
f_{2p} = p! (N-1)(N-2)\cdots (N-p)
\end{equation}

\subsection{$\mathbf{f_{2p+1}}$}

There are only two types of contractions possible, analogous to the first two types above. The recursion equation can therefore be written by inspection

\begin{equation}
F^{p;1}_{p+1,0;0,1} = (N(p+1) - p(p+1))F^{p-1,0,1}_{p,0,0;0,1}
\end{equation}
with the lauching condition $F^{0;1}_{1,0;0,1}=1$ which gives

\begin{equation}\label{f2pp1}
f_{2p+1} = (p+1)!(N-1)(N-2)\cdots (N-p)
\end{equation}

\subsection{$\mathbf{f_{2p}\rho_{2p}}$}

\noindent Here the first possibility involves contracting both the kernels of upper double cow-pie with lower cow-pies analogous to the computation of $f_{2p}$. This gives a factor of $Np - p(p-1) - p$. Another possibility involves contracting '$+$' of an upper double cow-pie with the single '$-$' in the lower row. This gives a factor of $-F^{p-1;0}_{p-1,0;1,1}$.

\begin{figure}[H]
	\begin{tikzpicture}
	
	\draw (5,-0.25) node[anchor=south] {\textbullet\quad\textbullet\quad\textbullet};
	
	\begin{scope}
	\draw (0,0) ellipse (1cm and 0.5cm);
	\node[draw=red] at (-0.4,0)  {$+$};
	\node[draw=red] at (0.4,0)  {$-$} ;
	\end{scope}

	\begin{scope}[xshift=3cm]
	\draw (0,0) ellipse (1cm and 0.5cm);
	\node[draw=red] at (-0.4,0)  {$+$};
	\node[draw=red] at (0.4,0)  {$-$} ;
	\end{scope}
	
	\begin{scope}[xshift=7cm]
	\draw (0,0) ellipse (1cm and 0.5cm);
	\node[draw=red] at (-0.4,0)  {$+$};
	\node[draw=red] at (0.4,0)  {$-$} ;
	\end{scope}

	\draw (5,-2.25) node[anchor=south] {\textbullet\quad\textbullet\quad\textbullet};
	
	\begin{scope}[yshift=-2cm]
	\draw (0,0) ellipse (1cm and 0.5cm);
	\node[draw=red] at (-0.4,0)  {$+$};
	\node[draw=red] at (0.4,0)  {$-$} ;
	\end{scope}
	
	\begin{scope}[xshift=3cm,yshift=-2cm]
	\draw (0,0) ellipse (1cm and 0.5cm);
	\node[draw=red] at (-0.4,0)  {$+$};
	\node[draw=red] at (0.4,0)  {$-$} ;
	\end{scope}
	
	\begin{scope}[xshift=7cm,yshift=-2cm]
	\draw (0,0) ellipse (1cm and 0.5cm);
	\node[draw=red] at (-0.4,0)  {$+$};
	\node[draw=red] at (0.4,0)  {$-$} ;
	\end{scope}
	
	\begin{scope}[xshift=9cm,yshift=-2cm]
	\draw (0,0) circle (0.5cm);
	\node[draw=red] at (0,0)  {$-$} ;
	\end{scope}
	
	\draw (-0.4,0) -- (9.0,-2);
	
	
	\end{tikzpicture}
\end{figure}
Other possibilities involve single contractions of upper kernels, which can be done in following ways: (a) '$+$' of the upper double cow-pie contracted with a '$-$' of the lower double cow-pie, (b) '$-$' from the upper double cow-pie with a $+$ from lower double cow-pie. One can see by explicit computation for the lower orders (\textit{i.e.} $p=2,3,\cdots$)  that their contribution is given by $-p F^{p-1,0,0}_{p-1,0,1;1,2}$. Notice that the coefficient is different from the naive expectation because not all single contractions are independent and we must be careful to avoid over-counting and to keep track of the index structure.

\begin{figure}[H]
	\begin{tikzpicture}
	
	\draw (5,-0.25) node[anchor=south] {\textbullet\quad\textbullet\quad\textbullet};
	
	\begin{scope}
	\draw (0,0) ellipse (1cm and 0.5cm);
	\node[draw=red] at (-0.4,0)  {$+$};
	\node[draw=red] at (0.4,0)  {$-$} ;
	\end{scope}

	\begin{scope}[xshift=3cm]
	\draw (0,0) ellipse (1cm and 0.5cm);
	\node[draw=red] at (-0.4,0)  {$+$};
	\node[draw=red] at (0.4,0)  {$-$} ;
	\end{scope}
	
	\begin{scope}[xshift=7cm]
	\draw (0,0) ellipse (1cm and 0.5cm);
	\node[draw=red] at (-0.4,0)  {$+$};
	\node[draw=red] at (0.4,0)  {$-$} ;
	\end{scope}
	
	\draw (5,-2.25) node[anchor=south] {\textbullet\quad\textbullet\quad\textbullet};
	
	\begin{scope}[yshift=-2cm]
	\draw (0,0) ellipse (1cm and 0.5cm);
	\node[draw=red] at (-0.4,0)  {$+$};
	\node[draw=red] at (0.4,0)  {$-$} ;
	\end{scope}
	
	\begin{scope}[xshift=3cm,yshift=-2cm]
	\draw (0,0) ellipse (1cm and 0.5cm);
	\node[draw=red] at (-0.4,0)  {$+$};
	\node[draw=red] at (0.4,0)  {$-$} ;
	\end{scope}
	
	\begin{scope}[xshift=7cm,yshift=-2cm]
	\draw (0,0) ellipse (1cm and 0.5cm);
	\node[draw=red] at (-0.4,0)  {$+$};
	\node[draw=red] at (0.4,0)  {$-$} ;
	\end{scope}
	
	\begin{scope}[xshift=9cm,yshift=-2cm]
	\draw (0,0) circle (0.5cm);
	\node[draw=red] at (0,0)  {$-$} ;
	
	\end{scope}
	
	\draw (-0.4,0) -- (3.4,-2);
	
	
	\end{tikzpicture}
\end{figure}

\begin{figure}[H]
	\begin{tikzpicture}
	
	\draw (5,-0.25) node[anchor=south] {\textbullet\quad\textbullet\quad\textbullet};
	
	\begin{scope}
	\draw (0,0) ellipse (1cm and 0.5cm);
	\node[draw=red] at (-0.4,0)  {$+$};
	\node[draw=red] at (0.4,0)  {$-$} ;
	\end{scope}

	\begin{scope}[xshift=3cm]
	\draw (0,0) ellipse (1cm and 0.5cm);
	\node[draw=red] at (-0.4,0)  {$+$};
	\node[draw=red] at (0.4,0)  {$-$} ;
	\end{scope}
	
	\begin{scope}[xshift=7cm]
	\draw (0,0) ellipse (1cm and 0.5cm);
	\node[draw=red] at (-0.4,0)  {$+$};
	\node[draw=red] at (0.4,0)  {$-$} ;
	\end{scope}
	
	\draw (5,-2.25) node[anchor=south] {\textbullet\quad\textbullet\quad\textbullet};
	
	\begin{scope}[yshift=-2cm]
	\draw (0,0) ellipse (1cm and 0.5cm);
	\node[draw=red] at (-0.4,0)  {$+$};
	\node[draw=red] at (0.4,0)  {$-$} ;
	\end{scope}
	
	\begin{scope}[xshift=3cm,yshift=-2cm]
	\draw (0,0) ellipse (1cm and 0.5cm);
	\node[draw=red] at (-0.4,0)  {$+$};
	\node[draw=red] at (0.4,0)  {$-$} ;
	\end{scope}
	
	\begin{scope}[xshift=7cm,yshift=-2cm]
	\draw (0,0) ellipse (1cm and 0.5cm);
	\node[draw=red] at (-0.4,0)  {$+$};
	\node[draw=red] at (0.4,0)  {$-$} ;
	\end{scope}
	
	\begin{scope}[xshift=9cm,yshift=-2cm]
	\draw (0,0) circle (0.5cm);
	\node[draw=red] at (0,0)  {$-$} ;
	
	\end{scope}
	
	\draw (0.4,0) --  (2.6,-2);
	
	
	\end{tikzpicture}
\end{figure}
Thus we get the recursion equation

\begin{equation}
F^{p,0,0}_{p,0,1;1,2} = p\left[N - p -1 \right]F^{p-1,0,0}_{p-1,0;1,2} - F^{p-1,0,1}_{p,0,0;1,1}
\end{equation}
$F^{p-1,0,1}_{p,0,0;1,1}$ can be again evaluated using the cow-pie formalism and its recursion equation is given by

\begin{equation}
F^{p;0}_{p,0;1,1} = (p+1)(N-p)F^{p-1,0}_{p,0;1,1}
\end{equation}
This system of recursion equations can be solved using the launching condition $F^{0,0,0}_{0,0,1;1,2}=0$ and $F^{0,0,0}_{1,0,0;1,1}=1$. Using the expression for $f_{2p}$ in \eqref{f2p} we get,

\begin{equation}\label{evenrho}
\rho_{2p} = -\frac{p}{N-1}
\end{equation}

\subsection{$\mathbf{f_{2p+1}\rho_{2p+1}}$}

We again have three cases to consider: (a) Both kernels of the upper cow-pie contracted with lower cow-pies, (b) Both kernel remain uncontracted, and (c) Only one of the kernels is contracted.\\

Case (a) is similar to the computation of $f_{2p+1}$ and gives a factor of $(p+1)(N-p)F^{p,0,0}_{p-1,0,1;1,2}$. Case (b) does not contribute as we do not obtain the desired operator. Case (c) is similar to the case of $f_{2p}\rho_{2p}$. Once again, by explicit computation for the lowest order, we can see that its contribution is $-(p+1)F^{p,0,0}_{p-1,0,1;1,2}$.

\begin{figure}[H]
	\begin{tikzpicture}
	
	\draw (5,-0.25) node[anchor=south] {\textbullet\quad\textbullet\quad\textbullet};
	
	\begin{scope}
	\draw (0,0) ellipse (1cm and 0.5cm);
	\node[draw=red] at (-0.4,0)  {$+$};
	\node[draw=red] at (0.4,0)  {$-$} ;
	\end{scope}

	\begin{scope}[xshift=3cm]
	\draw (0,0) ellipse (1cm and 0.5cm);
	\node[draw=red] at (-0.4,0)  {$+$};
	\node[draw=red] at (0.4,0)  {$-$} ;
	\end{scope}
	
	\begin{scope}[xshift=7cm]
	\draw (0,0) ellipse (1cm and 0.5cm);
	\node[draw=red] at (-0.4,0)  {$+$};
	\node[draw=red] at (0.4,0)  {$-$} ;
	\end{scope}
	
	\begin{scope}[xshift=9cm]
	\draw (0,0) circle (0.5cm);
	\node[draw=red] at (0,0)  {$-$} ;
	\end{scope}
	
	\draw (5,-2.25) node[anchor=south] {\textbullet\quad\textbullet\quad\textbullet};
	
	\begin{scope}[yshift=-2cm]
	\draw (0,0) ellipse (1cm and 0.5cm);
	\node[draw=red] at (-0.4,0)  {$+$};
	\node[draw=red] at (0.4,0)  {$-$} ;
	\end{scope}
	
	\begin{scope}[xshift=3cm,yshift=-2cm]
	\draw (0,0) ellipse (1cm and 0.5cm);
	\node[draw=red] at (-0.4,0)  {$+$};
	\node[draw=red] at (0.4,0)  {$-$} ;
	\end{scope}
	
	\begin{scope}[xshift=7cm,yshift=-2cm]
	\draw (0,0) ellipse (1cm and 0.5cm);
	\node[draw=red] at (-0.4,0)  {$+$};
	\node[draw=red] at (0.4,0)  {$-$} ;
	\end{scope}
	
	\begin{scope}[xshift=10cm,yshift=-2cm]
	\draw (0,0) ellipse (1cm and 0.5cm);
	\node[draw=red] at (-0.4,0)  {$+$};
	\node[draw=red] at (0.4,0)  {$-$} ;
	\end{scope}
	
	\draw (-0.4,0) -- (3.4,-2);
	
	
	\end{tikzpicture}
\end{figure}

\begin{figure}[H]
	\begin{tikzpicture}
	
	\draw (5,-0.25) node[anchor=south] {\textbullet\quad\textbullet\quad\textbullet};
	
	\begin{scope}
	\draw (0,0) ellipse (1cm and 0.5cm);
	\node[draw=red] at (-0.4,0)  {$+$};
	\node[draw=red] at (0.4,0)  {$-$} ;
	\end{scope}

	\begin{scope}[xshift=3cm]
	\draw (0,0) ellipse (1cm and 0.5cm);
	\node[draw=red] at (-0.4,0)  {$+$};
	\node[draw=red] at (0.4,0)  {$-$} ;
	\end{scope}
	
	\begin{scope}[xshift=7cm]
	\draw (0,0) ellipse (1cm and 0.5cm);
	\node[draw=red] at (-0.4,0)  {$+$};
	\node[draw=red] at (0.4,0)  {$-$} ;
	\end{scope}
	
	\begin{scope}[xshift=9cm]
	\draw (0,0) circle (0.5cm);
	\node[draw=red] at (0,0)  {$-$} ;
	\end{scope}
	
	\draw (5,-2.25) node[anchor=south] {\textbullet\quad\textbullet\quad\textbullet};
	
	\begin{scope}[yshift=-2cm]
	\draw (0,0) ellipse (1cm and 0.5cm);
	\node[draw=red] at (-0.4,0)  {$+$};
	\node[draw=red] at (0.4,0)  {$-$} ;
	\end{scope}
	
	\begin{scope}[xshift=3cm,yshift=-2cm]
	\draw (0,0) ellipse (1cm and 0.5cm);
	\node[draw=red] at (-0.4,0)  {$+$};
	\node[draw=red] at (0.4,0)  {$-$} ;
	\end{scope}
	
	\begin{scope}[xshift=7cm,yshift=-2cm]
	\draw (0,0) ellipse (1cm and 0.5cm);
	\node[draw=red] at (-0.4,0)  {$+$};
	\node[draw=red] at (0.4,0)  {$-$} ;
	\end{scope}
	
	\begin{scope}[xshift=10cm,yshift=-2cm]
	\draw (0,0) ellipse (1cm and 0.5cm);
	\node[draw=red] at (-0.4,0)  {$+$};
	\node[draw=red] at (0.4,0)  {$-$} ;
	\end{scope}
	
	\draw (0.4,0) --  (2.6,-2);
	
	
	\end{tikzpicture}
\end{figure}
So we have following recursion equation
\begin{equation}
F^{p,0,1}_{p+1,0,0;1,2} = (p+1)(N-p-1)F^{p-1,0,1}_{p,0,0;1,2}
\end{equation}
which can be solved with the launching condition $F^{1,0,0}_{0,0,1;1,2}=1$. Using \eqref{f2pp1} along with the recursion equations above, we get

\begin{equation}\label{oddrho}
\rho_{2p+1} = 1- \frac{p}{N-1}
\end{equation}
In the appendix we provide an alternate derivation of \eqref{evenrho} and \eqref{oddrho} using cow-pies.

\section{Matching with the Free Theory}
Having fixed the OPE coefficients of the free theory, we are now in a position to compute the anomalous dimensions of the interacting theory operators. This involves analyzing 3-point functions with $V_{2n}\times V^{A}_{2n+1\;a}$ OPEs in \eqref{intOPE} and demanding that in the $\e \rightarrow 0$, they go to corresponding quantities in free theory. In particular, we analyze 3-point correlators of the form

\begin{eqnarray}\label{int3pt1}
\langle V_{2n}(x_1)\;V_{2n+1\;a}^{A}(x_2)\;\bV^{B}_{1\;b}(x_3) \rangle &\rightarrow & \langle \phi_{2n}(x_1)\phi^{A}_{2n+1\;a}(x_2)\bphi^{B}_{1\;b}(x_3) \rangle \\ \nonumber  &\sim & f_{2n} (x^{2}_{12})^{-n} \langle \psi^{A}_{a}(x_2)\bpsi^{B}_{b}(x_3)
\end{eqnarray}
and

\begin{eqnarray}\label{int3pt2}
\langle V_{2n}(x_1)\;V_{2n+1\;a}^{A}(x_2)\;\bV^{B}_{3\;b}(x_3) \rangle &\rightarrow & \langle \phi_{2n}(x_1)\phi^{A}_{2n+1\;a}(x_2)\bphi^{B}_{3\;b}(x_3) \rangle \\ \nonumber & \sim & f_{2n}\rho_{2n} (x^{2}_{12})^{-n} (\fsl{x}_{12})_{ab} \langle \phi^{A}_{3\;a}(x_2)\bphi^{B}_{3\;b}(x_3)\rangle
\end{eqnarray}
The LHS of \eqref{int3pt2} can be evaluated, to the leading order, using $V_{2n}\times V^{A}_{2n+1\;a}$ OPE of \eqref{intOPE} and the fact that $V_{3}$ is a descendent of $V_1$, i.e,

\bea
\langle V^{A}_{1\;a}(x_1)\bV^{B}_{3\;b} \rangle &=& \alpha^{-1}(\e) \partial_{2\;\mu} \langle V^{A}_{1\;a}(x_1) \bV^{B}_{1\;c} \rangle (\gamma^{\mu})_{cb} \br  &=& \sigma\sqrt{(N-1)\gamma_1}\frac{\delta_{ab} \delta^{AB}}{(x^{2}_{12})^{\Delta_1 + \frac{1}{2}}}
\eea
Since this is proportional to $\sqrt{\gamma_1}$, it vanishes in the $\e \rightarrow 0$ limit. Therefore, to reproduce \eqref{freeOPE1} and \eqref{freeOPE2} we need $q_1$ to remain finite in this limit. We also need $q_2$ to blow up as $\e \rightarrow 0$ limit such that 

\bea
q^{i}_2 \alpha(\e) \rightarrow \rho_{i}, \quad i=2p,\;2p+1
\eea
The coefficients $q_i$ are determined by conformal symmetry whose details and explicit form can be found in Appendix. As alluded before, we find that $q_1$ is indeed finite in the free limit. The asymptotic behavior of $q_2$ is given by

\bea
q_{2}^{2n}\approx \frac{(\gamma_1 + \gamma_{2n}-\gamma_{2n+1})}{4 \gamma_1}, \qquad q_{2}^{2n+1}\approx \frac{(\gamma_1 + \gamma_{2n+1}-\gamma_{2n+2})}{4 \gamma_1}
\eea
Its evident that for $q_2$ to blow up $y_{1,1}$ has to vanish. This gives us following telescoping series

\begin{eqnarray}
y_{2n,1}-y_{2n+1,1} &=& 2\sigma \sqrt{(N-1)y_{1,2}}\;\rho_{2n}, \quad n = 1,2,\cdots \\ 
y_{2n+1,1}-y_{2n+2,1} &=& 2\sigma \sqrt{(N-1)y_{1,2}}\;\rho_{2n+1} \quad n = 0,1,\cdots
\end{eqnarray}
Together this can be written as

\bea
y_{i,1} - y_{i+1,1} = 2\sigma \sqrt{(N-1)y_{1,2}}\;\rho_{i}, \qquad i=1,2,\cdots
\eea
Summing the telescoping series gives

\begin{eqnarray}\label{anomdim}
y_{n,1} = K\sum^{n-1}_{m=1}\rho_{m}
\end{eqnarray}
This gives the anomalous dimensions of all the odd and even primaries in the theory once we fix the numerical value of $K$. To fix this we make use of \eqref{dimconstraint} which can be written as

\bea
2\delta + \gamma_3 = \gamma_1 + 1
\eea
This gives $y_{3,1}=-1$ which can now be used to fix $K$ by setting $n=3$ in \eqref{anomdim}.

\begin{equation}
y_{3,1} = K(\rho_1 + \rho_2)
\end{equation}
This gives $K=-\frac{(N-1)}{(N-2)}$ which fixes $\sigma=-1$ and furthermore fixes $y_{1,2}$ also. Thus we obtain

\begin{equation}\label{gamma1}
\gamma_1 = \frac{(N-1)}{4(N-2)^2}\e^2
\end{equation}
One can also compute the anomalous dimensions of $V_2$ which we obtain to be

\begin{equation}\label{gamma2}
\gamma_2 = -\frac{(N-1)}{(N-2)}\e
\end{equation}
which are in perfect agreement with the results of \cite{Derkachov,Gracey1,Gracey2}.

\section{Conclusion}
We have considered $O(N)$ Gross-Neveu model in $2+\e$ dimensions and computed the leading order anomalous dimension for a class of operators of the form $V_{2n}$, $V_{2n+1\; a}^A$ and $\bV_{2n+1\; a}^A$. In doing so, we have only appealed to the conformal symmetry and OPE, without getting bogged by Feynman diagrams. Our results are in agreement with the previous results in the literature obtained using Feynman diagrams. Our main result is \eqref{anomdim} along with the ``cow-pie'' formalism adapted to fermions, so our work provides an extension of \cite{Rychkov} and \cite{Chethan-Pallab} to fermions. 

It should also be noted that the leading order $\epsilon$-expansion coefficient is completely determined by conformal symmetry and does not depend on the dynamics. However, to go beyond the leading order, two and three point functions are not enough and one also needs information from the four point functions, as considered here \cite{Sen:2015doa}. 

\chapter{Summary \& Conclusions}
In this dissertation, we have explored various topics in $2+1$-D gravity and higher spin theories, conformal field theories and boundary conditions in gravity. The common theme that connects most of the work here is holography. \\

\section*{Higher Spin Theories}
Gravity and higher spin gauge theories in $2+1$ dimensions is substantially simple due to the lack of local dynamics. In our work, we have constructed the higher spin generalization of Kerr-de-Sitter and Quotient cosmologies for backgrounds with positive cosmological constant. A general spin-3 charged gauge field takes the form

\begin{eqnarray}
a &=& \left[\left(1 - \frac{L}{2l}\right)T_1 + i\left(1 + \frac{L}{2l}\right)T_2 - \frac{W}{8l}W_{-2} \right]d\omega \nonumber \\  &+& \mu \left[W_2 -\frac{L}{2l} W_0 + \frac{L^2}{16 l^2} W_{-2} - \frac{W}{l} (T_1 - i T_2) \right]d\bar{\omega} \nonumber
\end{eqnarray}
Thermodynamics of gravity solutions is an important testing ground for putative quantum gravity theories and it is therefore important to understand the thermodynamics of our solutions. We have demonstrated that the thermodynamics of higher spin generalizations of Kerr de Sitter universes are related to (higher spin) AdS black holes. A solution is said to be thermodynamically consistent if it obeys the first law of thermodynamics. The consistency of our (cosmological) solutions is ensured by imposing trivial holonomy on the time component of the gauge field. This is the higher spin equivalent of Gibbons-Hawking conical singularity argument and by a Euclidean continuation we show our solutions reproduce the thermodynamics of horizon (even though the notion of horizon is not gauge invariant). Higher spin theories share many approximate qualitative features with string theory. Thus the hope is that understanding higher spin theories can shed light on how string theory solves many of the puzzles in gravity such as singularity of the black holes and big bang. \\
For backgrounds with a negative cosmological constant, AdS/CFT provides a window to quantum gravity in those spaces. However, we live in a universe with a (tiny) positive cosmological constant and most of our particle physics are formulated in flat space. One would eventually like to understand quantum gravity is such scenario. In $2+1$ dimensions there exist a one way map based on Inonu-Wigner contraction of AdS$_3$ isometry algebra which relates quantities in AdS space with that of flat space. We provide a prescription to implement the Inonu-Wigner contraction, using a Grassmann parameter identified with the inverse AdS radius, $1/l \rightarrow \e, \e^2 =0$. Under this identification, a general AdS$_3$ solution (in a BMS like gauge) has the form

\begin{eqnarray}\nonumber
A &= -\frac{1}{\sqrt{2}}\left[\epsilon \left(\frac{\cM}{2}-1\right)du -\epsilon dr + \epsilon \cN d\phi + \left( \frac{\cM}{2}-1 \right) \right]T_0 \\ \nonumber &-\frac{1}{\sqrt{2}}\left[\epsilon \left(\frac{\cM}{2}+1\right)du -\epsilon dr + \epsilon \cN d\phi + \left( \frac{\cM}{2}+1 \right) \right]T_1 -\epsilon r d\phi T_2 \nonumber
\end{eqnarray}
which can be identified as the flat space solution. This maps the action, solutions and asymptotic charges of AdS into flat space, atleast at the classical level. We demonstrate that our prescription works even in higher spin theories and black hole singularity in higher spin theories. This is a piquing observation that need further investigation to see if there are deeper reasons why it works.\\
We have also generalized the recent work of Grumiller and Riegler on the most general AdS$_3$ boundary condition to higher spins. Taking their metric as our definition of asymptotically AdS space, we have constructed arguably the most general higher spin solution with the same metric asymptotically. Our solutions have the feature that the left moving sector has a Drinfeld-Sokolov reduced form while the right moving sector has completely unrestricted $sl(3)$ connection. Thus the asymptotic symmetry algebra of the left moving sector is a copy of $\cW_3$ and for the right moving sector we find the loop algebra $sl(3)_k$ with a central extension. The complete asymptotic symmetry algebra of these solutions is $\cW_3 \times sl(3)_k$ and we compare it with a previous work of Poojary and Suryanarayana where this asymptotic symmetry algebra has appeared before. Our solution is more general than that of Poojary-Suryanarayan since we find their metric also to be in the generalized Fefferman-Graham class.\\

\section*{Neumann Boundary Condition in Gravity and Holography}
We have also studied the issue of Neumann boundary condition in Einstein's gravity. In our work, we have constructed a boundary term for gravity that makes the variational problem well defined with Neumann boundary condition. This is the analogue of Gibbons-Hawking-York term for the Dirichlet boundary condition. To accomplish this, we sought a variational problem where the conjugate momentum associated with the boundary metric is held fixed. This gives a boundary  term which is similar to Gibbons-Hawking terms but with a different coefficient. Incidentally,  we find that this coefficient vanish in 4 dimensions, making Einstein-Hilbert action a well posed Nauemann problem in 4D. Our primary result is the explicit form of the action 

\begin{eqnarray}
S_{N} = \frac{1}{2\kappa}\int_{{\cal M}}d^{d+1}x \sqrt{-g}(R-2\Lambda) 
+ \frac{(3-d)}{2\kappa}\int_{\partial {\cal M}}d^{d}y \sqrt{|\gamma|}\varepsilon K
\end{eqnarray}
We also notice that in $d=2$ dimensions, our boundary term has appreared previously in the literature in other guises. The Neumann boundary condition is thermodynamically found to give a mixed ensemble and reproduce the first law of thermodynamics \cite{Pavan}. In asymptotically AdS spaces, our boundary condition lends itself to holographic interpretation as keeping the boundary stress tensor fixed. We have demonstrated that on-shell divergence of Neumann action can be eliminated by adding counter-terms that can be systematically computed by holographic renormalization. The renormalized action in various dimensions is 

\begin{eqnarray}
S^{d=2}_{ren} &=& \frac{1}{2\kappa}\int_{M}d^3x \sqrt{-g}(R-2\Lambda) + \frac{1}{2\kappa}\int_{\partial M}d^2x \sqrt{-\gamma}\Theta = S^{d=2}_N \br
S^{d=3}_{ren} &=& \frac{1}{2\kappa}\int_{M}d^4x \sqrt{-g}(R-2\Lambda) + \frac{1}{\kappa}\int_{\partial M}d^3x \sqrt{-\gamma}\left(1 - \frac{R[\gamma]}{4} \right) \br
S^{d=4}_{ren} &=& \frac{1}{2\kappa}\int_{M}d^5x \sqrt{-g}(R-2\Lambda) - \frac{1}{2\kappa}\int_{\partial M}d^4x \sqrt{-\gamma}\Theta +\frac{3}{\kappa}\int_{\partial M}d^4x \sqrt{-\gamma} \nonumber
\end{eqnarray}
where we have omitted the logarithmic term, responsible for trace anomaly. This term can be reinstated when required and do not contribute to on-shell action for simple solutions like AdS-Schwarzschild black holes. We find that when the trace anomaly vanish, the on-shell renormalized Dirichlet and Neumann action have same value. Our expectation is that there exist a version of holography with dynamical boundary metric whose investigation is deferred to future work.\\

\section*{$\varepsilon$-Expansion in the Gross-Neveu CFT}
In the third part of the thesis, we have studied $O(N)$ Gross-Neveu model in $2+\e$ dimension at the Wilson-Fisher fixed point. The aim of this chapter is to demonstrate the power of conformal symmetry in restricting the observables in the theory and in some cases, as considered here, in completely determining certain observable such as the anomalous dimension of composite operators. Here we generalized the earlier work of Rychkov and Tan to compute the anomalous dimension for a class of composite operators in the theory. In doing so we have employed the ``cow-pie'' contraction algorithm of Basu and Krishnan, extended to fermions. We find a general expression for the leading order anomalous dimension for the composite operators $V_{2n}$ and $V^{A}_{2n+1}$.

\begin{eqnarray}
y_{n,1} = -\frac{(N-1)}{(N-2)}\sum^{n-1}_{m=1}\rho_{m}
\end{eqnarray}
where $\rho_m$ is a specific OPE coefficient in the free theory. We compare our results to previous works in the literature, the leading order anomalous dimension of operators $\psi$ and $\bpsi \psi$, for instance, 

\begin{eqnarray}
\gamma_1 &=& \frac{(N-1)}{4(N-2)^2}\e^2, \\
\gamma_2 &=& -\frac{(N-1)}{(N-2)}\e,
\end{eqnarray}
match with the Feynman diagram computation. \\

\newpage

\appendix

\chapter{}
\section{$sl(3)$ algebra and generators}
For $sl(3)$ generators we use the principle embedding basis where the generators $\{ L_i, W_j \}$ satisfy the algebra

\begin{equation}
[L_i,L_j]  =  (i-j) L_{i+j}
\end{equation}
\begin{equation}
[L_i,W_j]  =  (2i-j) W_{i+j}
\end{equation}
\begin{equation}
[W_i, W_j] = -\frac{1}{3}(i-j)(2i^2 + 2j^2 -ij -8)L_{i+j}
\end{equation}

We work with the $3\times 3$ fundamental representation and the generators are explicitly given by

\begin{eqnarray}\label{generators}
L_{-1} &=& \left(
\begin{array}{ccc}
 0 & -2 & 0 \\
 0 & 0 & -2 \\
 0 & 0 & 0 \\
\end{array}
\right), \quad L_0 = \left(
\begin{array}{ccc}
 1 & 0 & 0 \\
 0 & 0 & 0 \\
 0 & 0 & -1 \\
\end{array}
\right), \quad L_1 = \left(
\begin{array}{ccc}
 0 & 0 & 0 \\
 1 & 0 & 0 \\
 0 & 1 & 0 \\
\end{array}
\right) \br 
W_{-2} &=& \left(
\begin{array}{ccc}
 0 & 0 & 8 \\
 0 & 0 & 0 \\
 0 & 0 & 0 \\
\end{array}
\right), \quad  W_{-1} = \left(
\begin{array}{ccc}
 0 & -2 & 0 \\
 0 & 0 & 2 \\
 0 & 0 & 0 \\
\end{array}
\right), \quad W_0 = \frac{2}{3}\left(
\begin{array}{ccc}
 1 & 0 & 0 \\
 0 & -2 & 0 \\
 0 & 0 & 1 \\
\end{array}
\right) \\
W_1 &=& \left(
\begin{array}{ccc}
 0 & 0 & 0 \\
 1 & 0 & 0 \\
 0 & -1 & 0 \\
\end{array}
\right), \quad W_2 = \left(
\begin{array}{ccc}
 0 & 0 & 0 \\
 0 & 0 & 0 \\
 2 & 0 & 0 \\
\end{array}
\right) \nonumber
\end{eqnarray} 
The Killing metric of $sl(3)$ is given by 

\begin{equation}
\gamma_{AB} = x\; \Tr[T_A,T_B], \quad x \in \mathbb{R}
\end{equation}
where $T_A \in \{L_i,W_j \},\; i=-1,\cdots,1,\;j=-2,\cdots,2$. We choose the constant $x$ such that

\begin{equation}
\gamma_{AB} = \left(
\begin{array}{cccccccc}
 0 & 0 & -1 & 0 & 0 & 0 & 0 & 0 \\
 0 & 1/2 & 0 & 0 & 0 & 0 & 0 & 0 \\
 -1 & 0 & 0 & 0 & 0 & 0 & 0 & 0 \\
 0 & 0 & 0 & 0 & 0 & 0 & 0 & 4 \\
 0 & 0 & 0 & 0 & 0 & 0 & -1 & 0 \\
 0 & 0 & 0 & 0 & 0 & 2/3 & 0 & 0 \\
 0 & 0 & 0 & 0 & -1 & 0 & 0 & 0 \\
 0 & 0 & 0 & 4 & 0 & 0 & 0 & 0 \\
\end{array}
\right)
\end{equation}

\section{Equations of Motion for General $sl(3,\bR)$ Connection}
For a general gauge field $a(t,\phi)$,

\begin{eqnarray}
a(t,\phi) &=& \left( \sum_{i=-1}^{1} \mui{i}(t,\phi)L_i + \sum_{j=-2}^{2} \nui{j}(t,\phi)W_j \right)dt \br &+& \left( \sum_{i=-1}^{1} \calL{i}(t,\phi)L_i + \sum_{j=-2}^{2} \calW{j}(t,\phi)W_j  \right)d\phi
\end{eqnarray}
equations of motion impose constraints on the chemical potentials and charges. The equation of motion is given by

\begin{equation}
F_{t\phi} = \partial_t a_{\phi} - \partial_{\phi}a_t + [a_t,a_{\phi}]
\end{equation}
which in the component form are given by

\begin{eqnarray}
\calL{1}\mui{0} - \calL{0}\mui{1} + 2\calW{1}\nui{0} &-& 2\calW{0}\nui{1} + 4\calW{-1}\nui{2} - 4\calW{2}\nui{-1} \br  && + \;\; \pp \mui{1} - \pt \calL{1} = 0
\end{eqnarray}
\begin{eqnarray}
-2\calL{-1}\mui{1} &+& 2\calL{1}\mui{-1} - 2\calW{-1}\nui{1} + 16\calW{-2}\nui{2} + 2\calW{1}\nui{-1} \br &-& 16\calW{2}\nui{-2} + \pp \nui{0} - \pt \calL{0} = 0 
\end{eqnarray}
\begin{eqnarray}
\calL{-1}\mui{0} - \calL{0}\mui{-1} + 2\calW{-1}\nui{0} &-& 4\calW{-2}\nui{1} + 2\calW{0}\nui{-1} + 4\calW{1}\nui{-2}\br && -\;\; \pp \nui{-1} + \pt \calL{-1} = 0
\end{eqnarray}
\begin{eqnarray}
2\calW{-2}\mui{0} - \calW{-1}\mui{-1} + \calL{-1}\nui{-1} + 2\calL{0}\nui{-2} - \pp \nui{-2} + \pt \calW{-2} = 0
\end{eqnarray}
\begin{eqnarray}
-\calW{-1}\mui{0} - 4\calW{-2}\mui{1} + 2\calW{0}\mui{-1} &-&  2\calL{-1}\nui{0} + \calL{0}\nui{-1} + 4\calL{1}\nui{-2} \br && +\;\; \pp \nui{-1} - \pt \calW{-1} = 0
\end{eqnarray}
\begin{eqnarray}
-3\calW{-1}\mui{1} + 3\calW{1}\mui{-1} - 3\calL{-1}\nui{1} + 3\calL{1}\nui{-1} + \pp \nui{0} - \pt \calW{0} = 0
\end{eqnarray}
\begin{eqnarray}
\calW{1}\mui{0} - 2\calW{0}\mui{1} + 4\calW{2}\mui{-1} &+& 2\calL{1}\nui{0} - \calL{0}\nui{1} - 4\calL{-1}\nui{2} \br && +\;\; \pp \nui{1} - \pt \calW{1} = 0
\end{eqnarray}
\begin{eqnarray}
2\calW{2}\mui{0} - \calW{1}\mui{1} + \calL{1}\nui{1} - 2\calL{0}\nui{2} + \pp \nui{2} - \pt\calW{2} = 0
\end{eqnarray}

\section{The Metric of Poojary-Suryanarayana}
In \cite{nemani}, the proposed gauge field solution \eqref{nemani_gauge1}, \eqref{nemani_gauge2} translates to the following metric in the Banados radial gauge $b = \exp(\rho L_0)$ that they work with:

\begin{eqnarray}
g_{\rho \rho} &=& 1,\quad g_{\rho t} = -\frac{1}{2}f^0, \quad g_{\rho \phi} = -\frac{1}{2}f^0 \br g_{tt} &=& \rme^{2\rho}(-1 + f^{-1}) + \left[ \frac{1}{4}f^{0^2} - f^1 f^{-1} + \frac{1}{3}g^{0^2} - g^{1}g^{-1} + \kappa + \tilde{\kappa} - f^{-1}\tilde{\kappa} + 4g^{-2}(g^2 + \tilde{\omega}) \right] \br &-& \rme^{-2\rho}\kappa(f^1 + \tilde{\kappa}) + 4\rme^{-4\rho}\omega(g^2 + \tilde{\omega}) \\
g_{t\phi} &=& \rme^{2\rho}f^{-1} + \left[ \frac{1}{4}f^{0^2} - f^1 f^{-1} + \frac{1}{3}g^{0^2} - g^{1}g^{-1} + \kappa - \tilde{\kappa}  + 4g^{-2}g^2 \right] - \rme^{-2\rho}f^{1}\kappa \br &+& 4\rme^{-4\rho}g^{2}\omega \br g_{\phi \phi} &=& \rme^{2\rho}(1+f^{-1}) + \left[ \frac{1}{4}f^{0^2} - f^1 (1+f^{-1}) + \frac{1}{3}g^{0^2} - g^{1}g^{-1} + \kappa + \tilde{\kappa} + f^{-1}\tilde{\kappa} + 4g^{-2}(g^2 - \tilde{\omega}) \right] \br &+& \rme^{-2\rho} \kappa(\tilde{\kappa} - f^1) + 4\rme^{-4\rho}\omega(g^2 - \tilde{\omega}) \nonumber
\end{eqnarray}
Notice that this metric does not fall under the usual Fefferman-Graham form because of the non-vanishing $g_{\rho t}$ and $g_{\rho \phi}$ components. However this metric is still allowed by the \textit{generalized Fefferman-Graham} class of metrics of \eqref{ads_metric}.

\chapter{}
\section{Asymptotic solution}

The relation between the various $g_i$'s (with $i<d$) in Fefferman-Graham expansion is determined by solving Einstein's equation iteratively. This was worked out in detail in \cite{deHaro} and here we collect some useful results for completeness. The indices below are raised with the metric $g_{0}$.

The determinant of induced metric on $\rho=\epsilon$ boundary can be expanded as follows

\begin{equation}\label{detexpansion}
\sqrt{-g} = \sqrt{-g_0}\left( 1 + \frac{1}{2}\epsilon \Tr (g_{0}^{-1}g_2) + \frac{1}{8}\epsilon^2 \left( (\Tr (g_{0}^{-1}g_2))^2 - \Tr (g_{0}^{-1}g_2)^2 \right) + O(\epsilon^3) \right)
\end{equation}

The leading coefficients $g_n$ for $n\neq d$ are given by \footnote{Our convention for Ricci tensor and Ricci scalar differ from \cite{deHaro} by a minus sign}

\begin{eqnarray}
g_{2\; ij} &=& -\frac{1}{(d-2)}\left( R_{ij} - \frac{1}{2(d-1)}R g_{0\;ij} \right) \\ \nonumber
g_{4\; ij} &=& \frac{1}{(d-4)}\left( \frac{1}{8(d-1)}D_i D_j R - \frac{1}{4(d-2)}D^k D_k R_{ij} + \frac{1}{8(d-1)(d-2)}g_{0\; ij}D^k D_k R \right. \\ \nonumber  &-& \left. \frac{1}{2(d-2)}R^{kl}R_{ikjl} + \frac{(d-4)}{2(d-2)^2}R_{i}^{\; k}R_{kj} + \frac{1}{(d-1)(d-2)^2}R R_{ij} \right. \\ &+& \left. \frac{1}{4(d-2)^2}R^{kl}R_{kl}g_{0\;ij} - \frac{3d}{16(d-1)^2 (d-2)^2}R^2 g_{0\;ij}   \right)
\end{eqnarray}

For $n=d$, one can obtain the trace and divergence of $g_n$ as well as the coefficient of logarithmic term $h_d$ from Einstein's equation and we refer the reader to Appendix A of \cite{deHaro}. On-shell $g_2$ is determined in terms of the induced metric $\gamma$ as \cite{deHaro}

\begin{eqnarray}\label{g2traces}
\Tr\; g_2 &=& \frac{1}{2\epsilon (d-1)}\left( -R[\gamma] + \frac{1}{(d-2)}\left(R_{ij}[\gamma]R^{ij}[\gamma] - \frac{1}{2(d-1)}R^2[\gamma] \right) + O[R^3[\gamma]] \right) \nonumber \\ \\ \nonumber
\Tr\; g_{2}^{2} &=& \frac{1}{(d-2)^2 \epsilon^2}\left( R_{ij}[\gamma]R^{ij}[\gamma] + \frac{4-3d}{4(d-1)^2}R^{2}[\gamma] + O[R^3[\gamma]] \right)
\end{eqnarray}

\chapter{}
\section{Legendre Transform Approach}

The Neumann action can be thought of as a boundary Legendre transform of the Dirichlet action. The Neumann and Dirichlet action are related by \cite{Pavan}

\begin{equation}\label{Neu_leg_action}
S^{ren}_N = S^{ren}_D - \int_{\partial \mathcal{M}}d^{d}x\;\pi^{ren}_{ab}\gamma^{ab}
\end{equation}
where $\pi^{ren}_{ab} = \frac{\delta S^{ren}_D}{\delta \gamma^{ab}}$. $\pi^{ren}_{ab}$ is further related to the renormalized boundary stress tensor as

\begin{equation}\label{pi-stress-tensor}
\pi^{ren}_{ab} = -\frac{\sqrt{-\gamma}}{2}T^{ren}_{ab}
\end{equation}
So, given the renormalized action and the boundary stress tensor for the Dirichlet case, we can use the above relations between the Dirichlet and Neumann action to obtain a renormalized action for the Neumann case. This serves as an independent check of the holographic renormalization of Neumann case and we will go through each case ($d=2,3,4$) separately here.

\subsection{AdS$_3$}
The renormalized Dirichlet action and stress-tensor for AdS$_3$ are given by \cite{Balasubramanian:1999re, deHaro}

\begin{eqnarray}
S^{ren}_D &=& \frac{1}{2\kappa}\int_{\cal M} d^3x \sqrt{-g}(R-2\Lambda)+ \frac{1}{\kappa}\int_{\partial \mathcal{M}}d^2x\sqrt{-\gamma}\Theta \\ \nonumber &-& \frac{1}{\kappa}\int_{\partial \mathcal{M}}d^2x\sqrt{-\gamma}
\end{eqnarray}
and

\begin{equation}
T^{ren}_{ab} = \frac{1}{\kappa}\left( \Theta_{ab} - \Theta \gamma_{ab} + \gamma_{ab} \right)
\end{equation}
where we have set $l=1$. Using \eqref{Neu_leg_action} and \eqref{pi-stress-tensor} we immediately see that

\begin{equation}
S^{ren}_{N} = \frac{1}{2\kappa}\int_{\cal M} d^3x \sqrt{-g}(R-2\Lambda)+ \frac{1}{2\kappa}\int_{\partial \mathcal{M}}d^2x\sqrt{-\gamma}\Theta
\end{equation}
which matches with the renormalized Neumann action obtained by holographic renormalization.

\subsection{AdS$_4$}
In AdS$_4$, the renormalized Dirichlet action and stress tensor is \cite{Balasubramanian:1999re, deHaro} (for $l=1$)

\begin{eqnarray}
S^{ren}_D &=& \frac{1}{2\kappa}\int_{\cal M} d^4x \sqrt{-g}(R-2\Lambda)+ \frac{1}{\kappa}\int_{\partial \mathcal{M}}d^3x\sqrt{-\gamma}\Theta \\ \nonumber &-& \frac{2}{\kappa}\int_{\partial \mathcal{M}}d^3x\sqrt{-\gamma}\left(1+ \frac{{}^{(3)}R}{4} \right)
\end{eqnarray}
and

\begin{equation}
T^{ren}_{ab} = \frac{1}{\kappa}\left( \Theta_{ab} - \Theta \gamma_{ab} + 2\gamma_{ab} - {}^{(3)}G_{ab} \right)
\end{equation}
Using \eqref{Neu_leg_action} and \eqref{pi-stress-tensor} we obtain

\begin{equation}
S^{ren}_N = \frac{1}{2\kappa}\int_{\cal M} d^4x \sqrt{-g}(R-2\Lambda) + \frac{1}{\kappa}\int_{\partial \mathcal{M}}d^3x\sqrt{-\gamma}\left(1- \frac{{}^{(3)}R}{4} \right)
\end{equation}
which is in agreement with the renormalized Neumann action obtained by holographic renormalization.

\subsection{AdS$_5$}
For the case of AdS$_5$ the renormalized action and stress-tensor are given by \cite{Balasubramanian:1999re, deHaro} (for $l=1$)

\begin{eqnarray}
S^{ren}_D &=& \frac{1}{2\kappa}\int_{\cal M} d^5x \sqrt{-g}(R-2\Lambda)+ \frac{1}{\kappa}\int_{\partial \mathcal{M}}d^4x\sqrt{-\gamma}\Theta \\ \nonumber &-& \frac{3}{\kappa}\int_{\partial \mathcal{M}}d^4x\sqrt{-\gamma}\left(1+ \frac{{}^{(4)}R}{12} \right)
\end{eqnarray}
and 

\begin{equation}
T^{ren}_{ab} = \frac{1}{\kappa}\left( \Theta_{ab} - \Theta \gamma_{ab} + 3\gamma_{ab} - \frac{1}{2}{}^{(4)}G_{ab} \right)
\end{equation}
Using \eqref{Neu_leg_action} and \eqref{pi-stress-tensor} we once again obtain the renormalized Neumann action which matches with the one obtained by holographic renormalization

\begin{eqnarray}
S^{ren}_N &=& \frac{1}{2\kappa}\int_{\cal M} d^5x \sqrt{-g}(R-2\Lambda)- \frac{1}{2\kappa}\int_{\partial \mathcal{M}}d^4x\sqrt{-\gamma}\Theta \\ \nonumber &+& \frac{3}{\kappa}\int_{\partial \mathcal{M}}d^4x\sqrt{-\gamma}
\end{eqnarray}

\chapter{}
\section{Gauss-Codazzi-Ricci relations}

Gauss-Codazzi relations helps us express the spacetime curvature tensors in terms of the intrinsic and extrinsic curvatures of the embedding hypersurface. They can be summarized as follows:

\begin{eqnarray}
R + 2R_{ab}u^a u^b &=& {}^{(2)}R - \hat{k}_{ab}\hat{k}^{ab} + \hat{k}^2 \\ \nonumber
\sigma_{ab}u_{c}R^{bc} &=& d_a \hat{k} - d_b \hat{k}^{b}_{a}  \\ \nonumber
\sigma_{ac}\sigma_{bd}R^{cd} &=& -\frac{1}{N}{\cal L}_{m}\hat{k}_{ab} - \frac{1}{N}d_a d_b N + {}^{(2)}R_{ab} + \hat{k}\hat{k}_{ab} - 2\hat{k}_{ac}\hat{k}^{c}_{b}
\end{eqnarray}
where ${\cal L}_m$ refers to the Lie derivative with respect to the vector $m^a = Nu^a$, $d_a$ is the covariant derivative w.r.t the metric $\sigma_{ab}$ and $\hat{k}_{ab}$ is the extrinsic curvature of $B$ embedded in ${\cal B}$. 

The last of these relations does not arise as commonly as the first two, we refer the reader to \cite{G-C-ref}. We need all three of them in our simplifications of the ADM version of the renormalized actions.

\chapter{}
\section{OPE coefficients from 3-point function}

As mentioned in Section \ref{sec1}, the OPE coefficients, $q_i$, are completely determined by the conformal symmetry \cite{Ferrara:1973eg}. Here we outline a procedure for obtaining these coefficients from an expansion of 3-point functions. For the case in hand, the coefficients are computed from a scalar-fermion-antifermion 3-pt correlator which takes following form \cite{Weinberg:2010fx,Fradkin:1996is}

\begin{eqnarray}\label{gen3ptfun}
\langle V_{2n}(x_1)V^{A}_{2n+1\;a}(x_2)\bV^{B}_{1\;b}(x_3) \rangle &=& C_{123} \frac{(\fsl{x}_{23})_{ab}\delta^{AB}}{(x^{2}_{12})^{l_3}\;(x^{2}_{23})^{l_1}\;(x^{2}_{31})^{l_2}} \\ \nonumber &+& C'_{123} \frac{(\fsl{x}_{12}\fsl{x}_{31})_{ab}\delta^{AB}}{(x^{2}_{12})^{l'_3}\;(x^{2}_{23})^{l'_1}\;(x^{2}_{31})^{l'_2}}
\end{eqnarray} 

\noindent where $l_1$, $l_2$ and $l_3$ and their primed counterparts are determined in terms of the scaling dimensions of the operators

\begin{eqnarray} \nonumber
l_1 &=& \frac{1}{2}\left[1-\Delta_{2n}+\Delta_{2n+1}+\Delta_1 \right] \\ 
l_2 &=& \frac{1}{2}\left[\Delta_{2n}-\Delta_{2n+1}+\Delta_1 \right] \\ \nonumber
l_3 &=& \frac{1}{2}\left[\Delta_{2n}+\Delta_{2n+1}-\Delta_1 \right] \\ \nonumber
\end{eqnarray}
and $l'_1=l_1-1/2$, $l'_2=l_2+1/2$ and $l'_3=l_3+1/2$. The functional form of the 3-pt function \eqref{gen3ptfun} is essentially fixed by imposing conformal symmetry, \textit{i.e.}

\begin{eqnarray}
0 &=& \langle G_I V_{2n}(x_1)V^{A}_{2n+1\;a}(x_2)\bV^{B}_{1\;b}(x_3) \rangle \\ \nonumber &=&  \langle [G_I,V_{2n}(x_1)]V^{A}_{2n+1\;a}(x_2)\bV^{B}_{1\;b}(x_3) \rangle + \langle V_{2n}(x_1)[G_I,V^{A}_{2n+1\;a}(x_2)]\bV^{B}_{1\;b}(x_3) \rangle \\ \nonumber &+& \langle V_{2n}(x_1)V^{A}_{2n+1\;a}(x_2)[G_I,\bV^{B}_{1\;b}(x_3)] \rangle 
\end{eqnarray}
where $G_I$ collectively stand for the generators of the conformal group and $V_{2n}(x_1)$, $V^{A}_{2n+1\;a}(x_2)$ and $\bV^{B}_{1\;b}(x_3)$ are all assumed to transform as primary under the action of $G_I$s.\\
\noindent The case of $n=1$ has to be treated separately since $V_3$ is not primary. For the time being, we consider the simpler case of all three operators being primary and return to $n=1$ case in the end. Now we imagine a scenario where the first two operators, $V_{2n}(x_1)$ and $V^{A}_{2n+1\;a}(x_2)$, are coming together such that $|x_{12}|\ll|x_{31}|$ and $|x_{12}|\ll|x_{23}|$.  This allows us to expand the 3-pt function \eqref{gen3ptfun} by eliminating $x_{31}$ using the relation

\bea
x_{31}^{2} = x^{2}_{23}\left( 1 + \frac{2 x_{12}.x_{23}}{x^{2}_{23}} + \frac{x^{2}_{12}}{x^{2}_{23}} \right)
\eea

\noindent Substituting this in \eqref{gen3ptfun} and keeping leading terms in $x_{12}$ we obtain following series

\begin{eqnarray}
&\;& \langle V_{2n}(x_1)V^{A}_{2n+1\;a}(x_2)\bV^{B}_{1\;b}(x_3) \rangle \equiv C_{123} \frac{(\fsl{x}_{23})_{ab}\delta^{AB}}{(x^{2}_{12})^{l_3}\;(x^{2}_{23})^{l_1}\;(x^{2}_{31})^{l_2}} \\ \nonumber &\approx & C_{123} (x^{2}_{12})^{-\frac{1}{2}\left[ \Delta_{2n} + \Delta_{2n+1}-\Delta_1 \right]} \left[ \frac{(\fsl{x}_{23})_{ab}}{(x^{2}_{23})^{\Delta_1 + \frac{1}{2}}} - 2l_2 \frac{(\fsl{x}_{23})_{ab}(x_{12}.x_{23})}{(x^{2}_{23})^{\Delta_1 + \frac{3}{2}}}  \right] \\ \nonumber &-&C'_{123} (x^{2}_{12})^{-\frac{1}{2}\left[ \Delta_{2n} + \Delta_{2n+1}-\Delta_1 \right]}\left[ \frac{(x^2_{12})^{3/2}\delta_{ab}}{(x^2_{23})^{\Delta_1 + \frac{1}{2}}} +\frac{(x^2_{12})^{1/2}(\fsl{x}_{12}\fsl{x}_{23})_{ab}}{(x^2_{23})^{\Delta+ \frac{1}{2}}}  \right]
\end{eqnarray}

\noindent Since the operators $V_{2n}$ and $V^{A}_{2n+1\;a}(x_2)$ are close, we may use OPE \eqref{intOPE}. Substituting this into the LHS of \eqref{gen3ptfun}, we obtain

\begin{eqnarray}
& \;& \langle  V_{2n}(x_1) V^{A}_{2n+1\;a}(x_2)\bV^{B}_{1\;b}(x_3) \rangle \approx \tilde{f}(x^{2}_{12})^{-\frac{1}{2}[\Delta_{2n}+\Delta_{2n+1}-\Delta_1]} \Big[ \langle V^{A}_{1\;a}(x_2)\bV^{B}_{1\;b}(x_3) \rangle \\ \nonumber  &+& q_1\; x_{12}^{\mu}\partial_{2\; \mu} \langle V^{A}_{1\;a}(x_2)\bV^{B}_{1\;b}(x_3)\rangle + q_2\;(\fsl{x}_{12}\fsl{\partial}_{2})_{ac}\langle V^{A}_{1\;c}(x_2)\bV^{B}_{1\;b}(x_3) \rangle \Big]
\end{eqnarray}

\noindent This evaluates to

\begin{eqnarray}\label{ope_exp}
& \;& \langle  V_{2n}(x_1) V^{A}_{2n+1\;a}(x_2)\bV^{B}_{1\;b}(x_3) \rangle \approx \tilde{f}(x^{2}_{12})^{-\frac{1}{2}[\Delta_{2n}+\Delta_{2n+1}-\Delta_1]} \left[ \frac{(\fsl{x}_{23})_{ab}}{(x^{2}_{23})^{\Delta_1 + \frac{1}{2}}} \right. \\ \nonumber &+& \left. q_1 \left( \frac{(\fsl{x}_{23})_{ab}}{(x^{2}_{12})^{\Delta_1 + \frac{1}{2}}}  - \frac{(2\Delta_1 + 1)(x_{12}.x_{23})(\fsl{x}_{23})_{ab}}{(x^{2}_{23})^{\Delta_1 + \frac{3}{2}}} \right) + q_2\;\frac{(d-2\Delta_1 -1)(\fsl{x}_{12})_{ab}}{(x^{2}_{23})^{\Delta_1 + \frac{1}{2}}}  \right]\delta^{AB}
\end{eqnarray}

\noindent Comparing this with the 3-pt expansion \eqref{gen3ptfun}, we see that the terms proportional to $C'_{123}$ does not match with the terms in \eqref{ope_exp}. Therefore we have $C'_{123}=0$ and

\begin{eqnarray}\label{coeff1}
q_1 &=& \frac{\Delta_{2n}-\Delta_{2n+1}+\Delta_1}{2\Delta_1 + 1} \\ \nonumber
q_2 &=& \frac{\Delta_{2n}-\Delta_{2n+1}+\Delta_1}{(2\Delta_1 + 1)(2\Delta_1 + 1 - d)} \\ \nonumber
\end{eqnarray}

\noindent The fermion-scalar-anti-fermion 3-point function is given by

\begin{eqnarray}\label{gen3ptfun2}
\langle V^{A}_{2n+1\;a}(x_1)V_{2n+2}(x_2)\bV^{B}_{1\;b}(x_3) \rangle = C_{123} \frac{(\fsl{x}_{12}\fsl{x}_{23})_{ab}\delta^{AB}}{(x^{2}_{12})^{m_3}\;(x^{2}_{23})^{m_1}\;(x^{2}_{31})^{m_2}} \\ \nonumber + C'_{123}\frac{(\fsl{x}_{31})_{ab}\delta^{AB}}{(x^{2}_{12})^{m'_3}\;(x^{2}_{23})^{m'_1}\;(x^{2}_{31})^{m'_2}}
\end{eqnarray}

\noindent with

\begin{eqnarray}
m_1 &=& \frac{1}{2}\left[1-\Delta_{2n+1}+\Delta_{2n+2}+\Delta_1 \right] \\ \nonumber
m_2 &=& \frac{1}{2}\left[\Delta_{2n+1}-\Delta_{2n+2}+\Delta_1 \right] \\ \nonumber
m_3 &=& \frac{1}{2}\left[1+\Delta_{2n+1}+\Delta_{2n+2}-\Delta_1 \right] \nonumber
\end{eqnarray}
and $m'_1=m_1 +1/2$, $m'_2=m_2 -1/2$ and $m'_3=m_3 -1/2$. Here we may set $C'_{123}=0$ which follows directly from the arguments of the previous case. Proceeding in a similar manner, the 3-point function expansion takes the form

\begin{eqnarray}\label{3pt2exp}
&\;& \langle V^{A}_{2n+1\;a}(x_1)V_{2n+2}(x_2)\bV^{B}_{1\;b}(x_3) \rangle \equiv C_{123} \frac{(\fsl{x}_{12}\fsl{x}_{23})_{ab}\delta^{AB}}{(x^{2}_{12})^{m_3}\;(x^{2}_{23})^{m_1}\;(x^{2}_{31})^{m_2}} \\ \nonumber &\approx & C_{123} (x^{2}_{12})^{-\frac{1}{2}\left[ \Delta_{2n+1}+\Delta_{2n+2}-\Delta_1 + 1 \right]}(\fsl{x}_{12})_{ac}\left[ \frac{(\fsl{x}_{23})_{cb}}{(x^{2}_{23})^{\Delta_1 + \frac{1}{2}}} - 2m_2 \frac{(x_{12}.x_{23})(\fsl{x}_{23})_{cb}}{(x^{2}_{23})^{\Delta_1 + \frac{3}{2}}}  \right]
\end{eqnarray}

\noindent On the other hand, OPE of the first two operators is given by

\begin{eqnarray}
V^{A}_{2n+1\;a}(x_1)\times V_{2n+2}(x_2) & \approx & \tilde{f}(x^{2}_{12})^{-\frac{1}{2}\left[ \Delta_{2n+1}+\Delta_{2n+2}-\Delta_1 + 1 \right]} (\fsl{x}_{12})_{ac} \\ \nonumber &\times & \Big[ \delta_{cd} + q_1\;\delta_{cd}x^{\mu}_{12}\partial_{2\;\mu} + q_2\;(\fsl{x}_{12}\fsl{\partial}_{2})_{cd} \Big]V^{A}_{1\;d}(x_2)
\end{eqnarray}

\noindent Substituting this in the LHS of \eqref{gen3ptfun2} and comparing with \eqref{3pt2exp}, we get

\begin{eqnarray}
q_1 &=& \frac{\Delta_{2n+1}-\Delta_{2n+2}+\Delta_1}{2\Delta_1 + 1} \\ \nonumber
q_2 &=& \frac{\Delta_{2n+1}-\Delta_{2n+2}+\Delta_1}{(2\Delta_1 + 1)(2\Delta_1 + 1 - d)} \\ \nonumber 
\end{eqnarray}

We now return to the case of $n=1$ in \eqref{gen3ptfun}. Here we have

\begin{eqnarray}
&\;&\langle V_2(x_1) V^{A}_{3\;a}(x_2)\bV^{B}_{1\;b}(x_3) \rangle =  -\frac{1}{\alpha(\e)}(\fsl{\partial}_2)_{ac}\langle V_2(x_1) V^{A}_{1\;c}(x_2)\bV^{B}_{1\;b}(x_3)\rangle \\ \nonumber &-&\frac{1}{\alpha(\e)} \left[ (d+2l_1)\delta_{ab} - 2l_3\frac{(\fsl{x}_{12}\fsl{x}_{23})_{ab}}{x^{2}_{12}}\right]\frac{1}{(x^{2}_{12})^{l_3}\;(x^{2}_{23})^{l_1}\;(x^{2}_{31})^{l_2}}
\end{eqnarray}

Expanding the above expression for $|x_{12}|\ll|x_{31}|$ and $|x_{12}|\ll|x_{23}|$ as previously, and comparing against the OPE gives the required coefficient

\begin{eqnarray}
q^{2}_2 \approx \frac{(\gamma_{2,1}+1)}{4\gamma_{1,2}\e}
\end{eqnarray}

where we have used the fact that $\gamma_1 = \gamma_{1,2}\e^2$ and $\gamma_2=\gamma_{2,1}\e$. Similarly, from the $\langle V^{A}_{1\;a}(x_1)V_4(x_2)\bV^{B}_{1\;b}(x_3)\rangle $ 3-pt function one obtains

\begin{equation}
q^{3}_{2} \approx -\frac{(1+\gamma_{4,1})}{4\gamma_{1,2}\e}
\end{equation}

\chapter{}
\section{Computing $f_{2p}\rho_{2p}$ and $f_{2p+1}\rho_{2p+1}$ from cow-pies}

In this appendix we give an alternate way to obtain $f_{2p}\rho_{2p}$ and $f_{2p+1}\rho_{2p+1}$ coefficients using cow-pie contractions. This works as a double check of our results, because there are not many results other than \eqref{gamma1} and \eqref{gamma2} that we can check in the literature.

\subsection{$\mathbf{f_{2p}\rho_{2p}}$}
For the ease of counting, we invert the cow-pie and start the contractions from the single kernel. There are two cases to consider,

\begin{enumerate}
\item[\textbf{Case I}]: The `$-$' kernel remains uncontracted. It is easy to see that its contribution is zero because rest of the contractions cannot give the desired operator.

\item[\textbf{Case II}]: The `$-$' kernel is contracted with `$+$' from the double cow-pie.

\begin{figure}[H]
\begin{tikzpicture}

\draw (5,-0.25) node[anchor=south] {\textbullet\quad\textbullet\quad\textbullet};

\begin{scope}
    \draw (0,0) ellipse (1cm and 0.5cm);
    \node[draw=red] at (-0.4,0)  {$+$};
    \node[draw=red] at (0.4,0)  {$-$} ;
\end{scope}

\begin{scope}[xshift=3cm]
\draw (0,0) ellipse (1cm and 0.5cm);
\node[draw=red] at (-0.4,0)  {$+$};
\node[draw=red] at (0.4,0)  {$-$} ;
\end{scope}
 
\begin{scope}[xshift=7cm]
\draw (0,0) ellipse (1cm and 0.5cm);
\node[draw=red] at (-0.4,0)  {$+$};
\node[draw=red] at (0.4,0)  {$-$} ;
\end{scope}

\begin{scope}[xshift=9cm]
\draw (0,0) circle (0.5cm);
\node[draw=red] at (0,0)  {$-$} ;
\end{scope}

\draw (5,-2.25) node[anchor=south] {\textbullet\quad\textbullet\quad\textbullet};

\begin{scope}[yshift=-2cm]
\draw (0,0) ellipse (1cm and 0.5cm);
\node[draw=red] at (-0.4,0)  {$+$};
\node[draw=red] at (0.4,0)  {$-$} ;
\end{scope}

\begin{scope}[xshift=3cm,yshift=-2cm]
\draw (0,0) ellipse (1cm and 0.5cm);
\node[draw=red] at (-0.4,0)  {$+$};
\node[draw=red] at (0.4,0)  {$-$} ;
\end{scope}
 
\begin{scope}[xshift=7cm,yshift=-2cm]
\draw (0,0) ellipse (1cm and 0.5cm);
\node[draw=red] at (-0.4,0)  {$+$};
\node[draw=red] at (0.4,0)  {$-$} ;
\end{scope}

\draw (9.0,0) -- (6.6,-2);


\end{tikzpicture}
\end{figure}
which contributes $-p F^{p-1,0,1}_{p,0,0;1,1}$ which can again be evaluated using cow-pies. Once again, we invert the cow-pie diagram and start the contractions from the uncontracted `$-$' in the double cow-pie. As can be readily seen, there are two cases to consider:

\begin{enumerate}
\item[\textbf{(a)}]: `$-$' remains uncontracted. This gives a contribution of $F^{p,0,0}_{p+1,0,0;1,1}$. 

\begin{figure}[H]
\begin{tikzpicture}

\draw (5,-0.25) node[anchor=south] {\textbullet\quad\textbullet\quad\textbullet};

\begin{scope}
    \draw (0,0) ellipse (1cm and 0.5cm);
    \node[draw=red] at (-0.4,0)  {$+$};
    \node[draw=red] at (0.4,0)  {$-$} ;
\end{scope}

\begin{scope}[xshift=3cm]
\draw (0,0) ellipse (1cm and 0.5cm);
\node[draw=red] at (-0.4,0)  {$+$};
\node[draw=red] at (0.4,0)  {$-$} ;
\end{scope}
 
\begin{scope}[xshift=7cm]
\draw (0,0) ellipse (1cm and 0.5cm);
\node[draw=red] at (-0.4,0)  {$+$};
\node[draw=red] at (0.4,0)  {$-$} ;
\end{scope}

\draw (5,-2.25) node[anchor=south] {\textbullet\quad\textbullet\quad\textbullet};

\begin{scope}[yshift=-2cm]
\draw (0,0) ellipse (1cm and 0.5cm);
\node[draw=red] at (-0.4,0)  {$+$};
\node[draw=red] at (0.4,0)  {$-$} ;
\end{scope}

\begin{scope}[xshift=3cm,yshift=-2cm]
\draw (0,0) ellipse (1cm and 0.5cm);
\node[draw=red] at (-0.4,0)  {$+$};
\node[draw=red] at (0.4,0)  {$-$} ;
\end{scope}
 
\begin{scope}[xshift=7cm,yshift=-2cm]
\draw (0,0) ellipse (1cm and 0.5cm);
\node[draw=red] at (-0.4,0)  {$+$};
\node[draw=red] at (0.4,0)  {$-$} ;
\end{scope}

\begin{scope}[xshift=9cm,yshift=-2cm]
\draw (0,0) circle (0.5cm);
\node[draw=red] at (0,0)  {$-$} ;
\end{scope}

\draw (6.6,0) -- (9.0,-2);


\end{tikzpicture}
\end{figure}
$F^{p,0,0}_{p+1,0,0;1,1}$ can in turn be evaluated using cow-pies. This is similar to the case of $f_{2p}$ and $f_{2p+1}$ and its recursion equation is given by

\begin{equation}\label{recursion2}
F^{p,0,0}_{p+1,0,0;1,1} = (p+1)(N-p)F^{p-1,0,0}_{p,0,0;1,1}
\end{equation}

\item[\textbf{(b)}]: `$-$' can be contracted with one of the double cow-pies. This gives a factor of $(p+1)F^{p,0,0}_{p,0,1;1,2}$

\begin{figure}[H]
\begin{tikzpicture}

\draw (5,-0.25) node[anchor=south] {\textbullet\quad\textbullet\quad\textbullet};

\begin{scope}
    \draw (0,0) ellipse (1cm and 0.5cm);
    \node[draw=red] at (-0.4,0)  {$+$};
    \node[draw=red] at (0.4,0)  {$-$} ;
\end{scope}

\begin{scope}[xshift=3cm]
\draw (0,0) ellipse (1cm and 0.5cm);
\node[draw=red] at (-0.4,0)  {$+$};
\node[draw=red] at (0.4,0)  {$-$} ;
\end{scope}
 
\begin{scope}[xshift=7cm]
\draw (0,0) ellipse (1cm and 0.5cm);
\node[draw=red] at (-0.4,0)  {$+$};
\node[draw=red] at (0.4,0)  {$-$} ;
\end{scope}

\draw (5,-2.25) node[anchor=south] {\textbullet\quad\textbullet\quad\textbullet};

\begin{scope}[yshift=-2cm]
\draw (0,0) ellipse (1cm and 0.5cm);
\node[draw=red] at (-0.4,0)  {$+$};
\node[draw=red] at (0.4,0)  {$-$} ;
\end{scope}

\begin{scope}[xshift=3cm,yshift=-2cm]
\draw (0,0) ellipse (1cm and 0.5cm);
\node[draw=red] at (-0.4,0)  {$+$};
\node[draw=red] at (0.4,0)  {$-$} ;
\end{scope}
 
\begin{scope}[xshift=7cm,yshift=-2cm]
\draw (0,0) ellipse (1cm and 0.5cm);
\node[draw=red] at (-0.4,0)  {$+$};
\node[draw=red] at (0.4,0)  {$-$} ;
\end{scope}

\begin{scope}[xshift=9cm,yshift=-2cm]
\draw (0,0) circle (0.5cm);
\node[draw=red] at (0,0)  {$-$} ;
\end{scope}

\draw (6.6,0) -- (9.0,-2);
\draw (7.4,0) --  (6.6,-2);


\end{tikzpicture}
\end{figure}

\end{enumerate}
\end{enumerate}
Putting all the pieces together, we have the following system of recursion equations

\begin{eqnarray}
F^{p,0,0}_{p,0,1;1,2} &=& -p F^{p-1,0,1}_{p,0,0;1,1} \\ \nonumber
F^{p,0,1}_{p+1,0,0;1,1} &=& F^{p,0,0}_{p+1,0,0;1,1} + (p+1)F^{p,0,0}_{p,0,1;1,2} \\ \nonumber
F^{p,0,0}_{p+1,0,0;1,1} &=& (p+1)(N-p)F^{p-1,0,0}_{p,0,0;1,1} \nonumber
\end{eqnarray}
which can be solved along with the launching conditions $F^{0,0,0}_{0,0,1;1,2}=0$, $F^{0,0,1}_{1,0,0;1,1}=1$ and $F^{0,0,0}_{1,0,0;1,1}=1$. Using the expression for $f_{2p}$ in \eqref{f2p}, we obtain

\begin{equation}
\rho_{2p} = - \frac{p}{N-1}
\end{equation}

\subsection{$\mathbf{f_{2p+1}\rho_{2p+1}}$}
We proceed analogous to the even case, \textit{i.e.} $f_{2p}\rho_{2p}$. As in the previous case, we start the contractions with the single kernel `$-$'. Again, we have two cases:

\begin{enumerate}
\item[\textbf{Case I}]: `$-$' remains uncontracted. This indeed contributes, with a factor of $F^{p,0,0}_{p+1,0,0;1,1}$ which can be further evaluated using cow-pies. It can be seen that the recursion equation for $F^{p,0,0}_{p+1,0,0;1,1}$ is given by

\begin{equation}
F^{p,0,0}_{p+1,0,0;1,1} = (p+1)(N-p)F^{p-1,0,0}_{p,0,0;1,1}
\end{equation}

\item[\textbf{Case II}]: `$-$' is contracted with one of the double cow-pies. This contributes a factor of $(p+1)F^{p,0,0}_{p,0,1;1,2}$, where $F^{p,0,0}_{p,0,1;1,2}$ can be furthermore evaluated using cow-pies. To this end, we invert the cow-pie diagram and once again consider two separate cases 

\begin{enumerate}
\item[\textbf{a}]: `$-$' of the double cow-pie remains uncontracted. It can be seen that this does not contribute as we do not get the desired operator.

\item[\textbf{b}]: `$-$' of the double cow-pie is contracted with one of the double cow-pies. This contributes $-pF^{p-1,0,1}_{p,0,0;1,2}$.

\begin{figure}[H]
\begin{tikzpicture}

\draw (5,-0.25) node[anchor=south] {\textbullet\quad\textbullet\quad\textbullet};

\begin{scope}
    \draw (0,0) ellipse (1cm and 0.5cm);
    \node[draw=red] at (-0.4,0)  {$+$};
    \node[draw=red] at (0.4,0)  {$-$} ;
\end{scope}

\begin{scope}[xshift=3cm]
\draw (0,0) ellipse (1cm and 0.5cm);
\node[draw=red] at (-0.4,0)  {$+$};
\node[draw=red] at (0.4,0)  {$-$} ;
\end{scope}
 
\begin{scope}[xshift=7cm]
\draw (0,0) ellipse (1cm and 0.5cm);
\node[draw=red] at (-0.4,0)  {$+$};
\node[draw=red] at (0.4,0)  {$-$} ;
\end{scope}

\begin{scope}[xshift=10cm]
\draw (0,0) ellipse (1cm and 0.5cm);
\node[draw=red] at (-0.4,0)  {$+$};
\node[draw=red] at (0.4,0)  {$-$} ;
\end{scope}


\draw (5,-2.25) node[anchor=south] {\textbullet\quad\textbullet\quad\textbullet};

\begin{scope}[yshift=-2cm]
\draw (0,0) ellipse (1cm and 0.5cm);
\node[draw=red] at (-0.4,0)  {$+$};
\node[draw=red] at (0.4,0)  {$-$} ;
\end{scope}

\begin{scope}[xshift=3cm,yshift=-2cm]
\draw (0,0) ellipse (1cm and 0.5cm);
\node[draw=red] at (-0.4,0)  {$+$};
\node[draw=red] at (0.4,0)  {$-$} ;
\end{scope}
 
\begin{scope}[xshift=7cm,yshift=-2cm]
\draw (0,0) ellipse (1cm and 0.5cm);
\node[draw=red] at (-0.4,0)  {$+$};
\node[draw=red] at (0.4,0)  {$-$} ;
\end{scope}

\begin{scope}[xshift=9.6cm,yshift=-2cm]
\draw (0,0) circle (0.5cm);
\node[draw=red] at (0,0)  {$-$} ;
\end{scope}

\draw (9.6,0) -- (9.6,-2);
\draw (10.4,0) --  (6.6,-2);


\end{tikzpicture}
\end{figure}
\end{enumerate}
\end{enumerate}

Thus we have following system of recursion equations, which can be solved along with the launching conditions $F{0,0,1}_{1,0,0;1,2}=1$, $F^{0,0,0}_{1,0,0;1,1}=1$, $F^{0,0,0}_{0,0,1;1,2}=0$.

\begin{eqnarray}
F^{p,0,1}_{p+1,0,0;1,2} &=& F^{p,0,0}_{p+1,0,0;1,1} + (p+1)F^{p,0,0}_{p,0,1;1,2} \\ \nonumber
F^{p,0,0}_{p+1,0,0;1,1} &=& (p+1)(N-p)F^{p-1,0,0}_{p,0,0;1,1}  \\ \nonumber
F^{p,0,0}_{p,0,1;1,2} &=& -p F^{p-1,0,1}_{p,0,1;1,2} \\ \nonumber
\end{eqnarray} 
Using \eqref{f2pp1}, along with above set of recursion equations gives

\begin{equation}
\rho_{2p+1} = 1- \frac{p}{N-1}
\end{equation}

\newpage

\bibliography{ref} 
\bibliographystyle{utphys}
 
\end{document}